\documentclass[universe,
article,
submit,moreauthors,pdftex]{Definitions/mdpi} 

\firstpage{1} 
\pubvolume{1}
\issuenum{1}
\articlenumber{0}
\pubyear{2021}
\copyrightyear{2020}
\datereceived{} 
\dateaccepted{} 
\datepublished{} 
\hreflink{https://doi.org/} 

\usepackage{graphicx}
\usepackage{epsfig}
\usepackage{rotating}
\usepackage{amssymb}
\usepackage{subfigure}
\usepackage{dsfont}
\usepackage{psfrag}
\usepackage{mathtools}
\usepackage{amsmath}
\usepackage{euscript}
\usepackage{array}
\usepackage{mathrsfs}
\usepackage{bbold}
\usepackage{bm}
\usepackage{bbm}
\usepackage{multirow}
\usepackage{dcolumn}
\usepackage{xcolor}

\newcommand{\beq}{\begin{equation}}
\newcommand{\eeq}{\end{equation}}
\newcommand{\beqs}{\begin{eqnarray}}
\newcommand{\eeqs}{\end{eqnarray}}
\newcommand{\lsim}{\mathrel{\raisebox{-
.6ex}{$\stackrel{\textstyle<}{\sim}$}}}

\newcommand{\Tr}{{\rm Tr}}

\Title{$Sp(2N)$ Lattice Gauge Theories and Extensions of the Standard Model of Particle Physics}

\TitleCitation{$Sp(2N)$ Lattice Gauge Theories and Extensions of the Standard Model of Particle Physics.\\
CTPU-PTC-23-09, 
PNUTP-23/A02}

\Author{Ed Bennett $^{1}$\orcidA{}, 
Jack Holligan $^{2,3}$\orcidB{}, 
Deog Ki Hong $^{4}$\orcidC{}, 
Ho Hsiao $^{5}$\orcidD{}, 
Jong-Wan Lee $^{4,6,7,*}$\orcidE{}, 
C.-J. David Lin $^{5,8,9}$\orcidF{}, 
Biagio Lucini $^{10,1}$\orcidG{}, 
Michele Mesiti $^{11}$\orcidH{}, 
Maurizio Piai $^{12}$\orcidI{}, 
and Davide Vadacchino $^{13,14}$\orcidJ{}}

\AuthorNames{Ed Bennett, Jack Holligan, Deog Ki Hong, Ho Hsiao, Jong-Wan Lee, C.-J. David Lin, Biagio Lucini, Michele Mesiti, Maurizio Piai, and Davide Vadacchino}

\AuthorCitation{Bennett, E.; Holligan, J.; Hong, D.~K.; Hsiao, H.; Lee, J.-W.; Lin, C.-J.~D.; Lucini, B.; Mesiti, M.; Piai, M.; Vadacchino, D.}

\address[5]{%
$^1$ \quad Swansea Academy of Advanced Computing, Swansea University (Bay Campus), Fabian Way, SA1 8EN Swansea, Wales, United Kingdom\\
$^2$ \quad Biomedical and Physical Sciences Building, Michigan State University, East Lansing, Michigan, USA, 48824\\
$^3$ \quad Physical Sciences Complex, University of Maryland, College Park, Maryland, USA, 20742\\
$^4$ \quad Department of Physics, Pusan National University, Busan 46241, Korea\\
$^5$ \quad Institute of Physics, National Yang Ming Chiao Tung University, 1001 Ta-Hsueh Road, Hsinchu 30010, Taiwan\\
$^6$ \quad Institute for Extreme Physics, Pusan National University, Busan 46241, Korea\\
$^7$ \quad Particle Theory  and Cosmology Group, Center for Theoretical Physics of the Universe, Institute for Basic Science (IBS), Daejeon, 34126, Korea\\
$^8$ \quad Center for High Energy Physics, Chung-Yuan Christian University, Chung-Li 32023, Taiwan\\
$^9$ \quad Centre for Theoretical and Computational Physics, National Yang Ming Chiao Tung University, 1001 Ta-Hsueh Road, Hsinchu 30010, Taiwan\\
$^{10}$ \quad Department of Mathematics, Faculty of Science and Engineering, Swansea University (Bay Campus), Fabian Way, SA1 8EN Swansea, Wales, United Kingdom\\
$^{11}$ \quad Steinbuch Centre for Computing, Karlsruher Institut für Technologie, Zirkel 2, 76131 Karlsruhe, Germany\\
$^{12}$ \quad Department of Physics, Faculty of Science and Engineering,
Swansea University (Singleton Park Campus), Singleton Park, SA2 8PP Swansea, Wales, United Kingdom\\
$^{13}$ \quad School of Mathematics and Hamilton Mathematics Institute, Trinity
College, Dublin 2, Ireland\\
$^{14}$ \quad Centre for Mathematical Sciences, University of Plymouth, Plymouth, PL4 8AA, United Kingdom}

\corres{Correspondence: j.w.lee@ibs.re.kr; 
}

\abstract{We review the current status of the long-term programme of numerical investigation of $Sp(2N)$ 
gauge theories with and without fermionic matter content. We start by
introducing the phenomenological as well as theoretical motivations
for this research programme, which are related to composite Higgs models, models of partial top  compositeness,
dark matter models, and in general to the physics of strongly coupled theories and their approach to the large-$N$ limit.
We  summarise the results of lattice studies conducted so far  in the $Sp(2N)$ Yang-Mills theories,
measuring the string tension, the mass spectrum of glueballs and the topological susceptibility,
and discuss their large-$N$ extrapolation.
We then focus our discussion on $Sp(4)$, and summarise numerical measurements of mass and decay constant
of  mesons in the theories with
fermion matter in either the fundamental or the antisymmetric representation, first in the quenched approximation,
and then with dynamical fermions. 
 We finally discuss the case of dynamical fermions in mixed representations, and exotic composite fermion states
 such as the chimera baryons.
 We conclude by sketching the future stages of the programme. 
 And we describe our approach to open access.
 }

\begin{document}

\tableofcontents

\section{Introduction}
\label{sec:introduction}

The past two decades have seen the publication of the first 
dedicated lattice studies of the four-dimensional $Sp(2N)$ gauge 
theories with $N>1$~\cite{Holland:2003kg,Bennett:2017kga,
Lee:2018ztv,Bennett:2019jzz,
Bennett:2019cxd,Bennett:2020hqd,
Bennett:2020qtj,Lucini:2021xke,Bennett:2021mbw,
Bennett:2022yfa,Bennett:2022gdz,Bennett:2022ftz,AS,
Lee:2022elf,Hsiao:2022kxf,
Maas:2021gbf,
Zierler:2021cfa,
Kulkarni:2022bvh}.
Large classes of $Sp(2N)$ gauge theories confine,  and, in the presence of matter 
fields, chiral symmetry breaking condensates govern the
long-distance dynamics.
The interest in these theories ultimately descends from 
 the nature of $Sp(2N)$ groups and their representations:
they possess symmetries and  (dynamically) yield symmetry-breaking patterns
that  are different from 
those of related $SU(N_c)$ theories~\cite{Peskin:1980gc}.
New opportunities for model-building and phenomenology
hence emerge, thanks to the peculiar  symmetries,
symmetry breaking patterns, spectroscopy, and  low-energy effective field theory (EFT)
description associated with $Sp(2N)$ gauge theories.
Yet, the microscopic dynamics of $Sp(2N)$ gauge theories is
not dissimilar from  $SU(N_c)$---in particular $SU(2)$ and $SU(3)$---theories.
Having implemented the necessary adjustments to the Monte Carlo update algorithms
that generate the ensembles~\cite{Bennett:2017kga,
Bennett:2019jzz,
Bennett:2020qtj,
Bennett:2022yfa},
as well as to the correlation functions used to measure spectral observables~\cite{Bennett:2019jzz,
Bennett:2019cxd,Bennett:2020qtj,
Bennett:2022yfa,Lee:2022elf,Hsiao:2022kxf},
 it is then possible to adapt
 the modern advancements of lattice gauge theories
to study the non-perturbative regime of the $Sp(2N)$ gauge theories.

The standard model (SM) of particle physics has been spectacularly successful at describing the strong and electroweak forces through which the known elementary particles interact among each other.  Yet, there is solid evidence that the SM is incomplete and must be extended to explain several astronomical and experimental observations, among which are the existence of dark matter, the matter-antimatter asymmetry, and non-zero masses of neutrinos. 
Furthermore, the SM is unnaturally fine tuned, since it does not provide a mechanism that explains why the Higgs boson has a mass at the electroweak scale, rather than receiving 
the expected large quantum corrections that would generate a mass at the Planck scale. 
To address these shortcomings, much effort has been devoted to developing models based on novel strongly coupled gauge theories as extensions of the standard model of particle physics

This review  summarises briefly the phenomenological and theoretical motivations 
to study $Sp(2N)$ gauge theories,
 and  then discusses at length the available (lattice) numerical results,
to facilitate their use by model-builders and phenomenologists.
We start by introducing in this first  section the main arguments
 why $Sp(2N)$ gauge theories 
are a promising topic of investigation.
These are further developed in the body of the paper.
 They include phenomenological consideration pertaining
 to composite Higgs, top (partial) compositeness,  dark matter physics,
and theoretical considerations about finite temperature phase transitions (and 
gravitational wave detection), as well as non-perturbative phenomena
in non-Abelian gauge theories, especially in relation to the large-$N_c$ extrapolation.
Within each such topic, we provide the context for the application of  $Sp(2N)$ theories, 
explaining the main ideas 
and their historical development. We complement the narrative by an ample  list
of references that
contain expanded explanations and technical details.

The discovery of the Higgs boson~\cite{Aad:2012tfa,Chatrchyan:2012xdj} has triggered a
revival of interest in  composite Higgs models (CHMs)~\cite{Kaplan:1983fs,
Georgi:1984af,Dugan:1984hq} 
(see, e.g., the reviews in Refs.~\cite{
Panico:2015jxa,Witzel:2019jbe,
Cacciapaglia:2020kgq}, the summary tables in Refs.~\cite{
Ferretti:2013kya,Ferretti:2016upr,Cacciapaglia:2019bqz},
and the selection of papers in Refs.~\cite{
Katz:2005au,Barbieri:2007bh,
Lodone:2008yy,Gripaios:2009pe,Mrazek:2011iu,Marzocca:2012zn,Grojean:2013qca,Cacciapaglia:2014uja,
Ferretti:2014qta,Arbey:2015exa,Cacciapaglia:2015eqa,Feruglio:2016zvt,DeGrand:2016pgq,Fichet:2016xvs,
Galloway:2016fuo,Agugliaro:2016clv,Belyaev:2016ftv,Csaki:2017cep,Chala:2017sjk,Golterman:2017vdj,
Csaki:2017jby,Alanne:2017rrs,Alanne:2017ymh,Sannino:2017utc,Alanne:2018wtp,Bizot:2018tds,
Cai:2018tet,Agugliaro:2018vsu,Cacciapaglia:2018avr,Gertov:2019yqo,Ayyar:2019exp,
Cacciapaglia:2019ixa,BuarqueFranzosi:2019eee,Cacciapaglia:2019dsq,Cacciapaglia:2020vyf,
Dong:2020eqy,Cacciapaglia:2021uqh,Banerjee:2022izw}
and 
Refs.~\cite{
Contino:2003ve,Agashe:2004rs,Agashe:2005dk,Agashe:2006at,
Contino:2006qr,Falkowski:2008fz,Contino:2010rs,Contino:2011np,Elander:2023aow}),
many of which also implement top (partial)  
compositeness~\cite{Kaplan:1991dc}
(see also Refs.~\cite{Grossman:1999ra,Gherghetta:2000qt,Chacko:2012sy}).
In this context, the lightest composite spin-$0$ and spin-$1/2$
states in a new strongly coupled  sector are
identified  with the heaviest particles in the Standard Model:
 the Higgs boson and the top quark.
The former emerges as one of the Pseudo-Nambu-Goldstone Bosons (PNGBs)
associated with spontaneous breaking of the global symmetry in the underlying microscopic theory.
The latter is a generalisation of the baryons, which plays the role of a top quark partner, and may involve fermions
in more than one representation of the gauge group, so that in the following we call them chimera baryons. 
These ideas admit a multitude of possible realisations with strikingly diverse phenomenological implications, 
as suggested by the vastness of the literature on this subject.
They can be tested by the future experimental programme of the 
Large Hadron Collider (LHC),
with aid from
computational techniques adapted to the study of
the non-perturbative nature of the underlying strong dynamics.

The natural choice of non-perturbative instrument for the investigation of strongly coupled gauge theories is  lattice
field theory.
Depending on the nature of the representations of the fermion matter field content,
three  different symmetry patterns emerge.  As in QCD-like theories with $N_f$ Dirac fermions in the
 fundamental representation, in the presence of complex representations 
 the global non-Abelian $SU(N_f)\times SU(N_f)$ is broken to the diagonal $SU(N_f)$ subgroup
 by the condensates forming in the theory.
 Real representations yield the spontaneous breaking of 
the enhanced $SU(2N_f)$ symmetry  to its $SO(2N_f)$ maximal subgroup.
Pseudo-real representations are characterised by the breaking of $SU(2N_f)$ to $Sp(2N_f)$.
Vacuum alignment arguments can be used to
 select the vacuum, on the basis of which deformations are 
 admissible~\cite{Peskin:1980gc}.
The resulting cosets, the PNGBs spanning them, and the masses induced by explicit breaking of the
global symmetries, are the starting point for the construction of CHMs.

A number of dedicated studies of the lattice $SU(2)$ gauge theories
relevant to CHMs have been performed~\cite{Hietanen:2014xca,Detmold:2014kba,
Arthur:2016dir,Arthur:2016ozw,Pica:2016zst,Lee:2017uvl,Drach:2017btk,Drach:2020wux,
Drach:2021uhl}, which with $N_f=2$ (Dirac) fermions transforming in the fundamental representation
of the gauge group yield the $SU(4)/Sp(4)$ 
coset relevant to the CHMs of interest in this paper. The low-energy theory
has five PNGBs, four of which are interpreted in terms of the SM Higgs doublet,
and one as a scalar SM singlet.

Studies of $SU(4)$ gauge theories have also been published~\cite{Ayyar:2017qdf,Ayyar:2018zuk,Ayyar:2018ppa,
 Ayyar:2018glg,Cossu:2019hse,Shamir:2021frg,DelDebbio:2021xlv}, 
 their field content consisting of mixed fermion representations, as required in models
 combining Higgs and top (partial) compositeness. Lattice studies consider matter consisting of Dirac fermions, while 
 the minimal model of this class would require odd numbers of Majorana fermions---with five 
 2-index antisymmetric Majorana
 fermions one can see the $SU(5)/SO(5)$ coset emerge, and the $14$ PNGBs
 can be reorganised into one scalar SM doublet and additional SM singlets and triplets~\cite{Ferretti:2014qta}. 
 
Lattice theories with an $SU(3)$ gauge group, in which the antisymmetric representation of the group
coincides with the (conjugate) fundamental, allow for an alternative way of combining 
 composite Higgs and top compositeness~\cite{Vecchi:2015fma}.
The chimera baryons, used as top quark partners, are actual baryons of the new  $SU(3)$ theory.
By exploiting  the dilaton EFT~\cite{Coleman:1985rnk,
Migdal:1982jp,Leung:1985sn,Bardeen:1985sm,Yamawaki:1985zg,Goldberger:2007zk}
in the new context of near conformal gauge theories and their EFT 
treatment~\cite{Matsuzaki:2013eva,Golterman:2016lsd,Kasai:2016ifi,
Hansen:2016fri,Golterman:2016cdd,Appelquist:2017wcg,Appelquist:2017vyy,
Golterman:2018mfm,Cata:2019edh,Appelquist:2019lgk,Golterman:2020tdq,Golterman:2020utm,Appelquist:2022mjb},
Refs.~\cite{Appelquist:2020bqj,
Appelquist:2022qgl} showed that 
it is possible to build new CHMs, based on $SU(N_f)\times SU(N_f)/SU(N_f)$ cosets
 (see also Refs.~\cite{Ma:2015gra,BuarqueFranzosi:2018eaj}), with input from 
 lattice data on the $SU(3)$ theory with $N_f=8$ fundamental fermions~\cite{Aoki:2014oha,
Appelquist:2016viq,
Aoki:2016wnc,
Gasbarro:2017fmi,
Appelquist:2018yqe}.

Gauge theories with $Sp(2N)$ group are special in this context. With
 $N_f=2$ (Dirac) fermions in the fundamental representation,
 they give rise to the same $SU(4)/Sp(4)$ coset as the aforementioned
$SU(2)=Sp(2)$ theories. In addition,  with $N>1$, $N_f=2$, and $n_f=3$ (Dirac) fermions 
transforming in the 2-index antisymmetric representation,
they yield bound states of two fundamental and one antisymmetric fermion (chimera baryons),
that can play the role of top partners, hence
 combining composite Higgs and top (partial) compositeness~\cite{Barnard:2013zea}. 
Progress has been made in studying the spectra of mesons~\cite{Bennett:2017kga,
Bennett:2019jzz,Bennett:2019cxd} and chimera baryons~\cite{Bennett:2022yfa}..\footnote{
We borrow the terminology and nomenclature associated with mesons and baryons from QCD,  when referring to the analogous composite states in new strongly coupled gauge theories.} 
These theories have been studied also with semi-analytical techniques~\cite{Bizot:2016zyu}, 
based on
replacing the fundamental dynamics with four-fermion interactions,
as in the Nambu-Jona-Lasinio model. The theories with $n_f=3$ and $N_f=0$ can also realise
alternative composite Higgs and dark matter models~\cite{Cacciapaglia:2019ixa}.

We must mention that an alternative way to study strongly coupled dynamics
is based upon gauge-gravity dualities; special
strongly-coupled field theories admit an equivalent description as weakly coupled theories of gravity
living in higher dimensions~\cite{Maldacena:1997re,Gubser:1998bc,Witten:1998qj,Aharony:1999ti}.
Indeed, the recent revival of interest on CHMs started before the Higgs discovery,
 driven by extra-dimensional models inspired by gauge/gravity dualities, and based on the minimal
$SO(5)/SO(4)$ coset~\cite{
Contino:2003ve,Agashe:2004rs,Agashe:2005dk,Agashe:2006at,
Contino:2006qr,Falkowski:2008fz,Contino:2010rs,Contino:2011np,Elander:2023aow}.
More recently, progress has been made towards building semi-realistic descriptions of the dynamics 
of the more complex  CHMs that are amenable to  lattice studies, but in the context of bottom-up holography~\cite{Erdmenger:2020lvq,
Erdmenger:2020flu,Elander:2020nyd,Elander:2021bmt}. Even the first steps towards 
embedding models with  $SO(5)/SO(4)$ coset into supergravity (and string theory) 
have been taken~\cite{Elander:2021kxk}.
The complementary role of these approaches to strong dynamics is actively being investigated.

A completely independent, compelling argument for new physics extending the standard model is that it does not provide
 an explanation for the nature and origin of dark matter. This could be explained by the existence of a new dark 
 sector---see Refs.~\cite{Strassler:2006im,Cheung:2007ut,Hambye:2008bq,Feng:2009mn,
Cohen:2010kn,Foot:2014uba} and the review in Ref.~\cite{Bertone:2016nfn}, for example.
This dark sector might consist of a new strongly coupled theory, with
matter consisting only of SM singlet fields. The new strong dynamics would lead to the formation 
of composite PNGBs and in general the spectroscopy resemble qualitatively that of a generalisation of
Quantum Chromo-Dynamics (QCD). These proposals go under the names of
 composite dark matter (CoDM), as in Refs.~\cite{DelNobile:2011je,
Hietanen:2013fya,Cline:2016nab,Dondi:2019olm,
Ge:2019voa,Beylin:2019gtw,Yamanaka:2019aeq,Yamanaka:2019yek,Cai:2020njb},
or strongly interacting dark matter (SIMP), as in 
Refs.~\cite{Hochberg:2014dra,Hochberg:2014kqa,Hochberg:2015vrg,Bernal:2017mqb,Berlin:2018tvf,
Bernal:2019uqr,Tsai:2020vpi,Kondo:2022lgg,Bernal:2015xba}.
$Sp(2N)$ gauge theories play a prominent role in many of these proposals,
and the first dedicated lattice studies of the spectroscopy of $Sp(4)$ with $N_f=2$, non-degenerate
(Dirac) fundamental fermions have  recently become available~\cite{Maas:2021gbf,
Zierler:2021cfa,
Kulkarni:2022bvh}.

The first  dedicated lattice  exploration of $Sp(2N)$ gauge theories
focused on the pure gauge dynamics, and its confinement/deconfinement phase
 transition at finite temperature~\cite{Holland:2003kg}. All $Sp(2N)$ Yang-Mills theories have
 centre symmetry ${\mathbb Z}_2$. The expectation value of the Polyakov loop
 behaves as the order parameter of the transition, vanishing at low temperature (${\mathbb Z}_2$-unbroken phase),  
 and becoming non-trivial  above some critical temperature $T_c$ (${\mathbb Z}_2$-broken phase).
In three spatial dimensions, while for $Sp(2)=SU(2)$ the phase transition is of second order,
when  $N>1$ there is evidence of a first-order phase transition. If originally this quest had
mostly a theoretical motivation, related to the general characterisation of phase transitions
in non-Abelian gauge theories, in recent times it has
acquired new phenomenological relevance, related to the aforementioned context of strongly interacting dark matter.

Such a dark $Sp(2N)$ sector might undergo a strong enough first order (dark confinement) phase
transition, in the early universe, to leave behind
 a relic stochastic background of gravitational 
waves~\cite{Witten:1984rs,Kamionkowski:1993fg,Allen:1996vm,Schwaller:2015tja,
 Croon:2018erz,Christensen:2018iqi}, potentially accessible  to 
 present and future gravitational-wave detectors~\cite{Seto:2001qf,
 Kawamura:2006up,Crowder:2005nr,Corbin:2005ny,Harry:2006fi,
 Hild:2010id,Yagi:2011wg,Sathyaprakash:2012jk,Thrane:2013oya,
 Caprini:2015zlo,
 LISA:2017pwj,
 LIGOScientific:2016wof,Isoyama:2018rjb,Baker:2019nia,
 Brdar:2018num,Reitze:2019iox,Caprini:2019egz,
 Maggiore:2019uih}.
For recent phenomenological studies, see for instance  Refs.~\cite{Huang:2020crf,Halverson:2020xpg,Kang:2021epo}, 
and references therein.
The finite-temperature behaviour of many gauge theories has been studied;
for examples of $SU(N_c)$ studies see Refs.~\cite{Lucini:2002ku,Lucini:2003zr,Lucini:2005vg,Panero:2009tv,Datta:2010sq,Lucini:2012wq},
for $Sp(N_c)$ see Ref.~\cite{Holland:2003kg},
and for $G_2$ see Ref.~\cite{Pepe:2005sz,Pepe:2006er,Cossu:2007dk,Bruno:2014rxa}.
A handful of dedicated  lattice calculations 
 focus on stealth dark matter 
with $SU(4)$ gauge dynamics~\cite{Appelquist:2015yfa,Appelquist:2015zfa,LatticeStrongDynamics:2020jwi}.
The recent Ref.~\cite{Borsanyi:2022xml} critically summarises the history of
  $SU(3)$ studies,
and the technical difficulties
intrinsic to current state-of-the-art lattice calculations.
It is hoped that by applying new ideas in lattice field theory, such as the Logarithmic Linear Relaxation (LLR) algorithm
~\cite{Langfeld:2012ah,Langfeld:2013xbf,Langfeld:2015fua},
some of these difficulties may be overcome---see in particular 
Ref.~\cite{Cossu:2021bgn} for zero-temperature studies of $SU(3)$,
and preliminary finite-temperature results for $SU(4)$ in Ref.~\cite{Springer:2021liy},
$SU(3)$ in Refs.~\cite{Mason:2022trc,Mason:2022aka}, and $SU(N_c)$ in Ref~\cite{Springer:2023wok}.
$Sp(2N)$ theories
can be explored with the LLR method, but
such lattice calculations are not available yet, and  we will not  discuss them further.

The final topic we touch upon in this introduction
is the observation that, while different in nature, the sequence of $Sp(N_c=2N)$ gauge theories shares (in the common sector
of the spectrum of bound states) the same large-$N_c$ limit as obtained with $SU(N_c)$ theories.
One can then study these theories as a complementary way of testing theoretical  expectations, for observables 
such as  the vacuum condensates and the mass spectra of bound states. And one can use the comparison 
between different sequences of theories to learn about commonalities and differences,
hence deducing general field-theoretical lessons.
In the case of pure Yang-Mills theories, the spectrum of glueballs can be computed, in the large-$N_c$ limit,
with the tools of gauge-gravity dualities---a selection of papers on the topic includes Refs.~\cite{Brower:2000rp,
Apreda:2003sy,Mueck:2004qg,Wen:2004qh,Kuperstein:2004yf,Elander:2013jqa,Athenodorou:2016ndx,
Elander:2018aub,Elander:2020csd}---or other semi-analytical approaches~\cite{Bochicchio:2016toi,Bochicchio:2013sra,
Hong:2017suj}. These can then be compared to the results of the lattice literature on $SU(N_c)$ 
Yang-Mills theories~\cite{Lucini:2001ej,Lucini:2004my,Lucini:2010nv,Lucini:2012gg,
Athenodorou:2015nba,Lau:2017aom,Hernandez:2020tbc,Athenodorou:2021qvs,Yamanaka:2021xqh,
Bonanno:2022yjr},  and $Sp(N_c)$ theories~\cite{Bennett:2017kga,Bennett:2020hqd,Bennett:2020qtj}.
The spectra of mesons and of fermion bound states 
are more challenging to compute on the lattice~\cite{Lucini:2012gg},
but equally interesting, and the quenched calculation may soften such difficulties,
while producing interesting results---for $Sp(2N)$ theories, see Ref.~\cite{Bennett:2019cxd}.
Other non-perturbative objects, such as the string tension (see Ref.~\cite{Aharony:2009gg}
and references therein) and the topological susceptibility of Yang-Mills theories---see the useful 
Refs.~\cite{Witten:1979vv,Veneziano:1979ec,Witten:1998uka,Vicari:2008jw}---are also accessible to the lattice~\cite{
Luscher:1981zq,Campostrini:1989dh,DelDebbio:2002xa,Lucini:2004yh,DelDebbio:2004ns,Luscher:2010ik,
Panagopoulos:2011rb,Bonati:2015sqt,Bonati:2016tvi,
Ce:2016awn,Alexandrou:2017hqw,Bonanno:2020hht,Borsanyi:2021gqg,Cossu:2021bgn,
Teper:2022mmj,Bonanno:2022vot,Bonanno:2022hmz}. Recently, the topological susceptibility of $Sp(2N)$ theories has been
the subject of dedicated studies summarised in Ref.~\cite{Bennett:2022ftz,Bennett:2022gdz}.

The paper is organised as follows.
In Sect.~\ref{sec:sp2n} we define the gauge actions of $Sp(2N)$  theories,  couple them to matter fields,
analyse the low-energy description---borrowing  ideas from the literature on
Chiral Perturbation Theory ($\chi$PT)   and
Hidden Local Symmetry (HLS)~\cite{
Bando:1984ej,Casalbuoni:1985kq,Bando:1987br,Casalbuoni:1988xm,
Harada:2003jx,Georgi:1989xy,Appelquist:1999dq,Piai:2004yb,
Franzosi:2016aoo}---and applications
in CHM, top compositeness, and SIMP contexts.
Significant parts of  this section follow Refs.~\cite{Bennett:2017kga,
Bennett:2019jzz,Bennett:2019cxd} and references therein.
Section~\ref{sec:latticesp2n} is a brief summary of lattice field theory numerical techniques
used  in Refs.~\cite{Bennett:2017kga,Lee:2018ztv,Bennett:2019jzz,Bennett:2019cxd,
Bennett:2020hqd,Bennett:2020qtj,Lucini:2021xke,Bennett:2021mbw,Bennett:2022yfa,
Bennett:2022gdz,Bennett:2022ftz,AS}, and we refer the reader to the original literature for  details.
We summarise in Sect.~\ref{sec:puresp4}  results obtained
in the (quenched) lattice  $Sp(2N)$ theory,  in which the only dynamical degrees of freedom 
correspond to the gluons. Besides  strings (or fluxtubes) 
and glueballs~\cite{Bennett:2017kga,Bennett:2020hqd,Bennett:2020qtj}, we discuss
quenched mesons~\cite{Bennett:2019cxd},
and  topological susceptibility~\cite{Bennett:2022gdz,Bennett:2022ftz}.
Section~\ref{sec:sp4} considers observables in lattice studies that implement  dynamical fermions~\cite{Bennett:2017kga,Lee:2018ztv,Bennett:2019jzz,Bennett:2022yfa,AS}.
After the summary and conclusion in Sect.~\ref{sec:conclusion}, 
we devote Appendix~\ref{sec:algebra} to a summary of technical details, and
the short Appendix~\ref{sec:open} to our open
access approach to data and analysis code.

\section{$Sp(2N)$ gauge theory and composite dynamics}
\label{sec:sp2n}

In this section we provide the microscopic description of the broad class of $Sp(2N)$ gauge theories of interest.
We discuss the field content and interactions, the symmetries and symmetry-breaking patterns
(including both explicit and spontaneous symmetry-breaking effects),  and some interesting results obtained by deploying
perturbation theory and  low-energy EFT arguments.
In the process, we fix the notation adopted in the paper.
We sketch the connection with  applications in the context of the phenomenology of
extensions of the standard model, focusing on composite Higgs models, on top partial compositeness, and on 
composite dark matter.
As a note of caution, we highlight that in this review we ignore almost completely
 the Abelian $U(1)$ global symmetry factors, except for occasionally mentioning the anomalous $U(1)_A\sim SO(2)_A$
 symmetry acting on the fermions.\footnote{In the presence of fermions transforming 
 in different representations of the gauge group, the triangle anomaly gives mass to only one 
linear combination of the PNGBs associated with the breaking of the chiral $U(1)$ symmetries acting on the different 
flavor species. Phenomenological implications are discussed for example in Ref.~\cite{Belyaev:2016ftv}.}
Lattice explorations of the flavor singlet mesons
 are in their early stages---see for instance
Ref.~\cite{Arthur:2016ozw}.

\subsection{Fields, symmetries and observables}
\label{sec:uvmodel}

We start by defining the short-distance dynamics in continuum field-theory terms. For convenience,
we write explicitly the Lagrangian density of the dynamical theory relevant to the CHM 
proposed in Ref.~\cite{Barnard:2013zea} (see also Ref.~\cite{Ferretti:2013kya}), but without coupling it to the SM fields.
This is an $Sp(4)$ gauge theory coupled to $N_f=2$ Dirac fermions $Q^{J\,a}$
transforming in the fundamental (f) representation of the gauge group, and $n_f=3$ Dirac fermions 
$\Psi^{j\,ab}$ transforming in the 2-index antisymmetric (as) representation.
All other gauge theories of relevance to this review can be obtained by either 
replacing the $Sp(4)$ gauge group by $Sp(N_c=2N)$ (with $N>1$) and/or by changing the number of  dynamical 
 fermion species $N_f$ and $n_f$. We follow the notation of Ref.~\cite{Bennett:2022yfa}---see also 
Refs.~\cite{Bennett:2019cxd,Bennett:2017kga} and references therein.

Here and in the following, we denote the color indices in the fundamental representation
by letters at the beginning of the Latin alphabet,
as in $a,b=1,\,\cdots,\,N_c=4=2N$. We capitalise the index to denote the adjoint representation,
so that $A=1,\,\cdots,\,N(2N+1)=10$ is used to denote the gauge bosons  of $Sp(4)$.
We reserve the letters in the middle of the Latin alphabet for flavor/family
indices in Dirac fermion notation, so that the capitalised 
$J,K=1,\,\cdots,\, N_f=2$ labels the Dirac species in the (f) representation,
while the lower-case $j,k=1,\,\cdots,\,n_f=3$ is used for the Dirac species in the (as) representation.
We find it useful also to denote by   characters taken  from the second half of the Latin alphabet
the flavor/family indices in 2-component spinor representation, so that
$M,N=1,\,\cdots,\,2N_f=4$ labels 2-component spinors transforming in the (f) representation,
while $m,n=1,\,\cdots,\,2n_f=6$ is reserved for the (as) representation of the gauge group.
We use letters taken from the second half of the Greek alphabet
to denote Lorentz indices, as in $\mu,\nu=0,\,1,\,2,\,3$. In different parts of the
text we use Minkowski (M) or Euclidean (E) space-time notation---when possible ambiguities cannot be 
resolved by the context, we will add the subscripts $(M)$ or $(E)$ to differentiate between the two.
Spinorial indices are denoted by the first letters of the Greek alphabet,
and we restrict their use to 2-component notation, for example by writing $\alpha,\beta=1,\,2$, but
we mostly omit writing them and leave them implicit instead.

The symplectic group $Sp(2N)$ is defined as the subgroup of $SU(2N)$
consisting of $2N\times 2N$ matrices $U$ that obey the defining relationship
\beqs
U\, \Omega \, U^T &=& \Omega\,,
\eeqs
where $\Omega$ is the $2N\times 2N$ symplectic matrix, which we can write in $N\times N$ blocks as
\beqs
{\Omega}&\equiv&\left(\begin{array}{c|c}
\mathbb{O}_{N \times N} & \mathbb{I}_{N \times N}\\
\hline
-\mathbb{I}_{N \times N} & \mathbb{O}_{N \times N}
\end{array}\right)\,.
\label{Eq:symplecticmatrix}
\eeqs
These matrices can also be written in the form
\beqs
U&=&
\left(\begin{array}{c|c}
\mathbb{A} & \mathbb{B}\\
\hline
-\mathbb{B}^{\ast} & \mathbb{A}^{\ast}
\end{array}\right)\,,
\eeqs
with the $N\times N$ matrices $\mathbb{A}$ and $\mathbb{B}$ satisfying the 
non-trivial relations
$\mathbb{A}^{\dagger} \mathbb{A} + \mathbb{B}^{\dagger} \mathbb{B} = \mathbb{I}_{N\times N}$ 
and $\mathbb{A}^T \mathbb{B} = \mathbb{B}^T \mathbb{A}$.\footnote{
This property is useful in defining the Cabibbo-Marinari~\cite{Cabibbo:1982zn} updating algorithm for 
$Sp(2N)$; see Appendix A of Ref.~\cite{Bennett:2020qtj} for technical details.}

In Minkowski space-time, with signature mostly $-$, the Lagrangian density is 
\beqs
{\cal L}&=& -\frac{1}{2} \Tr V_{\mu\nu} V^{\mu\nu}\,+\,\nonumber\\
&&
\,+\,\frac{1}{2}\sum_{J=1}^{2}\left(i\,\overline{Q^{J}}_a 
\gamma^{\mu}\left(D_{\mu} Q^J\right)^a
\,-\,i\,\overline{D_{\mu}Q^{J}}_a \gamma^{\mu}Q^{J\,a}\right)\,
-\,m^{(f)}\sum_{J=1}^{2}\overline{Q^J}_a Q^{J\,a}+\nonumber\\
&&
\,+\,\frac{1}{2}\sum_{j=1}^{3}\left(i\,\overline{\Psi^{j}}_{ab} \gamma^{\mu}\left(D_{\mu} \Psi^j\right)^{ab}
\,-\,i\,\overline{D_{\mu}\Psi^{j}}_{ab} \gamma^{\mu}\Psi^{j\,ab}\right)\,
-\,m^{(as)}\sum_{j=1}^{3}\overline{\Psi^j}_{ab} \Psi^{j\,ab}\,,
\label{eq:lagrangian}
\eeqs
where we have suppressed spinor indices, and summations over color and Lorentz indices are understood.
The irreducible 2-index antisymmetric representation of $Sp(4)$ is $\Omega$-traceless,
so that $\Tr\, \Omega \Psi =0$.
In this review, we take  the mass matrices for the two species of fermions
to be proportional to the identity matrix---see Ref.~\cite{Kulkarni:2022bvh} for the
generalisation to non-degenerate masses---and denote the masses as $m^{(f)}$ and $m^{(as)}$,
for the (f) and (as) representations, respectively.
The transformation 
properties under the action of an element $U$ of the 
$Sp(4)$ gauge group are $Q\rightarrow U Q$ and $\Psi \rightarrow U \Psi U^{\mathrm{T}}$.
Hence, the field-strength tensor, $V_{\mu\nu}$, and the covariant derivatives, are given by
\beqs
V_{\mu\nu}&\equiv& \partial_{\mu}V_{\nu}-\partial_{\nu}V_{\mu} + i g \left[V_{\mu}\,,\,V_{\nu}\right]\,,\\
D_{\mu} Q^J&=& \partial_{\mu} Q^J \,+\,i g V_{\mu} Q^{J}\,,\\ 
D_{\mu} \Psi^j&=& \partial_{\mu} \Psi^j \,+\,i g V_{\mu} \Psi^{j}\,+\,i g \Psi^{j} V_{\mu}^{\mathrm{T}}\,,
\eeqs
where $g$ is the gauge coupling, while $V_{\mu}=V_{\mu}^AT^A$ are matrix-values gauge fields---the  $T^A$ matrices are the generators of the gauge group, normalised  so that $\Tr\, T^AT^B=\frac{1}{2}\delta^{AB}$.

The Lagrangian density in Eq.~(\ref{eq:lagrangian})  is formally identical to that of 
the $SU(N_c)$ theories coupled to Dirac fermions. If the group is taken to be
$SU(3)$, then the equivalence of the 2-index antisymmetric representation and the 
(conjugate) fundamental implies that this would become an extension of QCD with two fermions with mass $m^{(f)}$
and three with mass $m^{(as)}$.
But the representations of $Sp(2N)$ are (pseudo-)real, which leads to an enhancement of the non-Abelian 
global symmetry from $SU(N_f)\times SU(N_f)$ and $SU(n_f)\times SU(n_f)$, acting
on the (f) and (as) fermions,  to $SU(2N_f)$ and $SU(2n_f)$, respectively. 
From here onwards, in the rest of this section we restrict attention to $N=2$, $N_f=2$, and 
$n_f=3$~\cite{Barnard:2013zea,Ferretti:2013kya}, as reinstating the general dependence on number of
colors and flavors is straightforward.

To demonstrate symmetry enhancement manifestly, we perform the following exercise.
First, we introduce 2-component spinors $q^{M\,a}$ and $\psi^{n\,ab}$, 
transforming in the (f) and (as) representations of the gauge group, respectively, with $M=1,\,\cdots,\,4$ and
$n=1,\,\cdots,\,6$. 
We then construct the four component spinors via the following definitions:
\beqs
Q^{J\,a}&\equiv&\left(
\begin{array}{c}
q^{J\,a} \cr \Omega^{ab}(-\tilde{C}q^{J+2\,\ast})_b
\end{array}
\right)\,,
~~~~
\Psi^{j\,ab}\,\equiv\,\left(
\begin{array}{c}
\psi^{j\,ab} \cr
\Omega^{ac}\Omega^{bd} (-\tilde{C}\psi^{j+3\,\ast})_{cd}
\end{array}
\right)\,,
\eeqs
where $\tilde{C}=-i \tau^2$ is the $2\times 2$ charge-conjugation matrix in spinor space, 
$\tau^2$ is  the second Pauli matrix, $J=1,\,2$ and $j=1,\,2,\,3$. Because of the contraction with the symplectic 
matrix $\Omega$, which raises and lowers the $Sp(4)$ index, the pseudo-real nature of the
(f) representation, and real nature of the (as) representation, what results are two Dirac fermions
of type (f) and three of type (as), which are those appearing in Eq.~(\ref{eq:lagrangian}).
By replacing the definitions in Eq.~(\ref{eq:lagrangian}),  after some tedious algebra one 
arrives at the identity
\beqs
{\cal L}&=& -\frac{1}{2} \Tr \,V_{\mu\nu} V^{\mu\nu}\,+\,\nonumber\\
&&
\,+\,\frac{1}{2}\sum_{M=1}^{4}\left(i\,(q^{M})^{\dagger}_{\,\,\,a}
\bar{\sigma}^{\mu}\left(D_{\mu} q^{M}\right)^a
\,-\,i\,(D_{\mu}q^{M})^{\dagger}_{\,\,\,a} \bar{\sigma}^{\mu}q^{M\,a}\right)\,+\,\nonumber\\
&&
\,-\,\frac{1}{2} m^{(f)} \sum_{M,N=1}^{4}\tilde{\Omega}_{MN}\left( q^{M\,a\, T} \Omega_{ab} \tilde{C} q^{N\,b} 
- (q^{M})^{\dagger}_{\,\,\,a}\Omega^{ab} \tilde{C} (q^{N\,\ast})_b\right)+\nonumber\\
&&
\,+\,\frac{1}{2} \sum_{m=1}^{6} \left(i\,(\psi^{m})^{\dagger}_{\,\,\,ab} \bar{\sigma}^{\mu}\left(D_{\mu} \psi^{m}\right)^{ab}
\,-\,i\,(D_{\mu}\psi^{m})^{\dagger}_{\,\,\,ab} \bar{\sigma}^{\mu}\psi^{m\,ab}\right)\,+\,\nonumber\\
&&
\,-\,\frac{1}{2}m^{(as)} 
\sum_{m,n=1}^{6} \omega_{mn}\left(\psi^{m\,ab\,\mathrm{T}} \Omega_{ac}\Omega_{bd} \tilde{C}\psi^{n\,cd}\,
-\,(\psi^{m\,})^{\dagger}_{\,\,\,ab}\Omega^{ac}\Omega^{bd} \tilde{C}(\psi^{n\,\ast})_{cd}\right)\,,
\label{eq:lagrangian2}
\eeqs
where the kinetic terms for the 2-component spinors are written by making use of the
$2 \times 2$ matrices $\bar{\sigma}^{\mu}\equiv \left(\mathbb{I}_2,\,\tau^i\right)$.
In these expressions, $\tilde{\Omega}=\Omega$, but notice that the former acts on the flavor space, while the latter
in the color space---the former is a $2N_f\times 2N_f$ matrix, while the latter is a $2N\times 2N$ one.
The matrix $\omega$ is defined to be symmetric, and we can use the explicit expression
\beqs
{\omega}&\equiv&\left(\begin{array}{c|c}
\mathbb{O}_{3 \times 3} & \mathbb{I}_{3 \times 3}\\
\hline
\mathbb{I}_{3 \times 3} & \mathbb{O}_{3 \times 3}
\end{array}\right)\,.
\label{Eq:symmetricmatrix}
\eeqs

With the Lagrangian density in the form of Eq.~(\ref{eq:lagrangian2}), it becomes manifest that the 
theory has a global $SU(4)\times SU(6)$ non-Abelian symmetry, and that the mass terms 
proportional to $m^{(f)}$ and $m^{(as)}$ introduce a (small) breaking effect, reducing 
 the exact symmetry to  the subgroups of
$SU(4)$ and $SU(6)$ that leave invariant, respectively, the matrices $\tilde{\Omega}$ and $\omega$.
Vacuum alignment arguments~\cite{Peskin:1980gc} suggest that 
fermion bilinear condensates form in the underlying dynamics, breaking spontaneously
the global symmetry in the same way, and hence PNGBs will emerge that describe the $SU(4)/Sp(4)$ coset
in the (f) sector, and the $SU(6)/SO(6)$ coset in the (as) sector.

We conclude this subsection with a set of  counting exercises and symmetry consideration, and
characterise the spectrum of lightest bound states of the theory, and the operators
that are used to define spectral observables from correlation functions.
Some of the bound states admit a weakly coupled  description 
as particles associated with fields in the low energy EFT description of the dynamics. 
More details and a broader set of considerations of this type can be found for example
in Appendix E and F of Ref.~\cite{Bennett:2019cxd}, in Appendix C of Ref.~\cite{Bennett:2022yfa},
in Section III.C of Ref.~\cite{Bennett:2020qtj},
and in the references therein.

Let us start with the glueballs. 
These are bound states that exist in the Yang-Mills theory, without matter fields, in the confined phase. 
They do not carry flavor, but they can have any (integer) spin $J$, and in general are 
characterised by $J^{PC}$, with $P$ the parity and $C$ the charge-conjugation eigenvalues,
except that, at odds with the $SU(N_c)$ cases, 
 in the  $Sp(2N)$ gauge theories $C=+$ for all glueballs.
  The interpolating operators sourcing the glueballs can be built from the Wilson loops, 
 traced path ordered products of links around closed spatial (contractible) loops, 
 along with appropriate projections to the states with desired spin and parity quantum numbers.
We will return in due time to the subtleties related to how the continuum rotation symmetry
is broken to the octahedral group $O_h$ on a hypercubic lattice theory.
Here we notice only the fact that in the presence of additional fermionic matter, one expects the
glueballs to mix with the flavor-singlet mesons.  Quantitative understanding of
these and related effects, which  involve disconnected diagrams, is an open problem on the lattice---an interesting
exploration  of this topic in the $SU(2)$ theory can be found in Ref.~\cite{Arthur:2016ozw}.

The flavored mesons made of (f) fermions can be classified by their spin $J$, the representation of the
unbroken $Sp(4)\sim SO(5)$ global symmetry group, and additional discrete quantum numbers, such 
the unbroken parity $P$---constructed by combining ordinary spatial parity and discrete internal symmetries.
As long as the mass terms are small, in appropriate units, the lightest states are going to be the PNGBs.
These have $J^P=0^-$, and transform as $5$ of $Sp(4)$, the $\Omega$-traceless antisymmetric representation.
In the language of 2-flavor QCD, the PNGBs are identified with the pions $\pi$. Their parity partner $J^P=0^+$ 
mesons transform as $5$ of $Sp(4)$, and are the analogous of the $a_0$ in QCD, in the sense that
if $U(1)_A=SO(2)_A$ were exact, $\pi$ and $a_0$ would be degenerate.
There are then four multiplets of spin-1 states. Two $J^P=1^-$ states transforming as the $10$ of $Sp(4)$ correspond to 
what in QCD are the $\rho$ and $\rho^{\prime}$ states, which have different properties in the global $SU(4)$,
but undergo mixing.
Two $J^P=1^+$ states exist, one of which transforms 
as a $5$ of $Sp(4)$, and is the analogue of the $a_1$, and one transforming as a $10$, related to the $b_1$ in QCD.
We summarise in Table~\ref{tab:mesons} the operators ${\cal O}_M$  sourcing these states
 (see also Ref.~\cite{Lewis:2011zb}), and 
their basic quantum numbers and properties. We label them as pseudoscalar (PS), scalar (S), vector (V), tensor (T),
axial-vector (AV), and axial-tensor (AT).
The (as) fermions give rise to a similar set of multiplets, but for the fact that the 
symmetric and antisymmetric representations are swapped. For example, the
$20$ PNGBs describing the $SU(6)/SO(6)$ coset are in the traceless symmetric representation.\footnote{We denote the set of PNGBs of $SU(6)/SO(6)$
as $20^{\prime}$, for consistency with the conventional notation of $SU(4)\sim SO(6)$,
as there are  three inequivalent representations with $20$ degrees of freedom, usually denoted as $20$, $20^{\prime}$,
and $20^{\prime\prime}$~\cite{Slansky:1981yr}.}

\begin{specialtable}[t]
\begin{center}
\caption{
Interpolating operators $\mathcal{O}_M$ built with Dirac fermions of types (f) and (as).
Colour and spinor indices are  implicit and summed over, and flavor combinations are denoted generically.
More details can be found in Ref.~\cite{Bennett:2019cxd}.
We also show the $J^P$ quantum numbers, the  corresponding QCD 
meson sourced by the operator with analogous quantum numbers, and
 the  irreducible representation  of  the unbroken global 
$Sp(4)\times SO(6)$ symmetry groups.
}
\label{tab:mesons}
\begin{tabular}{|c|c|c|c|c|c|}
\hline\hline
{\rm Label} & {\rm ~Interpolating operator~} & {\rm ~~~Mesons in~~~} 
& {\rm ~~~$J^{P}$~~~}
& $Sp(4)$ & $SO(6)$  \cr
 $M$ & $\mathcal{O}_M$& {\rm ~~~$N_f=2$ QCD~~~} 
& 
& &   \cr
\hline
PS & $\overline{Q^{I}}\gamma_5 Q^{J}$ & $\pi$ & $0^{-}$ & $5 $ & $1$\cr
S & $\overline{Q^{I}} Q^{J}$ & $a_0$ & $0^{+}$ & $5 $ & $1$\cr
V & $\overline{Q^{I}}\gamma_\mu Q^{J}$ & $\rho, \rho^{\prime}$ & $1^{-}$ & $10$ & $1$\cr
T & $\overline{Q^{I}}\sigma_{\mu\nu} Q^{J}$ & $\rho, \rho^{\prime}$ & $1^{-}$ & $10 $ & $1$ \cr
AV & $\overline{Q^{I}}\gamma_5\gamma_\mu Q^{J}$ & $a_1$ & $1^{+}$ & $5 $ & $1$\cr
AT & $\overline{Q^{I}}\gamma_5\sigma_{\mu\nu} Q^{J}$ & $b_1$ & $1^{+}$ & $10 $ & $1$\cr
\hline
ps & $\overline{\Psi^{k}}\gamma_5 \Psi^{j}$ & $\pi$ & $0^{-}$ & $1$ & $20^{\prime}$\cr
s & $\overline{\Psi^{k}} \Psi^{j}$ & $a_0$ & $0^{+}$ & $1$ & $20^{\prime}  $\cr
v & $\overline{\Psi^{k}}\gamma_\mu \Psi^{j}$ & $\rho, \rho^{\prime}$ & $1^{-}$ & $1$ & $15$ \cr
t & $\overline{\Psi^{k}}\sigma_{\mu\nu} \Psi^{j}$ & $\rho, \rho^{\prime}$ & $1^{-}$ & $1$  & $15 $\cr
av & $\overline{\Psi^{k}}\gamma_5\gamma_\mu \Psi^{j}$ & $a_1$ & $1^{+}$ & $1$ & $20^{\prime}  $\cr
at & $\overline{\Psi^{k}}\gamma_5\sigma_{\mu\nu} \Psi^{j}$ & $b_1$ & $1^{+}$ & $1$ & 
$15$\cr
\hline\hline
\end{tabular}
\end{center}
\end{specialtable}

\begin{specialtable}[H]
\begin{center}
\caption{
Interpolating operators $\mathcal{O}_{\rm CB}$ sourcing the lightest
chimera baryons, built with two Dirac fermions of types (f) and one of type (as),
with their
$Sp(4)\times SO(6)$ quantum numbers.
Details can be found in Ref.~\cite{Bennett:2022yfa}.
}
\label{tab:chimerabaryons}
\begin{tabular}{|c|c|c|c|}
\hline\hline
{\rm Label} & {\rm ~Interpolating operator~} & $Sp(4)$ & $SO(6)$  \cr
\hline
 ${\cal O}^{L,R}_{{\rm CB}, 1}$
 &
$\left(\overline{Q^{1\,a}} \gamma^5 Q^{2\,b}+\overline{Q^{2\,a}} \gamma^5 Q^{1\,b}\right)  \Omega_{bc} P_{L,R}\Psi^{k\,ca}$&
& \cr
${\cal O}^{L,R}_{{\rm CB}, 2}$
& $i\left(-\overline{Q^{1\,a}} \gamma^5 Q^{2\,b}+\overline{Q^{2\,a}} \gamma^5 Q^{1\,b}\right)\Omega_{bc} P_{L,R}\Psi^{k\,ca}$&
& \cr
${\cal O}^{L,R}_{{\rm CB}, 3}$
&
$\left(\overline{Q^{1\,a}}  \gamma^5 Q^{1\,b}-\overline{Q^{2\,a}}\gamma^5Q^{2\,b}\right)\Omega_{bc} P_{L,R}\Psi^{k\,ca}$&
$5$ & $6$ \cr
$ {\cal O}^{L,R}_{{\rm CB}, 4}$
&
$ -i\,\left(\overline{Q^{1\,a}}  Q^{2\,b}_{\,C}+\overline{Q_C^{2\,a}}  Q^{1\,b}\right)\Omega_{bc} P_{L,R}\Psi^{k\,ca}$&
& \cr
$ {\cal O}^{L,R}_{{\rm CB}, 5}$
&
$ i\,\left(-i\,\overline{Q^{1\,a}}  Q^{2\,b}_C+i\overline{Q^{2\,a}_C} Q^{1\,b}\right)\Omega_{bc} P_{L,R}\Psi^{k\,ca}$&
& \cr
\hline
$ {\cal O}^{\prime\,L,R}_{{\rm CB}, 1}$
 &
$ i\left(\overline{Q^{1\,a}}  Q^{2\,b}+\overline{Q^{2\,a}}  Q^{1\,b}\right) \Omega_{bc} P_{L,R}\Psi^{k\,ca}$&
   &\cr
${\cal O}^{\prime\,L,R}_{{\rm CB}, 2}$
 &
$
\left(\overline{Q^{1\,a}}  Q^{2\,b}-\,\overline{Q^{2\,a}}  Q^{1\,b}\right)\Omega_{bc} P_{L,R}\Psi^{k\,ca}$&
        &\cr
${\cal O}^{\prime\,L,R}_{{\rm CB}, 3}$
 &
$
i\left(\overline{Q^{1\,a}}   Q^{1\,b}-\overline{Q^{2\,a}}Q^{2\,b}\right)\Omega_{bc} P_{L,R}\Psi^{k\,ca}$&
  $5$ & $6$ \cr
$ {\cal O}^{\prime\,L,R}_{{\rm CB}, 4}$
 &
$
 \,\left(\overline{Q^{1\,a}}  \gamma^5 Q^{2\,b}_{\,C}+\overline{Q_C^{2\,a}}  \gamma^5 Q^{1\,b}\right)\Omega_{bc} P_{L,R}\Psi^{k\,ca}$&
    &\cr
$ {\cal O}^{\prime\,L,R}_{{\rm CB}, 5}$
 &
$
 i\,\left(\,\overline{Q^{1\,a}} \gamma^5  Q^{2\,b}_C-\overline{Q^{2\,a}_C} \gamma^5 Q^{1\,b}\right)\Omega_{bc} P_{L,R}\Psi^{k\,ca}$&
 &\cr
\hline\hline
\end{tabular}
\end{center}
\end{specialtable}

We list in Table~\ref{tab:chimerabaryons}  the explicit form of the
operators sourcing the  two sets of lightest chimera baryons in the theory, made of 
two (f) and one (as) elementary fermions.
The two sets we consider transform both as a $5$ of $Sp(4)$, and are one 
the $U(1)_A\sim SO(2)_A$ partner of the other, reproducing for these spin-$1/2$ states  the relation between
 PS and S mesons in the scalar sector of the spectrum. 
As conventional, the chiral projectors are
 \beqs
 P_{L,R}&\equiv&
 \frac{1}{2}\left(\frac{}{}\mathbb{I}_{4\times 4} \pm \gamma_5\frac{}{}
 \right)\,.
 \label{eq:chiral_projector}
 \eeqs
Other spin-$1/2$ and spin-$3/2$ states can be built systematically in a similar
 fashion~\cite{Hsiao:2022kxf}. (Table~1 of Ref.~\cite{Golterman:2017vdj} shows a classification
 of top partners for $SO(d)$ gauge theories.)
These operators also form multiplets of the global  $SU(6)$ symmetry and its unbroken $SO(6)$ subgroup,
and we will return to this part of the classification later in the paper.

\subsection{Perturbative considerations}
\label{sec:pert}

 The confining, QCD-like dynamics leading to the appearance of light PNGBs, that are essential to CHMs, can
be complemented by implementing the top (partial)  compositeness mechanism.
 Interacting near-conformal theories, with extended fermion matter content, 
  in which (chimera) baryon operators develop large anomalous dimensions
  are best suited to provide an origin for top partial compositeness, 
  for reasons we discuss in Sect.~\ref{sec:toppartner}.
The underlying strong interactions can be understood in full only with 
non-perturbative tools, such as lattice simulations. Yet,
 perturbative calculations, supplemented by other techniques,  provide useful insight into
  their infrared (IR) phase structure, and
guidance in identifying
 promising theories to be subjected to dedicated numerical studies. 
In this section, we briefly discuss the IR behaviour of non-Abelian gauge theories with fermions in 
the fundamental and/or two-index representations, 
and review existing analytical results relevant to $Sp(2N)$ gauge theories.

Yang-Mills  theories  are asymptotically free
at short distances. Their ultraviolet (UV) properties can be studied perturbatively,
as an expansion in the coupling $\alpha\equiv g^2/(4\pi)$.
When coupled to $N_f$ fundamental fermions, there is a maximum
 $N_f^{\rm AF}$ above which the theory loses asymptotic freedom. It can  be determined
from the renormalisation group (RG) analysis of the beta function 
$\beta(\alpha)\equiv \partial \alpha/\partial \log (\mu)$, estimated at the 1-loop order.
\footnote{
Although only  integer values of  $N_f$ are physically meaningful,  $N_f$ is treated as a continuous variable.
An alternative argument could be made by taking the large-$N_c$  (Veneziano) limit
while holding fixed the continuous ratio $x_f=N_f/N_c$.
}
If $N_f$ is sufficiently small,  the theory  confines in the IR, and breaks chiral symmetry, as in QCD.
For $N_f$ just below $N_f^{\rm AF}$, the theory admits the
Banks-Zaks fixed point~\cite{Caswell:1974gg,Banks:1981nn}, identified 
as a zero of the 2-loop beta function  at small coupling.
One therefore expects that asymptotically free gauge theories undergo
 a zero-temperature quantum phase transition, for a critical number of flavours $N_f^{\rm cr}$,
between the IR conformal and chirally broken phases. 
The interval $N_f^{\rm cr}<N_f<N_f^{\rm AF}$ is called {\it conformal window}, 
and has  been extensively studied by both analytical and numerical methods. 
For $N_f<N_f^{\rm cr}$, but in proximity of the conformal window,
  near-conformal dynamics has been suggested to display potential for phenomenological applications, 
in such contexts as (walking) technicolor, composite Higgs, and composite dark matter (e.g. see 
Refs.~\cite{Panico:2015jxa,Witzel:2019jbe,Cacciapaglia:2020kgq,Chivukula:2000mb,Lane:2002wv,
Hill:2002ap,Martin:2008cd,Sannino:2009za,Piai:2010ma}).

The determination of $N_f^{\rm cr}$ is notably difficult, because the coupling at the IR fixed point $\alpha_{\rm IR}$ grows in the approach to the lower end of the conformal window. 
As a first, crude approximation, one can identify $N_f^{\rm cr}$ as the number of flavors for 
which the zero of the 2-loop beta function disappears. 
This can be systematically improved to higher orders in $\alpha$, by solving  $0=\beta(\alpha)
 \equiv -2\alpha \sum_{\ell =1}^{\ell_{max}} b_\ell \left(\frac{\alpha}{4\pi}\right)^\ell$, 
where the coefficients $b_\ell$ are functions of $N_c$, $N_f$, and the fermion representation $R$, 
but suffer from the intrinsic limitation of perturbation theory. 
In particular, even if the existence of a fixed point is physical, and hence scheme independent, 
its determination and characterisation are affected by the scheme dependence 
of $\beta(\alpha)$ for $\ell\geq 3$. For example,
in the $\overline{\rm MS}$ scheme with $\ell_{max}=4$,  $N_f^{\rm cr}$ critical for various non-abelian gauge groups and representations can be found in Ref.~\cite{Pica:2010xq}.\footnote{
After the 5-loop beta function was computed \cite{Baikov:2016tgj,Herzog:2017ohr}, the conformal window has also been studied for $\ell_{max}=5$, and  in Ref.~\cite{Ryttov:2016ner}
the authors report on a strong instability of the perturbative expansion over a wide range of $N_f$ in the {\it would be} conformal window of $SU(3)$ gauge theories. 
}
Going beyond perturbation theory, several approaches intended to
 capture non-perturbative dynamics 
have been proposed in the literature,  such as the 
Schwinger-Dyson analysis in the ladder approximation~\cite{Appelquist:1988yc,Cohen:1988sq}, 
or a conjectured all orders beta function~\cite{Ryttov:2007cx,Pica:2010mt}
 inspired by the better controlled supersymmetric gauge theories---for
 the latter, see the review Ref.~\cite{Intriligator:1995au}. 

\begin{figure}[t]
\begin{center}
\includegraphics[width=.36\textwidth]{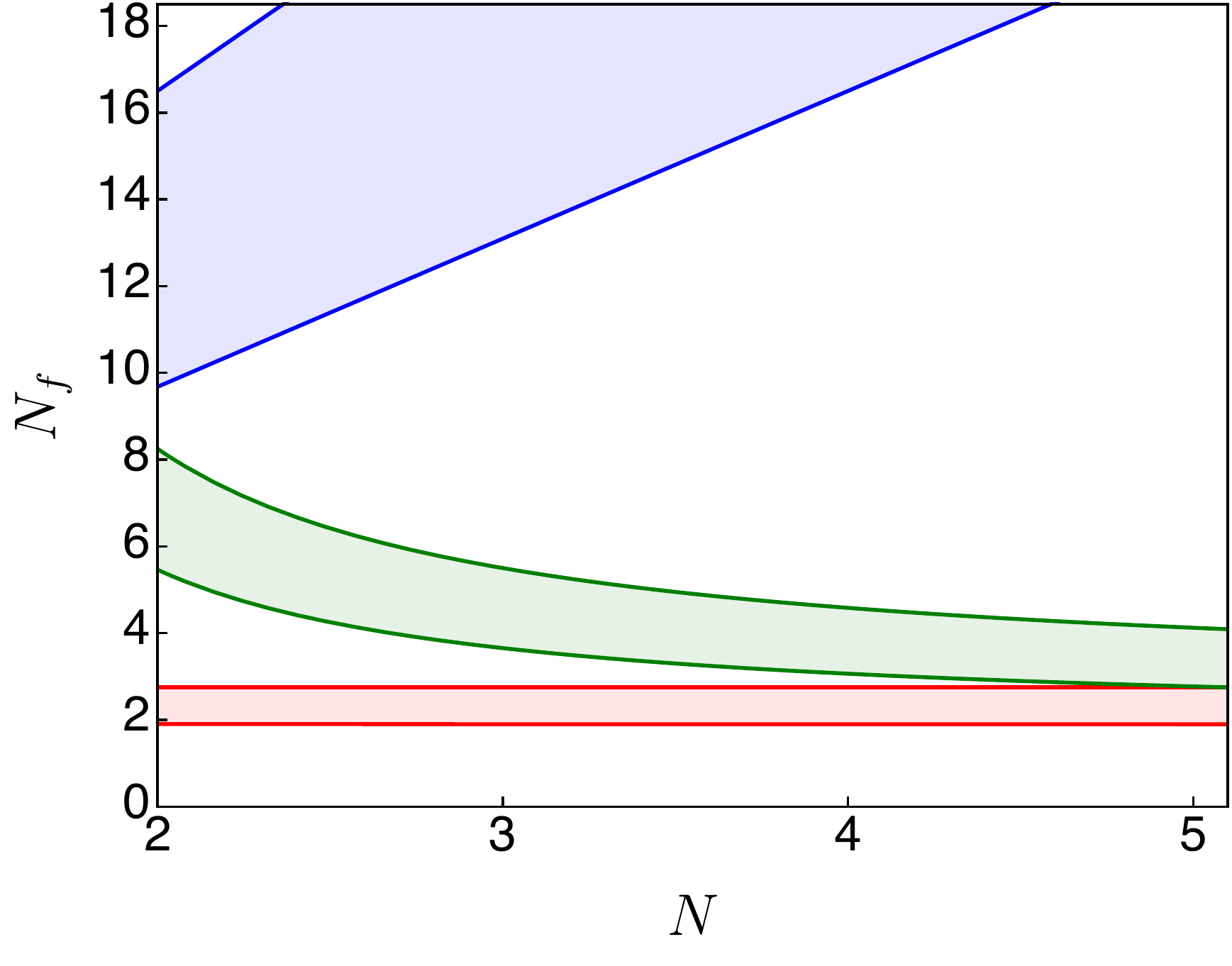}
\includegraphics[width=.36\textwidth]{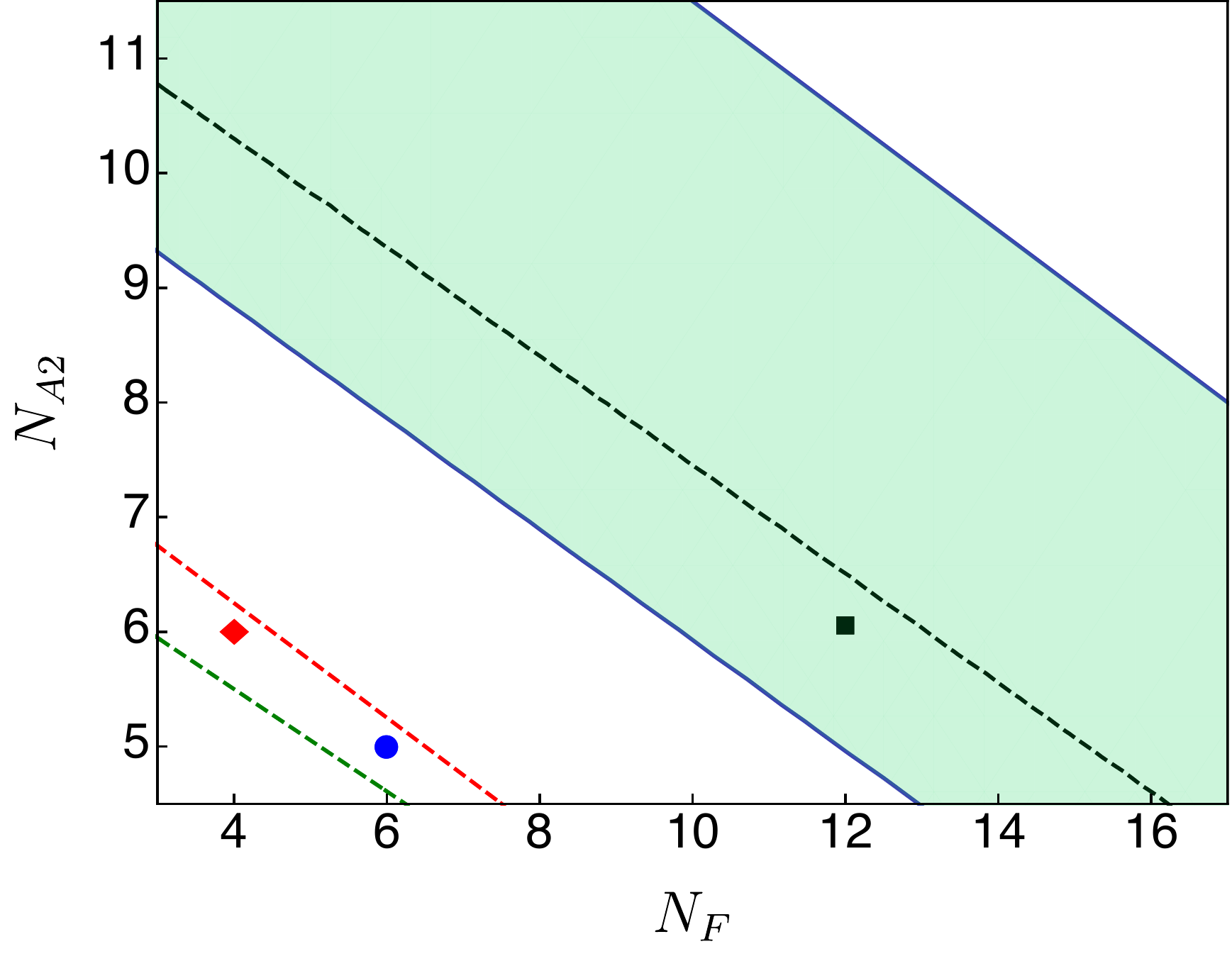}
\caption{%
Left panel: Conformal window of $Sp(2N)$ gauge theories with $N_R$ Dirac flavours in the fundamental (top, blue), 
antisymmetric (middle, green),
and symmetric (bottom, red) representations. 
Right panel: Conformal window of $Sp(4)$ gauge theories with $N_F=2N_f$ 
fundamental and $N_{A2}=2n_f$ antisymmetric Weyl fermions.
Red diamond, blue circle and black square indicate some representative CHMs, quoted in the main text. The left and right plots are from Refs.~\cite{Lee:2020ihn} and \cite{Kim:2020yvr}, respectively.
}
\label{Fig:cw}
\end{center}
\end{figure}

A number of recent studies~\cite{Ryttov:2016hdp,Ryttov:2016asb,Ryttov:2016hal,Ryttov:2017toz,
Ryttov:2017kmx,Ryttov:2017dhd,Gracey:2018oym,Ryttov:2018uue,Ryttov:2020scx}
discuss the determination of the conformal window in terms of a (Banks-Zaks) expansion
in the small physical parameter
 $\Delta_R \equiv N_R^{AF} - N_R$, where $N_R$ denotes the number of fermions in representation $R$.
Compared to the standard perturbative expansion, it has several salient features. 
First of all, it is scheme-independent, as the expansion parameter $\Delta_R$ is a physical quantity. 
Secondly, it has been found that its coefficients are positive, 
to the highest order known~\cite{Ryttov:2017toz,Ryttov:2017kmx},
which improves its convergence and stability.

A particularly interesting quantity, directly relevant for model-building considerations, 
is the anomalous dimension,  $\gamma^*$, of the fermion bilinear operator, measured at the IR fixed point.  
It has been suggested  that $\gamma^*_{\rm IR}=1$ 
at the lower edge of the conformal window~\cite{Cohen:1988sq,Kaplan:2009kr}. 
In Refs.~\cite{Kim:2020yvr,Lee:2020ihn}, it has further been shown that 
 the equivalent critical condition $\gamma^*_{\rm IR} (2-\gamma^{\ast}_{\rm IR})=1$ 
computed at a finite order in the BZ expansion results in a more rapidly convergent series expansion 
and thus can be used to improve the estimate for the critical value $N_R^{\rm cr}$.
\footnote{
This critical condition should agree with $\gamma^{\ast}_{\rm IR}=1$, 
yet it gives rise to different results at finite order in the  expansion. 
This critical condition reproduces the value of the critical coupling $\alpha^{\rm cr}$ 
obtained from the Schwinger-Dyson analysis in the ladder approximation~\cite{Appelquist:1998rb},
and furthermore $|1-\gamma^{\ast}_{\rm IR} |$ has a square-root singularity 
with respect to 
when the IR and UV fixed point merge~\cite{Lee:2020ihn}. 
}
The results of Refs.~\cite{Kim:2020yvr,Lee:2020ihn} are in excellent agreement with non-perturbative lattice results for $SU(2)$ and $SU(3)$ gauge theories coupled to the fundamental and two-index representations. 
The left panel of Fig.~\ref{Fig:cw}, borrowed from Ref.~\cite{Lee:2020ihn}, shows the conformal window of $Sp(2N)$ gauge theories coupled to fundamental, antisymmetric and symmetric Dirac fermions. According to this approach, 
$Sp(4)$ theories with either two fundamental or three antisymmetric flavours of fermions are in the confining phase. 
In the same paper, the uncertainties associated with the truncation of the BZ expansion, 
 which might capture some  non-perturbative effects, are also discussed, and their sizes estimated.

In the presence of fermions transforming in distinct representations, the properties of the IR fixed point
depend on all choices of  $N_R$.
The results for the theory of main  interest in this paper, namely  the $Sp(4)$ gauge theory coupled to fermions in the fundamental and antisymmetric representations, 
are presented in the right panel of Fig.~\ref{Fig:cw}~\cite{Lee:2020ihn}---here,
 $N_F=2N_f$ and $N_{A2}=2n_f$ 
count Weyl fermions. 
In the figure, the shaded region is the conformal window estimated from the critical condition $\gamma^* (2-\gamma^*)=1$ applied to the results in the $3^{\rm rd}$ order BZ expansion, 
while the dashed lines denote the analytical results obtained by the truncated Schwinger-Dyson
analysis (black), the all-orders beta function (red) and the 2-loop order beta function (green). 
The red diamond, blue circle and the black squares  indicate the UV-complete theory proposed in Refs.~\cite{Barnard:2013zea,Ferretti:2013kya,Ferretti:2014qta,Cacciapaglia:2019dsq}, in the CHM context.
The  $Sp(4)$ theory with $N_f=2$ and $n_f=3$ is expected to lie near the sill of the conformal window, 
which motivates further dedicated studies with non-perturbative lattice methods. 
As mentioned at the beginning of this section, the top partial compositeness mechanism is 
most effective with large anomalous dimensions 
of the chimera baryons. 
So far, this has only been estimated at the one-loop order in $\alpha$ in standard perturbation 
theory~\cite{BuarqueFranzosi:2019eee}.

\subsection{Low-energy EFT }
\label{sec:eft}

We focus here on the 
 flavored mesons. The PNGBs have masses that are expected to be suppressed, 
in respect to those of other  mesons transforming non-trivially 
 under the action of the unbroken $Sp(4)\times SO(6)$ symmetry.
 At least in principle, if the fermion masses are small with respect to the dynamical scale of the theory,
 this would create a hierarchy of scales in the spectrum, with the sole PNGBs being important 
in long-distance observables.
This (little) hierarchy is ultimately what drives interest in applications to CHMs.
 Generalising the chiral Lagrangian of QCD, 
one can write an EFT that captures
 the long-distance dynamics 
 within a weakly-coupled field theory description, by retaining only the fields associated with the PNGBs.
Following the notation in  Refs.~\cite{Bennett:2017kga,Bennett:2019jzz,Bennett:2019cxd,Bennett:2022yfa} 
(and references therein),
we recollect here the main properties of this EFT, and of its extension to include the lightest 
spin-1 flavored mesons.

We start by defining the relevant notation and conventions,
which for the most part follow Refs.~\cite{Bennett:2019cxd,Bennett:2022yfa}.  
An orthonormalised basis for the
 15 generators $\tilde{T}^A$ of the global $SU(4)$ can be chosen so that 
$A=1\,,\,\cdots\,,\,5$ denotes the broken generators and ${A}=6\,,\,\cdots\,,\,15$  the unbroken ones.
They obey the following relations:
\beqs
\label{Eq:broken}
\tilde{\Omega} \tilde{T}^A-\tilde{T}^{A\,\mathrm{T}} \tilde{\Omega} &=&0\,,
~~~~{\rm for}~~A=1\,,\,\cdots\,,\,5\,,\\
\label{Eq:unbroken}
\tilde{\Omega} \tilde{T}^A+\tilde{T}^{A\,\mathrm{T}} \tilde{\Omega} &=&0\,,
~~~~{\rm for}~~A=6\,,\,\cdots\,,\,15\,.
\eeqs
The same applies for the 35 generators $t^B$ of $SU(6)$, which we split in
$B=1\,,\,\cdots\,,\,20$ for the broken ones and  ${B}=21\,,\,\cdots\,,\,35$ for the unbroken ones. 
They satisfy the relations:
\beqs
\omega t^B-t^{B\,\mathrm{T}} \omega &=&0\,,
~~~~{\rm for}~~B=1\,,\,\cdots\,,\,20\,,\\
~~~~\omega t^B+t^{B\,\mathrm{T}} \omega &=&0\,,
~~~~{\rm for}~~B=21\,,\,\cdots\,,\,35\,.
\eeqs

We introduce two non-linear sigma-model fields. 
The matrix-valued $\Sigma_6$ transforms in the same way as the bilinear operator 
in the underlying dynamics
$\Omega_{ab}q^{M\,a\,T}\tilde{C} q^{N\,b}$, in the antisymmetric
representation of the global  $SU(4)$.
Namely, for any $\tilde{U} \in SU(4)$, $\Sigma_6 \rightarrow \tilde{U} \Sigma_6 \tilde{U}^{\mathrm{T}}$.
 $\Sigma_{21}$ has the quantum numbers 
of $-\Omega_{ab}\Omega_{cd}\psi^{m\,ac\,T}\tilde{C} \psi^{n\,bd}$,
and transforms in the symmetric representation of the $SU(6)$ global symmetry:
$\Sigma_{21} \rightarrow u \Sigma_{21} u^{\mathrm{T}}$ for any $u \in SU(6)$.
In the vacuum, the 2-index antisymmetric representation of $SU(4)$
decomposes as $6=1\oplus 5$ of the unbroken $Sp(4)$,
and the 2-index, symmetric representation of $SU(6)$ as $21=1\oplus 20^{\prime}$ of $SO(6)$.
We parametrise the non-linear sigma-model fields $\Sigma_6$ and $\Sigma_{21}$ 
in terms of
the PNGB fields  $\pi_5$ and $\pi_{20}$ as
\beqs
\Sigma_6&\equiv&e^{\frac{i \pi_5}{f_5}}\tilde{\Omega}e^{\frac{i \pi_5^{\mathrm{T}}}{f_5}}=
e^{\frac{2i \pi_5}{f_5}}\tilde{\Omega}=\tilde{\Omega} e^{\frac{2i \pi_5^{\mathrm{T}}}{f_5}}\,,\\
\Sigma_{21}&\equiv&e^{\frac{i \pi_{20}}{f_{20}}}{\omega}e^{\frac{i \pi_{20}^{\mathrm{T}}}{f_{20}}}=
e^{\frac{2i \pi_{20}}{f_{20}}}\omega=\omega e^{\frac{2i \pi_{20}^{\mathrm{T}}}{f_{20}}}\,.
\label{Eq:Sigmas}
\eeqs
The shorthands $\pi_5\equiv\sum_{A=1}^5 \pi_5(x)^A\tilde{T}^A$
and $\pi_{20}\equiv\sum_{B=1}^{20} \pi_{20}(x)^B{t}^B$ are used to lighten the notation.
The decay constants  are denoted by $f_{5}$ and $f_{20}$, and are introduced to make
the exponents dimensionless.{We choose the conventions used in
 this parameterisation and in the Lagrangian density 
so that, when applied to the QCD chiral Lagrangian, 
 the decay constant 
is $f_{\pi}\simeq 93$ MeV.} 
These relations are equivalent to imposing (and solving) the non-linear constraints 
$\Sigma_6 \Sigma_6^{\dagger}=\mathbb{I}_{4\times 4}$
and $\Sigma_{21} \Sigma_{21}^{\dagger}=\mathbb{I}_{6\times 6}$.
With the specific choice of $SU(4)$ basis in Ref.~\cite{Lee:2017uvl} (which we reproduce in Appendix~\ref{sec:algebra}),
the five PNGBs in the $SU(4)/Sp(4)$ coset are written as follows~\cite{Bennett:2019cxd}:
\beqs
\pi_5\hspace{-2pt}=\hspace{-2pt}\frac{1}{2\sqrt{2}}
\hspace{-2pt}
\left(\hspace{-4pt}
\begin{array}{cccc}
 \pi_5^{\,\,3} & \pi_5^{\,\,1}-i \pi_5^{\,\,2} & 0 & -i \pi_5^{\,\,4}+\pi_5^{\,\,5} \\
 \pi_5^{\,\,1}+i \pi_5^{\,\,2} & -\pi_5^{\,\,3} & i \pi_5^{\,\,4}-\pi_5^{\,\,5} & 0 \\
 0 & -i \pi_5^{\,\,4}-\pi_5^{\,\,5} & \pi_5^{\,\,3} & \pi_5^{\,\,1}+i \pi_5^{\,\,2} \\
 i \pi_5^{\,\,4}+\pi_5^{\,\,5} & 0 & \pi_5^{\,\,1}-i \pi_5^{\,\,2} & -\pi_5^{\,\,3}
\end{array}
\hspace{-4pt}
\right)\,,
\label{Eq:pion}
\eeqs
where we have omitted  the explicit dependence on the space-time coordinates.
A similar expression  holds for $\pi_{20}$, given a choice of basis for $SU(6)$.

The symmetry breaking effects due to the fermion masses in the
underlying dynamical theory are captured in the
EFT Lagrangian density with the introduction of (non-dynamical) spurion fields
$M_6 \equiv m^{(f)}\, \tilde{\Omega}$ and $M_{21} \equiv - m^{(as)} \,\omega$.
Formally, they transform as $M_6\rightarrow U^{\ast} M_6 U^{\dagger}$
and $M_{21}\rightarrow u^{\ast} M_{21} u^{\dagger}$ under the action of the $SU(4)\times SU(6)$ 
global symmetry transformations---but they are not fields, they are constants.
The Lagrangian density describing the PNGBs of the $SU(4)/Sp(4)$ coset is
\beqs
\label{Eq:Lpi6}
{\cal L}_6&=&\frac{f_{5}^2}{4}\Tr\left\{\frac{}{}\partial_{\mu}\Sigma_6 (\partial^{\mu}\Sigma_6 )^{\dagger}\frac{}{}\right\}
\,-\,\frac{v_6^3}{4}\Tr \left\{\frac{}{}M_6 \Sigma_6 \frac{}{}\right\}\,+\,{\rm h.c.}\\
&=&\Tr\left\{\frac{}{}\partial_{\mu}\pi_5\partial^{\mu}\pi_5\frac{}{}\right\}\,+
\,\frac{1}{3f_{5}^2}\Tr\left\{\frac{}{}
\left[\partial_{\mu}\pi_5\,,\,\pi_5\right]\left[\partial^{\mu}\pi_5\,,\,\pi_5\right]\frac{}{}\right\}\,+\,\cdots\nonumber\,+\\
&&\,+\,\frac{1}{2}\,m^{f} v_6^3\,\Tr (\Sigma_6\Sigma_6^{\dagger})  \,-\, \frac{m^{(f)} v_6^3}{f_5^2}\Tr \pi_5^2
 \,+\,\frac{m^{f} v_6^3}{3 f_5^4}\Tr \pi_5^4 \,+\,\cdots\,,
 \label{eq:hls_larg}
\eeqs
where $v_6$ parameterises the condensate, and where we include only the leading-order terms in both
the derivative and mass expansions.
The expansion for the $SU(6)/SO(6)$ PNGBs is
formally identical:
\beqs
\label{Eq:Lpi21}
{\cal L}_{21}&=&\frac{f_{20}^2}{4}\Tr\left\{\frac{}{}\partial_{\mu}\Sigma_{21} (\partial^{\mu}\Sigma_{21} )^{\dagger}\frac{}{}\right\}
\,-\,\frac{v_{21}^3}{4}\Tr \left\{\frac{}{}M_{21} \Sigma_{21} \frac{}{}\right\}\,+\,{\rm h.c.}\,.
\eeqs
Notice the opposite sign in the definition of $M_{21}$, which combines with the defining property
 $\tilde{\Omega}^2=-\mathbb{I}_{4\times 4}$ (as opposed to
$\omega^2=\mathbb{I}_{6\times 6}$), so that by just replacing the condensates
$v_6\rightarrow v_{21}$ one can recover the same expressions for the physical observables.

By perturbatively expanding the Lagrangian density, one can extract the propagator and the couplings
of the EFT, and compute observable quantities. The definitions 
and conventions are such that the Gell-Mann-Oakes-Renner (GMOR)~\cite{Gell-Mann:1968hlm}
 relation can be recovered, in both meson sectors:
\beqs
\label{eq:gmor}
m_{\pi_5}^2f_{\pi_5}^2&=&m^{(f)}\,v_6^3\,,\\
m_{\pi_{20}}^2f_{\pi_{20}}^2&=&m^{(as)}\,v_{21}^3\,,
\eeqs
 relating the pion masses $m_{\pi_5}$ and $m_{\pi_{20}}$
to the decay constants $f_{\pi_5}=f_{5}$ and $f_{\pi_{20}}=f_{{20}}$.
One can then add subleading corrections, following the same process applied for the chiral 
Lagrangian---the only technicality worth noting is that the normalisations of multi-trace deformations
depend on the dimension of the matrices, and hence on the number of fermion species.

\subsubsection{Hidden Local Symmetry }
\label{sec:hls}

Refs.~\cite{Bennett:2017kga,Bennett:2019cxd} report an extension of
 the  EFT description
 to include   the lightest 
V and AV states (corresponding to the $\rho$ and $a_1$ in 2-flavor QCD), besides the PNGBs.
This is done within the framework of Hidden Local Symmetry (HLS)~\cite{
Bando:1984ej,Casalbuoni:1985kq,Bando:1987br,Casalbuoni:1988xm,Harada:2003jx} 
(see also~\cite{Georgi:1989xy,Appelquist:1999dq,Piai:2004yb,Franzosi:2016aoo}).
We report here the basic construction, and comment about the applicability of such approach.

 \begin{figure}
\begin{center}
\begin{picture}(330,100)
\put(0,0){\includegraphics[width=.265\textwidth]{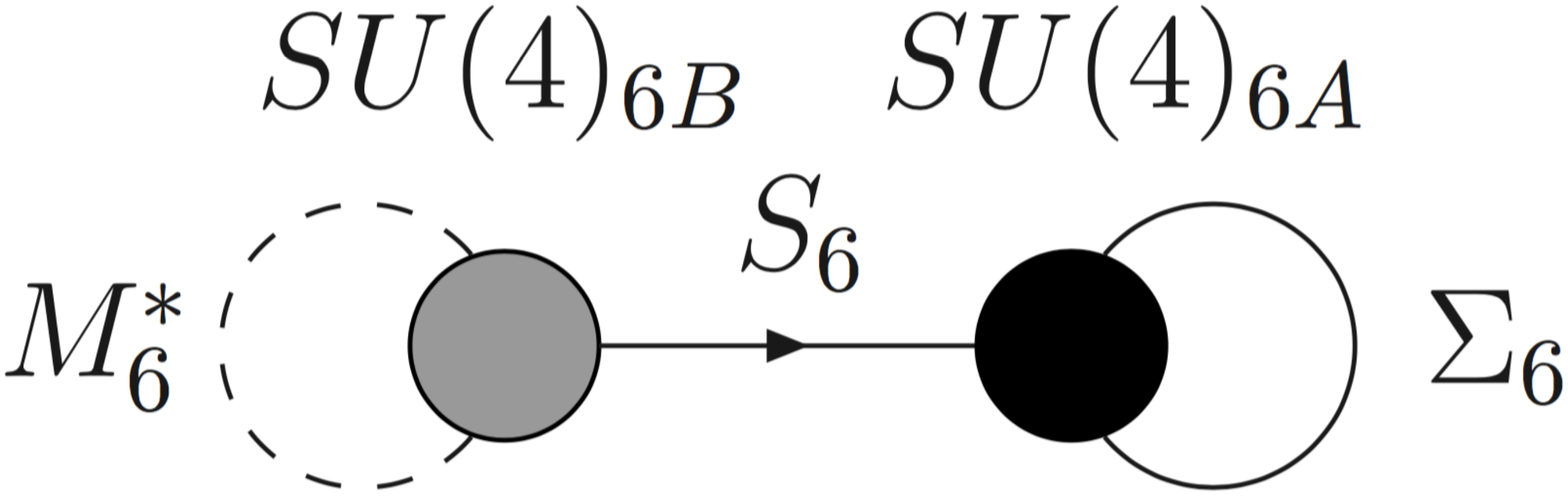}}
\put(165,0){\includegraphics[width=.265\textwidth]{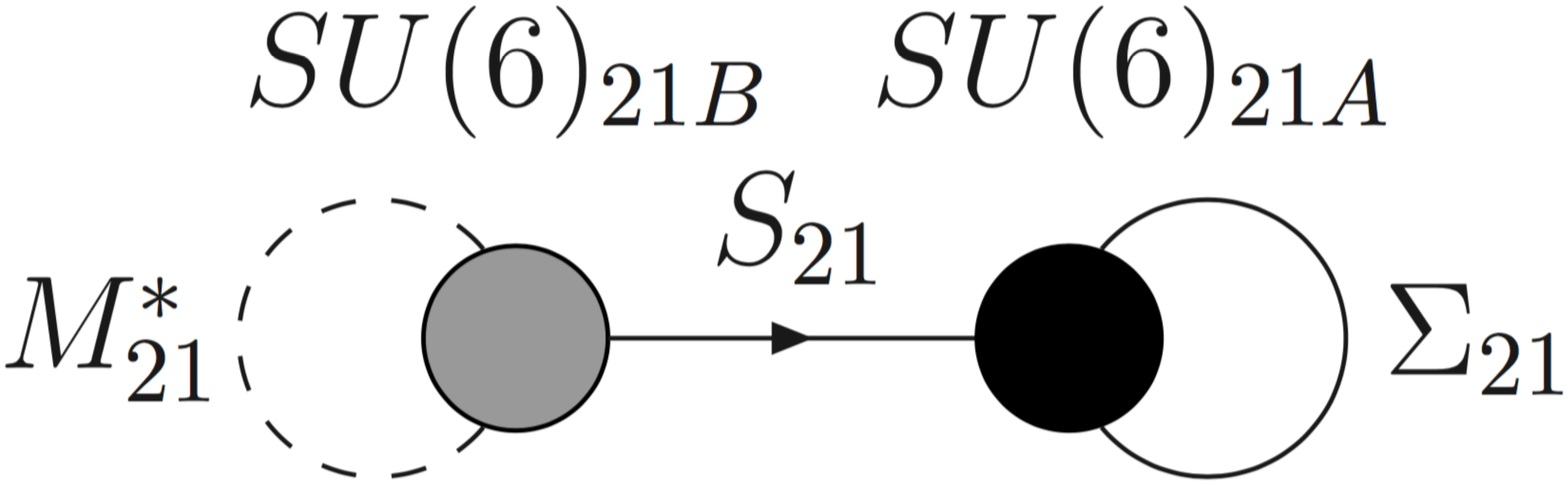}}
\end{picture}
\caption{From Ref.~\cite{Bennett:2017kga}, 
 the symmetries and their representations in the  low-energy 
EFT descriptions based on HLS. In the left panel, $SU(4)_{6A}$ is gauged, while $SU(4)_{6B}$ 
is a global symmetry and the fact that  $S_6$ and $\Sigma_6$ 
are non-trivial in the vacuum  breaks the symmetry 
to $Sp(4)$. As a result, all the gauge bosons are heavy, and in addition one has  five PNGBs. 
In the right panel, the same construction
is  applied to $SU(6)_{21B}\times SU(6)_{21 A}$ 
and to its breaking to the $SO(6)$ subgroup.}
\label{Fig:EFTHLS}
\end{center}
\end{figure}

We introduce two meson sectors that are completely independent from one another,
which is a reasonable approximation as long as one allows only for single-trace 
operators~\cite{Bennett:2017kga,Bennett:2019cxd}.
One starts by promoting the unbroken $SU(4)$ and $SU(6)$
global symmetries acting on the (f) and (as) fermions
to $SU(4)_{6B}\times SU(4)_{6A}$ and $SU(6)_{21B}\times SU(6)_{21A}$, respectively.
One then introduces two sets of sigma-model, matrix-valued fields;  $\Sigma_6$ transforms in the antisymmetric 
2-index representation of
$SU(4)_{6A}$ and $\Sigma_{21}$  in the symmetric 2-index
representation of
$SU(6)_{21A}$;
$S_6$ transforms in the $(4,\bar{4})$ bifundamental representation 
 of $SU(4)_{6B}\times SU(4)_{6A}$, and
 $S_{21}$  in the $(6,\bar{6})$ of
  $SU(6)_{21B}\times SU(6)_{21A}$.
Hence, the transformation rules are as follows:
\beqs
S_6 &\rightarrow& U_{6B} S_6 U_{6A}^{\dagger} \,,~~~~~~~~
\Sigma_6 \,\rightarrow\, U_{6A} \Sigma_{6} U_{6A}^{\mathrm{T}}
\,,\\
S_{21} &\rightarrow& U_{21B} S_{21} U_{21A}^{\dagger} \,,~~~~~~~~
\Sigma_{21} \,\rightarrow\, U_{21A} \Sigma_{21} U_{21A}^{\mathrm{T}}
\,,
\eeqs
where $U_{6A}$ and $U_{6B}$ are group elements  of $SU(4)_{6A}$ and $SU(4)_{6B}$, respectively,
 while $U_{21A}$ and $U_{21B}$ are in $SU(4)_{21A}$ and $SU(4)_{21B}$.
These fields are subject to  the nonlinear constraints
$\Sigma_6\,\Sigma_6^{\dagger} = \mathbb{I}_{4\times 4} = S_6\,S_6^{\dagger}$,
which are solved by parameterising $S_6 =e^{\frac{2i \sigma_6}{F}}$ 
and $\Sigma_6=e^{\frac{2i \pi_6}{f}}\tilde{\Omega}=\tilde{\Omega} e^{\frac{2i \pi_6^{\mathrm{T}}}{f}}$.
Analogous expressions apply to the $SU(6)$ sector.
This process  yields a parametrisation for the exactly massless Nambu-Goldstone bosons describing the cosets
$SU(4)_{6B}\times SU(4)_{6A}/Sp(4)$ and $SU(6)_{21B}\times SU(6)_{21A}/SO(6)$, respectively.

As a next step  one gauges (weakly) the $SU(4)_{6A}$ and $SU(6)_{21A}$ symmetries,
by introducing the appropriate gauge fields, covariant derivatives, field-strength tensors,
and gauge couplings $g_6$ and $g_{21}$.
The  Higgs mechanism turns $15+35$ of the exact  Nambu-Goldstone bosons 
into  the longitudinal components of the resulting massive vectors, which have the quantum numbers of the 
states identified with the $\rho$ and $a_1$ particles in QCD.
In order for the remaining $5+20$ pseudoscalars to acquire a physical mass 
 as PNGBs do, one must add sources of explicit symmetry breaking.
This is done by writing the 
$M_6$ and $M_{21}$ matrices as spurions that,
under the action of the global $SU(4)_{6B}$ and $SU(6)_{21B}$,
transform as follows:
  $M_6^{\ast} \,\rightarrow\, U_{6B} M_{6}^{\ast} U_{6B}^{\mathrm{T}}$
  and 
    $M_{21}^{\ast} \,\rightarrow\, U_{21B} M_{21}^{\ast} U_{21B}^{\mathrm{T}}$.
One then uses $\Sigma_6$,  $\Sigma_{21}$,  and their derivatives,  as well as $M_6$ and $M_{21}$, to build 
all possible operators allowed by the symmetries, organises them
as an expansion in derivatives (momenta $p^2$) and explicit mass terms, 
and writes a Lagrangian density that includes all
such operators up to a given order in the expansion.
We also restrict attention to operators that can be written as single traces, as repeatedly anticipated.

The Lagrangian density for the $SU(4)/Sp(4)$ mesons is  Eq.~(2.16) of Ref.~\cite{Bennett:2019cxd}:
\beqs
{\cal L}_6&=& 
-\frac{1}{2}\Tr\,A_{\,\mu\nu}A^{\mu\nu}
-\frac{\kappa}{2}\Tr\left\{ A_{\mu\nu} \Sigma (A^{\mu\nu})^{\mathrm{T}} \Sigma^{\ast}\right\}\nonumber+\\
&&+\frac{f^2}{4}\Tr\left\{\frac{}{}D_{\mu}\Sigma\,(D^{\mu}\Sigma)^{\dagger}\right\}\,+\,\frac{F^2}{4}\Tr\left\{\frac{}{}D_{\mu}S\,(D^{\mu}S)^{\dagger}\right\}\nonumber+\\
&&+b \frac{f^2}{4}\Tr\left\{D_{\mu}(S\Sigma )\left(D^{\mu}(S\Sigma)\right)^{\dagger}\right\}\,+\,
c\frac{f^2}{4} \Tr\left\{D_{\mu}(S\Sigma S^{\mathrm{T}})\left(D^{\mu}(S\Sigma S^{\mathrm{T}})\right)^{\dagger}\right\}\,+\nonumber\\
&&-\frac{v^3}{8}\Tr\left\{\frac{}{}M\, S\, \Sigma\,S^{\mathrm{T}} \right\}\,+\,{\rm h.c.} +\label{Eq:L6}\\
&&- \frac{v_1}{4} \Tr\left\{\frac{}{} M\, (D_{\mu} S)\, \Sigma \, (D^{\mu} S)^{\mathrm{T}} \right\}\,
 -\frac{v_2}{4} \Tr\left\{\frac{}{} M\, S\,(D_{\mu}  \Sigma) \, (D^{\mu} S)^{\mathrm{T}} \right\}\,+\,{\rm h.c.}\,+\nonumber\\
 &&
 -\frac{y_3}{8}\Tr\left\{A_{\mu\nu}\Sigma\left[(A^{\mu\nu})^{\mathrm{T}}S^{\mathrm{T}} M S-S^{\mathrm{T}} M S A^{\mu\nu}\right]\right\}\,+\,{\rm h.c.}\nonumber+\\
  &&
 -\frac{y_4}{8}\Tr\left\{A_{\mu\nu}\Sigma\left[(A^{\mu\nu})^{\mathrm{T}}S^{\mathrm{T}} M S+S^{\mathrm{T}} M S A^{\mu\nu}\right]\right\}\,+\,{\rm h.c.}\,\nonumber+
 \\
&&
+\frac{v_5^2}{32}\Tr \left\{\frac{}{}M S \Sigma S^{\mathrm{T}} M S \Sigma S^{\mathrm{T}}\frac{}{}\right\}\,+\, {\rm h.c.}\,.\nonumber
\eeqs
To lighten the notation, we suppressed the subscript ``$6$" on all fields and  parameters, and
multi-trace operators have been omitted~\cite{Bennett:2017kga}.
The covariant derivatives contain the parameter $g_{6}$, that controls the 
strength of the coupling of the spin-1 states.
We write
\beqs
D_\mu S &\equiv & \partial_\mu S - i  S g_{6}   A_\mu\,
\eeqs
and
\beqs
D_\mu \Sigma &\equiv & \partial_\mu \Sigma + i \left[
(g_{6} A_\mu) \Sigma + \Sigma (g_{6} A_\mu)^T
\right]\,.
\eeqs

The Lagrangian density in Eq.~(\ref{Eq:L6}) can be adapted to the $SU(6)/SO(6)$ sector. 
One replaces $\Sigma_6$ by $\Sigma_{21}$,  $S_6$ by $S_{21}$, $M_6$ by 
$M_{21}\equiv -m \omega$,  $A^A_{6\,\mu}$ by $A^A_{21\,\mu}$ ,
$g_6$ by $g_{21}$, and furthermore changes the sign of the second term in the first line
  $\kappa_{6} \rightarrow -{\kappa}_{21}$.
With these conventions, 
masses and decay constants are given by the same relations
as in Ref.~\cite{Bennett:2017kga}, to which we refer the reader for further details.

One has to adopt some caution in the way one uses Eq.~(\ref{Eq:L6}).
In particular, one has to ensure that the parameters are all within a range of values 
such that the EFT is weakly coupled.
The main source of concern here is the size of the gauge couplings $g_{6,21}$, and the related
effective couplings ( $g_{\rho\pi\pi}$).
The mass of the vector mesons
can be estimated as $M_{\rho}^2 \sim {\cal O}(\frac{1}{4}g_{6,21}^2f_{6,21}^2)$, up to a
 complicated functional dependence on all the parameters~\cite{Bennett:2017kga}.
Hence, if in comparing to lattice data one finds that $M_{\rho}\gg f_{\pi}$,
this might imply that the coupling is not small---barring  the possibility of cancellations and fine-tuning.
In practice, for the $Sp(4)$ theory, as in 2-flavor QCD, real data seem to sit half-way between
the extremes of trustable perturbative and uncontrolled strong-coupling regimes:
the self-couplings are perturbative,
but not small enough that one can do precision measurements and calculations
 with the tree-level Lagrangian and its couplings.
On the other hand, this is a non-renormalisable EFT, in which the number of independent couplings proliferates
going to higher-order in the loop expansion, making unappealing the in principle viable programme of systematic expansion
beyond the leading order.
Nevertheless, the organisational principles, order-of-magnitude estimates,
 and general lessons associated with  Eq.~(\ref{Eq:L6}) have general value.

\subsection{Phenomenological applications}
\label{sec:pheno}

In this subsection we consider three examples of applications of the $Sp(2N)$ gauge theories of interest: a model
of composite Higgs, a realisation of top partial compositeness,
and two opportunities arising in the context of  strongly interacting dark matter.
For the most part, we  make explicit reference to
 the $SU(4)/Sp(4)$ model with the field content discussed in Refs.~\cite{Barnard:2013zea,Ferretti:2013kya}, 
 but, where appropriate we also 
 highlight considerations that have more general validity, applicable to larger classes of models.

\subsubsection{Composite Higgs }
\label{sec:comphiggs}

We start by recalling the basic properties of the standard model, and the motivations for compositeness.
For concreteness,  we 
 postulate the existence of three right-handed neutrinos, singlet under the SM gauge group.
All the fermions are then Dirac particles, and can be classified in terms of their quantum numbers
under the symmetry
\beqs
SU(3)_c \times SU(2)_L \times SU(2)_R \times U(1)_{B-L}\,.
\eeqs
The $SU(3)_c$ symmetry is gauged, with coupling $g_s$, and describes the strong nuclear forces.
The $SU(2)_L$ and  the hypercharge subgroup $U(1)_Y \subset SU(2)_R \times U(1)_{B-L}$
are also gauged, with couplings $g_L$ and $g_Y$, respectively, 
in the electroweak (EW) theory.
The hypercharge generator is
$Y=\tilde{T}^3_R+\frac{1}{2}(B-L)$, 
where $\tilde{T}^3_R$ is the diagonal generator of $SU(2)_R$, and  $B-L$ is anomaly free;
quarks have baryon number  $B=+\frac{1}{3}$ and no lepton number,
while leptons have no baryon number, and lepton number $L=+1$.
The field content of the standard model consists of three copies (families) of (Dirac) 
quarks transforming as $(3,2,2,1/3)$
of $SU(3)_c \times SU(2)_L \times SU(2)_R \times U(1)_{B-L}$, and three families of leptons 
transforming as $(1,2,2,-1).$
The chiral symmetry acting on the left-handed and right-handed projections of the fermions
admits the local isomorphism
$SU(2)_L\times SU(2)_R \sim SO(4)_{EW}$,
which plays an important role in the following.  

In the minimal version of the standard model, electroweak symmetry breaking (EWSB) is implemented by adding
to the field content a scalar (Higgs)
 transforming as $\Phi \sim (1,2,2,0)$. The Lagrangian density for $\Phi$ consists of its
kinetic term, with appropriate covariant derivatives, coupling it to the $SU(2)_L\times U(1)_Y$ gauge fields,
a renormalizable potential with $SO(4)_{EW}$ global symmetry, and Yukawa couplings to the fermions
which break explicitly the $SU(2)_R$ global symmetry (as does the hypercharge coupling).
It is customary to write $\Phi$ in terms of  a doublet of complex scalars transforming as $H\sim (1,2,+1/2)$ under the 
SM gauge group $SU(3)_c\times SU(2)_L\times U(1)_Y$, 
and define the conjugate field $\tilde{H} \equiv i\tau^2 H^{\ast}$, so that the $2\times 2$ complex matrix
\beqs
\Phi&\equiv&\left(\frac{}{}\tilde{H}\,,\,H\frac{}{}\right)
\eeqs
transforms under the action of $SU(2)_L\times SU(2)_R$ as
$
\Phi\,\rightarrow\,U_L \Phi U_R^{\dagger}
$.  It is at times useful also to write the Higgs fields in real components
$h\equiv (h_1,\,h_2,\,h_3,\,h_4)$, defined by
 \beqs
 H&=&\frac{1}{\sqrt{2}}\left(\begin{array}{c} h_3+i\,h_4 \cr h_1+ i\,h_2 \end{array}\right)\,.
 \eeqs
The potential, ${\cal V}$, can be written as 
\beqs
{\cal V}&=&\lambda\left(H^{\dagger}H \,-\,\frac{v_W^2}{2}\right)^2\,=\,
\frac{\lambda}{4}\left(\Tr\,\Phi^{\dagger}\Phi \,-\,{v_W^2}\right)^2
\,=\, \frac{\lambda}{4}\left(h^T\,h\, -\,{v_W^2}\right)^2\,.
\eeqs
 The $SO(4)_{EW}$ global symmetry of ${\cal V}$ is manifest in the last expression.
The minimisation of the potential yields  a vacuum expectation value (VEV) for the scalar, 
that induces EWSB. In turn,  because of the coupling of $H$ to gauge bosons and fermions, it also 
provides them with a  mass.
With these conventions,  the  electroweak VEV $v_W$ is related
 to the Fermi constant $G_F$ by
 $v_W\equiv \frac{1}{\sqrt{\sqrt{2}\,G_F}}\sim 246~{\rm GeV}$,
the mass of the Higgs boson is given by the relation $m_h^2=2  \lambda v_W^2
 \simeq (125~{\rm GeV})^2$~\cite{Aad:2012tfa,Chatrchyan:2012xdj},
the $W^{\pm}$ bosons have mass $M_W =\frac{1}{2} g_L v_W \simeq 80~{\rm GeV}$,
and for the $Z$ bosons $M_Z^2 =\frac{1}{4} (g_L^2+g_Y^2) v_W^2 \simeq (91~{\rm GeV})^2$.

The standard model has passed successfully  countless experimental tests.
Yet, it is not a complete theory: several of its interactions (the Yukawa couplings, the $U(1)_Y$
gauge coupling, and the scalar self-coupling $\lambda$) are not asymptotically free,
and most likely require ultraviolet (UV) completion above some new physics scale $\Lambda$.
One way to show how this may lead to a general problem is by considering quantum corrections
on the Higgs potential.
Following  Coleman and Weinberg in Ref.~\cite{Coleman:1973jx}, the (divergent part of the)
1-loop effective potential computed (perturbatively)  in the
external field formalism, in the presence of a hard cutoff scale $\Lambda$, can be written as follows:
\beqs
\delta {\cal V}& = & \frac{\Lambda^2}{32 \pi^2} {\cal ST}r\,{\cal M}^2 \,
+\,\frac{1}{64\pi^2} {\cal ST}r\,\left({\cal{M}}^4\ln\left(\frac{{\cal M}^2}{\Lambda^2}\right)\right)\,,
\label{Eq:CW}
\eeqs
where $\cal{M}$ is the mass matrix of all the fields in the classical external field background,
and ${\cal ST}r$ denotes the super-trace, a trace in which bosons enter with positive sign, 
while fermions count with negative sign.
For example, the contributions of the top quark, that has mass $m_t \simeq (173~~{\rm GeV})^2$, and
the $W^{\pm}$,
 $Z$,
and  Higgs boson, 
 to the quadratically divergent part of this potential
are estimated to be~\cite{Einhorn:1992um}
\beqs
\delta {\cal V}& = &\frac{\Lambda^2}{32\pi^2}
\left[\frac{}{}{3}\left({2M_W^2}{}+{M_Z^2}{}\right)+{3m_h^2}{}
-{12m_t^2}\frac{}{}\right] \frac{2 H^{\dagger}H}{v_W^2}\,.
\eeqs
If $\Lambda \gg {\cal O}({\rm TeV})$, for example
if $\Lambda \sim {\cal O}(M_P)$, with $M_P$ the Planck scale characterising quantum gravity,
then the experimental value of the Higgs
boson mass is reproduced only at the price of fine-tuning 
 loop effects against appropriately chosen counter-terms.

This fine-tuning phenomenon is usually referred to as (big) hierarchy problem.
It can avoided by replacing the Higgs sector with a new strongly-coupled dynamical theory.
In the potential in Eq.~(\ref{Eq:CW}), $\Lambda$ is the 
characteristic scale of the new physics sector, above which new particles and interactions appear.
Given that the Higgs sector has the same $SU(2)_L\times SU(2)_R\sim SO(4)_{EW}\rightarrow SU(2)_V\sim SO(3)$
global non-Abelian symmetry breaking pattern as in 2-flavor QCD, it is intuitive to model the new sector as 
a generalisation of  QCD itself. The new gauge theory, with appropriate matter fields, 
is asymptotically free in the far UV, but at scale $\Lambda$  strongly coupling
induces the formation of composite condensates, EWSB appears, and the theory confines. This
idea predates most of modern particle physics, and goes under the name of technicolor (TC).
We are not going to explore further this topic, but rather refer the reader to the original 
papers on technicolor~\cite{Weinberg:1975gm,Susskind:1978ms}, 
walking technicolor (WTC)~\cite{Holdom:1984sk,Yamawaki:1985zg,Appelquist:1986an}, 
and extended technicolor (ETC)~\cite{Dimopoulos:1979es,Eichten:1979ah},
 as well as to more recent reviews in Refs.~\cite{Chivukula:2000mb,Lane:2002wv,
Hill:2002ap,Martin:2008cd,Sannino:2009za,Piai:2010ma}.
To the present purposes, it suffices to notice that in its original, QCD-like formulation,
the spectrum of TC would not contain a light state identifiable with the Higgs boson.
Furthermore, generic TC models would struggle to satisfy indirect bounds from electroweak precision 
physics, encoded in the $S$ and $T$ parameters of Peskin and Takeuchi~\cite{Peskin:1991sw},
and their generalisations as in Ref.~\cite{Barbieri:2004qk}, or in the subleading terms of the
 electroweak chiral Lagrangian~\cite{Appelquist:1980vg,Longhitano:1980iz,
Longhitano:1980tm,Appelquist:1993ka,Appelquist:1994qz}.

The solution provided by CHMs~\cite{Kaplan:1983fs,
Georgi:1984af,Dugan:1984hq}  relies on the engineering of
a two-stage symmetry-breaking pattern, which introduces 
a little hierarchy of scales.
At the strong coupling scale $\Lambda$, an approximate global symmetry $G$ is spontaneously broken
to a subgroup $H$. While all other composite  particles have mass  ${\cal O}(\Lambda)$,
the PNGBs have suppressed mass, and decay constant $f_{\pi}$.
The PNGBs admit an EFT description in terms of weakly coupled fields, and 
by  embedding  the SM gauge group into $G$, and coupling the PNGBs 
to the SM fermions, one can introduce a perturbative instability,
which triggers EWSB in the vacuum. A hierarchy $v_W\ll f_{\pi}$ emerges, between $f_{\pi}$,
which originates in the strong-coupling dynamics, and the 
electroweak VEV, which has a weak-coupling origin,
as a destabilising perturbation of the vacuum.
This  vacuum misalignment phenomenon relies on a special modification
of the vacuum alignment arguments ubiquitous in the theory of phase transitions (and exploited in Ref.~\cite{Peskin:1980gc}).

Let us now return to  the $SU(4)/Sp(4)$ model of Refs.~\cite{Barnard:2013zea,Ferretti:2013kya}.
We have already established that with $N_f=2$ Dirac fermions transforming as the fundamental representation of $Sp(2N)$,
the strong dynamics gives rise to the spontaneous breaking of  $SU(4)$ to $Sp(4)$. 
Working in the basis, in flavor space, in which $\tilde{\Omega}=\Omega$ in 
Eq.~(\ref{Eq:symplecticmatrix}), in Sect.~\ref{sec:eft} we chose a parametrisation of the five PNGBs, 
in Eq.~(\ref{Eq:pion}), and we will present a choice of generators for $SU(4)$ 
in Eqs.~(\ref{Eq:basisSU(4)}) of Appendix~\ref{sec:algebra}.
We  now discuss the embedding of 
$SO(4)_{EW}$.\footnote{By preserving the whole $SO(4)_{EW}$, the model preserves custodial symmetry, 
 suppressing new physics contributions to the $T$ parameter~\cite{Peskin:1991sw}.}
In Appendix~\ref{sec:algebra} we define a first embedding $SO(4)_0$, in Eqs.~\eqref{Eq:SU2L} and~(\ref{Eq:SU2R}),
so that the vacuum $\Sigma_{6}\propto \tilde{\Omega}$ leaves $SO(4)_0$ unaffected.
We then define a second embedding in Eqs.~(\ref{Eq:SU2LTC}) and~(\ref{Eq:SU2RTC}), denoted as $SO(4)_{TC}$.
$SO(4)_{TC}$ is broken at scale $f_{\pi}$ to the $SO(3)_{TC}\sim SU(2)_{V,TV}$ subgroup;
 this is the embedding one would use in a 
traditional technicolor theory, in which the strong coupling and EWSB scale coincide.
In practice, by doing so one establishes that the EFT field $\Phi$ describes light particles that originate
in the fundamental theory as composite excitations sourced by the operator $\overline{Q_R} Q_L$,
with $Q_{L,R}$ the chiral projections of the two Dirac fields transforming in the fundamental representation of $Sp(2N)$.

We write the generators of $SO(4)_{EW}$ as a linear combination of the two:
\beqs
\tilde{T}^i_{\chi,EW}&\equiv& \sin(\theta) \,\tilde{T}^i_{\chi,TC}\,+\,\cos(\theta) \tilde{T}^{i}_{\chi,0}\,,
\eeqs
for $i=1,\,2,\,3$ and $\chi=L,\,R$.
The vacuum (mis-)alignment angle, $\theta\equiv  \frac{v_W}{f_{\pi}}$, is determined 
dynamically by the interplay between symmetry-breaking terms 
that stabilise the EW vacuum, and hence favor $\theta=0$ and $SO(4)_{EW}\rightarrow SO(4)_0$,
and symmetry-breaking interactions that destabilise it, and trigger EWSB.
A nice discussion of the typical, generic potential one expects to arise from combining such symmetry breaking terms
(which originate from the masses of the fundamental fermions,  the gauging of the $SU(2)_L\times U(1)_Y$ 
subgroup, and the coupling to the SM fermions) can be found in Eq.~(125) of Ref.~\cite{Contino:2010rs}, which 
  studies  the potential for $|h|\equiv\sqrt{h^Th}$,
\beqs
V_{\rm eff} &=& \alpha \cos\left(\frac{|h|}{f_{\pi}}\right) \,-\,\beta \sin^2\left(\frac{|h|}{f_{\pi}}\right)\,.
\eeqs
The coefficients $\alpha$ and $\beta$ are model dependent, and determined by the non-trivial interplay between 
weakly coupled effects encoded in the EFT,
and strongly coupled effects that can in principle be extracted from matrix elements in the strongly coupled sector.

Other model-dependent quantities are the number, masses, and couplings of the additional  PNGBs, besides
 $H$; models with $SU(4)/Sp(4)$ coset predict an additional singlet, while other CHMs
have richer spectra.
Precision electroweak (and Higgs boson) observables are affected by the additional light scalars, 
and the spin-1 bound states. 
Except for  the PNGBs, bound states have masses of order the scale $\Lambda$, 
and carry EW quantum numbers; they can be detected in direct searches at colliders.
As anticipated, we do not discuss in this review  the singlet sector,
though it may have important phenomenological implications both for collider and 
dark matter physics---for broader phenomenological 
considerations see Refs.~\cite{Panico:2015jxa,Witzel:2019jbe,Cacciapaglia:2020kgq,
Ferretti:2016upr,Cacciapaglia:2019bqz} and references therein.
The feasibility of direct and indirect searches depends on dynamical information from the 
underlying microscopic theory, which requires non-perturbative methods.
Lattice studies 
can measure, in increasing order of difficulty: masses of some bound states (relevant to direct searches), 
decay constants (entering for instance precision electroweak observables) and 
other matrix elements (relevant for example for vacuum misalignment),
and couplings between bound states (determining width, production and decay rates of new particles).

\subsubsection{Top partial compositeness }
\label{sec:toppartner}

This subsection is devoted to the mechanism producing the mass of the SM fermions.
We start with the standard model, in which the Yukawa couplings take the form
\beqs
{\cal L}_Y&=&
-Y^{(\mathfrak{u})}_{ij}\overline{\mathfrak{q}^i_L}\,\tilde{H}\, \mathfrak{u}^j_R
-Y^{(\mathfrak{d})}_{ij}\overline{\mathfrak{q}^i_L}\,H \, \mathfrak{d}^j_R
-Y^{(\mathfrak{n})}_{ij}\overline{\mathfrak{\ell}^i_L}\,\tilde{H}\, \mathfrak{n}^j_R
-Y^{(\mathfrak{e})}_{ij}\overline{\mathfrak{\ell}^i_L}\,H \,\mathfrak{e}^j_R\,+\,{\rm h.c.}\,,
\label{Eq:Y}
\eeqs
where $\mathfrak{q}^i_L\equiv \left(\begin{array}{c} \mathfrak{u}^i \cr \mathfrak{d}^i \end{array}\right)_L$ and 
$\mathfrak{\ell}^i_L\equiv \left(\begin{array}{c} \mathfrak{n}^i \cr \mathfrak{e}^i \end{array}\right)_L$ are the quark and lepton 
$SU(2)_L$ doublets, respectively, while $\mathfrak{u}^i_R$, $\mathfrak{d}^i_R$, $\mathfrak{n}^i_R$, and $\mathfrak{e}^i_R$
are the right-handed up- and down-type quark,   neutrino, and charged lepton $SU(2)_L$ singlets. 
The index $i=1,\,2,\,3$ labels the three families, and the $3\times 3$ complex Yukawa matrices $Y^{(\mathfrak{u})}_{ij}$,
$Y^{(\mathfrak{d})}_{ij}$,
$Y^{(\mathfrak{n})}_{ij}$, and 
$Y^{(\mathfrak{e})}_{ij}$ are proportional to the mass matrices in the EWSB vacuum,
via the relations $m^{(\phi)}_{ij}=Y^{(\phi)}_{ij}\frac{v_W}{\sqrt{2}}$, for $\phi=
\mathfrak{u},\mathfrak{d},\mathfrak{n},\mathfrak{e}$.
Not only does Eq.~(\ref{Eq:Y}) provide masses for all the fermions, after EWSB, but 
it also automatically implements the Glashow-Iliopoulos-Maiani
(GIM) mechanism, suppressing Flavor Changing Neutral Current (FCNC) processes~\cite{Glashow:1970gm}.

Let us now discuss what changes if  the Higgs field $H$ is composite.
Broadly speaking, there are two ways to couple elementary fermions to a strongly-coupled vector-like theory
that yields EWSB---we find it useful to refer the reader
 to the discussions in Ref.~\cite{Chacko:2012sy}, although a vast literature on the subject predates it.
For concreteness, we refer to the $Sp(4)$ gauge theory
in Ref.~\cite{Barnard:2013zea,Ferretti:2013kya}, using the  conventions introduced in Sect.~\ref{sec:sp2n}.

 The first possibility arises because, in the EFT, the quantum numbers of $\Phi$ 
 do not depend on whether it is an elementary SM field, or  describes a mesonic composite state.
Equation~(\ref{Eq:Y})  originates 
in the coupling to the meson operator  $\overline{Q_R}Q_L$ (transforms as $\Phi$):
 \beqs
 {\cal L}_{\rm ETC}&=&-\frac{1}{\Lambda_{ETC}^2}
 \frac{}{}\left(\frac{}{}
  c^{(\mathfrak{u})\ast}_{ij} \overline{\mathfrak{u}^j_R}
 \,,\,
  c^{(\mathfrak{d})\ast}_{ij} \overline{\mathfrak{d}^j_R}
  \frac{}{}\right)\,
Q_L\frac{}{}
\,
 \frac{}{}
   \overline{Q_R}\,{\mathfrak{q}_L^i}\,
 \frac{}{}
 \,+\,{\rm h.c.}\,,
 \eeqs
 and a similar term for the leptons.
 The dimensionless  parameters $c^{(\phi)}_{ij}$ are  proto-Yukawa couplings.
These interactions involve  four-fermion operators, 
 have engineering dimension $6$, 
 spoil asymptotic freedom, and
force us to introduce the new physics scale $\Lambda_{ETC}$.
This is the construction adopted in the ETC literature, and above  $\Lambda_{ETC}$
 a further, more fundamental theory unifies family/generation 
physics with the strong dynamics, in such a way that 
 all the $c^{(\phi)}_{ij}$ have  a dynamical origin.
New physics also produces  other four-fermion interactions, 
involving only SM fermions, which spoils the GIM mechanism,
 so that the experimentally verified suppression of FCNC processes 
 requires that $\Lambda_{ETC} \gg \Lambda$.
An example of valiant effort at producing a semi-realistic implementation of this challenging
model-building programme
can be found in Refs.~\cite{Appelquist:1993sg,Appelquist:2002me,Appelquist:2003uu,
Appelquist:2003hn, Appelquist:2004mn,Appelquist:2004es, Appelquist:2004ai}.

The magnitude of the
Yukawa couplings one is likely to generate in this fashion may be too small.
In matching to the low energy EFT at the scale $\Lambda$, one  replaces
$\overline{Q_R}Q_L \rightarrow 4\pi  \kappa \Phi  \Lambda^2$,\footnote{The important difference 
between CHMs and TC is that $v_W\ll f_{\pi}$, so that $\Lambda$ can naturally be larger than the TeV scale.}
so that $Y^{(\phi)}_{ij} = 4 \pi \kappa \frac{\Lambda^2}{\Lambda_{ETC}^2}\,c^{(\phi)}_{ij}$
 is suppressed by the ratio $\Lambda^2/\Lambda_{ETC}^2\ll 1$.
The top quark Yukawa coupling $y_t\equiv \sqrt{2}\,m_t/v_W\sim 1$ is particularly problematic,
as on the basis of Naive Dimensional Analysis (NDA)~\cite{Georgi:1992dw}, one expects the strong dynamics to yield
$\kappa \sim {\cal O} (1)$. Hence, one would need an unreasonably low scale $\Lambda_{ETC} \sim 3 \Lambda$, 
in order to reproduce a large enough top mass.

If the underlying dynamics is quasi-conformal in proximity of the strong
coupling scale $\Lambda$, up to some scale $\Lambda_W$, 
and if the scaling dimension of the $\overline{Q_R}Q_L$ operator is $y<3$ in this regime, then
the constant $\kappa$ receives an anomalous enhancement 
$\eta\sim{\cal O}\left(\left(\frac{\Lambda_{W}}{\Lambda}\right)^{3-y}\right)$.
For example, if $\Lambda_W=\Lambda_{ETC}$, and $y=2$~\cite{Cohen:1988sq,Leung:1989hw},
then it might be possible to arrive at a reasonable estimate, provided $\Lambda_{ETC}/\Lambda\lsim 4\pi$.
(If $y\sim 1$ were admissible,
 concerns about the ratio $\Lambda_{ETC}/\Lambda$ would be superseded~\cite{Luty:2004ye}.)

Top  partial compositeness (TPC)~\cite{Kaplan:1991dc} is an alternative way to 
generate the top mass.
In the microscopic theory,  one couples the top fields  to strongly coupled operators, 
${\cal B}_{L,R}$ with spin-$1/2$, scaling
dimensions $\Delta_{L,R}$, and carrying appropriate quantum numbers to preserve the 
SM gauge symmetry. Schematically, one writes:
\beqs
{\cal L}_{TPC}&=&-\frac{c_L}{\Lambda_{TPC}^{\Delta_L-5/2}}\overline{\mathfrak{q}_L}\,{\cal B}_L 
-\frac{c_R}{\Lambda_{TPC}^{\Delta_R-5/2}}\overline{\mathfrak{u}_R}\,{\cal B}_R\,+\,{\rm h.c.}\,.
\label{Eq:TPC1}
\eeqs
 The scale $\Lambda_{TPC}\gg \Lambda$ is 
 introduced to compensate for the fact that  ${\cal B}_{L,R}$ are composite operators, and
 Eq.~(\ref{Eq:TPC1}) introduces higher-dimension, non-renormalisable operators.
Matching to the low energy EFT  leads to effective Yukawa couplings
of the form
\beqs
{\cal L}_{t} &=&-4 \pi {\kappa}_t c_L c_R\left(\frac{\Lambda}{\Lambda_{TPC}}\right)^{\Delta_L+\Delta_R-5}\,\,
\overline{\mathfrak{q}^3_L}\,
\tilde{H}\, \mathfrak{u}^3_R\,,
\label{Eq:Lt}
\eeqs
with $\kappa_t\sim {\cal O}(1)$, another parameter that has its origin in the strong dynamics.

Generically, $\Delta_{L,R}$ is expected to be large, suppressing 
 the top mass. For example, in the $SU(N_c)$ theory, with odd $N_c$,
 and with fermion matter fields in the fundamental representation,  the
baryons have engineering dimension $\Delta_{L,R}=\frac{3}{2}N_c$, so that
$\Delta_L+\Delta_R-5=3N_c-5 \gg 1$.
But this needs not to be so. First, if the theory is approximately scale invariant, 
the dimensions $\Delta_{L,R}$ may be smaller, thanks to non-perturbative effects.
For $\Delta_{L,R}\simeq \frac{5}{2}$,  the suppression factor in Eq.~(\ref{Eq:Lt}) would
 depend logarithmically on $\Lambda/\Lambda_{TPC}$.
Second, ${\cal B}_{L,R}$ may have a different composition, as is
 the case for the $Sp(4)$ theory with $N_f=2$ and $n_f=3$ that we introduced in 
earlier in this section~\cite{Barnard:2013zea,Ferretti:2013kya},
where ${\cal B}_{L,R}$ are  identified with the 
chimera baryons in the top part of Table~\ref{tab:chimerabaryons}.
Whether the former is also true, and under what conditions,
are highly non-trivial questions about the strong dynamics,
that future dedicated lattice studies can in principle try to answer (see also Sect.~\ref{sec:pert}).

In our prototype CHM, the presence of $n_f=3$ Dirac fermions transforming in the 2-index antisymmetric 
representation $\Psi^{j\,ab}$ introduces an $SU(6)$ global symmetry,  explicitly and 
spontaneously broken to $SO(6)$. It also defines a natural $SU(3)_L \times SU(3)_R\subset SU(6)$ 
subgroup, itself explicitly and spontaneously broken to the diagonal $SU(3)_c$. This coincides with
the SM gauge group describing strong nuclear interactions.
A specific basis of $SU(3)_c\subset SU(6)$ generators
is given  in Appendix~C of Ref.~\cite{Bennett:2019cxd}.
The traceless, diagonal generator
of $SU(6)$ that commutes with $SU(3)_c$, is also unbroken.
As explained in Ref.~\cite{Ferretti:2013kya}, a linear combination of this $U(1)_X$ generator
and of the unbroken $\tilde{T}^3_R$ yields the SM hyperchange $Y$. ($X$, appropriately normalised, is related to $B-L$.)
In the same way in which
the set $(\pi_5^1,\,\pi_5^2,\,\pi_5^4,\,\pi_5^5)$ transforms as a $4$ of $SO(4)_{EW}$, and hence we can identify it
with the Higgs field $\Phi$,  the $({\cal O}_{CB,1},\,{\cal O}_{CB,2},\,{\cal O}_{CB,4},\,{\cal O}_{CB,5})$
operators have the same transformation properties under $SU(2)_L\times SU(2)_R$. Furthermore, because of the
presence of $\Psi$ in the constituents, the chimera baryons transform as $SU(3)_c$ triplets,
and the hypercharge $Y$ is such
 to match all transformation properties of the quarks, aside from the 
 fact that the field content is vector-like, rather than chiral.
 In the literature, sometimes
 these operators are said to source the top partners.

Lattice studies of chimera baryons  in strongly coupled theories are non-trivial.
(See Ref.~\cite{Ayyar:2018zuk} for lattice  study  in a $SU(4)$ theory
 with multiple fermion representations.)
Even in ordinary QCD, the study of  the baryons is  resource intensive,
and produces noisy numerical signals.
Additional difficulties arise  with
fermions in different representations, which require developing  dedicated software, and a
 complicated scanning of the multi-dimensional parameter space of the lattice theory~\cite{Bennett:2022yfa}.
So far, $Sp(4)$ results are restricted to the masses of the lightest such states.
Measuring scaling dimensions of  chimera baryon operators is an ambitious long-term goal.

\subsubsection{Composite dark matter }
\label{sec:compDM}

We anticipated in Sect.~\ref{sec:introduction} that strongly coupled dark matter sectors have much phenomenological potential.
We extend the discussion in this subsection, focusing on two examples.
First, we follow Refs.~\cite{Hochberg:2014dra,Hochberg:2014kqa,Hochberg:2015vrg,Bernal:2017mqb,
Berlin:2018tvf};
relic cold dark matter (CDM)
made  of  self-interacting particles yields predictions for dark-matter distribution profiles
in the small scale structure of astrophysical objects---for example
the centers of  dark-matter  halos have cores with spherical symmetry~\cite{Spergel:1999mh}.
Second, we follow Refs.~\cite{Caprini:2019egz,Halverson:2020xpg,Huang:2020crf}, and discuss how 
the presence of a first-order phase transition in a dark sector
could, in principle, be detected in  experiments such a the LISA~\cite{Caprini:2019egz}, 
that are sensitive to  relic stochastic gravitational wave backgrounds.

In Ref.~\cite{Hochberg:2014dra}, SIMP models were proposed, in which 
 a strongly self-interacting dark sector is feebly coupled
to the SM particles, but in thermal equilibrium with the visible sector, 
and $3\rightarrow 2$ annihilation processes
are strong enough to resolve the {\it `core vs. cusp'}~\cite{deBlok:2009sp} and
{\it `too big to fail'}~\cite{Boylan-Kolchin:2011qkt} problems in small scale
structures, while reproducing the same successful predictions of  
weakly interacting massive particles (WIMPs) 
 in large-scale structures.
Combination of numerical and observational studies  of rotational velocities 
in spiral galaxies indicate
the existence of a spherical core, in sharp qualitative contrast with the 
generic expectation from collisionless CDM models, leading to power-law
dark matter distributions and  cusp profiles.
 Similarly, the highly peaked distribution of dark matter expected within the WIMP-based CDM 
 paradigm would predict the existence of massive, satellite subhaloes, which have not been observed experimentally.

Realisations of the SIMP scenario were identified in general strongly-coupled dark sectors~\cite{Hochberg:2014kqa} 
that, in analogy with ordinary QCD, 
admit a Wess-Zumino-Witten (WZW) interaction term~\cite{Wess:1971yu,Witten:1983tw,Witten:1983tx}.
The $Sp(4)$ gauge theory with $N_f=2$ fermions transforming in the fundamental representation is the minimal
model realising this paradigm.
In studying the phenomenology of such models, Ref.~\cite{Hochberg:2015vrg} highlighted the
importance of having non-perturbative information about the
spin-1 bound states in the strongly-coupled dark sector, for example because it 
determines the phenomenology of models in which 
a dark photon mediates the interactions between visible and hidden sector that 
keep them in thermal equilibrium at freeze out.
This suggestion was further developed in Ref.~\cite{Berlin:2018tvf}, by noticing that in the presence of 
symmetry-breaking, large masses for the PNGBs, the physics of the vector mesons can have a dominant effect in determining 
the CDM relic abundance.
Furthermore, in the presence 
of a small mass splitting within the dark PNGB multiplet,
the observed
dark matter density may result from
other depletion mechanisms that rely on the exchange of dark vectors,
but do not assume that dark and visible sectors are at thermal equilibrium with one another
 (see for instance Ref.~\cite{Bernal:2017mqb}).

This brief, incomplete, collection of thoughts about the
phenomenological and model-building developments taking place over the past ten years
of dark matter studies
is yet sufficient to support three points of general validity.
\begin{itemize}

\item Gauge theories with $Sp(2N)$ group, coupled to  $N_f$ families of fundamental 
matter, might play a central role in SIMP model building, and it is hence a priority to study them on the lattice,
both in the minimal $N=2=N_f$ realisation and  its extensions.

\item In dark matter models, the mass of the lightest spin-1 composite state lies between that of the
PNGBs, and about twice of it. 
This is to be contrasted with the CHM context, where addressing the little hierarchy problem
requires a scale separation between PNGBs and heavier states.  And this is diametrically opposite to TC, where 
 gauge invariance forbids fermion masses.
For lattice practitioners, this observation
makes the quenched calculations into a reasonable approximation of the true dynamics in the phenomenologically relevant 
region of parameter space.
 
 \item Many variations of the mechanism yielding  the SIMP 
 dark matter relic abundance exist, including SIMP adaptations~\cite{Bernal:2015xba}
 of the freeze-in mechanism~\cite{McDonald:2001vt, 
 Hall:2009bx,Yaguna:2011qn,Campbell:2015fra,Kang:2015aqa},
 and more are likely to be proposed
 in the near future. This observation suggests to carry out broad, unprejudiced explorations
 of the whole parameter space. 
As high precision
 measurements are not yet a priority, 
 feasible investigation strategies for these explorations make reasonable use of available 
 computing resources.

\end{itemize}

A quite significant amount of information can already be found in 
Refs.~\cite{Bennett:2017kga,
Bennett:2019jzz,
Bennett:2019cxd,
Bennett:2020qtj}, that report   the masses and decay constants of PNGBs and  
other light mesons in $Sp(4)$ theories
with $N_f=2$  (quenched or dynamical) fermions, 
as well as the masses of glueballs in $Sp(2N)$
with $N=1,\,2,\,3,\,4$.
Future studies of the spectrum of mesons with dynamical matter transforming in the antisymmetric representation
will further contribute. A systematic study of the mesons in the low dimensional representations of $Sp(2N)$,
for varying $N$, inside the regime of validity of the quenched approximation, 
will tabulate  dynamical input that is essential to SIMP
model builders.
The ongoing programme of study of $Sp(4)$ theories in the presence of 
small mass splitting within the fermion sector  provides complementary strong-coupling input for
 phenomenology~\cite{Maas:2021gbf,Zierler:2021cfa,Kulkarni:2022bvh}.

Without  
 pretence of encyclopaedic completeness, we also discuss  the potential 
implications for the early universe of  first-order phase transitions.
At the transition, the formation of bubbles of true vacuum, their growth, collisions, 
and the resulting sound waves and friction
source a relic  stochastic GW background, which is detectable,
in principle, in future experiments.
The original motivation to consider such scenarios comes from 
the miscellaneous environment of hidden sectors and strongly coupled dark matter models.
A broad portfolio of tools has been optimised to analyse 
the specific  reach of future experimental programmes,
and test  broad classes of new physics models.

One such tool is the online software package PTPlot~\cite{Caprini:2019egz}.
Developed with the specifications of LISA, PTPlot provides the gravitational wave power spectrum $h^2\Omega_{\rm GW}(f)$
predicted for a given choice of input parameters, as a function of the frequency $f$,
and compares it to the sensitivity curves, 
 determined by the experiment configuration 
and its expected noise level.
Sound waves are the main source of gravitational waves, and following Ref.~\cite{Caprini:2019egz}
(to which we refer the reader, as to the original literature, for details) 
we ignore other sources.
  The power spectrum is computed from
(model-dependent) knowledge of the following five parameters.
\begin{itemize}

\item The  (percolation) temperature $T_{\ast}$ (or Hubble parameter $H_{\ast}$) at which 
the phase transition ends. The  phase transition starts at the critical temperature $T_c>T_{\ast}$.

\item The inverse duration  of the transition, measured by the bubble nucleation rate $\beta$ computed at $T_{\ast}$,
defined in terms of $S(T)$, the 3-dimensional action of the system:
\beqs
\frac{\beta}{H_{\ast}} & \equiv& T\frac{\partial }{\partial T}\left.\left(\frac{S(T)}{T}\right)\right|_{T_{\ast}}\,.
\eeqs

\item The parameter $\alpha$, determining the strength of the transition, depends on
 $\Delta\theta$, the jump at the transition
 in trace of the stress-energy tensor $\theta\equiv e-3p$, and the enthalpy $\omega_+=e_++p_+$ in the high-$T$ phase:
\beqs
\alpha&\equiv &\frac{\Delta\theta}{3\omega_+}\,.
\eeqs

\item The bubble wall speed $v_W$---the efficiency parameter $\kappa$ (the ratio of bulk kinetic energy to vacuum energy) depends
 on $\alpha$ and $v_W$~\cite{Espinosa:2010hh}.

\item The number of degrees of freedom $g_{\ast}$ after the phase transition.

\end{itemize}

We specify a dark, strongly coupled gauge theory.
Following Refs.~\cite{Huang:2020crf,Halverson:2020xpg}, we assume the transition
to be very fast,  so that  $T_*\simeq T_c$ and ${\beta}/{H_{\ast}}\gg 1$.
Furthermore, we  assume a relativistic bubble wall velocity 
$
v_W  \simeq  1
$;
a precise determination of the wall velocity would require dedicated studies of the bubble wall dynamics, and as shown
in Ref.~\cite{Huang:2020crf} the signal strength depends only mildly on this parameter.
We borrow from Ref.~\cite{Panero:2009tv}
the  lattice  indication that $p\ll e \simeq \theta$ for $SU(N_c)$ Yang-Mills theories near $T_c$,
and that $p$ varies smoothly across $T_c$. As a result
\beqs
\alpha &\simeq&\frac{1}{3}\,.
\eeqs

The value of $\beta/H_{\ast}$ can be obtained by modelling and measuring the effective action,
or with detailed  knowledge about the surface tension of the bubbles. For $SU(N_c)$ Yang-Mills theories,
Refs.~\cite{Huang:2020crf,Halverson:2020xpg} agree in indicating the range 
\beqs
10^{4} & \lsim & \frac{\beta}{H_{\ast}} \,\,\,\lsim\,\,\, 10^5\,,\label{Eq:beta}
\eeqs
which is affected by large uncertainties, for all $N_c$. Finally, 
the number of relativistic degrees of freedom is the sum of the SM ones and the new dark  sector ones.
For example, for a $SU(N_c)$ dark sector coupled to the SM fields (no right-handed neutrinos):
\beqs
g_{\ast}&=&n_B^{(SM)} +\frac{7}{8}n_F^{(SM)}+n_B^{(N_c)} +\frac{7}{8}n_F^{(N_c)}=106.75+2(N_c^2-1)\,,
\eeqs
while if we treat the SM neutrinos as Dirac particles, then
$
g_{\ast}=112+2(N_c^2-1)
$.

By making use of the online interface of PTPlot~\cite{Caprini:2019egz}, one can compare the 
GW power spectrum, $h^2 \Omega_{GW}(f)$, as a function of the frequency $f$, 
to the predicted reach of LISA (3-year exposure). Assuming $v_W=1$, 
$\alpha=0.33$, and $g_{\ast}=142$, one finds empirically that holding fixed the product
$T_{\ast} \beta/H_{\ast}=10000\,\,{\rm GeV}$
the peak of the GW signal appears at frequencies close to $f\simeq 0.001$ Hz, near the best reach of LISA.
The GW signal could be detected by LISA for $\beta/H_{\ast}\leq 100$, which can be compared to 
the inequalities~(\ref{Eq:beta}).

Coming back to the  topic of this review, 
 the percolation temperature $T_{\ast}$ is essentially a free parameter, and additional GW experiments
are being planned~\cite{Seto:2001qf,
 Kawamura:2006up,Crowder:2005nr,Corbin:2005ny,Harry:2006fi,
 Hild:2010id,Yagi:2011wg,Sathyaprakash:2012jk,Thrane:2013oya,
 Caprini:2015zlo,
 LISA:2017pwj,
 LIGOScientific:2016wof,Isoyama:2018rjb,Baker:2019nia,
 Brdar:2018num,Reitze:2019iox,Caprini:2019egz,
 Maggiore:2019uih},
that will be sensitive to higher frequencies and lower values of 
$h^2 \Omega_{GW}$.
Hence,  it is possible that dark sectors based on $Sp(2N)$ theories that undergo first
 order phase transitions in the early universe are testable 
 via their relic stochastic GW background.
Furthermore, the quantities $\alpha$ and $\beta$ have not yet been  computed for $Sp(2N)$
theories with $N>1$. (The $Sp(2)=SU(2)$ case is trivial, as the transition is believed to be of second order.)
Large-$N$ universality suggests that  similar results should hold for $Sp(2N)$ as for $SU(N_c)$
 theories, in which the thermodynamics  depends mildly on $N_c >2$.
 Some interesting work in this direction,
based on gauge-gravity dualities and their relation to large-$N_c$ theories, can be
found in Refs.~\cite{Bigazzi:2020phm,Ares:2020lbt,Bea:2021zsu,Bigazzi:2021ucw,
Henriksson:2021zei,Ares:2021ntv,Ares:2021nap,Morgante:2022zvc}.
But dedicated, non perturbative studies of $Sp(2N)$ theories at finite 
temperature are needed, for which the LLR method~\cite{Langfeld:2012ah, 
Langfeld:2013xbf,Langfeld:2015fua,Cossu:2021bgn} offers
an intriguing opportunity, as argued in Refs.~\cite{Springer:2021liy,Mason:2022trc,Mason:2022aka,Springer:2023wok}.

\section{$Sp(2N)$ lattice gauge theories}
\label{sec:latticesp2n}

This section introduces the lattice treatment of the theories of interest. 
We start by describing the lattice action, for bosons and fermions, in Sect.~\ref{sec:action}, 
and the numerical Monte Carlo algorithms adopted in Sect.~\ref{sec:simulation}.
Section~\ref{sec:gradient_flow} discusses scale setting and topology.
Section~\ref{sec:measure} introduces the strategy employed in data analysis, 
focusing mostly on the two-point functions used for spectroscopy measurements.
Additional information on the lattice theory and its systematic effects are presented in Sect.~\ref{sec:systematics},
which discusses the bulk phase structure and finite volume
effects.

\subsection{Lattice action}
\label{sec:action}

For the numerical calculations we first rewrite  Eq.~(\ref{eq:lagrangian}) in four-dimensional Euclidean space-time, then discretise the lattice action, that contains the gauge-field term $S_g$ and the fermion matter-field term $S_f$, 
\beq
S=S_g+S_f.
\label{eq:lattice_action}
\eeq
We use the standard Wilson plaquette action for the gauge fields. With the bare lattice coupling $\beta = 4N/g^2$, it gives
\beq
S_g\equiv\beta \sum_x \sum_{\mu <\nu} 
\left(
1-\frac{1}{2N}\textrm{Re}\,{\Tr} \,\mathcal{P}_{\mu\nu}
\right)\,,
\label{eq:gauge_action}
\eeq
where the plaquette $\mathcal{P}_{\mu\nu}$ is defined as
\beq
\mathcal{P}_{\mu\nu}(x)\equiv U_\mu (x)U_\nu(x+\hat{\mu})U^\dagger_\mu (x+\hat{\nu})U^\dagger_\nu(x)\,.
\label{eq:plaquette}
\eeq
The gauge link $U_{\mu}(x) \in Sp(2N)$  transforms in the fundamental representation. The massive Wilson-Dirac action for fermionic fields is
\beq
S_f \equiv a^4 \sum_{j=1}^{N_f}\sum_x \overline{Q}^j(x) D^{(f)}_m Q^j(x)+
a^4 \sum_{j=1}^{n_f}\sum_x \overline{\Psi}^j(x) D^{(as)}_m \Psi^j(x),
\label{eq:fermion_action}
\eeq
with the definition of the massive Wilson-Dirac operator
\beqs
D^{R}_m \psi^R_j(x) &\equiv& (4/a+m^{R}_0) \psi^{R}_j(x)\label{eq:DiracF} \\
& & -\frac{1}{2a}\sum_\mu \nonumber
\left\{(1-\gamma_\mu)U^{R}_\mu(x)\psi^{R}_j(x+\hat{\mu})
+(1+\gamma_\mu)U^{R,\,\dagger}_\mu(x-\hat{\mu})\psi^{R}_{j}(x-\hat{\mu})\right\}\,,
\eeqs
where we denote as $a$ the lattice spacing, $R$ the representation, with $(f)$ and $(as)$ being the fundamental and antisymmetric, respectively, and $m_0^{R}$ the (degenerate) bare masses of the fermion fields $\psi^{R}_i$. The link variable for fundamental fermions, $U_{\mu}^{(f)}(x)$, is the same as $U_{\mu}(x)$ in Eq.~(\ref{eq:plaquette}). For the antisymmetric fermions, the link variable, $U_{\mu}^{(as)}(x)$, is obtained by the construction
\beqs
\left(U^{(as)}_\mu\right)_{(ab)(cd)}(x)\equiv
{\rm Tr}\left[
(e_{(as)}^{(ab)})^\dagger U_\mu(x) e_{(as)}^{(cd)} U^{\mathrm{T}}_\mu(x)\right],~~~{\rm with}~a<b,~c<d.
\label{eq:U_AS}
\eeqs
The basis matrices are defined as 
\beqs
(e_{(as)}^{(ab)})_{c,N+c}\equiv -(e_{(as)}^{(ab)})_{N+c,c}\equiv 
\left\{\begin{matrix}
&\frac{1}{\sqrt{2\,a\,(a-1)}},~~~\textrm{for}~c<a,\\
&\frac{-(a-1)}{\sqrt{2\,a\,(a-1)}},~~~\textrm{for}~c=a,\\
\end{matrix}\right.
\label{eq:eas_diag}
\eeqs
for $b=N+a$ with $2\leq a\leq N$, and
\beqs
(e_{(as)}^{(ab)})_{cd}\equiv \frac{1}{\sqrt{2}}(
\delta_{ad}\delta_{bc}
-\delta_{ac}\delta_{bd}
)\,
\label{eq:eas}
\eeqs
for $b\neq N+a$. 
We assign the multi-index pairs $(ab)$ with the order $1 \leq a < b \leq 2N$.   In this work, the spatial extents, $L_x/a$, $L_y/a$ and $L_z/a$, of the lattice are taken to be the same, while the temporal extent, $T/a$, can be different.
Periodic boundary conditions are imposed for all fields in the spatial directions.
For the temporal direction, we use periodic and anti-periodic boundary conditions for the gauge and fermion fields, respectively. 
Using the lattice actions described above, we generate gauge-field ensembles with Monte Carlo (MC) methods, as described in the next section. 

The lattice theory with $N_f=2$ and $n_f=3$ (massive) Dirac flavours 
is expected to exhibit the same global (flavour) symmetry breaking pattern of the continuum theory discussed in
 Sec.~\ref{sec:uvmodel}.  Namely, the breaking patterns are $SU(4) \longrightarrow Sp(4)$ and $SU(6) \longrightarrow SO(6)$, for the fundamental and antisymmetric sectors, respectively. 
This information is encoded 
in the spectrum of the Dirac operator, which can be modelled 
by  chiral random matrix theory (ChRMT) \cite{Verbaarschot:1994qf}. 
In particular, ChRMT predicts that the distribution of the unfolded density 
of spacings, $s$, between subsequent Dirac eigenvalues, $P(s)$, is described by the Wigner 
surmise with a Dyson index different for the symmetry breaking patterns. 
In Ref.~\cite{Bennett:2022yfa}, 
we computed the Dirac eigenvalues for fermions in the fundamental and antisymmetric representations from a quenched ensemble  with  lattice size $4^4$, 
and found that the numerical results are in good agreement 
with the ChRMT predictions of $P(s)$. 
We hence confirmed that fermions are correctly implemented in the code 
used for numerical simulations and measurements.

\subsection{Simulation strategies}
\label{sec:simulation}

In the lattice studies reported in Refs.~\cite{Bennett:2017kga,
Lee:2018ztv,Bennett:2019jzz,
Bennett:2019cxd,Bennett:2020hqd,
Bennett:2020qtj,Lucini:2021xke,Bennett:2021mbw,
Bennett:2022yfa,Bennett:2022gdz,Bennett:2022ftz,AS,
Lee:2022elf,Hsiao:2022kxf}, numerical calculations are carried out by 
using the HiRep code~\cite{DelDebbio:2008zf,hirep-upstream}, with bespoke software implementation of  $Sp(2N)$ gauge groups~\cite{hirep-repo}. For pure $Sp(2N)$ gauge theories, gauge configurations are generated with the heat bath (HB) algorithm, and decorrelation between configurations is improved by micro-canonical over-relaxation (OR) updates. Similar to the case of $SU(N_c)$~\cite{Cabibbo:1982zn}, the gauge links evolve with the minimal set of $SU(2)$ subgroups covering the whole $Sp(2N)$ group to ensure ergodicity. 
A variant of the (modified) Gram-Schmidt algorithm allows to correct the link variables 
and keep them in the desired group manifold over the updates.  This re-symplectisation procedure is important for correcting for numerical errors arising from the limit of machine precision.  

Simulations with dynamical fermions are performed using the hybrid Monte Carlo (HMC) algorithm for even number of Dirac flavours.  For simulating odd number of Dirac flavours, we resort to the rational HMC (RHMC) algorithm~\cite{Clark:2003na}.  Contrary to the HB algorithm, the explicit form of the group generators of $Sp(2N)$ enters the definition of the molecular dynamics (MD) update (see also Refs.~\cite{Takaishi:2005tz,DeGrand:1990dk} for the relevant choice of integrators and conditioning of the fermion matrices). Again, the link variables are re-symplectised to correct for machine-precision errors. Beside the Gram-Schmidt method mentioned above, this can also be achieved by carrying out projections with the quaternion basis, as described in Appendix C of Ref.~\cite{Bennett:2017kga}.

Correlations between consecutive trajectories (Monte-Carlo steps) exist in the algorithms mentioned above. 
In order to obtain independent gauge-field configurations, we monitor the average value of the plaquette along Monte-Carlo steps, and investigate its autocorrelation time in all our simulations.  
We find that it is sufficient to perform measurements for every 12 trajectories in quenched simulations, and for every 8 to 28 trajectories for dynamical calculations.
Furthermore, we typically discard a few hundred  initial trajectories for the purpose of thermalisation, which is monitored via the plaquette value.  Statistical analysis  employs the standard bootstrap method.

\subsection{Scale setting and topology}
\label{sec:gradient_flow}

The raw data obtained from lattice calculations are all expressed in lattice units---each
ensemble with a given set of lattice parameters defines a lattice theory
at some value of the lattice spacing, $a$, which depends on the chosen couplings.
It is therefore necessary to set a common scale to convert all the lattice results to the same continuum theory in a consistent way,
using a procedure of {\it scale setting}.
The gradient flow method for the scale setting
is particularly suitable for lattice studies of novel strongly coupled theories,
as the ones considered here.
The lattice version of the gradient flow for the gauge fields, the {\it Wilson flow}, is nowadays 
common practice in the field.
Thus we do not venture into a complete treatment of this technique here, referring the reader to Refs.~\cite{Luscher:2010iy,Luscher:2013vga} for further details;
we instead briefly define the gradient flow scheme used for this work
and discuss the key numerical results.

\begin{figure}
\begin{center}
\includegraphics[width=.6\textwidth]{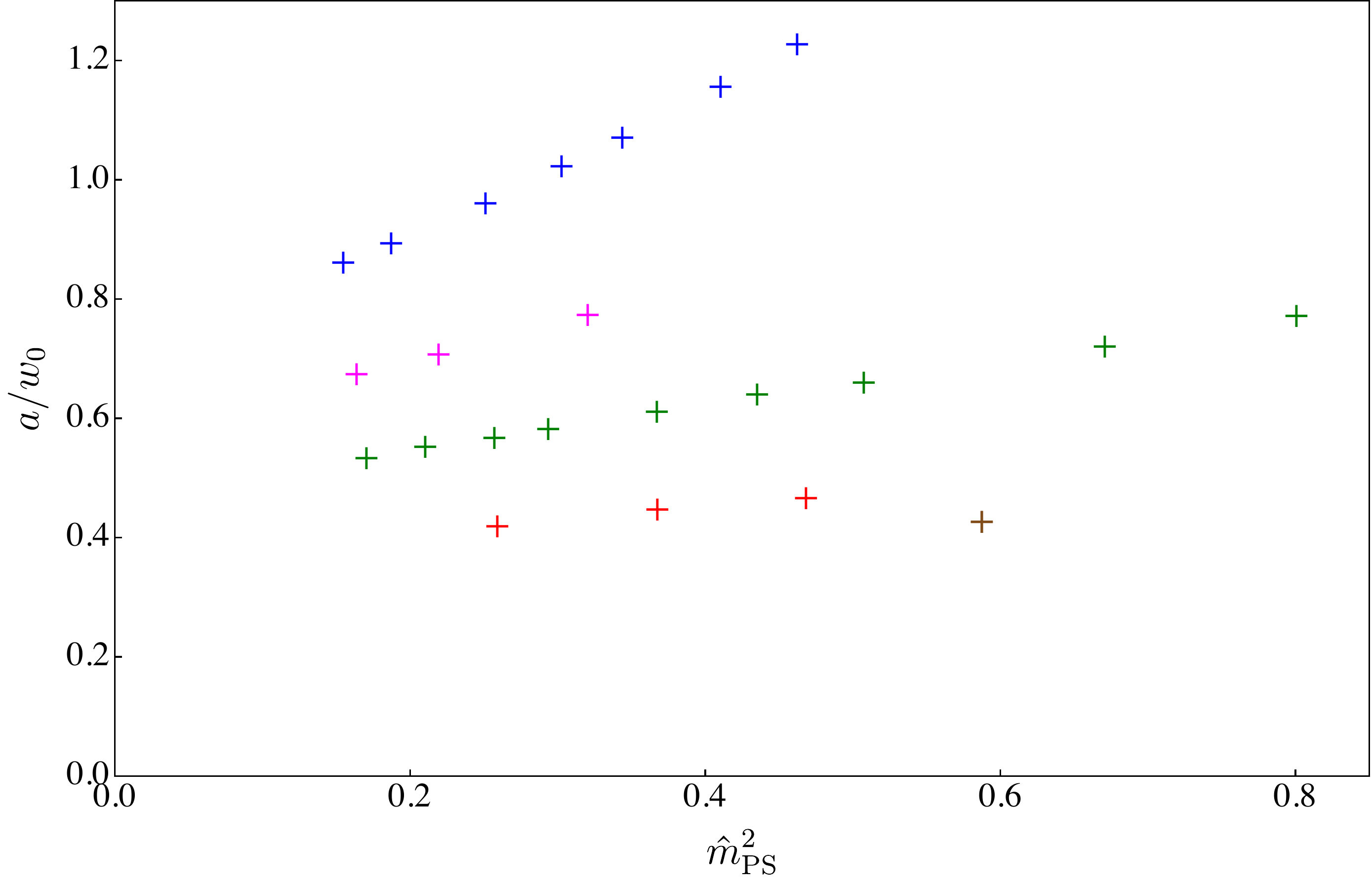}
\caption{%
Inverse of the gradient flow scale, $a/w_0$, in the $Sp(4)$ gauge theory coupled to $N_f=2$ fundamental fermions,
as a function to the mass of the lightest pseudoscalar $\hat{m}_{\rm PS}={m}_{\rm PS}w_0$.
Different colours denote different $\beta$ values; from top to bottom: $6.65$ (blue), $7.05$ (magenta), $7.2$ (green), $7.4$ (red) and $7.5$ (brown). 
The plot is taken from Ref.~\cite{Bennett:2019jzz}. 
}
\label{Fig:inv_w0}
\end{center}
\end{figure}

The gradient flow is defined via a diffusion equation in which a new gauge field $B_\mu(t,x)$ at a fictitious flow time $t$ (having length dimension two)
is defined from the four-dimensional gauge field $A_\mu(x)$ as
\beq\label{eq:GF_equation}
\frac{d B_\mu(t,x)}{dt}=D_\nu G_{\nu\mu}(t,x),~\textrm{with}~B_\mu(0,x)=A_\mu(x),
\eeq
where $D_\nu$ is the covariant derivative and $G_{\nu\mu}$ is the field-strength tensor.
For $t>0$ any gauge invariant observables built out of $B_\mu(t,x)$ are renormalised \cite{Luscher:2011bx}.
An observable that does not generate new operators along the flow time is the action density,
\beq
\label{Eq:EE}
E(t,x)=-\frac{1}{2}\Tr G_{\mu\nu}(t,x) G_{\mu\nu}(t,x).
\eeq
After defining a dimensionless quantity using the expectation value of $E(t,x)$,
\beq
  \mathcal{E}(t) \equiv t^2 \langle E(t,x)\rangle,
\eeq
one can obtain the scale $t_0$ by imposing the condition
\beq
\mathcal{E}(t)|_{t=t_0} = \mathcal{E}_0.
\eeq
Here, the renormalisation scale can be identified with the diffusion radius $\mu=1/\sqrt{8t}$.
The reference scale $\mathcal{E}_0$ is  chosen empirically  so that  lattice artefacts are minimised. 
Two further choices are made: firstly, rather than taking the simple plaquette operator $G_{\mu\nu}=\mathcal{P}_{\mu\nu}$,  in Eq.~(\ref{Eq:EE})
one can replace $G_{\mu\nu}$ with a four-plaquette clover, denoted by $\mathcal{C}_{\mu\nu}$, that will also be used  to define the topological
charge density.
Second, rather than $\mathcal{E}(t)$, one can consider
\begin{equation}
  \mathcal{W}(t)\equiv t \frac{d}{dt}\left\{ \mathcal{E}(t) \right\}~,
\end{equation}
and define the scale $w_0$~\cite{Borsanyi:2012zs} by the relation
\begin{equation}
      \mathcal{W}(t)|_{t=w_0^2}=\mathcal{W}_0\equiv 0.35~.
\end{equation}
Since $w_0$ and $t_0$ are affected differently by 
discretisation effects,
their comparison allows for an assessment
of their magnitude.

\begin{figure}[t]
\centering
\includegraphics[width=0.6\textwidth]{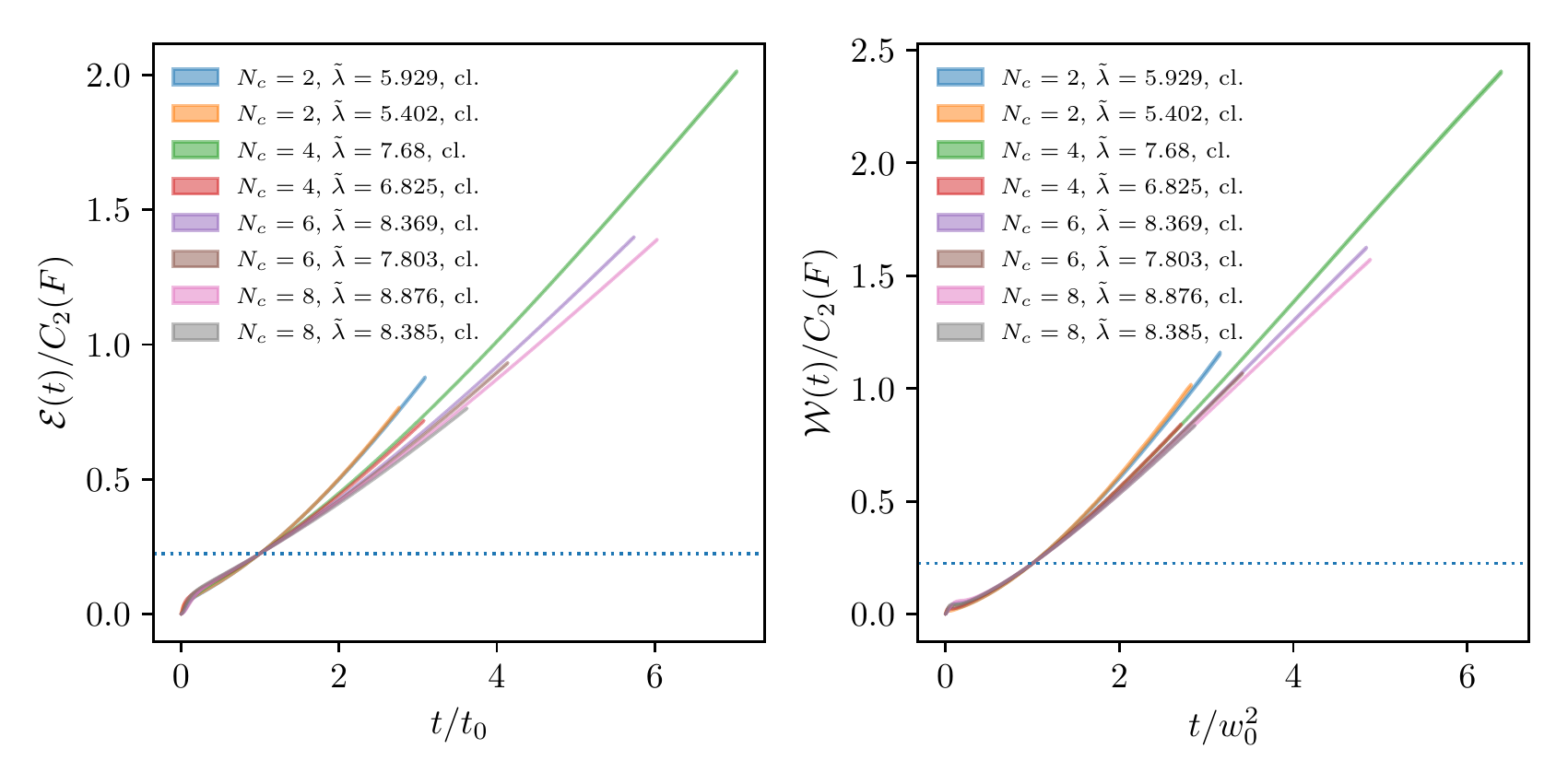}
\caption{Left panel: $\mathcal{E}(t)/C_2(F)$
    as a function of the rescaled flow time, $t/t_0$. Right
    panel: $\mathcal{W}(t)/C_2(F)$ as a function of the rescaled
    flow time $t/w_0^2$.  
    Both quantities are computed using the four-plaquette
    clover-leaf discretisation on the
    ensembles corresponding to the finest and coarsest available
    lattices for each $N_c$, with 
    $C_2(F)=(N_c+1)/4$.   
    The figure adopts the 
    choice  $c_e=c_w=0.225$ (horizontal dashed line).
    The plots are taken from Ref.~\cite{Bennett:2022ftz}.
\label{fig:scaled_flows_0.5}}
\end{figure}

While the flow scale shows mild quark-mass dependence in a typical lattice calculation for QCD with light quarks \cite{Borsanyi:2012zs},
for the $Sp(4)$ theory involving dynamical fermions considered here it turns out to significantly depend on the fermion mass, as shown in Fig.~\ref{Fig:inv_w0}.
Notice that the mass dependence is milder on finer lattices.
We introduce the hatted notation to present  physical quantities in units of the Wilson flow scale $w_0$, e.g.\ $\hat{m} \equiv m w_0 = m^{\rm latt} w_0^{\rm latt}$ with $m^{\rm latt} \equiv m a$ and $w_0^{\rm latt} \equiv w_0/a$.

When studying $Sp(2N)$ pure gauge theories on the lattice, it is convenient 
to define a way to relate the value of the 
scales obtained at different values of $N$. It can be shown that the following relation holds true,
\begin{equation}
\mathcal{E}(t) = \frac{3 \lambda}{64 \pi^2} C_2(F)
\end{equation}
at leading order in a perturbative expansion in the 't Hooft coupling,
defined as $\lambda = 4\pi N_c \alpha$, with 
$\alpha(\mu)$ the renormalized coupling in the
Wilson Flow scheme and $C_2(F)=(2N+1)/4$ the quadratic Casimir of the fundamental
representation of the $Sp(2N)$ group. It is then natural, 
especially in the context of
studies about the large-$N_c$ limit of gauge theories, to set
\begin{equation}
\mathcal{E}_0 = c_e C_2(F) \,,\quad
\mathcal{W}_0 = c_w C_2(F)~.
\end{equation}
where $c_e$ and $c_w$ are empirically chosen constants. The usefulness of
this scaling law outside of perturbation theory can 
be assessed numerically.
The behaviours of $\mathcal{E}(t)$ and $\mathcal{W}(t)$, rescaled
with $C_2(F)$, as a function of the rescaled flow times $t/t_0$ and
$t/w_0^2$, respectively, are displayed 
in Fig.~\ref{fig:scaled_flows_0.5}. Notice the approximate superposition of the 
curves corresponding to different values of $N$ and similar values
of the 't Hooft coupling, which holds beyond perturbation theory.

\begin{figure}[t]
\centering
\includegraphics[width=.45\textwidth]{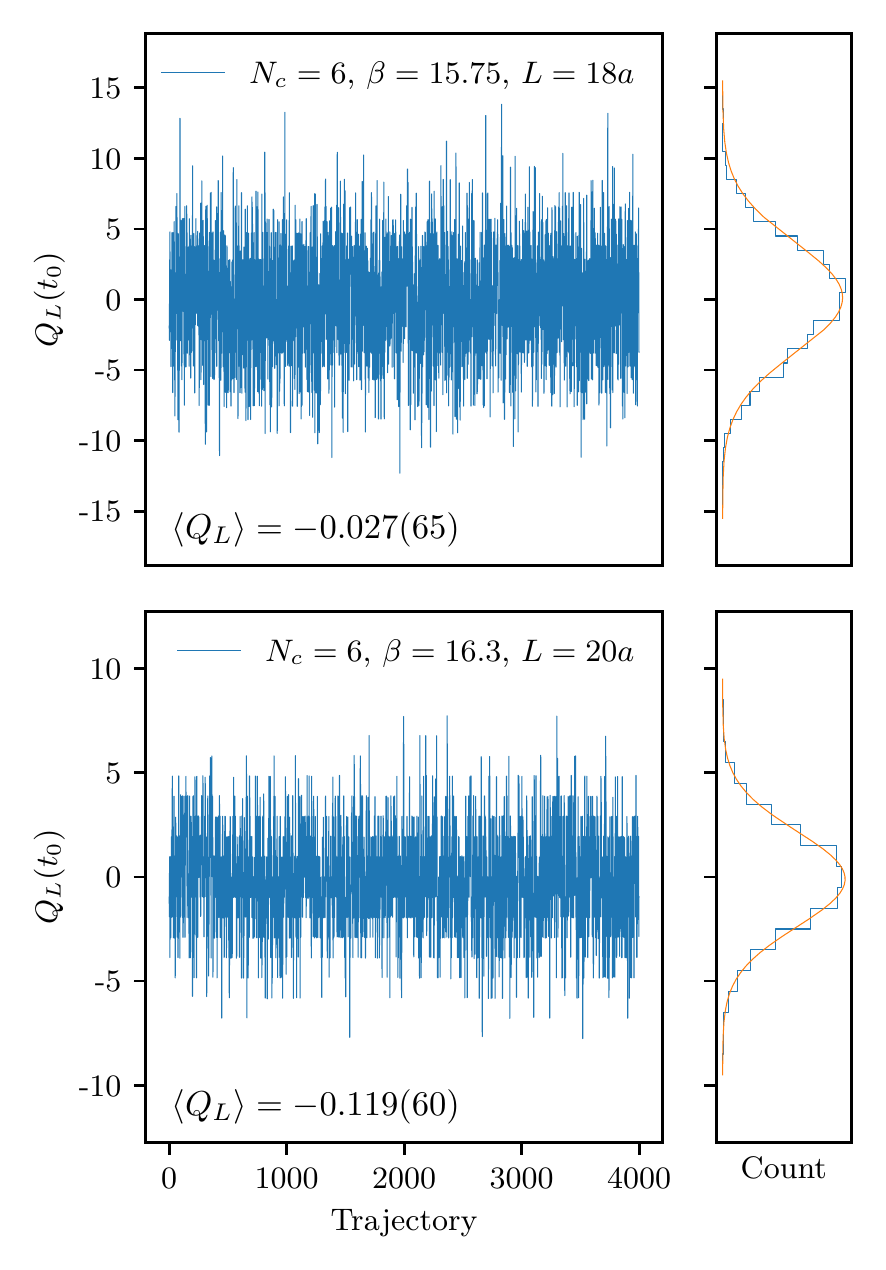}
\caption{The topological charge $Q_L$  as a function of simulation 
time (trajectory) for the ensembles corresponding to the coarsest (top) and finest (bottom) lattice
with $N_c=6$. 
The value of $Q_L$ is computed at $t=t_0$, where the value of $t_0$ is obtained from $c_e=0.225$. 
The average value of the topological charge along the trajectory
is reported in the bottom left-hand side of the plot. 
The side panel contains the cumulative histogram 
of the values of $Q_L(t_0)$.
The orange curve is a gaussian fit to the cumulative 
histogram. The plots are taken from Ref.~\cite{Bennett:2022ftz}.
\label{fig:TC_hist_6}}
\end{figure}

We close this section with a brief discussion on the topological charge $Q$. The discretisation of this observable is not unique. As $a\to 0$, valid lattice definitions differ by terms proportional to $a^4$. Regardless of the definition, lattice measurements of $Q$ are dominated by UV fluctuations. 
In order to extract the value of the topological charge at finite lattice spacing, an efficient strategy is to compute $Q$ on configurations that have
been evolved according to the Wilson flow, Eq.~(\ref{eq:GF_equation}), up to a finite flow-time $t$.

\clearpage

In the lattice studies on $Sp(2N)$ we review, the definition of the topological charge
$Q_L$ is
\beq
Q_L(t)=\sum_x q(t,\,x)\,,\quad
q_L(t,\,x) = \frac{1}{32\pi^2}\varepsilon_{\mu\nu\rho\sigma} \textrm{Tr}\left\{
\mathcal{C}_{\mu\nu}(t,\,x)\mathcal{C}_{\rho\sigma}(t,\,x)
\right\}\,.
\eeq
where $q_L(t,\,x)$ and $\mathcal{C}_{\mu\nu}(t,\,x)$ are, respectively, the 
topological charge density and the four-plaquette clover-leaf 
operator computed at space-time site $x$ and flow time $t$~\cite{Sheikholeslami:1985ij,Hasenbusch:2002ai}. 
This observable is used for monitoring the simulations. Specifically, for each of the lattice settings, 
the topological charge is computed
and its value as a function of simulation time inspected to ascertain absence of
 topological freezing, so that Monte Carlo configuration are not stuck in particular values of $Q_L$, which would
 indicate one is not sampling correctly the space of the configurations---more sophisticated  
 ideas exist to address topological freezing~\cite{Luscher:2011kk,Endres:2015yca, Luscher:2017cjh},
 and might be implemented in the future.
 As an example, the Monte Carlo time history of the topological charge $Q_L$ is reported 
in Fig.~\ref{fig:TC_hist_6} for the case $N_c=2N=6$.
The trajectory of $Q_L$ does not display any sign of
topological freezing. In the side panel, the frequency
histograms of $Q_L$ are reported. The distribution of $Q_L$
is compatible with a Gaussian centered at $Q_L=0$, as
expected from theoretical considerations.

When the topological charge plays a quantitative role in the physics observables of interest, for instance in the measurement of its susceptibility, its $\alpha$-rounded version is used in order to further reduce discretisation effects, see Sect.~\ref{sec:topology} for details.

\subsection{Measurements: two-point functions, masses and decay constants}
\label{sec:measure}

Spectroscopy studies are a crucial component in understanding gauge theories. These studies involve the computation of masses and decay constants of the low-lying hadronic states, such as those listed in Tables~\ref{tab:mesons} and ~\ref{tab:chimerabaryons}. 

Two-point correlation functions are a central tool for these calculations.  
For mesons, the structure of a generic two-point correlator  (using the notation $x\equiv(t,\vec{x})$) is
\beq
\label{eq:meson_2pt}
C_{M,M^\prime} (t) \equiv \sum_{\vec{x}} \left < \mathcal{O}_M(x) \mathcal{O}^\dagger_{M^\prime} (0) \right>\,,
\eeq
where $M$ and $M^{\prime}$ are labels appearing in the first column of Table~\ref{tab:mesons}, with $\mathcal{O}_{M}$ and $\mathcal{O}_{M^{\prime}}$ being the corresponding interpolating operators.  These operators overlap with the lowest-lying mesonic states with zero spatial momentum. Carrying out the Wick contraction for the fermion fields in Eq.~(\ref{eq:meson_2pt}), the correlation function is 
\beqs\label{Eq:cMM}
C_{M,M^\prime} (t) = - \sum_{\vec{x}} \Tr \left [ \Gamma_M S^{R}(x) \Gamma_{M^\prime} \gamma_5 S^{R\, \dagger}(x) \gamma_5 \right]\,,
\eeqs
where the trace is taken in both  spinor and colour spaces, with $\Gamma_{M^{(\prime)}}$ being the relevant Dirac matrix in $\mathcal{O}_{M^{(\prime)}}$.  In Eq.~(\ref{Eq:cMM}), the symbol $S^{R}$ denotes the fermion propagator in the representation $R$. We define respectively the $(f)$ and $(as)$ fermion propagators as 
\beq
S^{\,i\,a}_{Q\,\,\,\,b\,\alpha\beta}(x) = 
\langle Q^{i\,a}_{\,\,\,\,\,\alpha}(x) \overline{Q^{i\,b}}_{\beta}(0) \rangle
~{\textrm{and}}~S^{\,k\,ab}_{\Psi\,\,\,\,\,\,\,\,cd\,\alpha\beta}(x) =
 \langle \Psi^{k\,ab}_{\,\,\,\,\,\,\,\,\,\,\alpha}(x) 
\overline{\Psi^{k\,cd}}_{\beta}(0) \rangle\,,
\label{eq:fermion_prop}
\eeq
where $a,\,b,\,c,\,d$ are colour indices while $\alpha$ and $\beta$ are spinor indices. 
In the case of a point source, the meson interpolating operator is constructed at one space-time point, and the fermion propagator is computed by solving the Dirac equation
\beq\label{eq:source}
D^{R}_{a\alpha,b\beta}(x,y)S^{R\,b\beta}_{\,\,\,\,c\gamma}(y) = \delta_{x0}\delta_{\alpha\gamma}\delta_{ac}\,,
\eeq
with $D^{R}$ referring to the Dirac operator in representation $R$. 
Using $Z_2\times Z_2$ single-time stochastic wall sources~\cite{Boyle:2008rh} (with number of hits $3$, in our case) improves the signal by increasing the overlap of interpolating operators and the lowest-lying physical state. 
At large Euclidean time $t\rightarrow \infty$, the correlator with  $M=M^\prime$ behaves as
\beq\label{eq:meson_c}
C_{M,M}(t) \rightarrow \frac{\left | \left <  0 \left | \mathcal{O}_M \right | M \right > \right |^2 }{2m_{M}} \left [ e^{-m_{M}t} + e^{-m_{M}(T-t)} \right]\,,
\eeq
where $|M\rangle$ denotes the lowest-lying mesonic state that overlaps with $\mathcal{O}_{M}$, with $m_{M}$ being its mass, and $T$  the temporal extent of the lattice. The combination $M=\text{PS}$ and $M^\prime=\text{AV}$ is used to determine the pseudoscalar meson decay constant, as the correlator reads
\beq
C_{\textrm{PS},\textrm{AV}}(t) \rightarrow \frac{ \left <  0 \left | \mathcal{O}_{\text{AV}} \right | \text{PS} \right > \left <  0 \left | \mathcal{O}_{\text{PS}} \right | \text{PS} \right >^*  }{2m_{\text{PS}}} \left [ e^{-m_{\text{PS}}t }- e^{-m_{\text{PS}}(T-t)} \right]\,.
\eeq
The decay constants of the PS, V, and AV mesons are extracted from the matrix elements:
\beqs
\left <0 \left | \mathcal{O_{\text{AV}}} \right | \text{PS} \right > &\equiv& \sqrt{2} f_{\text{PS}} p^{\mu},\\
\left <0 \left |\mathcal{O_{\text{V}}} \right | \text{V} \right > &\equiv& \sqrt{2} f_{\text{V}} m_{\text{V}} \epsilon^{\mu},\\
\left <0 \left |\mathcal{O_{\text{AV}}} \right | \text{AV} \right > &\equiv& \sqrt{2} f_{\text{AV}} m_{\text{AV}}\epsilon^{\mu},
\eeqs
where $p^{\mu}$ and $\epsilon^{\mu}$ are the momentum and polarisation four-vectors, respectively. The PS decay constant, $f_{\text{PS}}$, is normalised by adopting the convention which yields $f_{\text{PS}} \simeq 93$ MeV in QCD. 
Furthermore, we renormalise the decay constants using the renormalisation constants obtained in lattice perturbation theory for Wilson fermions at the one-loop level 
with tadpole improvement~\cite{Martinelli:1982mw}.

The zero momentum two-point function of a chimera baryon, after the Wick contractions, takes the form,
\beqs
\label{eq:chimera_corr}
C_{\text{CB}}(t) &\equiv& \sum_{\vec{x}} \langle \mathcal{O}_{\rm CB}(x) \overline{\mathcal{O}_{\rm CB}}(0) \rangle  \nonumber \\
&=& - \sum_{\vec{x}} \left ( \Gamma^2 {S_{\Psi}^{k\,cd }}_{c^\prime d^\prime} (x,0)  \overline{\Gamma^{2}} \right)\Omega_{cb}\Omega^{b^\prime c^\prime} \Omega_{ad}\Omega^{d^\prime a^\prime}
 \nonumber \\
&& ~~~
 \times\Tr \left [ \Gamma^{1}  S^{b}_{Q\,\,\,b^{\prime}}(x,0) \overline{\Gamma^{1}} S_{Q\,\,\,\,a^\prime}^{a}(x,0) \right ]\,,
\eeqs
where we define $\overline{\Gamma} \equiv \gamma^0 \Gamma^\dagger \gamma^0$, with $\Gamma^{1,2}$ being the Dirac matrices appearing in the second column of Table~\ref{tab:chimerabaryons}.  The trace is over the spinor indices.
Unlike mesonic correlators, the chimera-baryon two-point function in Eq.~(\ref{eq:chimera_corr}) contains contributions from both even and odd parity states. The asymptotic behaviour of such a correlator at $t\rightarrow \infty$ is thus,
\beq
C_{\textrm{CB}}(t) \rightarrow \mathcal{P}_e \left[ c_ee^{-m_et} + c_oe^{-m_o(T-t)} \right] - \mathcal{P}_o \left[ c_oe^{-m_ot} + c_ee^{-m_e(T-t)} \right]\,,
\eeq
where $\mathcal{P}_{e,\,o} \equiv (1\pm \gamma^0)/2$ are the parity projectors in the non-relativistic limit. We denote as $m_e$ and $m_o$ the masses of the baryons in parity even and odd states, respectively, while $c_e$ and $c_o$ are coefficients related to matrix elements of the interpolating operator between the baryon states and the vacuum. 
By combining the correlators of both parity projections, 
$C_{e} \equiv \mathcal{P}_{e} C_{\textrm{CB}}$ and $ C_{o}\equiv \mathcal{P}_{o} C_{\textrm{CB}}$, we obtain
\beq\label{eq:CB_c}
\tilde{C}_{\textrm{CB}}(t) = \frac{1}{2} \left [ C_{e}(t) - C_{o}(T-t) \right ]
\xrightarrow{t\rightarrow \infty} \frac{1}{2}\left [ c_{e} e^{-m_et} + c_{o} e^{-m_o(T-t)} \right ] .
\eeq
The masses are extracted by fitting Eq.~(\ref{eq:meson_c}) for a meson and Eq.~(\ref{eq:CB_c}) for a chimera baryon.

Glueballs and torelons are color-singlet states of the system. Their existence descents from the confining nature of the theory. These states transform according to the irreducible representations of the spacetime symmetries of the system, which identify classification \emph{channels}. In the continuum,
the symmetry channels are the $J^P$ representations of the Poincar\'e group. The lattice is governed by the octahedral group, which is the symmetry group (rotations and parity transformation) of the cube. Near the continuum limit, degeneracies of states arise that restore  Poincar\'e invariance. The masses of the low-lying glueball states in all $J^P$ channels and of the ground state torelon were determined in ${Sp}(2N)$ theories 
for $N=1$, $2$, $3$, and $4$---see Ref.~\cite{Bennett:2020qtj} and references therein. In the rest of this section, we provide an overview of the methodology that underpins Ref.~\cite{Bennett:2020qtj}, with the  results reviewed in Sect.~\ref{sec:glueball}. 

On the lattice, states are generated from the vacuum by the action of gauge-invariant operators. These are defined as the trace of path-ordered, $\mathrm{P}$, products of link variables along closed space-like lattice paths $\mathcal{C}$,
\begin{equation}\label{eq:op_c}
U_{\mathcal{C}}(t,\,\vec{x}) = \mathrm{Tr}~\mathrm{P} \prod_{(x,\mu)\in\mathcal{C}}
U_\mu(x)~,
\end{equation}
where $x=(t,\,\vec{x})$ are the coordinates of any site that belongs to the path.
Elementary paths can be linearly combined with suitably chosen weights that preserve the symmetry channel. This fact can be exploited to optimise the signal-over-noise ratio, for instance using a variational approach involving the combination of multiple operators for each given symmetry channel. This observation underpins efficient methods of extraction of masses from lattice data, such as {\em smearing} and {\em blocking}. 

Glueballs are sourced by operators defined on contractible paths. They transform
in the trivial representation of the center of the group. As mentioned above, on the lattice the spacetime symmetries are described by the octahedral group, which has five irreducible  representations, each with two parity sectors. These ten channels are labelled by $A_1^\pm$, $A_2^\pm$, $E^\pm$, $T_1^\pm$, $T_2^\pm$, where $\pm$ indicate the parity $P$ and $A_1$, $A_2$ etc. the irreducible representations $R$ in standard crystallographic notation. 
The ground state mass in channel $R^P$ is determined
variationally. Among all the possible linear
combinations of operators defined as in Eq.~(\ref{eq:op_c}), the ones with the maximal
overlap with the ground state, denoted by $\tilde{O}^{R^P}$, are found. The large euclidean-time 
behaviour the two-points correlation functions of these operators then allows 
to extract the mass in channel $R^P$,  for different choices of the lattice spacing, at each value
of $N$.

Torelons are sourced by operators defined on paths that wind around the lattice
along a compactified direction. They transform  non-trivial representation
under the action of the centre of the group. From the ground state energy of the torelon, 
 the string tension, $\sigma$, can be extracted. 
The string tension is defined as the energy per unit length of a fluxtube
winding around a compactified direction of the system.
If the length of the
winding direction is $L$ and $m$ is the mass of torelon, then in general
\begin{equation}
\label{eq:string_expansion}
m=\sigma L\left( 1+ \sum_{k=1}^{\infty} \frac{d_k}{(\sigma L^2)^k}\right) \ ,
\end{equation}
where $d_k$ are dimensionless coefficients. The first three subleading terms in this expansion have been computed and have been shown to be constrained by symmetries  (i.e., they are {\em universal})~\cite{Luscher:1980fr,Luscher:1980ac,
Polchinski:1991ax,Luscher:2004ib,Aharony:2009gg,
Drummond:2004yp,HariDass:2006sd,Drummond:2006su,Dass:2006ud,Aharony:2013ipa,
Dubovsky:2015zey}. For a winding direction of sufficiently large length,
\begin{equation}
m=\sigma L \ , 
\end{equation}
in agreement with the classical picture of fluxtubes as strings of constant energy per unit length.
In Ref.~\cite{Bennett:2020qtj}, the value of the string tension $\sigma$ has been obtained from fits of the Nambu-Goto formula
\begin{equation}
m_\mathrm{NG}(L) = \sigma L \sqrt{1-\frac{\pi}{\sigma L^2}}~ \ ,
\end{equation}
which can be shown to reproduce the universal terms of Eq.~(\ref{eq:string_expansion}), to order $L^{-5}$. 

\subsection{Bulk phase structure and finite volume effects}
\label{sec:systematics}

The lattice action in Eq.~(\ref{eq:lattice_action}) involves at most three bare parameters: the lattice coupling $\beta$ and two  bare  masses $m_{0,\,{\rm latt}}^{(f)} \equiv am_0^{(f)}$ and $m_{0,\,{\rm latt}}^{(as)} \equiv am^{(as)}_0$, as we restrict attention to mass matrices that are flavour  degenerate.
The continuum and massless counterpart of this lattice theory can be obtained by taking the zero limit of $1/\beta$ and $m_{\rm latt}^R$ (after accounting for the additive renormalisation to the bare mass of the Wilson-Dirac fermions).
Understanding  the phase space of the lattice theory is necessary to choose  appropriate values of $\beta$, for which numerical simulations are doable on lattices of realistic size, without severe finite size effects, 
and yet such as to still be in the weak coupling regime. 
The latter condition is particularly important when the strong and weak coupling regimes are separated a first order bulk phase transition: 
the dynamics of the strong coupling regime could systematically differ from the continuum theory.   

The average plaquette value is an order parameter for  lattice bulk phase transitions. 
By measuring the ensemble average of the plaquette with  initial configuration either unity or
 random, on a small lattice (e.g.\ $4^4$), 
one associates the presence of  (strong) hysteresis as a sign of  first order phase transitions. 
By computing the plaquette susceptibility and using different sizes of lattice, the study of the volume dependence 
can confirm the order of phase transition and  pin down the location of the phase boundaries. 

\begin{figure}
\begin{center}
\includegraphics[width=.6\textwidth]{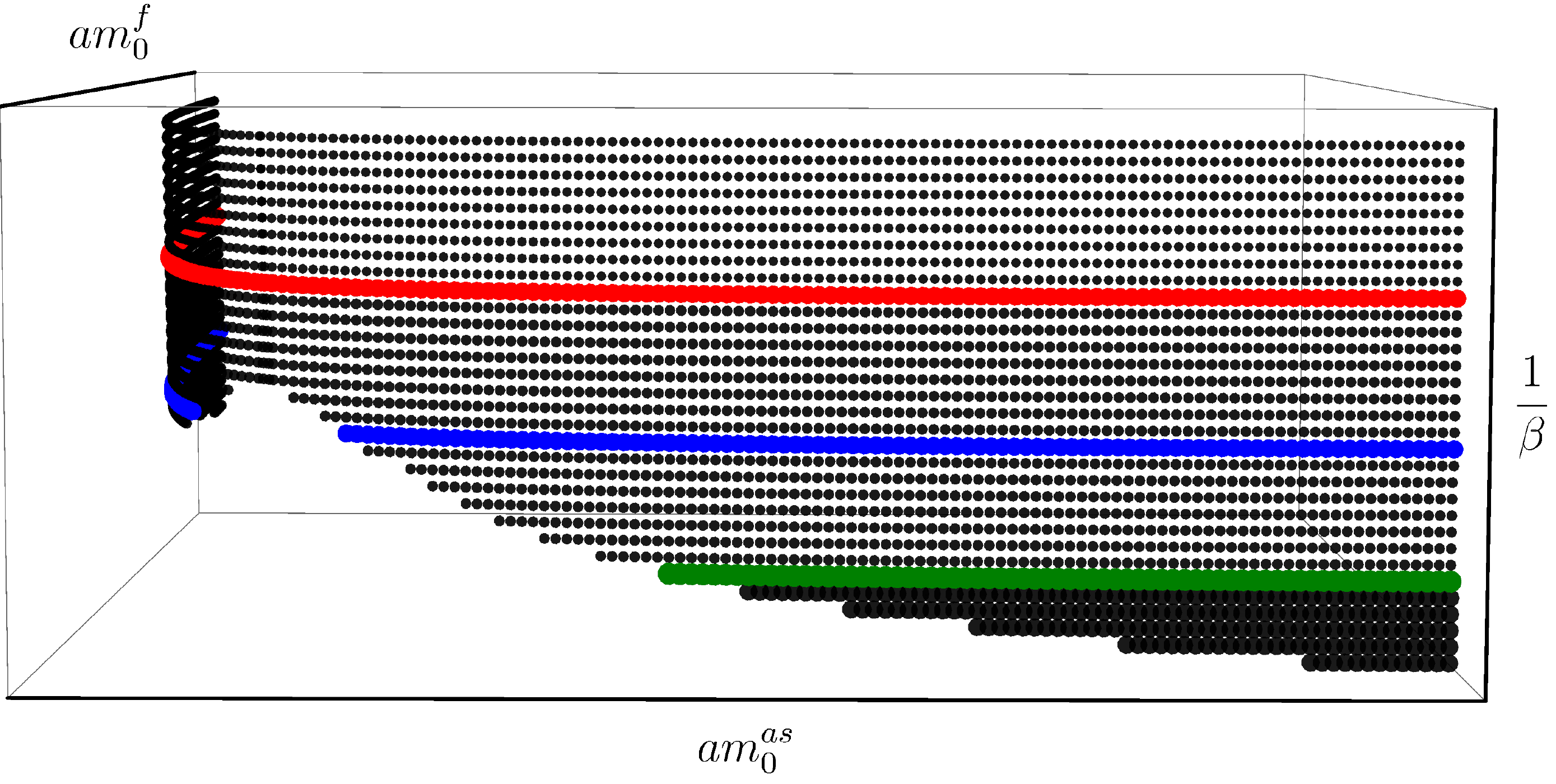}
\caption{%
Schematic diagram of the phase space of $Sp(4)$ lattice gauge theory with $N_f=2$ and $n_f=3$ Wilson Dirac fermions.
The black surface denotes the 1st order phase transition.
Coloured solid lines represent three distinct cases with fixed $\beta$ values:
the transition is always 1st order (red), becomes crossover for an interval with small masses (blue),
and is only 1st order with a large mass of antisymmetric fermions (green). 
The image is taken from Ref.~\cite{Bennett:2022yfa}.
}
\label{Fig:phase_diagram}
\end{center}
\end{figure}

The phase structure of $Sp(4)$ Yang-Mills has first been studied with (unimproved) Wilson plaquette action in Ref.~\cite{Holland:2003kg}, and later in Ref.~\cite{Bennett:2017kga}.
The bulk phase transition disappears above $\beta \gtrsim 7.5$. 
With degenerate fermions in a given representation the parameter space extends to a two-dimensional plane 
that can be scanned by measuring the average plaquette values. 
References~\cite{Bennett:2017kga,Lee:2018ztv} show that the bulk transition is of first order at strong coupling 
in the $Sp(4)$ theory with $N_f=2$ fundamental and $n_f=3$ antisymmetric Wilson-Dirac fermions, respectively. 
The weak coupling regime is $\beta \gtrsim 6.7$ for the former and $\beta \gtrsim 6.5$ for the latter. 
The critical beta value associated with the phase boundary decreases as more fermionic degrees of freedom are involved. 
Finally, the phase space of the $Sp(4)$ theory with fermions in both representations, two fundamental and three antisymmetric Dirac flavours, 
has been explored in Ref.~\cite{Bennett:2022yfa}---see Fig.~\ref{Fig:phase_diagram}---and
the weak coupling region  extends to smaller beta values  $\beta \gtrsim 6.3$. 
The infinite mass limit of either $a m_0^{(f)}$ or $a m_0^{(as)}$ recovers the phase structure of the theory with the same number of dynamical Dirac fermions 
in the fundamental or antisymmetric representation, which is asymmetric as represented by the green solid line in Fig.~\ref{Fig:phase_diagram}.

\begin{figure}[t]
\begin{center}
\includegraphics[width=.33\textwidth]{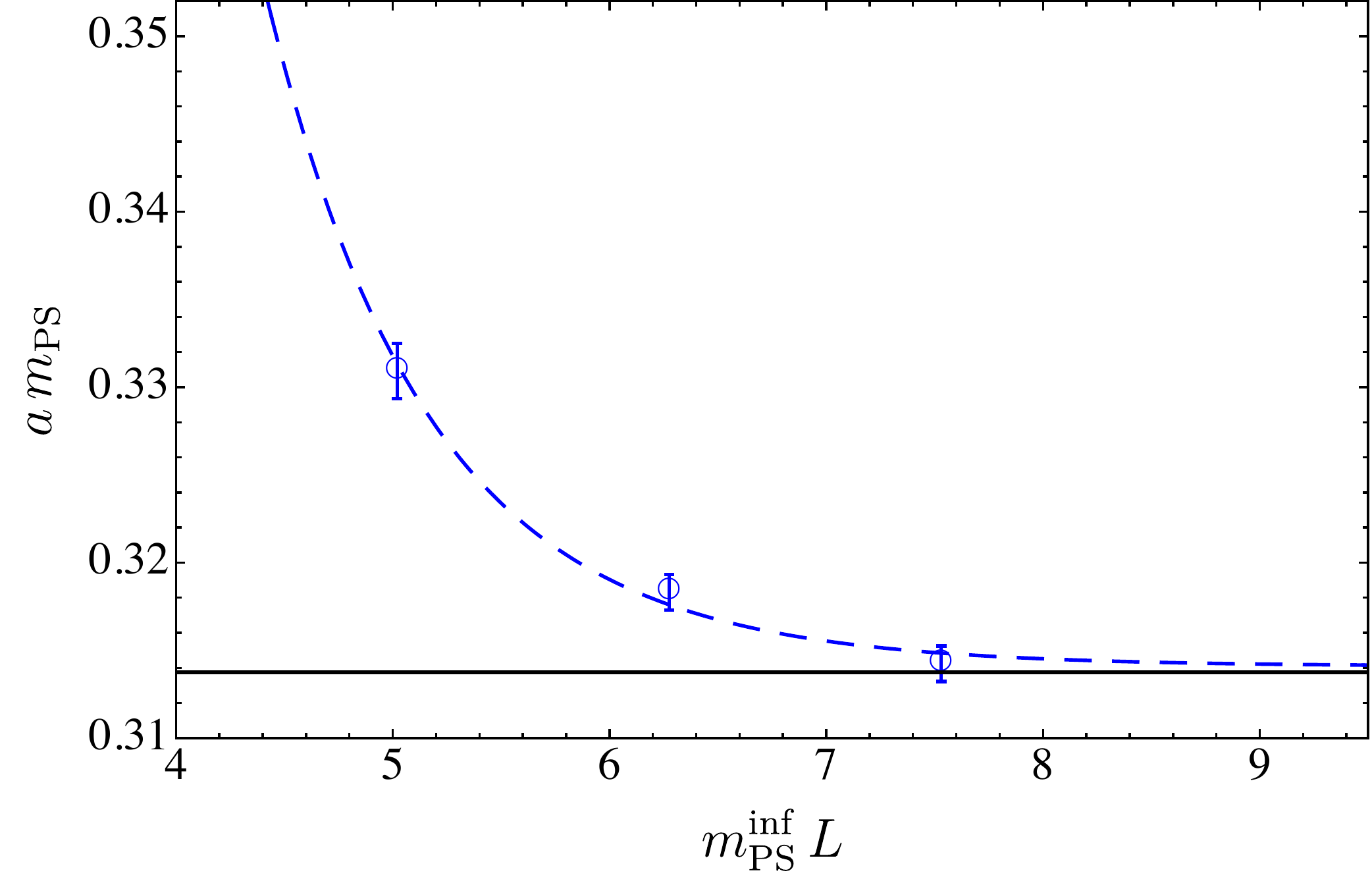}
\includegraphics[width=.33\textwidth]{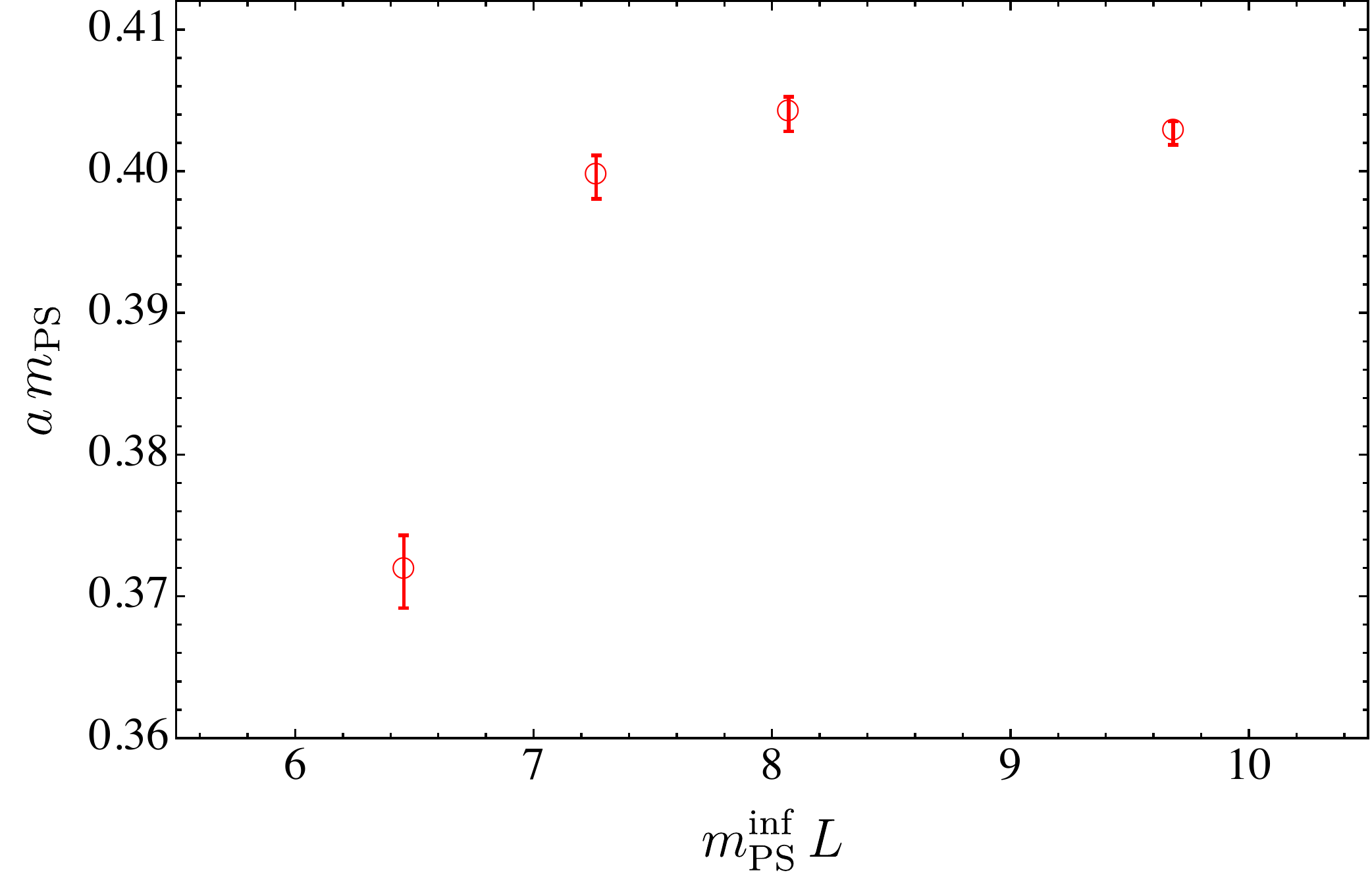}
\includegraphics[width=.33\textwidth]{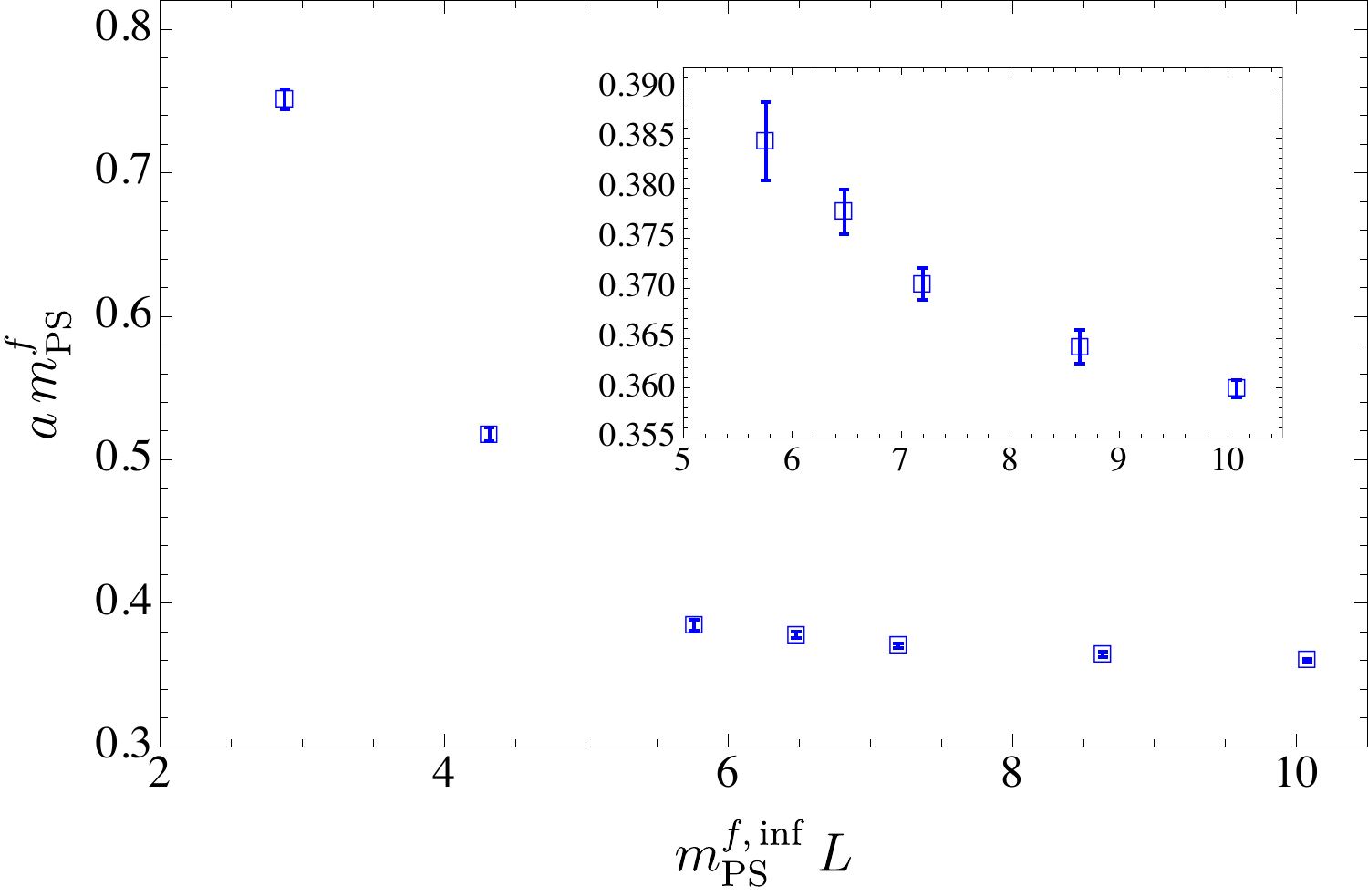}
\includegraphics[width=.33\textwidth]{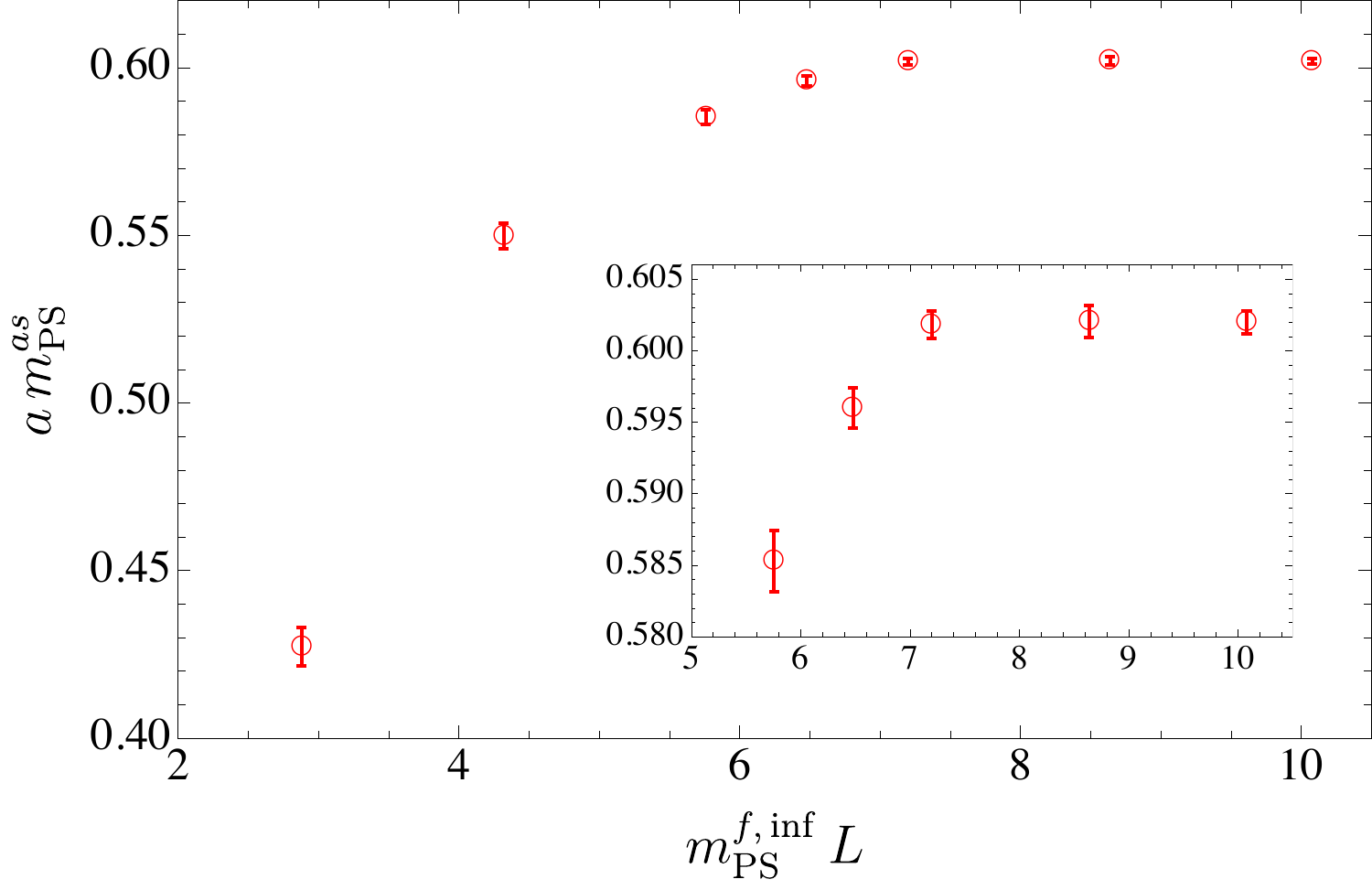}
\caption{%
Volume dependence of  pseudoscalar meson masses.
Top-left and top-right panels show the results for the $Sp(4)$ theories containing $N_f=2$ fundamental and $n_f=3$ antisymmetric fermions, respectively. 
Bottom-left and bottom-right  panels display the masses of the PS meson composed of fundamental and antisymmetric fermions, respectively,
but measured in the dynamical $Sp(4)$ theory containing both $N_f=2$ fundamental and $n_f=3$ antisymmetric fermions.
The lattice parameters are $\beta=7.2$, $a m_0^{(f)} = -0.79$ (top-left), $\beta=6.8$, $a m_0^{(as)}=-1.03$ (top-right),
and $\beta=6.5$, $a m_0^{(f)}=-0.71$, $a m_0^{(as)}=-1.01$ (bottom panels).
The mass in the infinite volume limit, $am_{\rm PS}^{\inf}$,
is estimated from  the largest available lattice, except for the top-left panel
in which it is determined from  infinite volume extrapolation (dashed line). 
Plot derived  from Refs.~\cite{Bennett:2019jzz,Bennett:2021mbw,Bennett:2022yfa}. 
}
\label{Fig:FV}
\end{center}
\end{figure}

Finite volume (FV) effects are an inherent source of systematic errors in lattice calculations. 
In confining theories,  they are expected to be exponentially suppressed 
or quantities that do not involve scattering states,
if the volume is larger than the longest (intrinsic) scale of the theory, 
e.g.\ the Compton wavelength of the lightest state---usually, the pseudoscalar meson, for which one requires $m_{\rm PS}^{\rm inf} \,L \gg 1$. 
To quantify the size of FV effects, we compute $m_{\rm PS}$ by varying the spacial lattice extent $L$ and 
investigate its  dependence on $m_{\rm PS}^{\rm inf} \,L$. 
Illustrative examples for different dynamical theories are shown in Fig.~\ref{Fig:FV}: 
the top-left and top-right panels are for the $Sp(4)$ theories with $N_f=2$ fundamental \cite{Bennett:2019jzz} and $n_f=3$ antisymmetric fermions \cite{Bennett:2021mbw}, respectively, 
while the bottom panels are for the theory with both $N_f=2$ fundamental and $n_f=3$ antisymmetric fermions \cite{Bennett:2022yfa}. 
We find that the FV effects can be safely neglected if the condition $m_{\rm PS}^{\rm inf} \,L \gtrsim 7.0$ is satisfied, 
except for $am_{\rm PS}^{f}$ in the two-representation theory, in which the condition becomes more stringent as $m_{\rm PS}^{\rm inf} \,L \gtrsim 8.5$. 
Such conditions are sufficient to ensure that FV effects are within a percent level.

We highlight  that the FV corrections to $am_{\rm PS}$ have  opposite sign for mesons composed of fundamental and of antisymmetric fermions. 
This  can be understood within the low-energy description of chiral perturbation theory ($\chi$PT), 
as  FV corrections are dominated by the contribution of PS states wrapping around each lattice spatial direction. 
The NLO expression of the PS mass squared at finite volume in the continuum theory is
\beqs
m_{\rm PS}^2= M^2\left(
1+a_M \frac{A(M)+A_{\rm FV}(M)}{F^2}+b_M(\mu)\frac{M^2}{F^2}+\mathcal{O}(M^4)
\right),
\eeqs
where $M$ and $F$ are the mass and decay constant of the PS meson defined at the leading order in the $\chi$PT, and $\mu$ is the renormalisation scale.  
$A(M)$ is the chiral logarithm arising from the one-loop integral at infinite volume, 
while $A_{\rm FV}(M)$ is the FV
 contribution obtained by replacing integrals with discrete sums on a cubic box of size $L$. 
 $A(M)$ and $A_{\rm FV}(M)$ are independent of the details of the theory, encoded in the coefficient $a_M$~\cite{Bijnens:2009qm}
\beqs
a_M = \begin{cases}
-\frac{1}{2}-\frac{1}{N_f}\,,~~&{\rm for}~SU(2N_f) \rightarrow Sp(2N_f)\,, \\
-\frac{1}{N_f}\,,~&{\rm for}~~SU(N_f)\times SU(N_f)\rightarrow SU(N_f)\,, \\
\frac{1}{2}-\frac{1}{2N_f}\,,~&{\rm for}~~SU(2N_f)\rightarrow SO(2N_f)\,.
\end{cases}
\label{eq:coeff_a}
\eeqs
For the two fundamental and three antisymmetric flavours, corresponding to the first and the third classes, one finds that $a_M=-1$ and $+1/3$. 
The resulting FV corrections would have an opposite sign and thus agree with our findings in Fig.~\ref{Fig:FV}.

\begin{figure}[!t]
    \centering
    \includegraphics[width=0.5\textwidth]{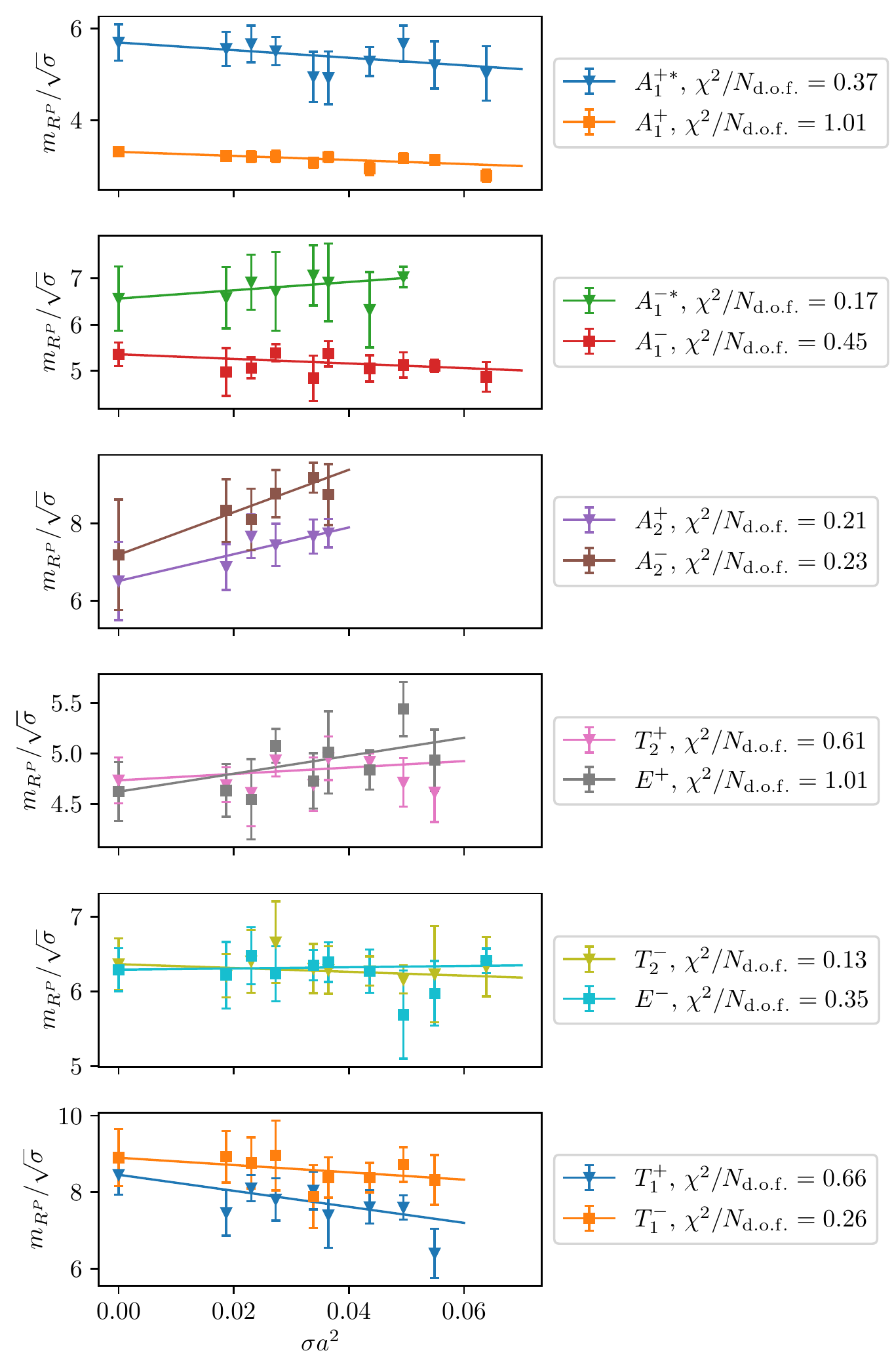}
      \caption{Glueball masses in units of $\sqrt{\sigma}$ in each channel 
      $R^P$ of the $ Sp(2N)$ theory with $N=4$, 
      as a function of $\sigma a^2$. The value at $\sigma a^2\rightarrow 0$ is obtained,
      for each symmetry channel $R^P$, by a likelihood analysis of the measurements with 
      Eq.~(\ref{eq:glue_cont_lim})---see the solid lines. The plots are taken from Ref.~\cite{Bennett:2020qtj}.}
    \label{fig:glue_extrapolation}
\end{figure}

\section{Numerical investigation I: Pure $Sp(2N)$}
\label{sec:puresp4}

We summarise in this section the main results 
for the measurement of physical observables obtained 
in $Sp(2N)$ lattice gauge theories in which only gauge dynamics is included in 
generating the ensembles.
Section~\ref{sec:glueball} focuses on string tension
and glueball masses~\cite{Bennett:2017kga,Bennett:2020hqd,Bennett:2020qtj},
Sect.~\ref{sec:quenchedmeson} reports a selection of measurements of 
meson masses in the quenched approximation~\cite{Bennett:2017kga,Bennett:2019cxd},
and Sect.~\ref{sec:topology} reports
on the topological susceptibility of the $Sp(2N)$ theories~\cite{Bennett:2022gdz,Bennett:2022ftz}.
We only reproduce some illustrative examples, and refer the reader to the original publications
 for  more extensive selections of numerical results, and for technical details about the calculations.

\subsection{Glueballs and string tension}
\label{sec:glueball}

\begin{figure}[!t]
\includegraphics[width=0.6\textwidth]{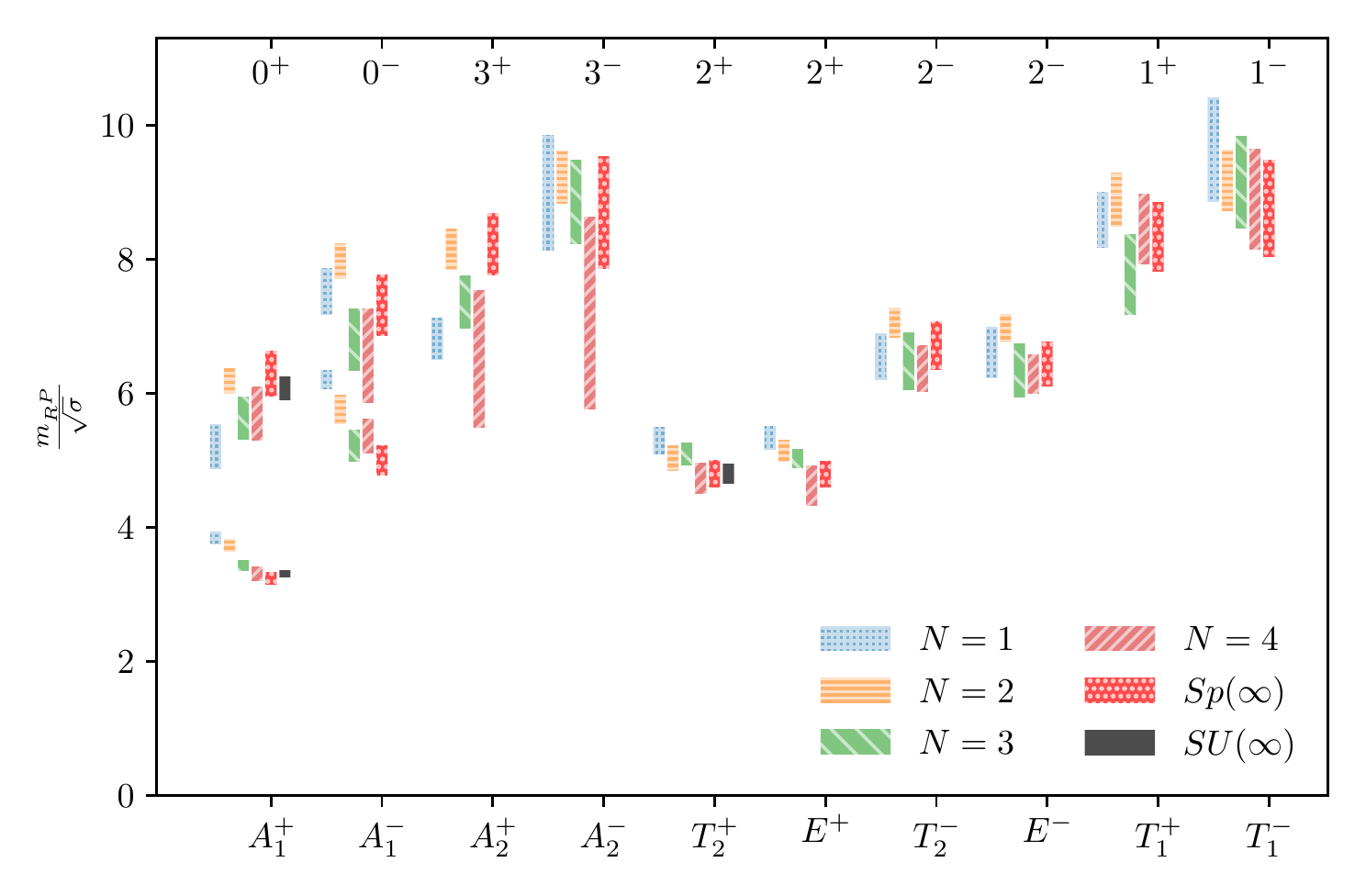}
\caption{Continuum limit of the  glueball spectrum of $Sp(2N)$ 
    gauge theories in units of $\sqrt{\sigma}$ for 
    $N=1,\,2,\,3,\,4$ and $N=\infty$, for each $R^P$ channel (bottom horizontal axis),
    and continuum channels (top horizontal axis). 
    The spectrum $A_1^{++}$, $A_1^{++*}$, and $E^{++}$ states for
    $SU(\infty)$ is also reported, for comparison~\cite{Lucini:2004my}. 
    The boxes represent 1$\sigma$ statistical errors. The plot is taken from Ref.~\cite{Bennett:2020qtj}.}
    \label{fig:glueballs_spectrum}
\end{figure}

\begin{specialtable}[!b]
\caption{Continuum limit extrapolations of ${m_{R^P}}/{\sqrt{\sigma}}$. 
  For $N=2$, these values are the
  weighted average of
  those in Ref.~\cite{Bennett:2017kga} and Ref.~\cite{Bennett:2020qtj}. 
  In the case of $SU(N_c \to \infty)$, we have $m/\sqrt{\sigma}=3.307(53)$ 
  for the $A_1^{++}$ channel, $6.07(17)$
  for the $A_1^{++*}$ channel, and $4.80(14)$ for the $E^{++}$ channel~\cite{Lucini:2004my}.
The table is taken from Ref.~\cite{Bennett:2020qtj}.} 
\label{tab:glue_cont_spectrum}
\centering
\begin{tabular}{|c|c|c|c|c|c|} 
\hline
& $1$ & $2$ & $3$ & $4$ & $\infty$ \\
\hline
$R^P$ & $m_{R^P}/\sqrt{\sigma}$ &  $m_{R^P}/\sqrt{\sigma}$ & $m_{R^P}/\sqrt{\sigma}$ & $m_{R^P}/\sqrt{\sigma}$ & $m_{R^P}/\sqrt{\sigma}$ \\
\hline
$A_1^+$  & $ 3.841(84)$  & $ 3.577(49)      $  &  $ 3.430(75)$  & $ 3.308(98)$ 	&  $ 3.241(88)$   \\
$A_1^{+*}$ & $ 5.22(33) $  & $ 6.049(40)      $  &  $ 5.63(32) $  & $ 5.58(44)$ &  $ 6.29(33) $   \\
$A_1^-$  & $ 6.20(14) $  & $ 5.69(16)       $  &  $ 5.22(23) $  & $ 5.36(26) $ 	&  $ 5.00(22) $   \\
$A_1^{-*}$  & $ 7.37(72) $  & $ 7.809(79)      $  &  $ 6.59(49) $  & $7.76(85)$ &  $ 7.31(45) $   \\
$A_2^+$  & $ 6.81(31) $  & $ 7.91(16)       $  &  $ 7.36(39) $  & $ 6.5(1.0) $ 	&  $ 8.22(46) $   \\
$A_2^-$  & $ 8.99(86) $  & $ 9.30(38)       $  &  $ 8.60(67) $  & $ 7.2(1.4) $ 	&  $ 8.69(83) $   \\
$T_2^+$  & $ 5.29(20) $  & $ 5.050(88)      $  &  $ 5.09(16) $  & $ 4.73(23) $ 	&  $ 4.80(20) $   \\
$T_2^-$  & $ 6.55(34) $  & $ 6.879(88)      $ &  $ 6.47(43) $  & $ 6.36(35) $  	&  $ 6.71(35) $   \\
$E^+$   & $ 5.33(18)  $ & $ 5.05(13)       $  &  $ 5.03(13) $  & $ 4.62(29) $ 	&  $ 4.79(19) $   \\
$E^-$   & $ 6.61(37)  $ & $ 6.65(12)       $  &  $ 6.34(40) $  & $ 6.29(29) $ 	&  $ 6.44(33) $   \\
$T_1^+$  & $ 8.58(41) $  & $ 8.67(28)       $  &  $ 7.77(59) $  & $ 8.45(52) $ 	&  $8.33(51) $   \\
$T_1^-$  & $ 9.63(77) $  & $ 9.24(33)       $  &  $ 9.15(69) $  & $ 8.90(75) $ 	&  $8.76(72) $   \\
\hline
\end{tabular}
\end{specialtable}

Numerical results for glueballs and string tension are available 
for several values of the lattice spacing. 
For each $Sp(2N)$ group, and for each representation $R^P$, the extrapolation for 
the ratio $m_{R^P}/\sqrt{\sigma}$ is performed with the relation
\begin{equation}\label{eq:glue_cont_lim}
\frac{m_{R^P}}{\sqrt{\sigma}}(a) = \frac{m_{R^P}}{\sqrt{\sigma}} ( 1+ c_{R^P} \sigma a^2 )\,.
\end{equation}
The leading-order linear behaviour
in $a^2$ in Eq.~(\ref{eq:glue_cont_lim}) describes the data well for all channels,
as attested by the values of the $\chi^2/N_\mathrm{d.o.f.}$ reported in the figure.
As an example, Figure~\ref{fig:glue_extrapolation} shows the extrapolations to the continuum limit for all the 
channels in the case $N=4$. 
Similar results are obtained for $N=1$, $2$, and $3$~\cite{Bennett:2020qtj}.
The values of the masses in the spectrum extrapolated to the continuum limit
are reported in Table~\ref{tab:glue_cont_spectrum}, and displayed in 
Fig.~\ref{fig:glueballs_spectrum}.
The masses in the $E^\pm$ and $T_2^\pm$ channels have 
degenerate continuum limit, as expected by  rotational 
invariance. Because the masses are degenerate even
at non-zero values of $a$, we infer that discretisation effects are small in all
the ensembles. 
The lightest glueball states in the spectrum are found in the channels 
$0^+$, $2^+$, and $0^-$, for every value of $N$, consistently with
the pattern observed in gauge theories
with $SU(N_c)$ groups~\cite{Lucini:2010nv}.

\begin{figure}[!t]
    \centering
    \includegraphics[width=0.5\textwidth]{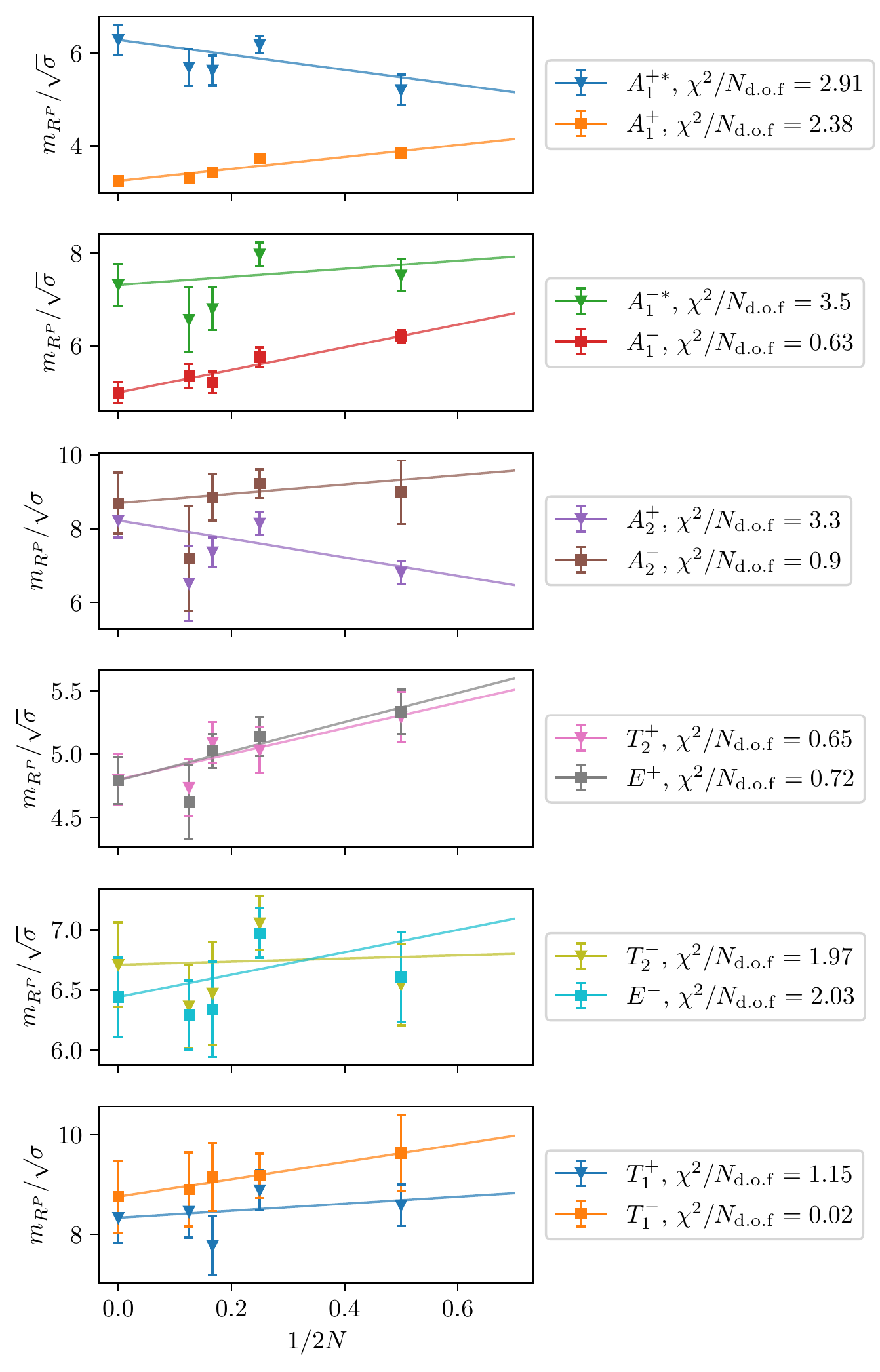}
    \caption{Glueballs masses in units of $\sqrt{\sigma}$ for
    each channel $R^P$, as a function of $1/2N$.
    The value of $m_{R^P}/\sqrt{\sigma}(\infty)$ is  obtained from the best 
    fit of Eq.~(\ref{eq:glueballs_large_N}) to the numerical measurements
      at fixed $N$. The plots are taken from Ref.~\cite{Bennett:2020qtj}.}
    \label{fig:glueballs_large_N}
\end{figure}

The leading-order, finite-$N$ correction to  glueball masses 
near the $N\to\infty$ limit is 
\begin{equation}\label{eq:glueballs_large_N}
\frac{m_{R^P}}{\sqrt{\sigma}}(N) = \frac{m_{R^P}}{\sqrt{\sigma}}(\infty) +  \frac{c_{R^P}}{N}\,,
\end{equation}
and is used to perform the large-$N$ limit extrapolation
in each $R^P$ channel. The results are displayed in Fig.~\ref{fig:glueballs_large_N},
 for all symmetry channels. The numerical results
are also reported in the last column of Table~\ref{tab:glue_cont_spectrum}.

Figure~\ref{fig:glueballs_spectrum} displays also
large-$N_c$ extrapolations in the $SU(N_c)$ family of gauge groups~\cite{Lucini:2004my},
for comparison, showing the compatibility of the results obtained for the two different group sequences. This is in agreement with the expectation that, in the
large-$N$ limit, the gauge theories based on 
the $Sp(2N)$ and the $SU(N_c)$ families of groups agree in their common sector.

\begin{figure}[!t]
\centering
\includegraphics[width=0.55\textwidth]{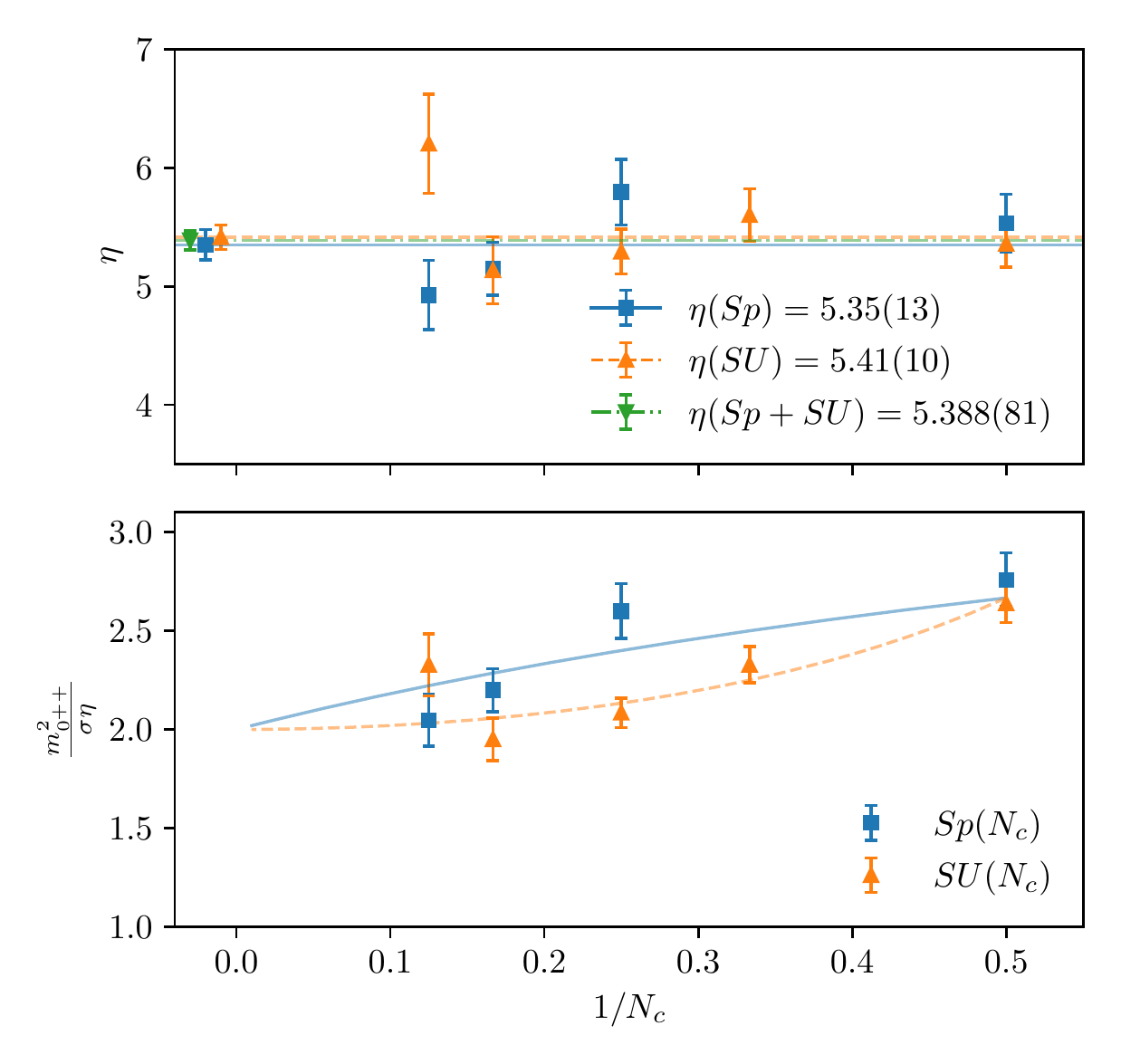}
\caption{Top panel: the ratio $\eta$ for  $SU(N_c)$ 
and $Sp(N_c=2N)$ theories in $d=3+1$ space-time dimensions. 
Fits of $\eta$ are  shown for 
the $Sp(N_c)$ family, the $SU(N_c)$ family and 
their combination. 
 Bottom panel:  ratio
$m^2_{0^{++}}/\sigma \eta$,	plotted as 
a function of $1/N_c$ in $d=3+1$; 
the lines are the ratios of the
quadratic Casimir operators, ${C_2(A)}/{C_2(F)}$, of the adjoint 
representation over the	corresponding ones 
of the fundamental representation. The 
plots are taken from Ref.~\cite{Bennett:2020qtj}.
}  
\label{fig:eta}
\end{figure}

\begin{figure}[!h]
    \centering
    \includegraphics[width=0.45\textwidth]{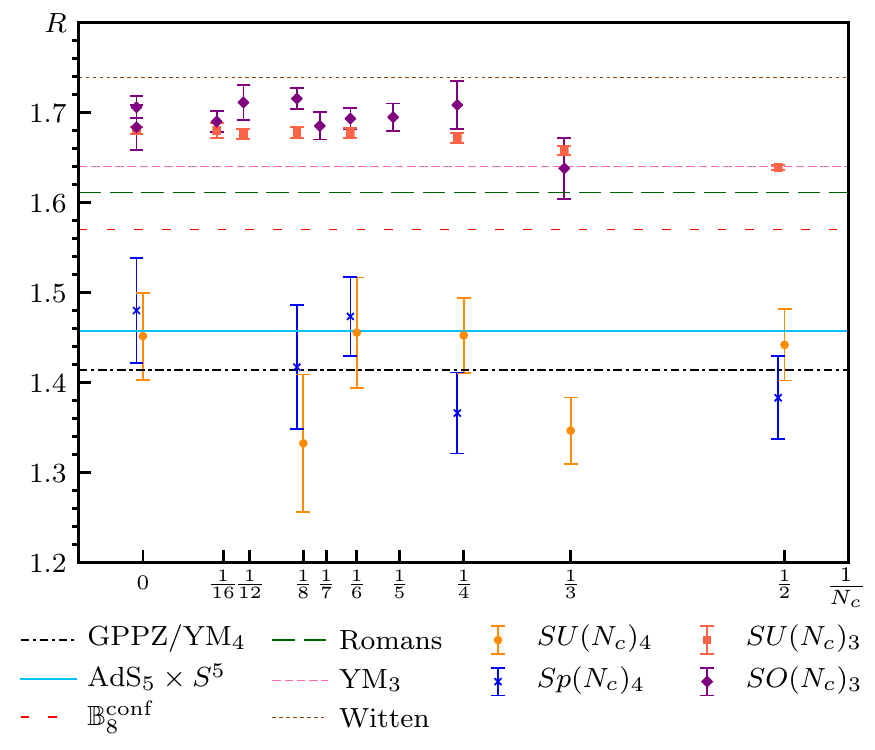}
    \caption{Numerical and semi-analytical results for the ratio $R$.     Different
 markers denote lattice continuum extrapolations in $3+1$ dimensions for $Sp(N_c)$  
 and $SU(N_c)$~\cite{Lucini:2004my}, as well as in $2+1$ dimensions for 
 $SO(N_c)$~\cite{Lau:2017aom} and $SU(N_c)$~\cite{Athenodorou:2016ebg}. 
 Extrapolations to the $N_c\rightarrow \infty$ limit are also included. 
 Differently rendered lines at $R=\sqrt{2}, 1.46, 1.57,  1.61,  1.74$, are the holographic 
 calculations in the GPPZ model~\cite{Mueck:2004qg}, the circle reduction of AdS$_5\times S^5$~\cite{Brower:2000rp,Elander:2020csd}, 
 the holographic model $\mathbb{B}_8^{\rm conf}$ in Ref.~\cite{Elander:2018gte},
the Witten model~\cite{Brower:2000rp,Elander:2013jqa}, and
the circle reduction of Romans supergravity~\cite{Wen:2004qh,Elander:2013jqa}, respectively. 
 With $R=\sqrt{2}, 1.64$ we report the field theoretical results from 
 Refs.~\cite{Bochicchio:2013sra} and~\cite{Leigh:2006vg}, for YM theories in $3+1$ and $2+1$ dimensions, respectively. 
    The plot is taken from Ref.~\cite{Bennett:2020hqd}.}
    \label{fig:RYangMills}
\end{figure}

Measurements of glueball masses in $Sp(2N)$
and $SU(N_c)$ gauge theories can be used to test conjectured
universal  behaviours in Yang-Mills theories. 
We mention two such tests here, 
referring to  Ref.~\cite{Hong:2017suj} and 
Ref.~\cite{Bennett:2020hqd} for details.
Ref.~\cite{Hong:2017suj} suggested that the ratio of the mass of the lightest $0^{++}$ glueball to the string tension,
normalised to the ratio ${C_2(F)}/{C_2(A)}$ of the Casimir operator for the fundamental ($F$) and adjoint ($A$) representation, 
might be a universal quantity in Yang-Mills theories, denoted as $\eta$, dependent only on the space-time dimensionality.
By fitting a constant to the numerical results for Yang-Mills theories in
 $d=2+1$ and $d=3+1$ dimensions~\cite{Bennett:2020qtj,Hong:2017suj} yields:
\begin{equation}\label{eq:eta}
\eta \equiv \frac{m^2_{0^{++}}}{\sigma} \frac{C_2(F)}{C_2(A)}
=\begin{cases}
               5.388(81)(60),\quad (d= 3 + 1)\\
               8.440(14)(76),\quad (d= 2 + 1)
            \end{cases}
~.
\end{equation}
Here, first and second parentheses denote  statistical 
and systematic uncertainties, respectively,  the latter estimated as the difference between 
the two sequences of gauge groups. 
Although none
of them is conclusive, several arguments, based on 
Bethe-Salpeter equations, scale anomaly, and sum rules,
might be able to explain the striking agreement between this conjecture and numerical
results displayed in 
Fig.~\ref{fig:eta} for $d=3+1$ dimensions---see Ref.~\cite{Hong:2017suj} for similar results in $d=2+1$ dimensions . In a similar spirit, we borrow  Fig.~\ref{fig:RYangMills} from
Ref.~\cite{Bennett:2020hqd}, to highlight
a regular pattern in the ratio $R\equiv\frac{m_{2^{++}}}{m_{0^{++}}}$, a quantity that can also 
be compared with a plethora of predictions obtained with non-perturbative instruments alternative to 
lattice techniques.

\subsection{Quenched mesons}
\label{sec:quenchedmeson}

\begin{figure}[t]
\begin{center}
\includegraphics[width=.35\textwidth]{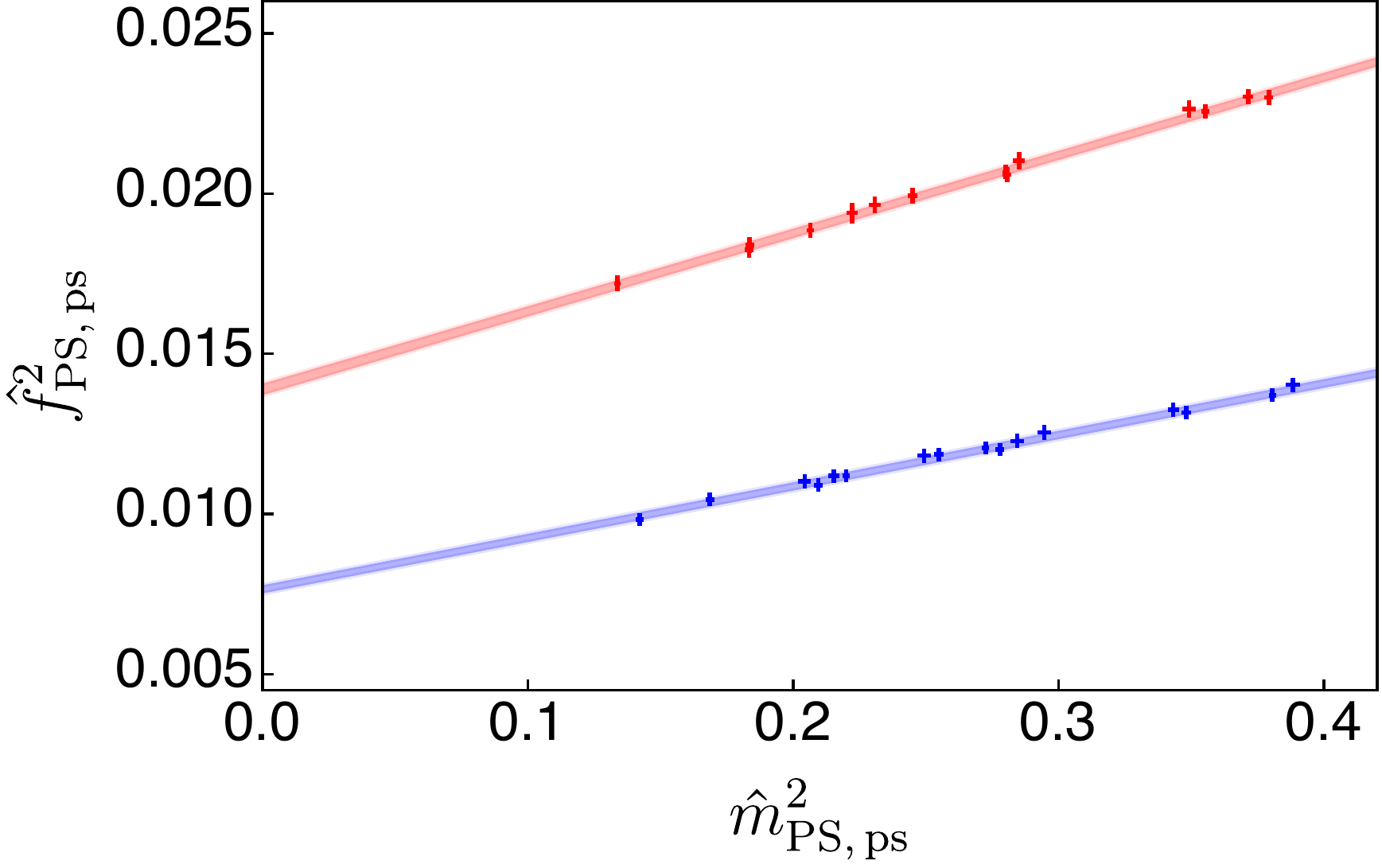}
\includegraphics[width=.35\textwidth]{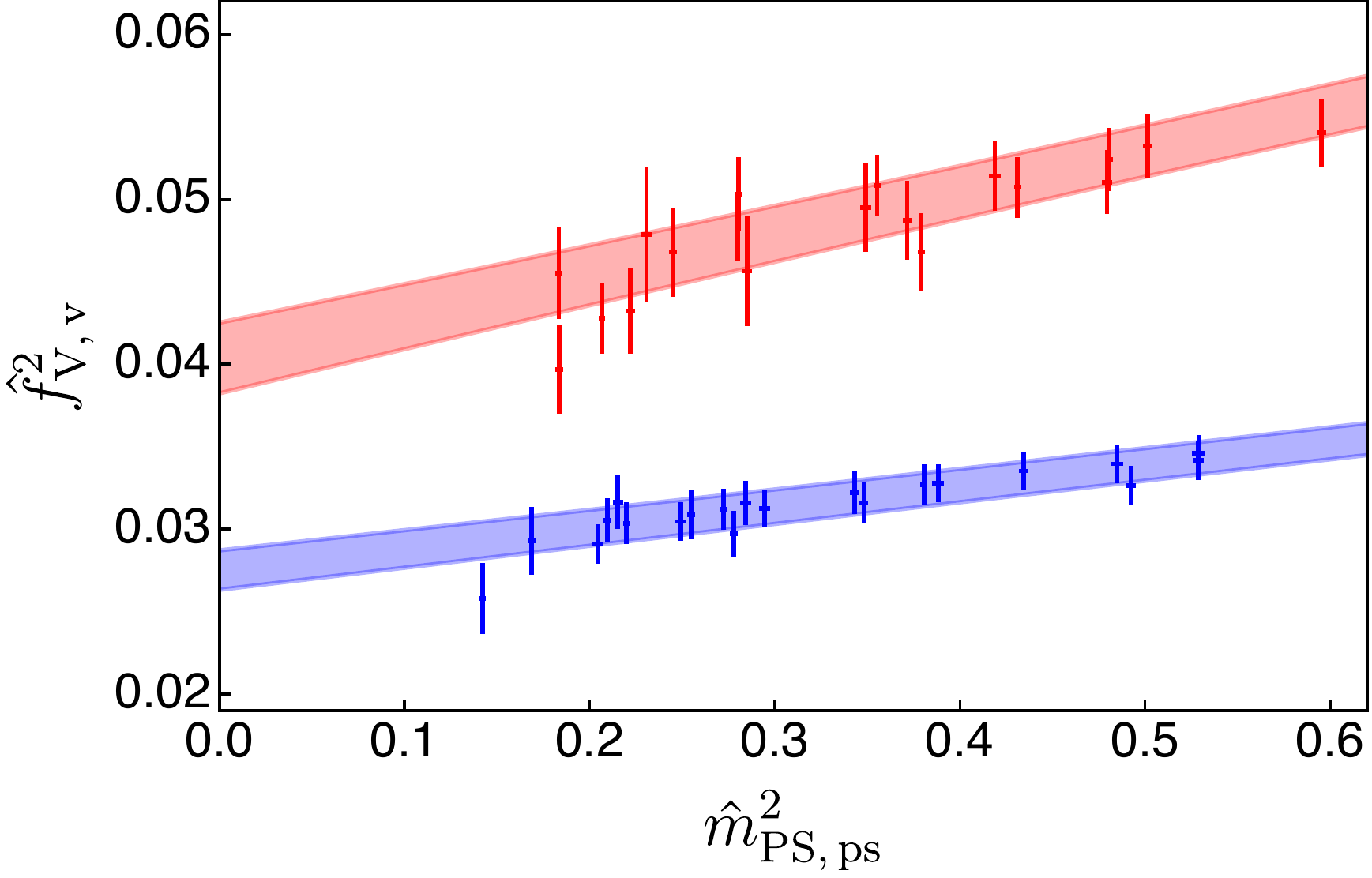}
\includegraphics[width=.35\textwidth]{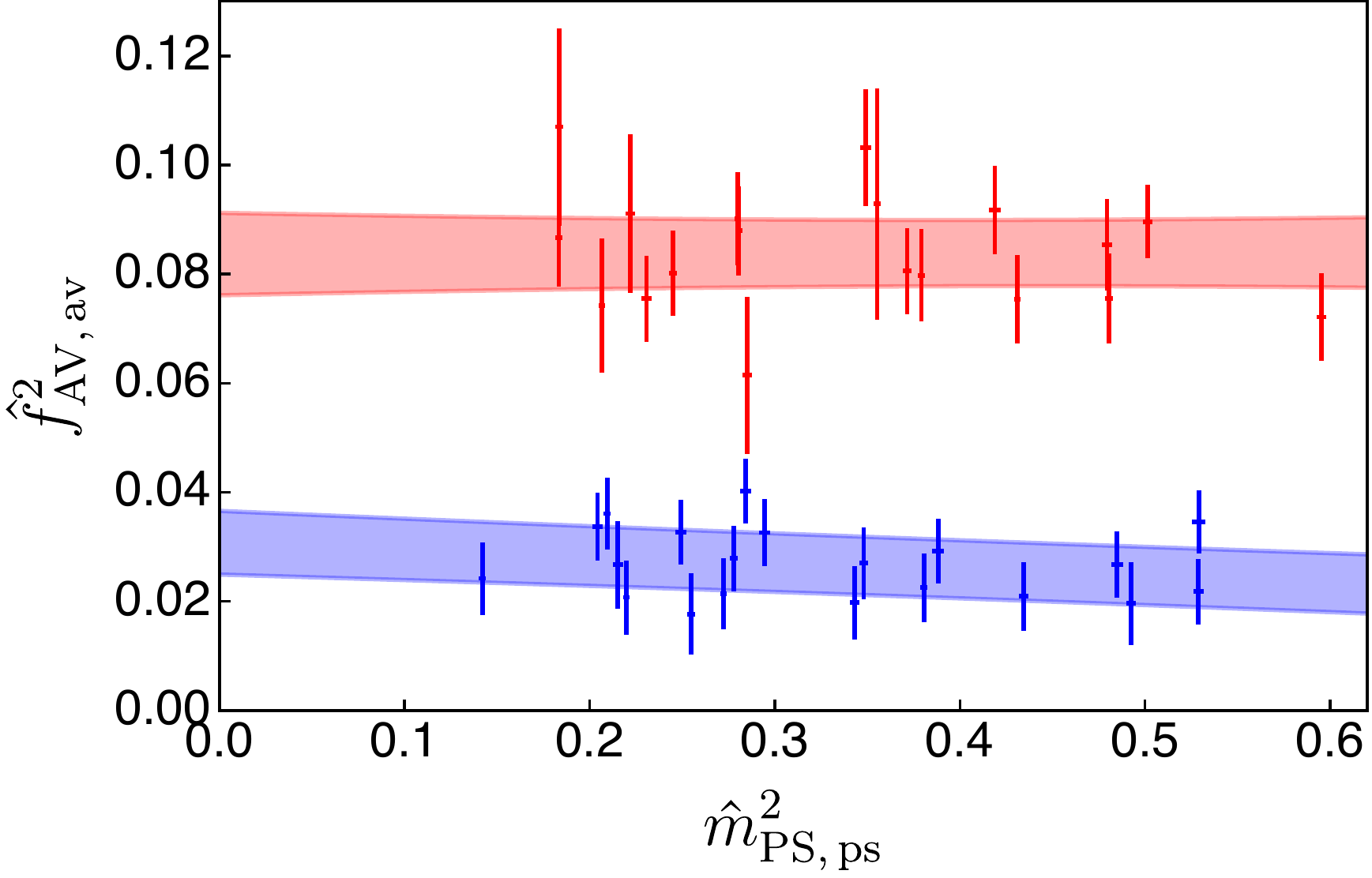}
\caption{%
Decay constant squared, as a function of the pseudoscalar meson mass squared, of flavoured mesons
composed of fermion constituents in the fundamental (blue) and antisymmetric (red) representations in the quenched approximation.  The plots are taken from Ref.~\cite{Bennett:2019cxd}.
}
\label{Fig:Qf}
\end{center}
\end{figure}

The first step in the study of any new gauge theory with fermion matter content
 is the measurement of the spectrum of mesons 
in the quenched approximation, as it sets 
a reference framework for subsequent dynamical fermion simulations.
Furthermore,  this exercise already provides useful information in the mass regime
that is  interesting for model-building purposes; for example, both for CHMs and for SIMPs
based on the $SU(4)/Sp(4)$ coset, which are microscopically realised by $Sp(2N)$ theories with $N_f=2$
fundamental fermions, the masses of the underlying fermions
are not small, so that the quenched approximation already provides useful estimates of 
the meson spectrum, which can then be refined with dynamical simulations.
Ref.~\cite{Bennett:2019cxd} performed the quenched analysis for $Sp(4)$, restricted to flavored mesons,
both for fermions transforming in the fundamental 
as well as the 2-index antisymmetric representations.
Further research will extend these studies in the future, 
by performing the calculations for chimera baryons composed of fermions in these two representations~\cite{Hsiao:2022kxf}, 
as well as considering mesons composed of fermions in the symmetric representation, 
and finally  by extending the study  to theories with larger groups~\cite{Bennett:2021mbw}. 

A complete description of the ensembles, and the measurements they are used for,
 can be found in Ref.~\cite{Bennett:2019cxd}.  $200$ thermalised configurations are 
 generated for each value of the coupling
 used for the glueballs in Ref.~\cite{Bennett:2020qtj}, $\beta=7.62,\, 7.7,\, 7.85,\, 8.0,\,8.2$, but 
on larger lattices, with $N_t\times N_s^3=48\times 24^3$ for $\beta=7.62$,
and  $N_t\times N_s^3=60\times 48^3$ for the other ensembles. 
In order to ensure that finite-volume effects are negligible, in comparison with statistical uncertainties,
the fermion masses in the propagators are chosen so that $m_{\rm PS, \,ps} L \gtrsim 7.5$. 
By inspection, one finds that  $f_{\rm PS} L \gtrsim 1.6$ and $f_{\rm ps} L \gtrsim 2.3$ for the fundamental and antisymmetric representation fermions, respectively, large enough to ensure applicability of the Chiral Perturbation Theory.
All the measurements have $m_{\rm V, \,v}/m_{\rm PS, \,ps} < 2$, so that the vector bound states cannot decay.

\begin{figure}[t]
\begin{center}
\includegraphics[width=.35\textwidth]{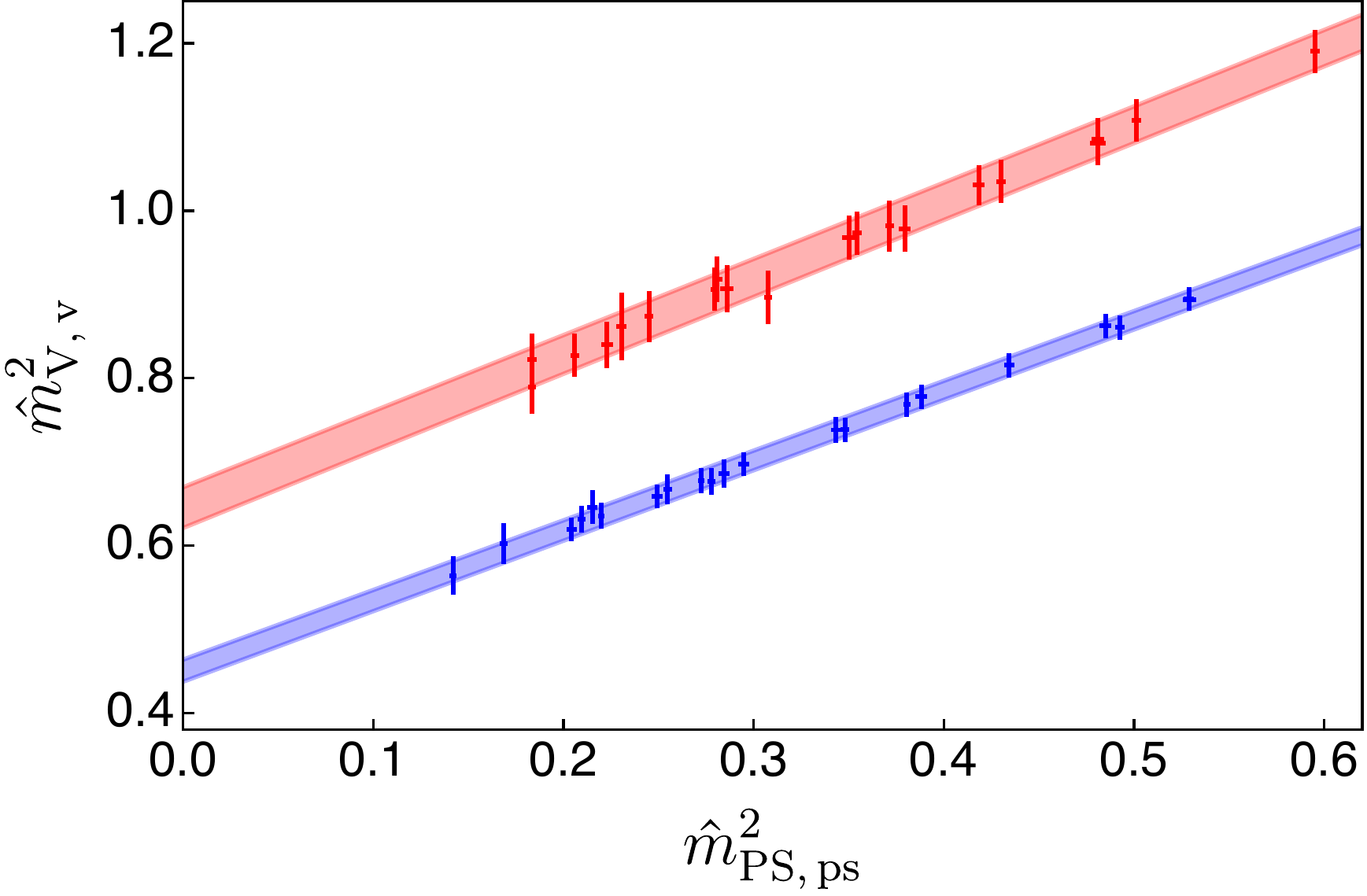}
\includegraphics[width=.35\textwidth]{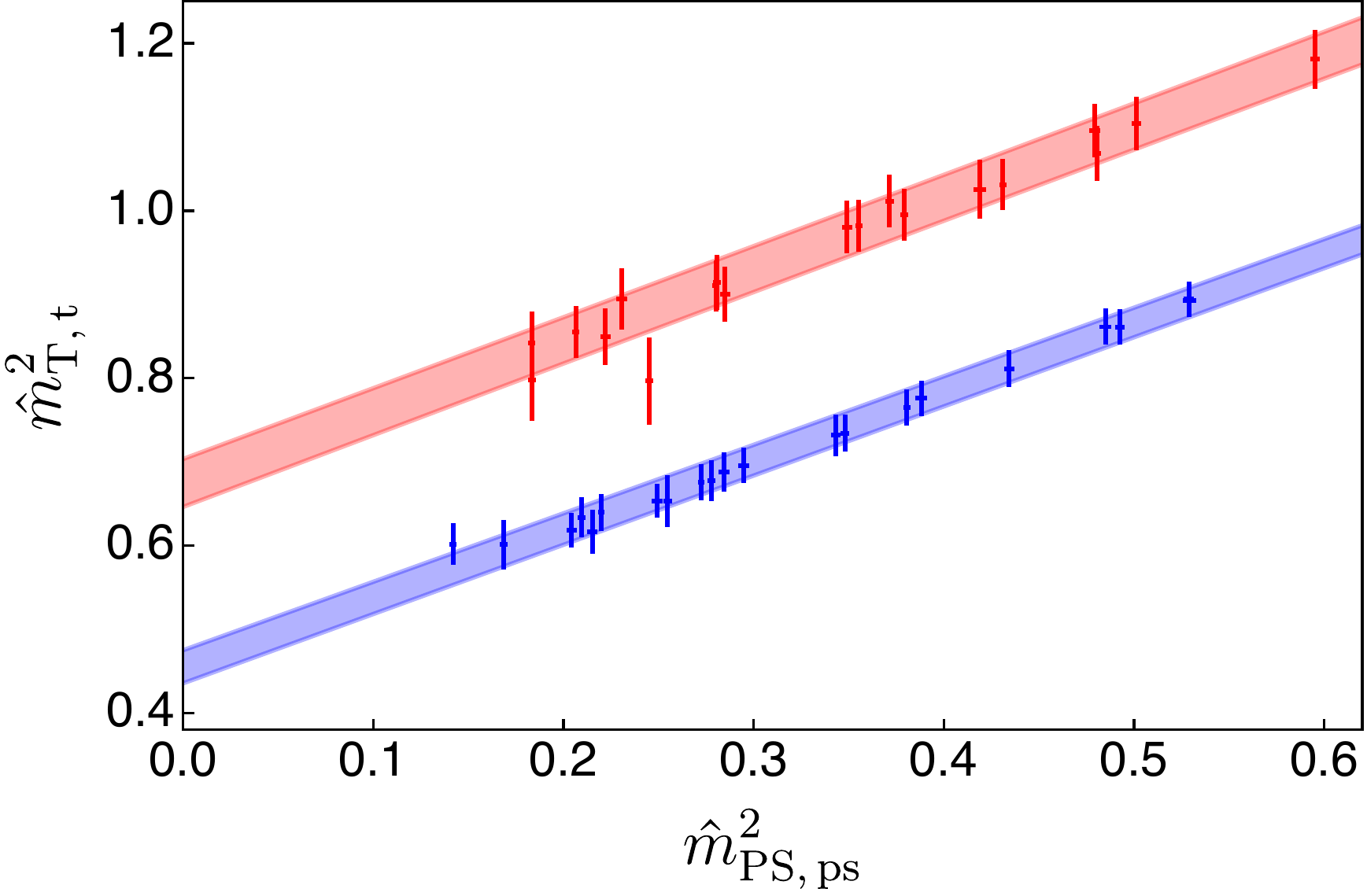}
\includegraphics[width=.35\textwidth]{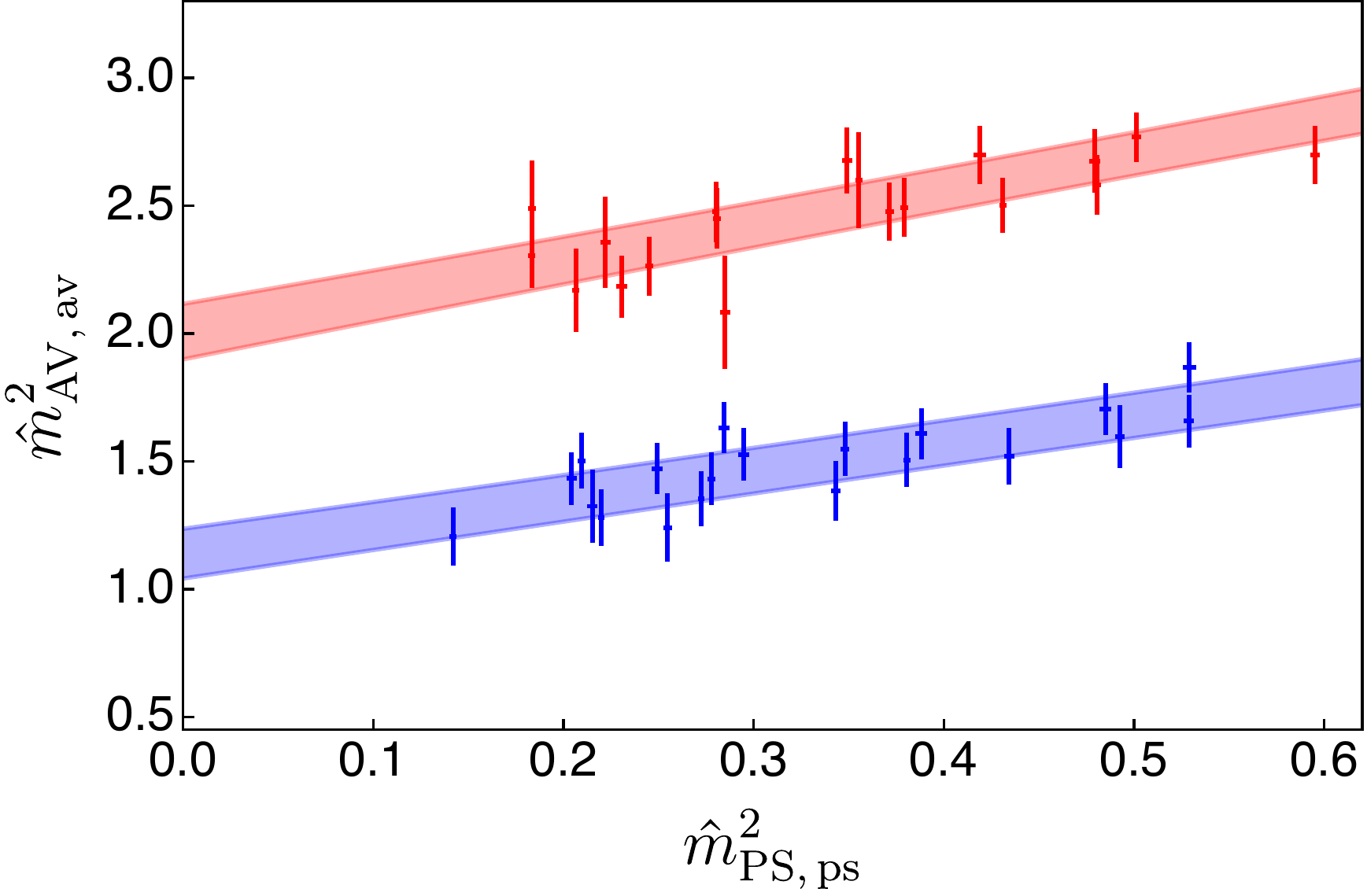}
\includegraphics[width=.35\textwidth]{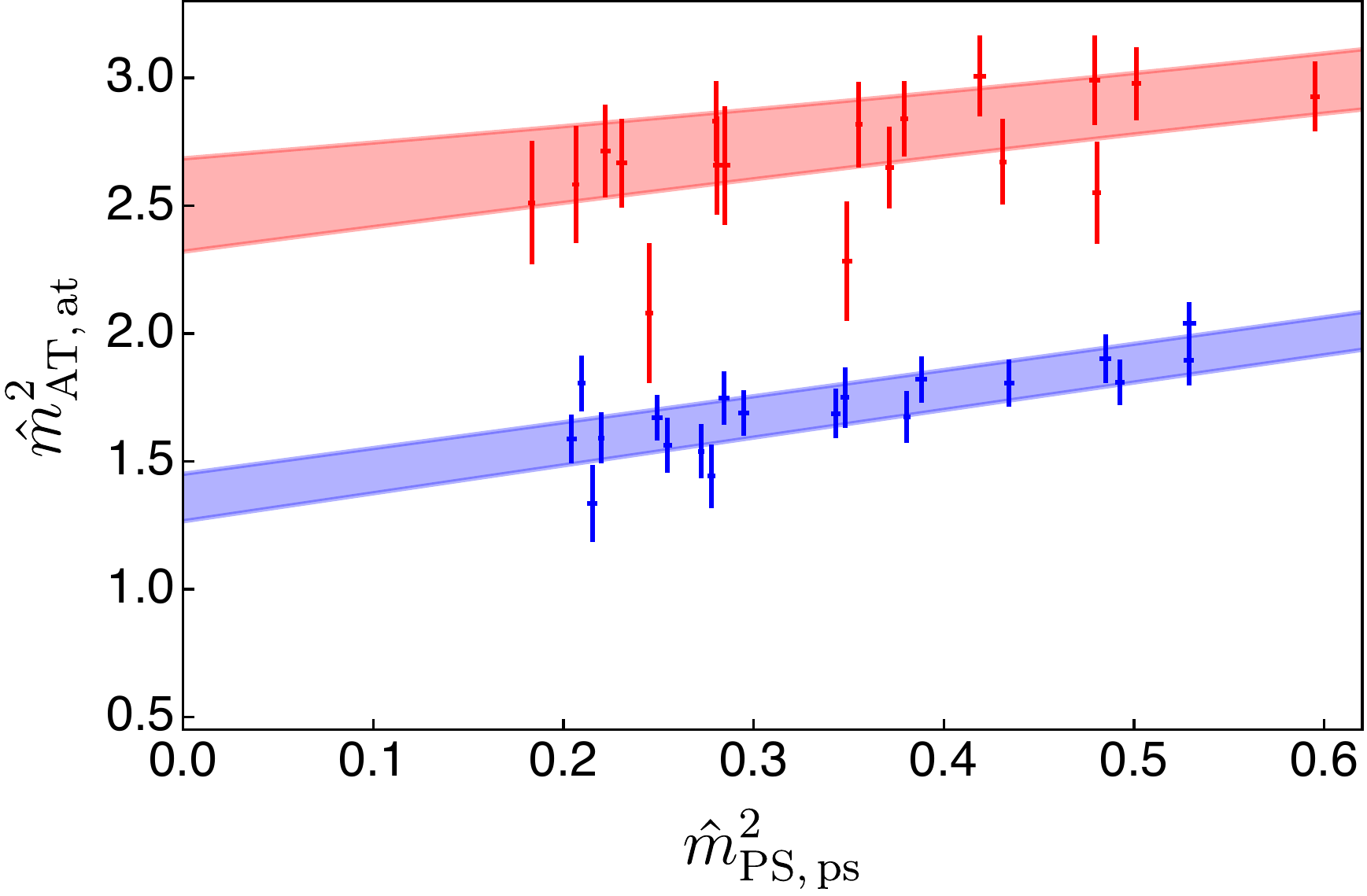}
\includegraphics[width=.35\textwidth]{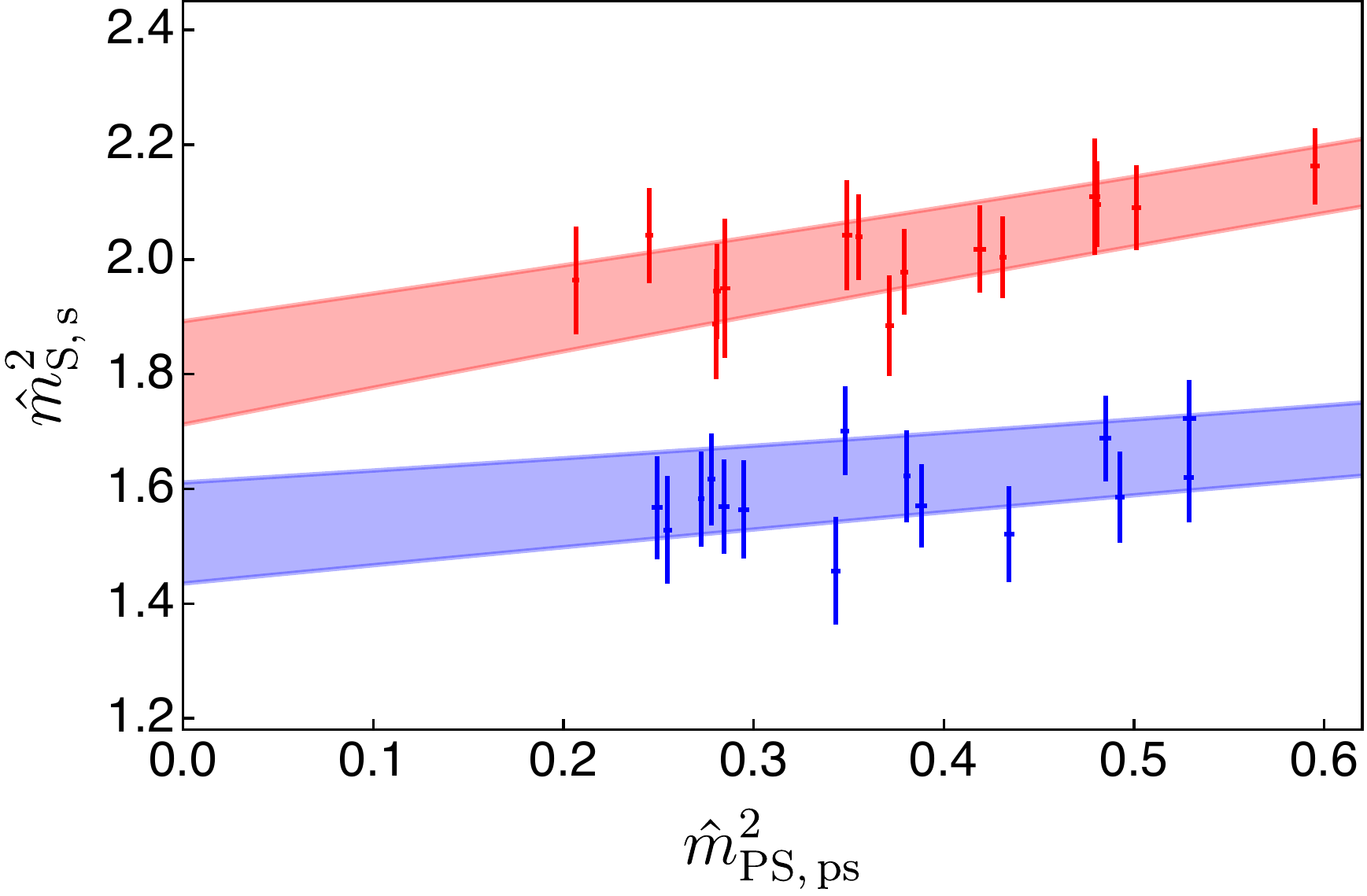}
\caption{%
Mass squared of flavour non-singlet mesons, as a function of the pseudoscalar meson mass squared, 
composed of fermion constituents in the fundamental (blue) and antisymmetric (red) representations in the quenched approximation. The plots are taken from Ref.~\cite{Bennett:2019cxd}.
}
\label{Fig:Qm}
\end{center}
\end{figure}

Figures~\ref{Fig:Qf} and~\ref{Fig:Qm} are taken from Ref.~\cite{Bennett:2019cxd}, 
and show the massless and continuum extrapolations of the lattice measurements of the flavoured-meson decay constants and the masses, respectively.  
Lattice measurements  are combined by making use of a  double expansion in small $\hat{m}_{\rm PS}^2$ and $\hat{a}$---we recall that the hatted notation uses  the gradient flow scale $w_0$, as discussed in Sect.~\ref{sec:systematics}, so that $\hat{m}\equiv m\, w_0$, 
for example---by adopting tree-level NLO Wilson chiral perturbation theory (W$\chi$PT)~\cite{Sheikholeslami:1985ij,Rupak:2002sm} (see also Ref.~\cite{Sharpe:1998xm}, and 
Refs.~\cite{Symanzik:1983dc,Luscher:1996sc} on improvement),
and writing 
\beqs
\hat{f}^{2,{\textrm{NLO}}}_M &\equiv& \hat{f}^{2,\chi}_M (1+L^0_{f,M}\hat{m}^2_{\textrm{PS}} ) + W^0_{f,M} \hat{a}\,, 
\label{eq:f_nlo}
\\
\hat{m}^{2,{\textrm{NLO}}}_M &\equiv& \hat{m}^{2,\chi}_M (1+L^0_{m,M}\hat{m}^2_{\textrm{PS}} ) + W^0_{m,M} \hat{a}\,, 
\label{eq:m_nlo}
\eeqs
where $\hat{f}^\chi$ and $\hat{m}^\chi$ are the decay constant and the mass in the chiral limit, while
 $L^0$ and $W^0$ are  low-energy constants to be determined from the fits to the numerical data. 
Implicit in this formalism is the replacement of the pseudoscalar mass squared for the fermion mass,
which is justified at this order of the chiral expansion, as long as the relation  $\hat{m}_{\rm PS}^2=2B \hat{m}_f$ holds.

In Figs.~\ref{Fig:Qf} and~\ref{Fig:Qm}, 
each data point has been obtained by subtracting the finite lattice-spacing correction from the raw data, 
and the bands denote the results of the fit obtained after removing the last terms in Eqs.~(\ref{eq:f_nlo}) and~(\ref{eq:m_nlo}). 
The width  of the bands represents the statistical uncertainties. 
With present accuracy, there is no evidence of corrections beyond linear order in $\hat{m}_{\rm PS,\,ps}^2$, 
to $\hat{f}_{\rm PS,\,ps}^2$ for $\hat{m}_{\rm PS,\,ps}^2 \lesssim 0.4$ and to all the other observables for $\hat{m}_{\rm PS,\,ps}^2 \lesssim 0.6$, in agreement with 
 Eqs.~(\ref{eq:f_nlo}) and~(\ref{eq:m_nlo}). 
The masses and decay constants of mesons composed of fermions transforming in the antisymmetric representation
are always larger than those of fundamental ones, 
for equal values of the pseudoscalar masses. 
A particularly important quantity in the context of CHM and top partial compositeness
 is the pseudoscalar decay constant, which shows the hierarchy $\hat{f}_{\rm ps}^2/\hat{f}_{\rm PS}^2=1.81(4)$,
 in the massless limit.
The masses of vector and tensor mesons are consistent with each other, 
as the two channels contain the same states, although  these two
measurements are affected by comparatively large discretisation effects~\cite{Bennett:2019cxd}.

\begin{figure}[t]
\begin{center}
\includegraphics[width=.6\textwidth]{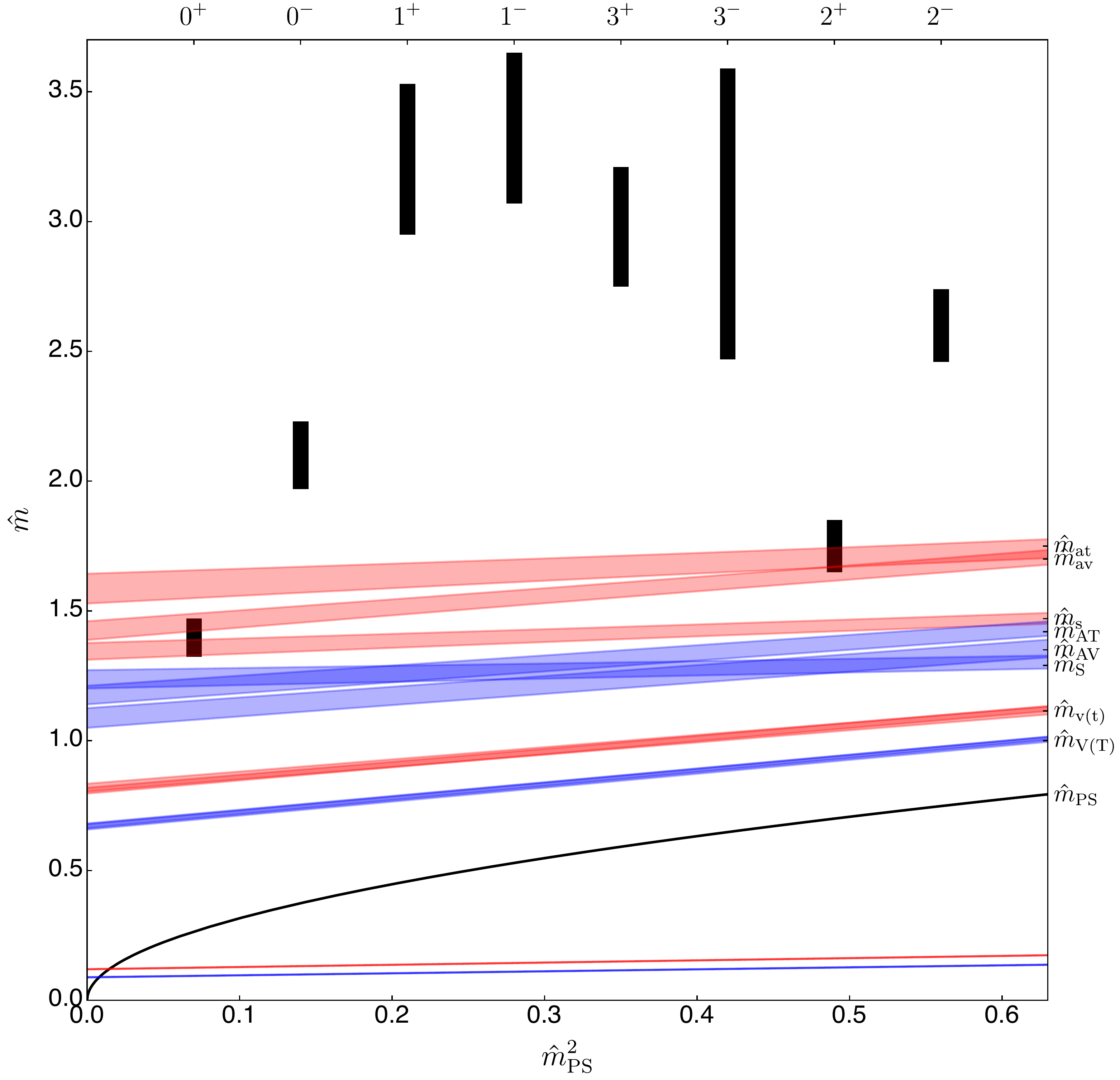}
\caption{%
A summary plot of quenched mass spectra of $Sp(4)$ gauge theory.
The red and blue colours denote the mesons composed of fermions in the
 fundamental and antisymmetric representations, respectively.
The black coloured data are for the glueballs in various channels classified by the quantum number $J^P$. 
The plot is taken from Ref.~\cite{Lucini:2021xke}.
}
\label{Fig:Qspec}
\end{center}
\end{figure}

Figure~\ref{Fig:Qspec} summarises the mass spectra of the ground 
states for (flavoured) mesons and glueballs in the quenched $Sp(4)$ theory.
The meson masses are shown as a function of the pseudoscalar mass squared, chosen to be the same
for the fundamental and antisymmetric representations. 
We also include the pseudoscalar decay constants, for completeness.
Glueball masses are denoted by their quantum numbers $J^P$.
As seen in the figure, the mass dependence of mesons in the two different representations are similar to each other,
but the antisymmetric ones are heavier than the fundamental ones, in all individual channels.
The lightest, $0^+$, glueball has  mass of the order of that of heavy mesons in the antisymmetric representation.

We anticipate, as a closing comment, some of the results of  Sect.~\ref{sec:fund}, 
obtained with dynamical calculations for mesons in the theory with $N_f=2$ fermions in the fundamental representation.
The comparison between dynamical and quenched calculations of continuum and chiral extrapolations show a  discrepancies of about $25\%$ for $\hat{m}_{\rm S}^2$,
 $20\%$ for $\hat{f}_{\rm PS}^2$, 
$10\%$ for $\hat{m}_{\rm V}^2$, and smaller for the other measurements. 
In the case of the two-index antisymmetric representation, only preliminary
 results for mesons have been reported recently~\cite{Lee:2022elf}, 
but the massless extrapolation has not been made, and thus the analogous comparison is not yet possible. 
Dedicated investigations are undergoing and the results will be published in the near future~\cite{AS}.

\begin{figure}[t]
\centering
\begin{picture}(380,340)
\put(0,170){\includegraphics[width=.34\textwidth]{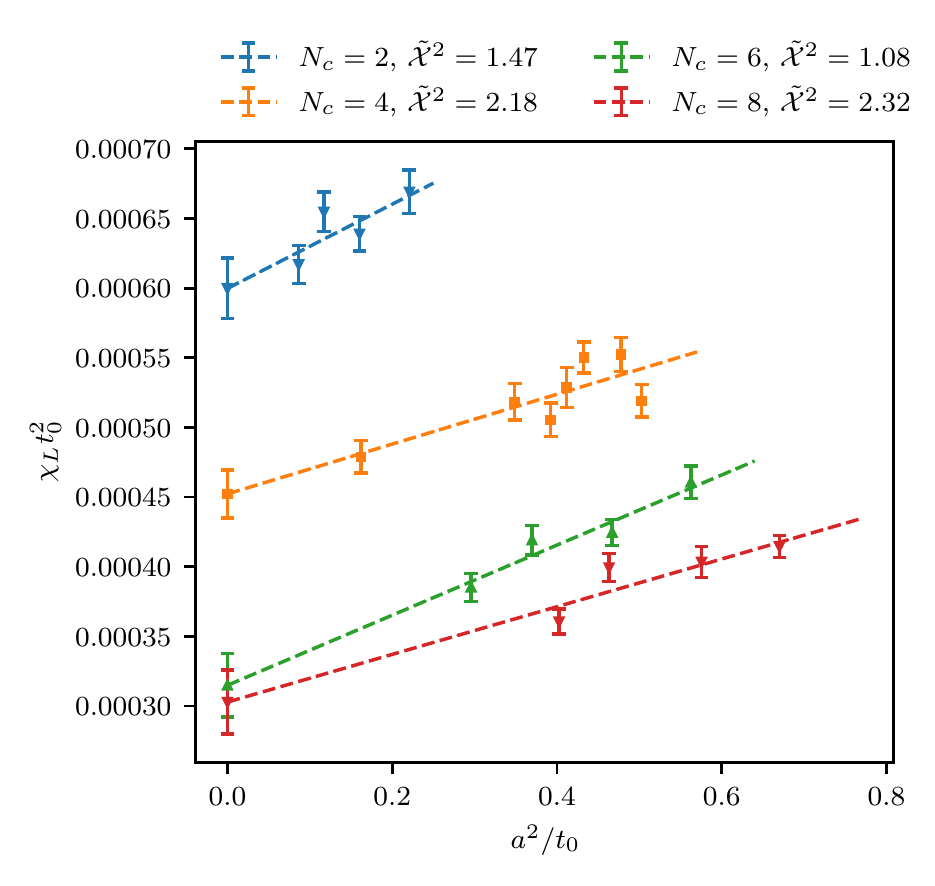}}
\put(190,170){\includegraphics[width=.34\textwidth]{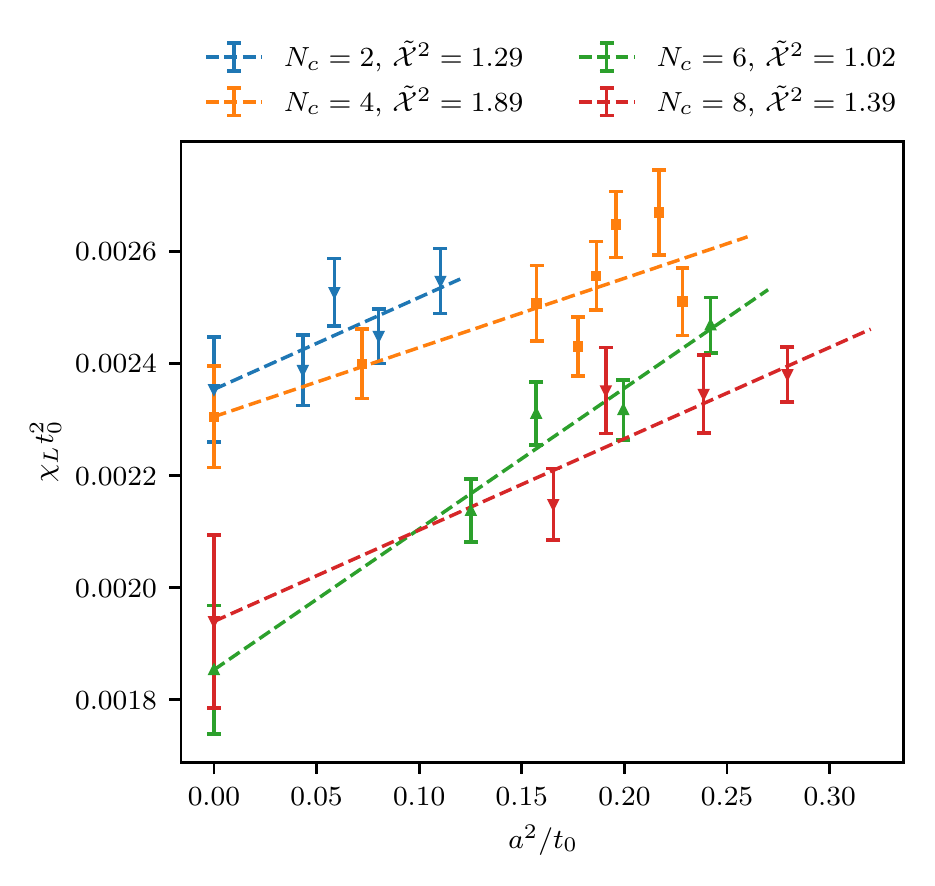}}
\put(0,0){\includegraphics[width=.34\textwidth]{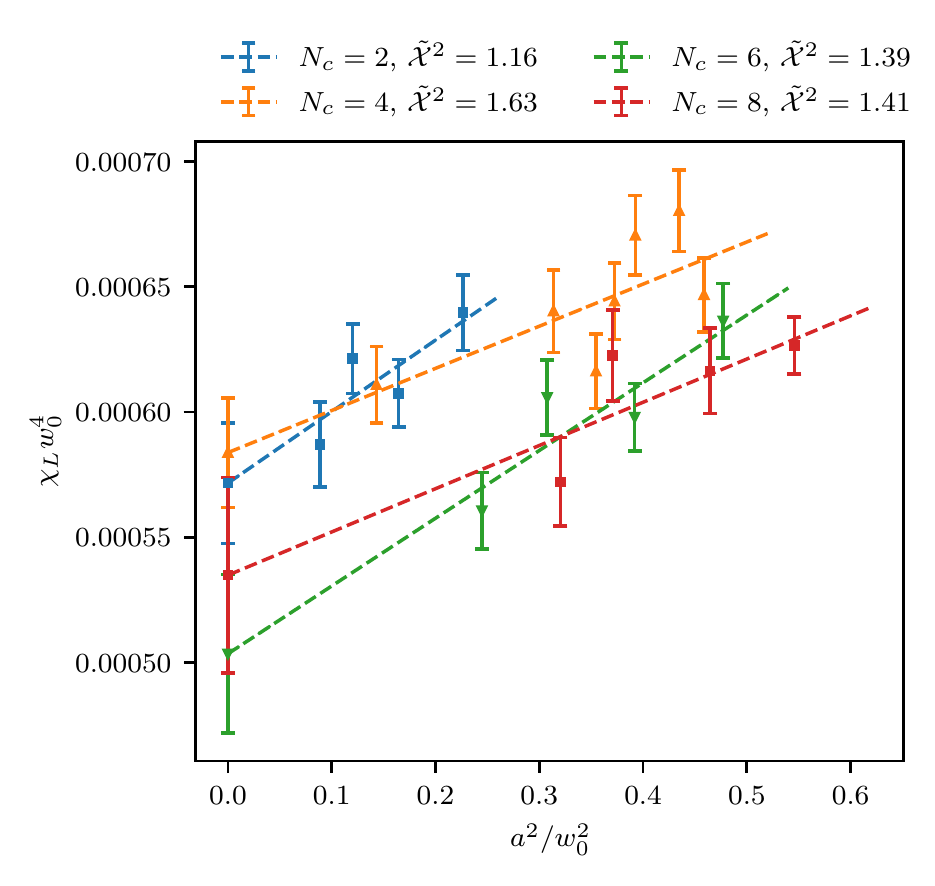}}
\put(190,0){\includegraphics[width=.34\textwidth]{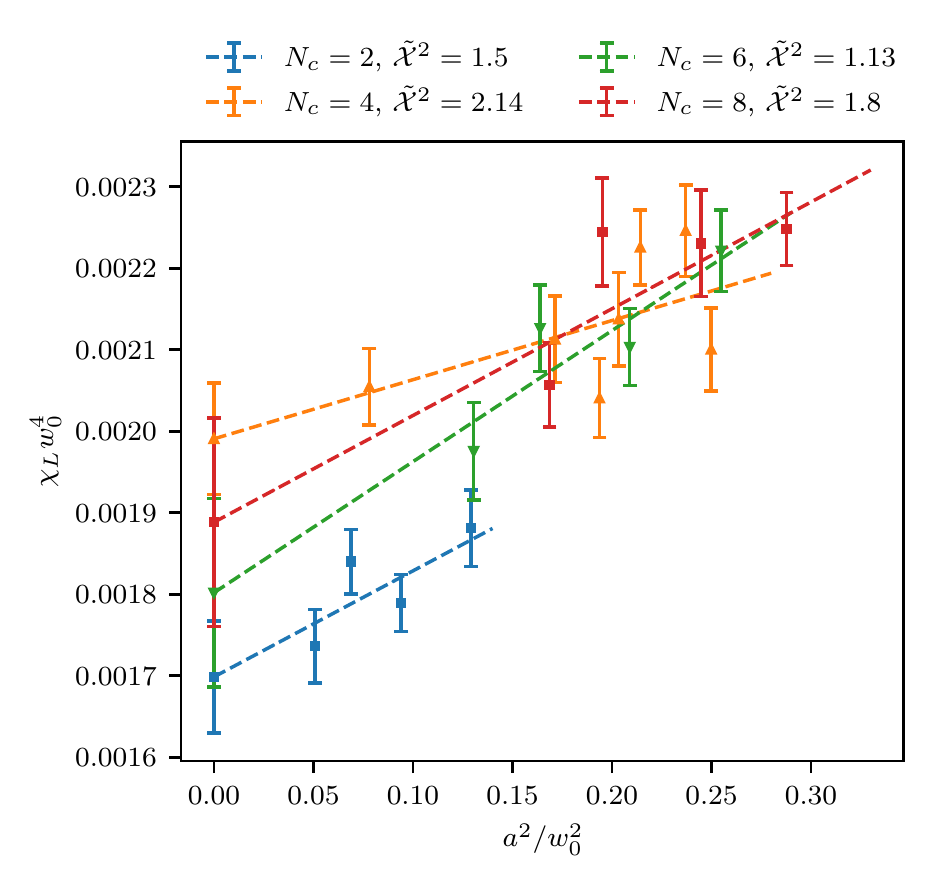}}
\end{picture}
\caption{Topological susceptibility per unit volume $\chi_L t_0^2$ as a 
function of $a^2/t_0$ (top panels) and $\chi_L w_0^4$ as a 
function of $a^2/w_0^2$ (bottom), in $Sp(N_c)$ Yang-Mills theories 
with $N_c=2,\,4,\,6,\,8$. We adopt reference values  $c_e=c_w=0.225$ (left panels) and
$c_e=c_w=0.5$ (right).
Our continuum
extrapolations  are represented as 
dashed lines. The plots are taken from Ref.~\cite{Bennett:2022ftz}.
\label{fig:top_susc_contlim}}
\end{figure}

\begin{figure}[!h]
\centering
\begin{picture}(300,210)
\put(30,0){\includegraphics[width=.40\textwidth]
{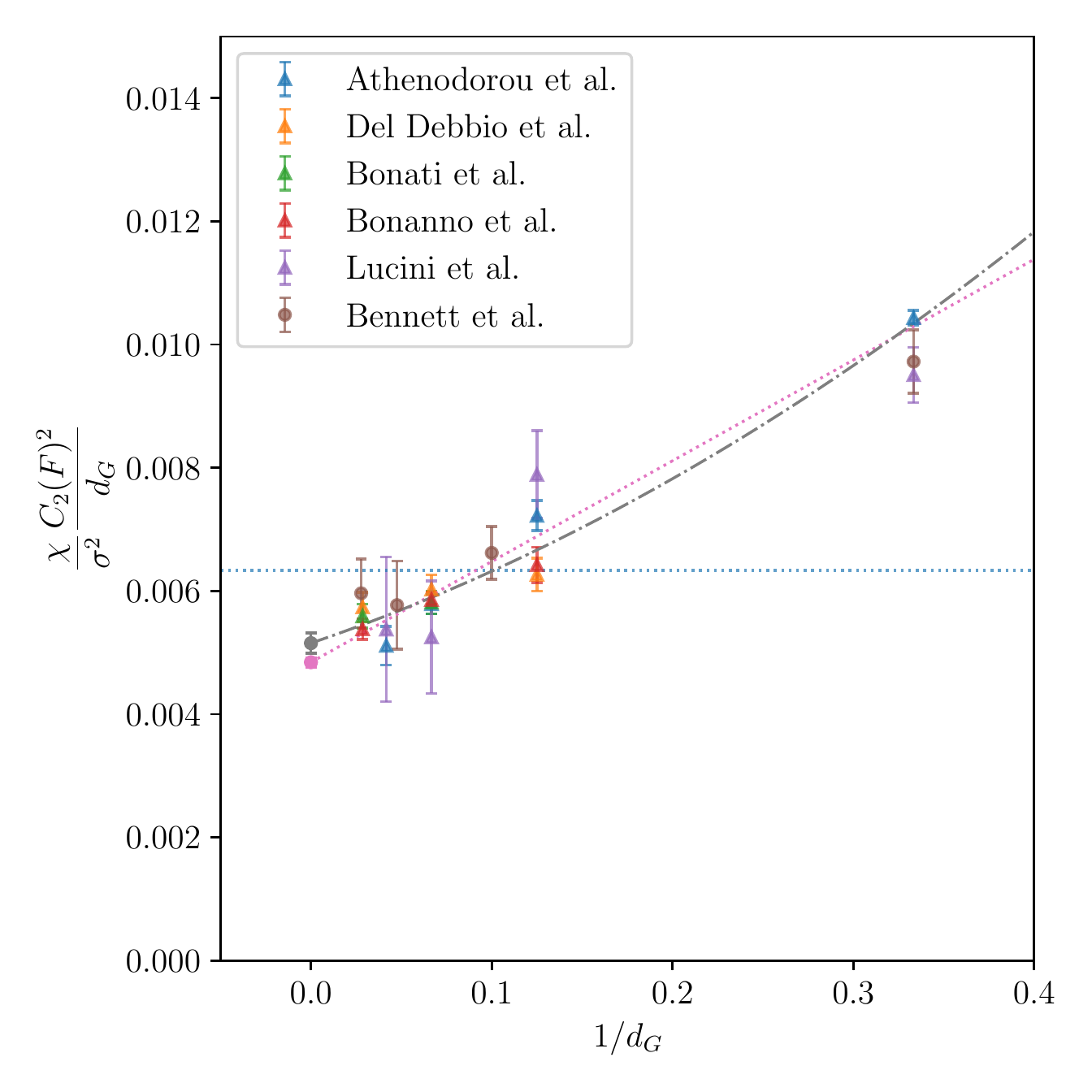}}
\end{picture}
\caption{Ratio of topological susceptibility 
and square of the string tension, rescaled 
by the group factor $C_2(F)^2/d_G$, 
as a function of $1/d_G$.
Dotted and dashed-line are results of 
a 2-parameter (dotted line), and 3-parameter fit (dashed line) 
including $O(1/d_G^2)$ corrections.
The horizontal dashed line is the na\"ive dimensional analysis
estimate $1/(4\pi)^2$. The plot is taken from Ref.~\cite{Bennett:2022gdz}.
\label{fig:scaled_top_susc}}
\end{figure}

\subsection{Topology }
\label{sec:topology}

We report here a selection of results taken  from Ref.~\cite{Bennett:2022ftz}, 
in which the ($\alpha$-rounded) topological charge~\cite{DelDebbio:2002xa}, denoted 
$\tilde{Q}_L$, is studied for
several values of $\beta$ and groups $Sp(2N)$. 
The topological susceptibility is then
obtained for each value of $N$ and $\beta$ from
\begin{equation}\label{eq:flowed_topo_suscept}
\chi_L (t) a^4 \equiv \frac{1}{L^4}\left\langle \frac{}{}\tilde{Q}_L(t)^2 \frac{}{}\right\rangle~.
\end{equation}
The continuum extrapolations can be obtained with the Wilson flow scale-setting procedure using the 
relation
\begin{equation}
\chi_L(a) t_0^2 = \chi_L(a=0) t_0^2 + c_1 \frac{a^2}{t_0}
\end{equation}
where $c_1$ is a dimensionless coefficient. Alternatively, one can adopt
$w_0$ to set the scale, and use the same formula, but replacing $t_0$ with $w_0^2$.
These extrapolations are displayed
in Figure~\ref{fig:top_susc_contlim}.

One would like to compare the value of the topological susceptibility in $Sp(2N)$ 
and  $SU(N_c)$ gauge theories. Scaling arguments
(see for instance Ref.~\cite{Bennett:2022gdz}) suggest to rescale the topological
susceptibility in units of the squared string tension as follows:
\begin{equation}
\eta_\chi \equiv \frac{\chi}{\sigma^2}\frac{C_2(F)^2}{d_G}~,
\end{equation}
where $d_G$ is the dimension of the gauge group, and test whether it captures universal
feature of Yang-Mills theories. In the  large-$N$ regime, one expects that
\begin{equation}
\lim_{N\to\infty} \frac{\chi}{\sigma^2}\frac{C_2(F)^2}{d_G}
 = b \frac{\chi_\infty}{\sigma_\infty^2} = \eta_\chi(\infty)\,,
 \end{equation}
 where $b=1/4$ for $Sp(2N)$ and $b=1/8$ for $SU(N_c)$.
 A compilation of results from the literature on $SU(N_c)$ gauge theories, 
 along with the results for $Sp(2N)$~\cite{Bennett:2022gdz,Bennett:2022ftz}, on the 
 rescaled topological susceptibility, is displayed
 as a function of $1/d_G$ in Fig.~\ref{fig:scaled_top_susc}. A combined fit
yields
 \begin{equation}
 \lim_{N_c\rightarrow \infty} 
\eta_{\chi}
= \left(48.42 \pm 0.77 \pm 3.31\right)\times 10^{-4}\,,
\end{equation}
where the first error is the statistical error from a 2-parameters linear 
fit in $1/d_G$. The second error is the difference between the result of a
2-parameters fit ${\cal O}(1/d_G)$, and  a 3-parameter ${\cal O}(1/d_G^2)$, performed on the same data. 
Both fits are displayed in Fig.~\ref{fig:scaled_top_susc}. We observe that  the
na\"ive dimensional analysis estimate  for $\eta_\chi(\infty)$ is of
the same order of magnitude as the numerical results.

\section{Numerical investigations II:
Dynamical fermions in $Sp(4)$}
\label{sec:sp4}

This section contains a selection of numerical results obtained in
 $Sp(4)$ gauge theories with dynamical matter fields.
In the case of $N_f=2$ fundamental Dirac fermions, we show 
in Sect.~\ref{sec:fund} 
the results for the masses and decay constants of flavoured mesons in various spin-0 and spin-1 channels, 
and discuss their implications for low-energy dynamics. 
More complete information can be found in Refs.~\cite{Bennett:2017kga,Lee:2018ztv,Bennett:2019jzz}.
For theories with other fermion field content,
we discuss in general terms the spectrum of mesons and (chimera) baryons 
in Sect.~\ref{sec:beyond_fund}. We refer the reader to Refs.~\cite{Bennett:2022yfa,
Lee:2022elf,Hsiao:2022kxf,AS} for extended selections of numerical results.

\subsection{$N_f=2$ fundamental fermions}
\label{sec:fund}

\begin{figure}
\begin{center}
\includegraphics[width=.38\textwidth]{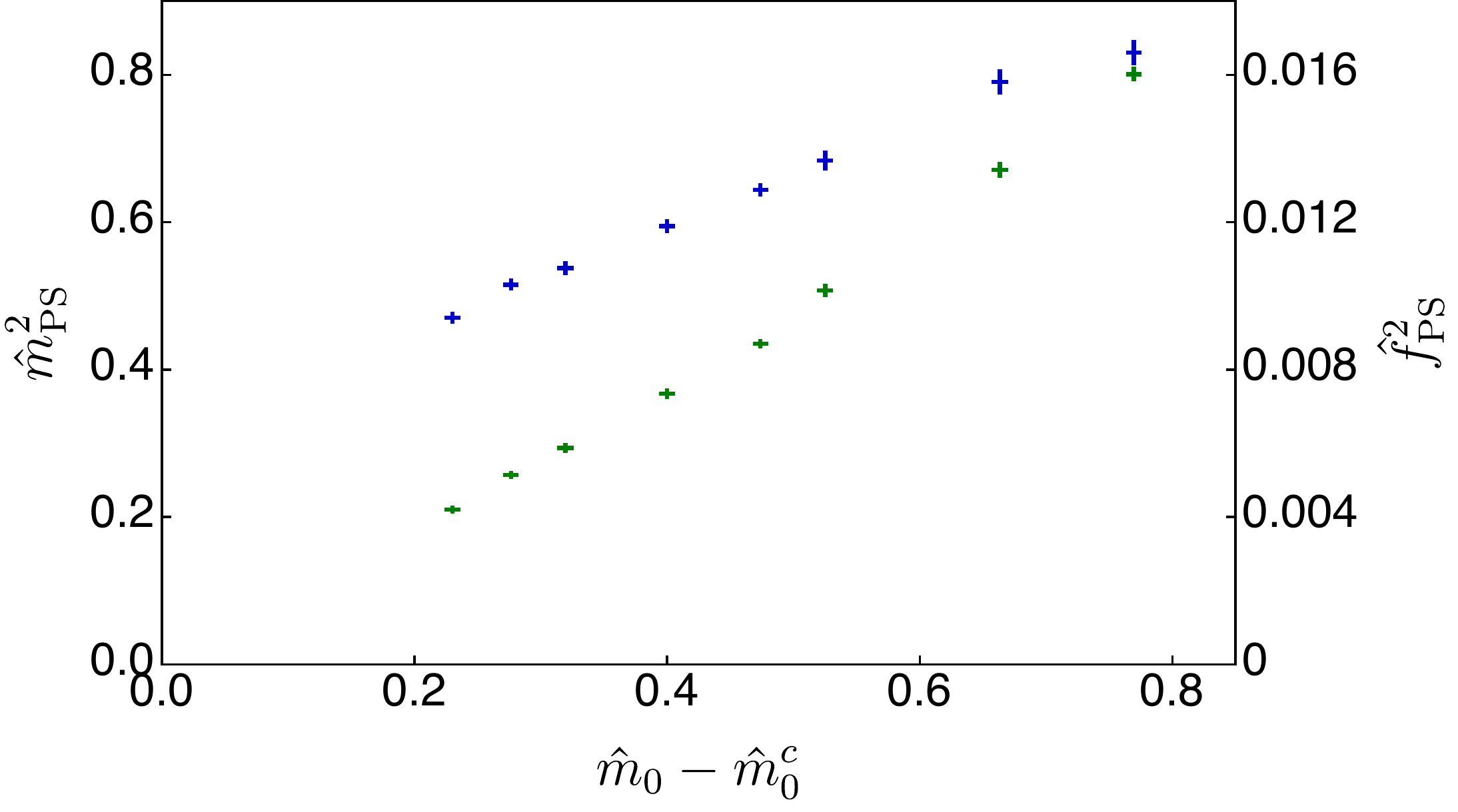}
\includegraphics[width=.32\textwidth]{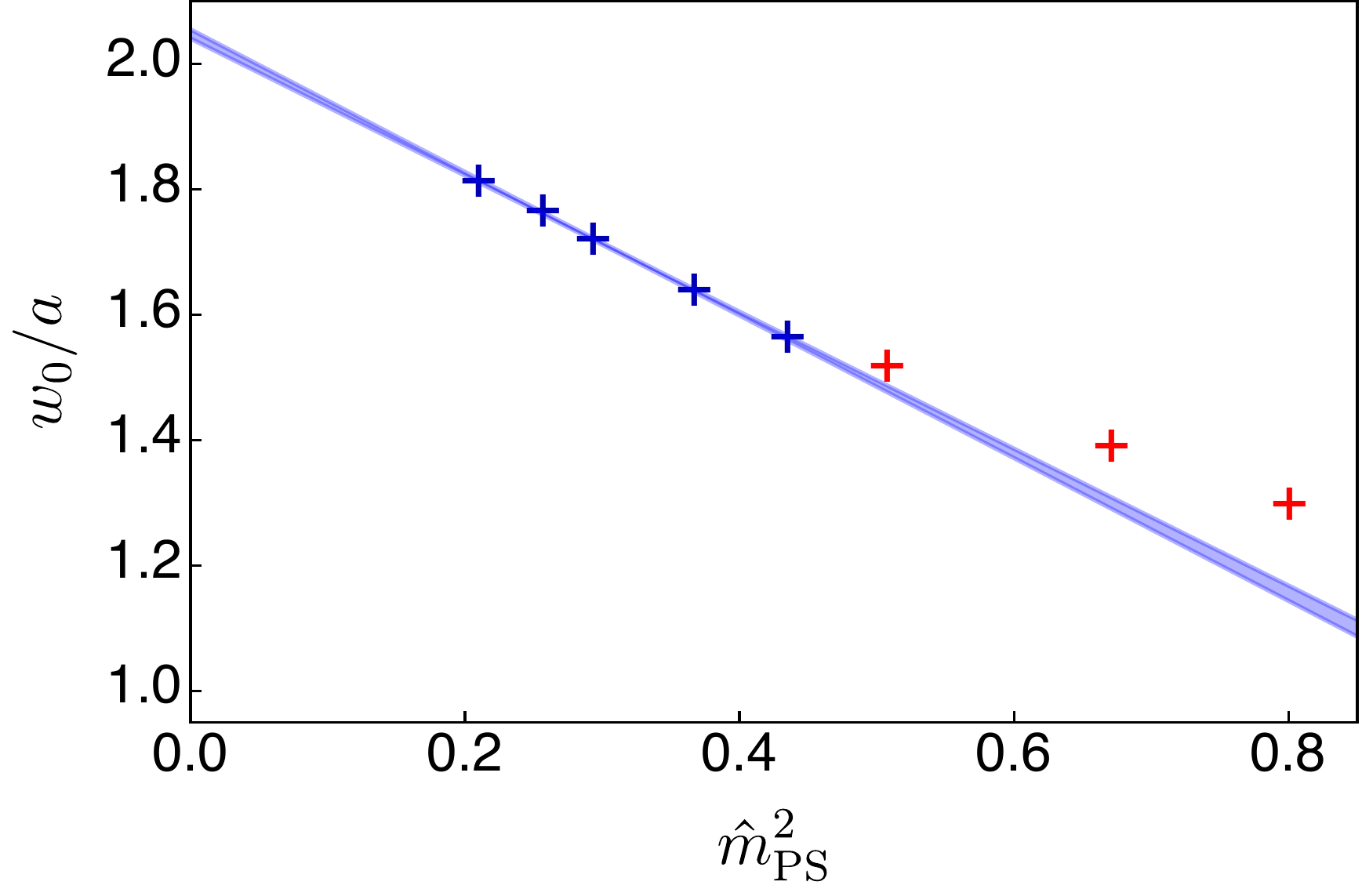}
\caption{%
The square of masses (green) and decay constants (blue) of the
 pseudoscalar mesons as a function of the bare fermion masses (left panel),
  and the relation between gradient flow scale $w_0/a$ and 
  squared pseudoscalar masses (right), for $\beta=7.2$. 
The plots are taken from Ref.~\cite{Bennett:2019jzz}.
}
\label{Fig:D_chiral}
\end{center}
\end{figure}

The  $Sp(4)$ gauge theory with $N_f=2$ dynamical fermions transforming in the fundamental representation
is treated with the Wilson-Dirac formulation and HMC algorithm, as discussed in Sects.~\ref{sec:action} and~\ref{sec:simulation}.
Careful analysis of the average plaquette value and its susceptibility indicates the
 existence of a first-order bulk phase transition~\cite{Bennett:2017kga}, 
that can be avoided for $\beta\gtrsim 6.8$.
Reference~\cite{Bennett:2019jzz} hence discusses five values of the coupling: $\beta=6.9,\,7.05,\,7.2,\,7.4,\,7.5$.
The bare fermion mass, $m_0$, is chosen so that the (pseudoscalar and vector meson) composite states
are lighter than the cut-off scale, identified with the inverse of the lattice spacing, $1/a$.

The gauge ensembles used for the measurements reported in Ref.~\cite{Bennett:2019jzz} 
consist typically of $100\sim 150$ 
thermalised configurations, separated by at least one autocorrelation time.
In order for the size of finite-volume effects,
as discussed in Sect.~\ref{sec:systematics}, 
to be negligibly small,  in respect to the statistical uncertainties, 
the stringent bound $m_{\rm PS} L \gtrsim 7.5$ is imposed,
and ensembles that do not satisfy this criterion are discarded.

\begin{figure}[t]
\begin{center}
\includegraphics[width=.35\textwidth]{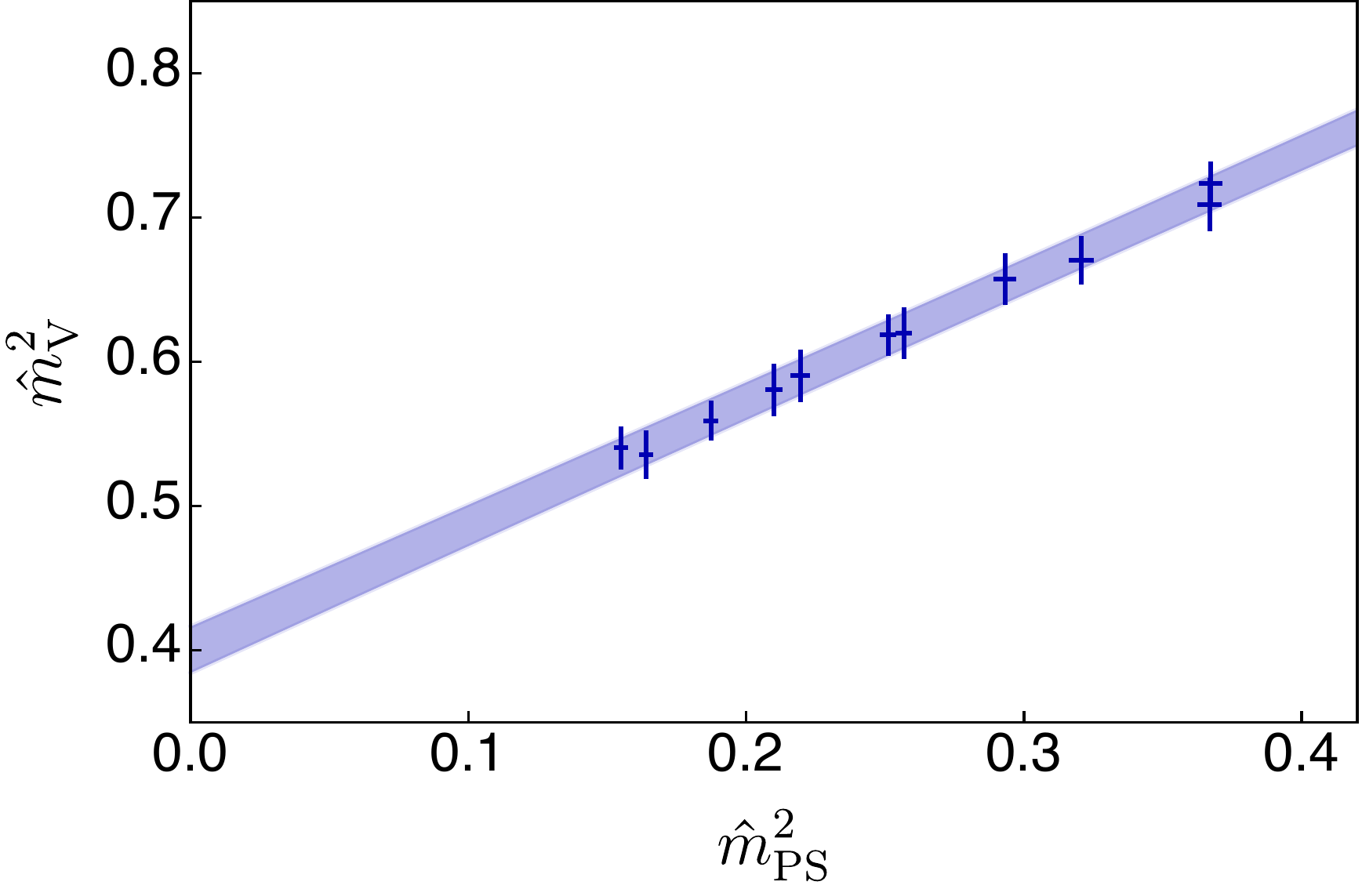}
\includegraphics[width=.35\textwidth]{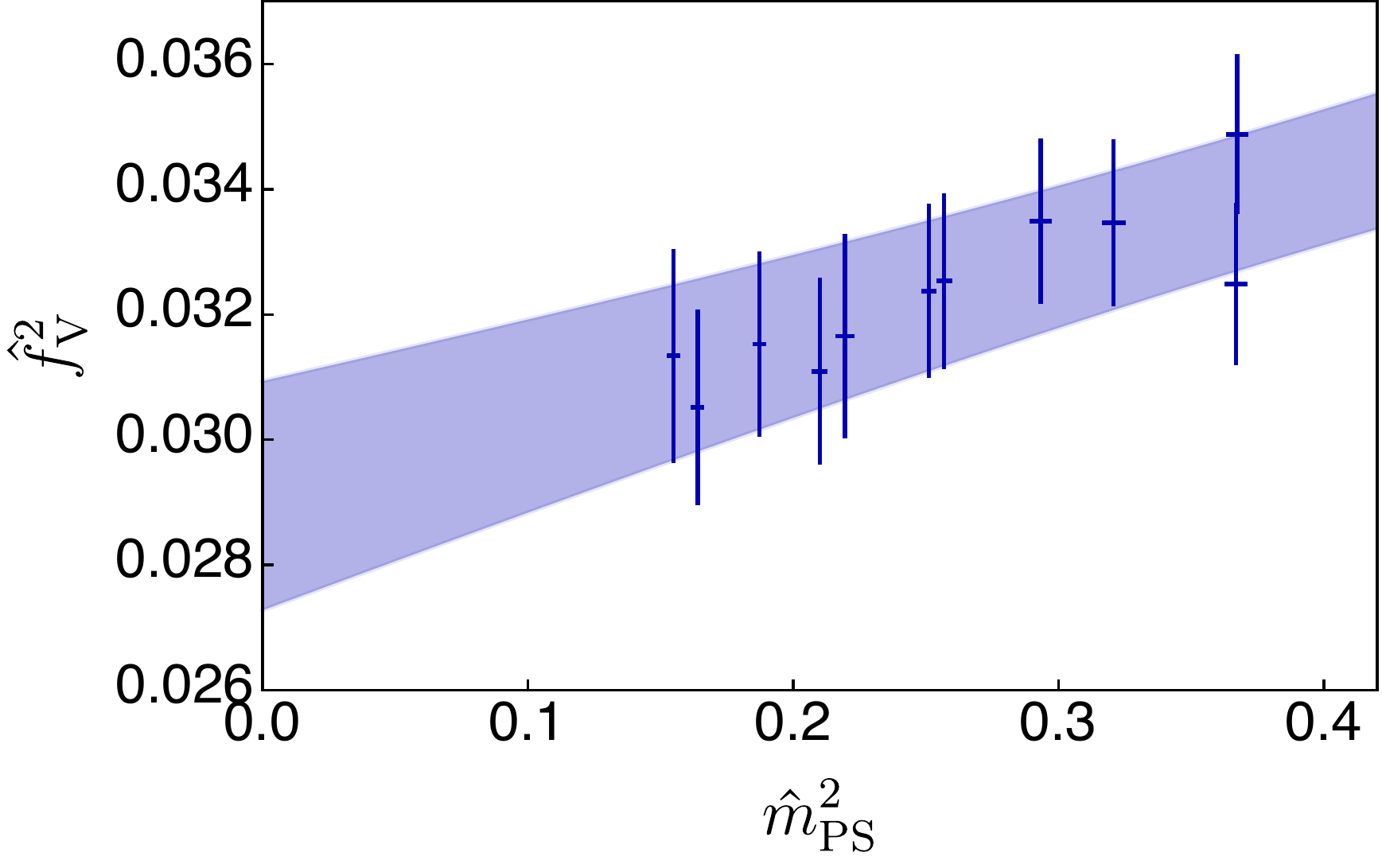}
\includegraphics[width=.35\textwidth]{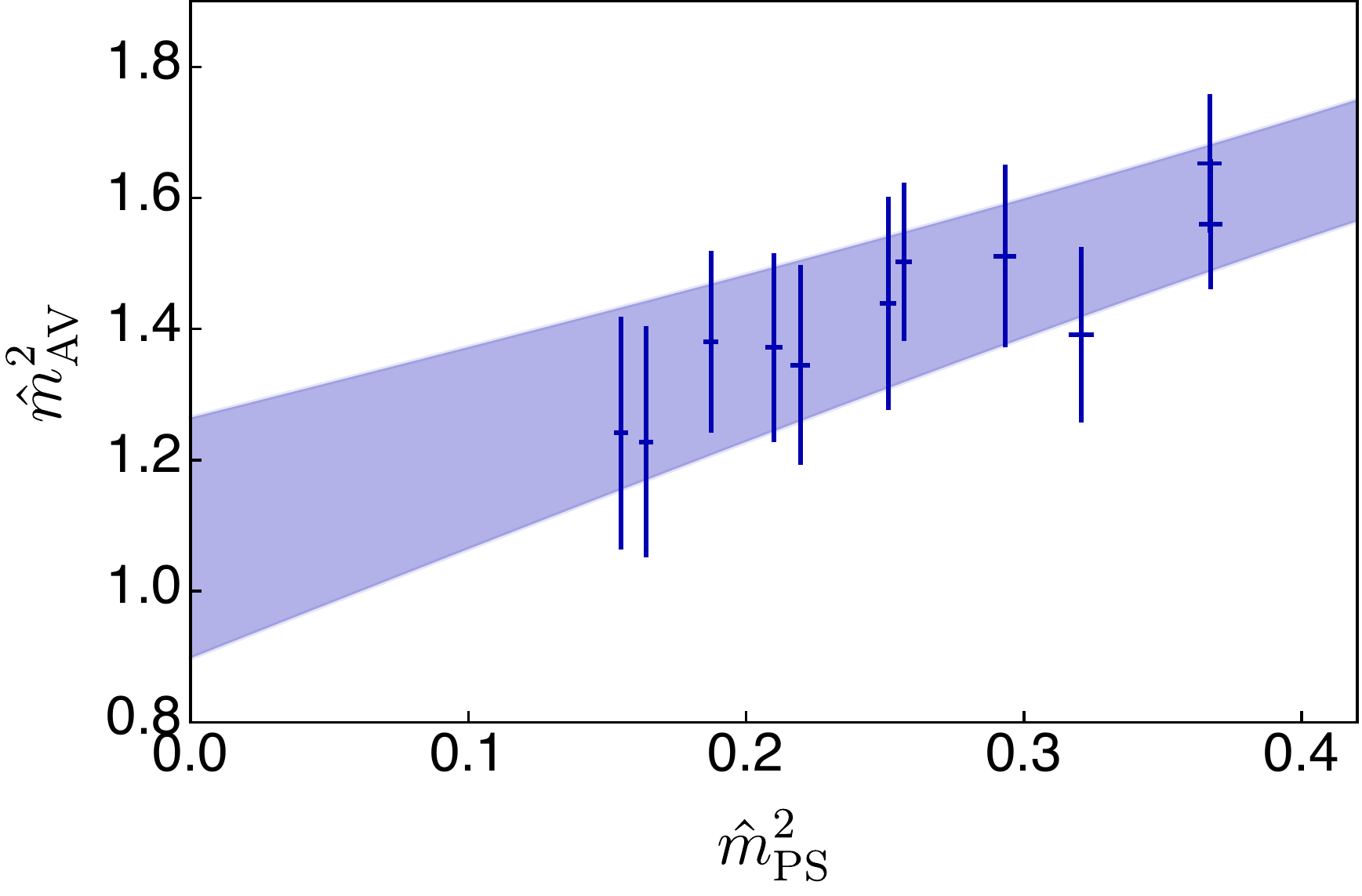}
\includegraphics[width=.35\textwidth]{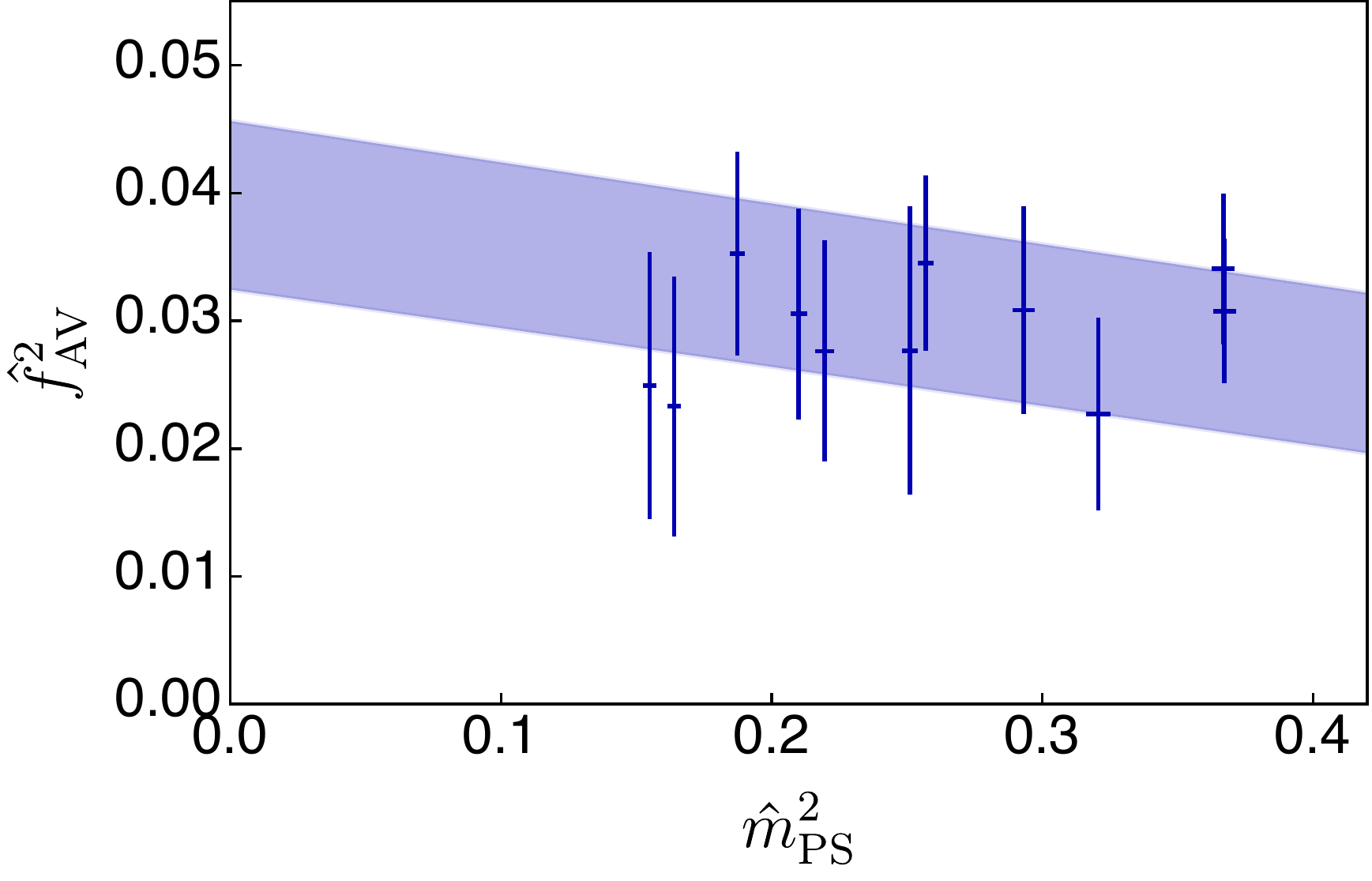}
\includegraphics[width=.35\textwidth]{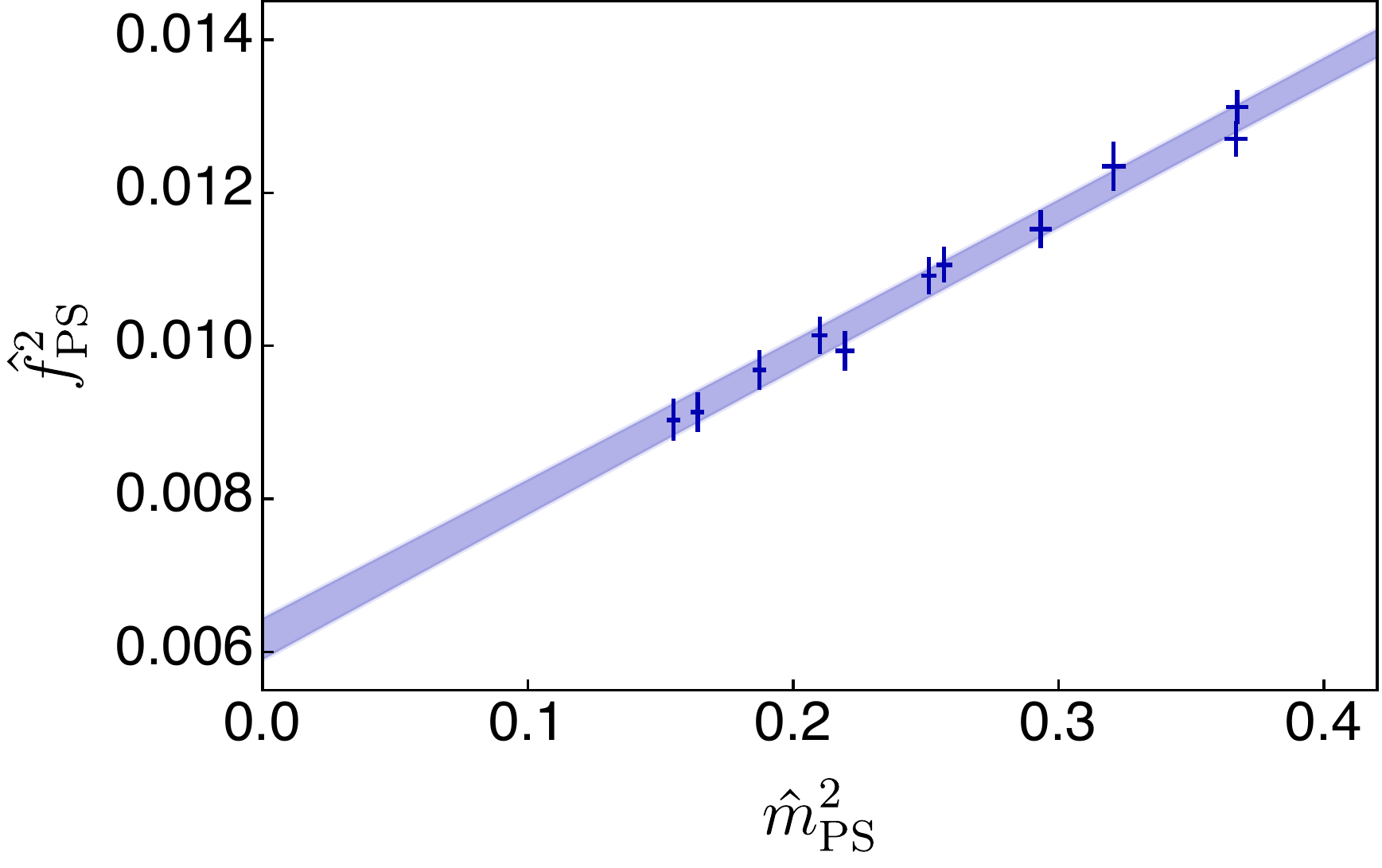}
\caption{%
Global fit and continuum extrapolation of masses and decay constants of flavour non-singlet spin-$0$ and  spin-$1$ mesons, 
based upon the low-energy EFT considerations based on hidden local symmetry (HLS). 
The plots are taken from Ref.~\cite{Bennett:2019jzz}.
}
\label{Fig:global_fit_dyn}
\end{center}
\end{figure}

All dimensional quantities are expressed in terms of the gradient flow scale, $w_0$, discussed in Sec.~\ref{sec:systematics},
in line with the treatment of quenched measurements. While the gradient flow depends itself
 non-trivially on both $\beta$ and $m_0$,
yet a {\it mass-dependent} scheme is adopted in the massless and continuum limit extrapolations,  in which the gradient flow scale is measured at a given fermion mass, as in Ref.~\cite{Ayyar:2017qdf}.
This approximation neglects corrections appearing only in higher-order terms of
 the effective field theory.

As the $Sp(4)$ theory with $N_f=2$ fundamental Dirac fermions is expected to lie deep inside of the chiral symmetry broken phase,
classical results such as the GMOR relation in Eq.~(\ref{eq:gmor}) should hold.
The left panel of Fig.~\ref{Fig:D_chiral} shows how the pseudoscalar mass squared, $\hat{m}^2_{\rm PS}$,
and decay constant, $\hat{f}^2_{\rm PS}$,  depend on $\hat{m}_0\equiv (m_0 a) (w_0/a)$ and  the critical 
value  $\hat{m}_0^c$---identified numerically by performing a linear fit to the lightest five data points, and extrapolating to the limit  $\hat{m}^2_{PS}\rightarrow 0$---for the choice $\beta=7.2$.
The decay constant $\hat{f}_{\rm PS}$ extrapolates to a finite value in the massless limit.
Both $\hat{f}_{\rm PS}$ and $\hat{m}^2_{\rm PS}$  are linear in the fermion mass when $\hat{m}^2_{\rm PS}\lesssim 0.4$.

The right panel of Fig.~\ref{Fig:D_chiral} shows
 the relation between $w_0/a$ and $\hat{m}^2_{\rm PS}$, for the same ensemble
 with $\beta=7.2$. One expects it to obey the next-to-leading-order (NLO) result~\cite{Bar:2013ora}:
\beq
\hat{w}_0^{\rm NLO} (\hat{m}_{\rm PS}^2) = 
\hat{w}_0^{\chi} \left(1+k_1 \frac{\hat{m}_{\rm PS}^2}{(4\pi \hat{f}_{\rm PS})^2}\right)\,,
\label{Eq:w0_EFT}
\eeq
and the lightest five points exhibit this linear behaviour. 
A fit to the data using Eq.~(\ref{Eq:w0_EFT}) yields  $\chi^2/N_{\rm d.o.f} \simeq 0.5$~\cite{Bennett:2019jzz},
supporting the adoption of W$\chi$PT, as in Eqs.~(\ref{eq:f_nlo}) and~(\ref{eq:m_nlo}).

\begin{figure}
\begin{center}
\includegraphics[width=.35\textwidth]{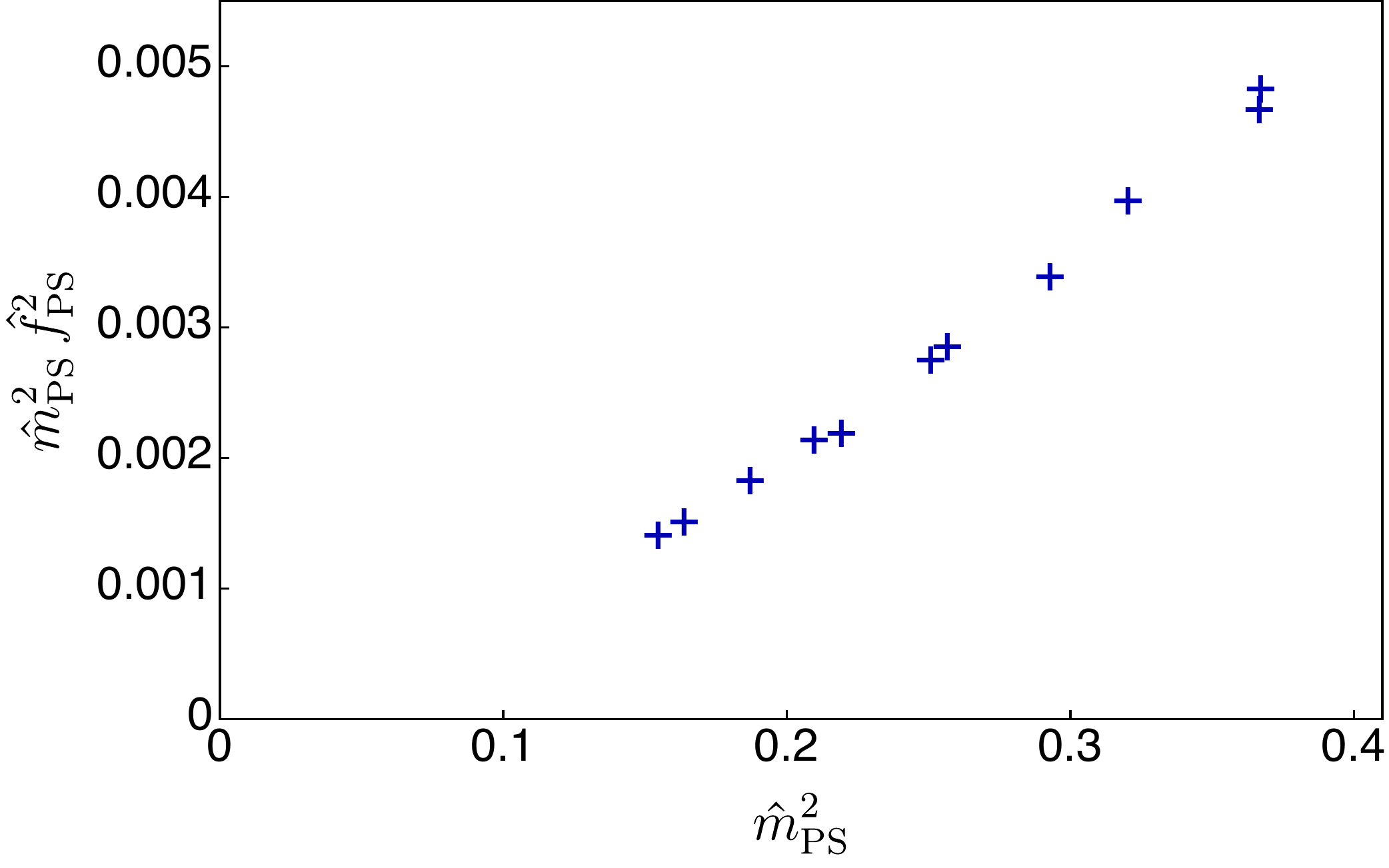}
\includegraphics[width=.35\textwidth]{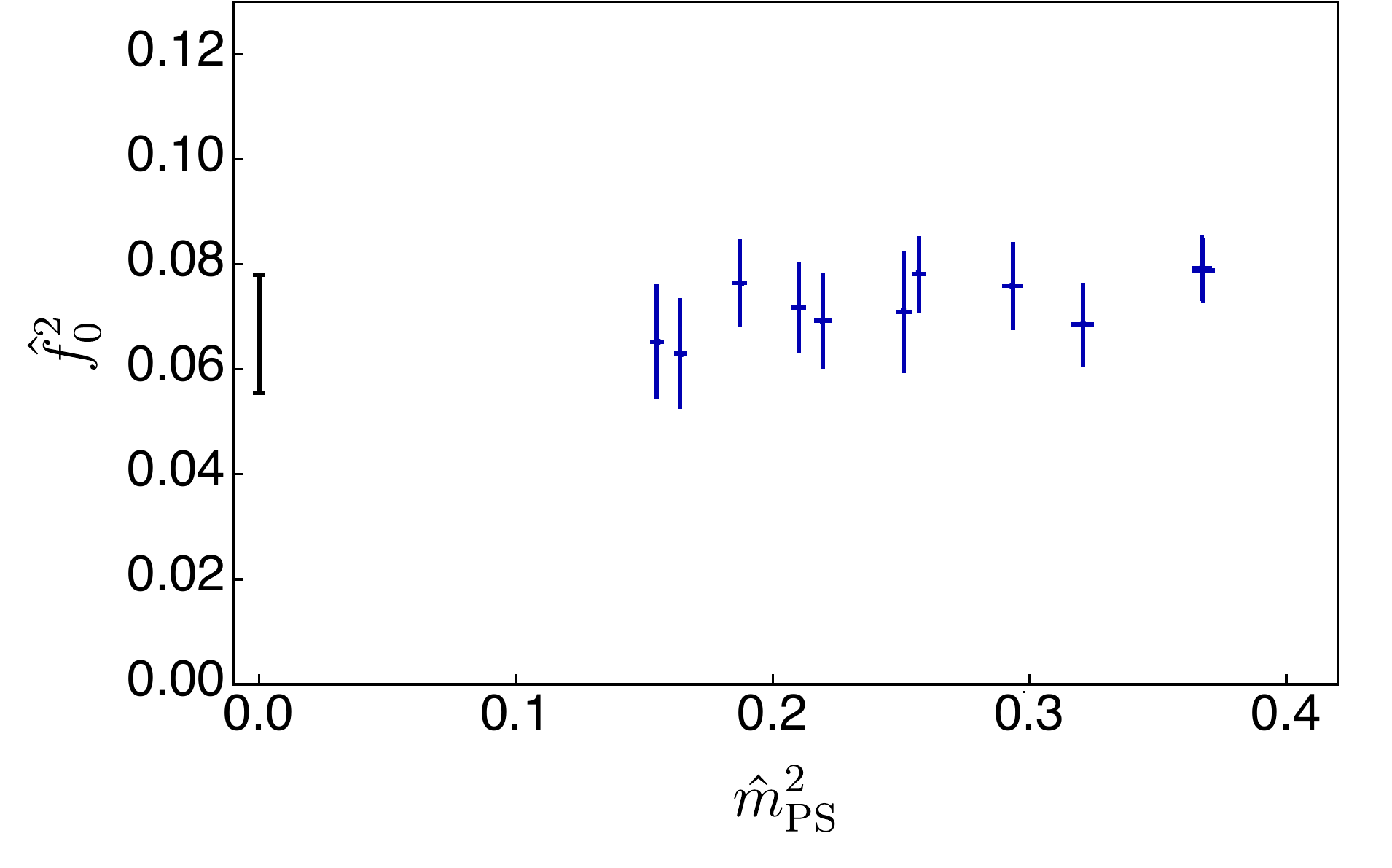}
\includegraphics[width=.35\textwidth]{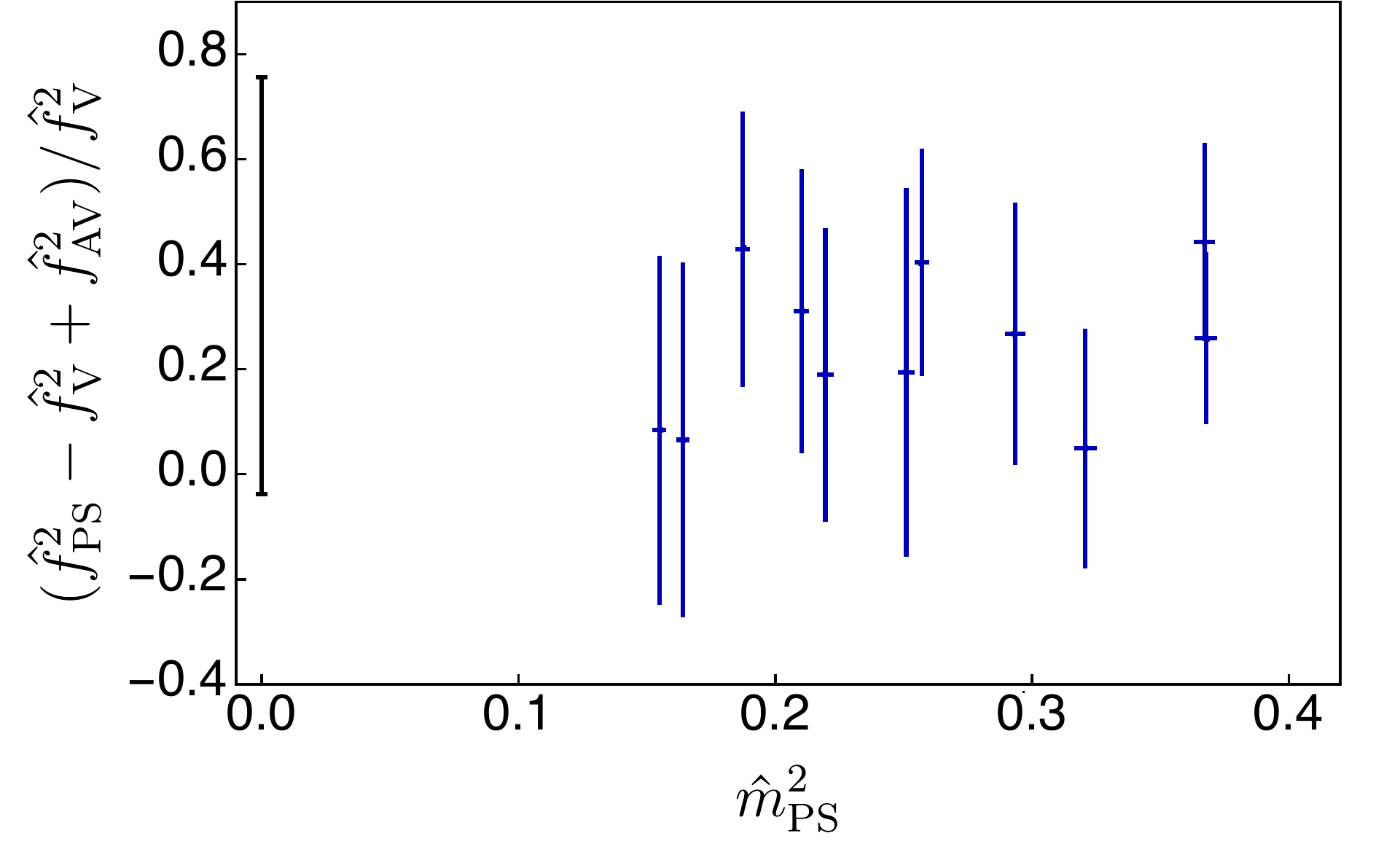}
\includegraphics[width=.35\textwidth]{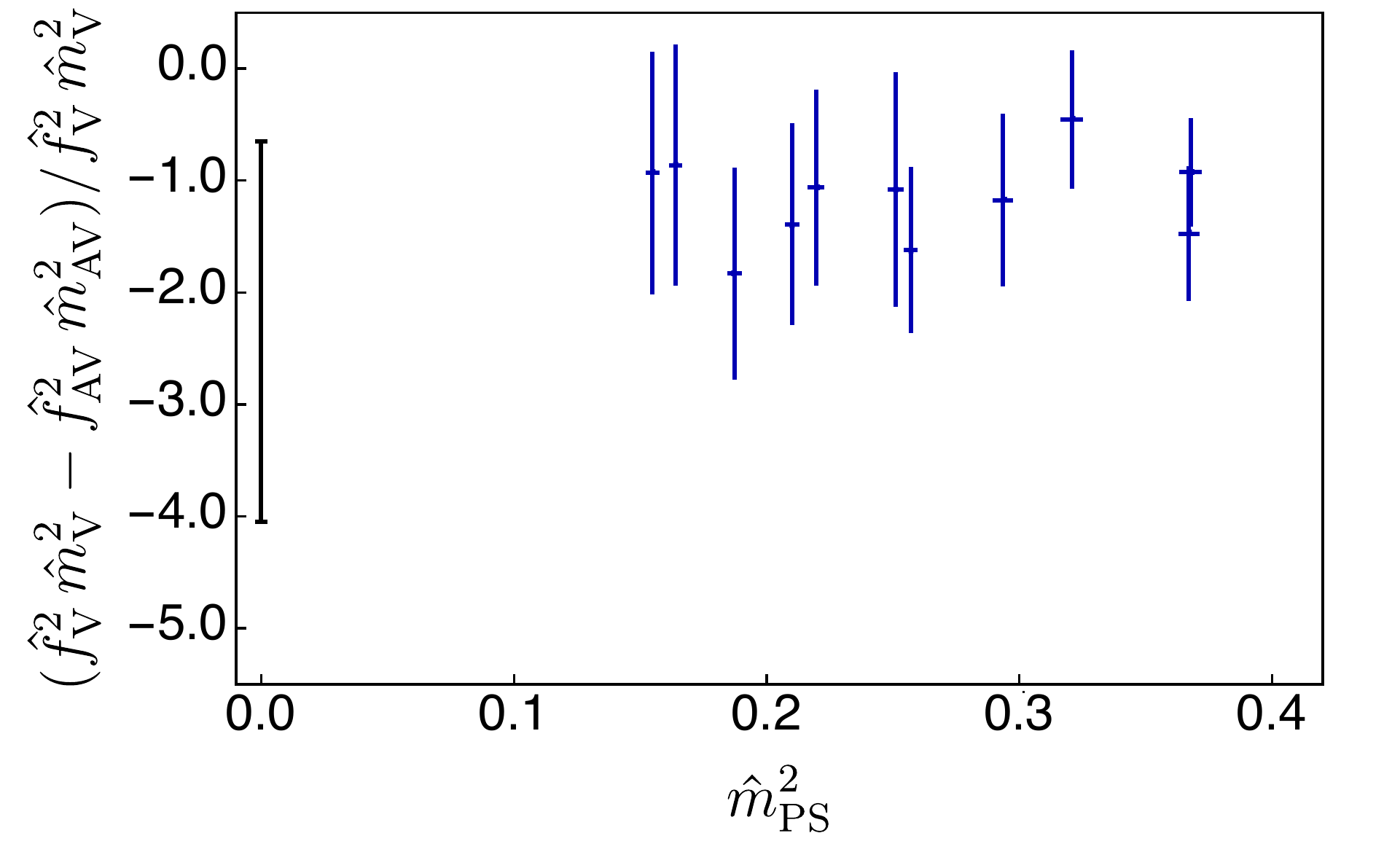}
\caption{%
Continuum extrapolation of GMOR relation and Weinberg sum rules in the $Sp(4)$ gauge theory 
with $N_f=2$ fundamental Dirac fermions.
The plots are taken from Ref.~\cite{Bennett:2019jzz}.
}
\label{Fig:sum_rules}
\end{center}
\end{figure}

In contrast to the quenched theory, however,
in the case of dynamical fermions the $w_0$ scheme is mass-dependent, as discussed above:
$\hat{m}_{\rm PS}^2$ and $\hat{a}$ are measured in units of $\hat{w}_0 (\hat{m}_{\rm PS}^2)$,
with the replacement of $\hat{w}_0 (\hat{m}_{\rm PS}^2)$ by $\hat{w}_0^\chi$.
The key requirements for  the validity of W$\chi$PT can hence be summarised as follows:
\beq
\frac{m_{\rm PS}}{\Lambda_\chi}\,,\,a \Lambda_\chi < 1~\textrm{and}~f_{\rm PS} \, L> 1,
\label{Eq:chipt_condition}
\eeq
where $\Lambda_\chi$ is the symmetry breaking scale, roughly estimated as $\Lambda_\chi=4 \pi f_{\rm {PS}}$.
By restricting attention to $\hat{m}_{\rm PS}^2 \lesssim 0.4$ for pseudoscalar mesons
(extended to $\hat{m}_{\rm PS}^2 \lesssim 0.6$ for all the other mesons),
the first condition is automatically satisfied.
The second condition is satisfied restricting the acceptable lattice spacing to  $\hat{a} <1$,
which is also needed in the expansions in Eqs.~(\ref{eq:f_nlo}) and~(\ref{eq:m_nlo}).
The ensembles satisfying these two conditions also have $f_{\rm PS}\, L \gtrsim 1.5$, satisfying the third one.
Continuum and massless extrapolations are restricted to ensembles satisfying all of these conditions,
making use of Eqs.~(\ref{eq:f_nlo}) and~(\ref{eq:m_nlo}), as for the quenched theory.
We refer the reader to Ref.~\cite{Bennett:2019jzz} for details of the  fits,
including the values of $\chi^2/N_{\rm d.o.f}$.

As discussed in Sec.~\ref{sec:hls}, HLS further extends the EFT to include 
the spin-1 states.
Reference~\cite{Bennett:2019jzz} focuses on the eleven lightest and finest  ensembles with $\hat{m}_{\rm PS}^2 \lesssim 0.4$,
in which range one is allowed to replace the fermion mass by the pseudoscalar mass squared.  
The resulting expressions involve $10$ of the $12$ unknown parameters in Eq.~(\ref{eq:hls_larg}).
The final results of the global (uncorrelated) fit are presented by blue bands in Fig.~\ref{Fig:global_fit_dyn}, 
along with the continuum values of the masses and decay constants. 
The value of  $\chi^2/N_{\rm d.o.f}\sim 0.4$ supports the EFT fit,
and, despite the weak constraints on some other combinations of the parameters,
one finds that  $g_{\rm VPP}^\chi=6.0(4)(2)$---first and second parentheses denote statistical and 
systematic errors in the numerical fits.

The  EFT based on HLS incorporates several striking, testable predictions. 
The first one is the GMOR relation extended to include NLO corrections: 
\beq
m_{\rm PS}^2 f_{\rm PS}^2=m_f (v^3 + m_f v_5^2)\,,
\eeq
where $v$ and $v_5$ are  associated with the spurion mass terms in Eq.~(\ref{eq:hls_larg})---see the 
top-left panel in Fig.~\ref{Fig:sum_rules}.

Within this truncated  EFT treatment,  reasonable assumptions lead to the omission 
of certain operators,  and
one finds that the sum of the decay constant squared for PS, V and AV, 
\beq
f_0^2 \equiv f_{\rm PS}^2 +f_{\rm V}^2 +f_{\rm AV}^2,
\eeq
is independent of $m_f$~\cite{Bennett:2017kga}.
The top-right panel of Fig.~\ref{Fig:sum_rules} shows the measurements of $f_0$ at finite mass and the massless extrapolation,
providing  strong evidence of the mass independence of $f_0^2$.
Also the  violations of Weinberg's sum rules are independent of the fermion mass,
as shown in
 the bottom panels in Fig.~\ref{Fig:sum_rules}.
 
\begin{figure}
\begin{center}
\includegraphics[width=.5\textwidth]{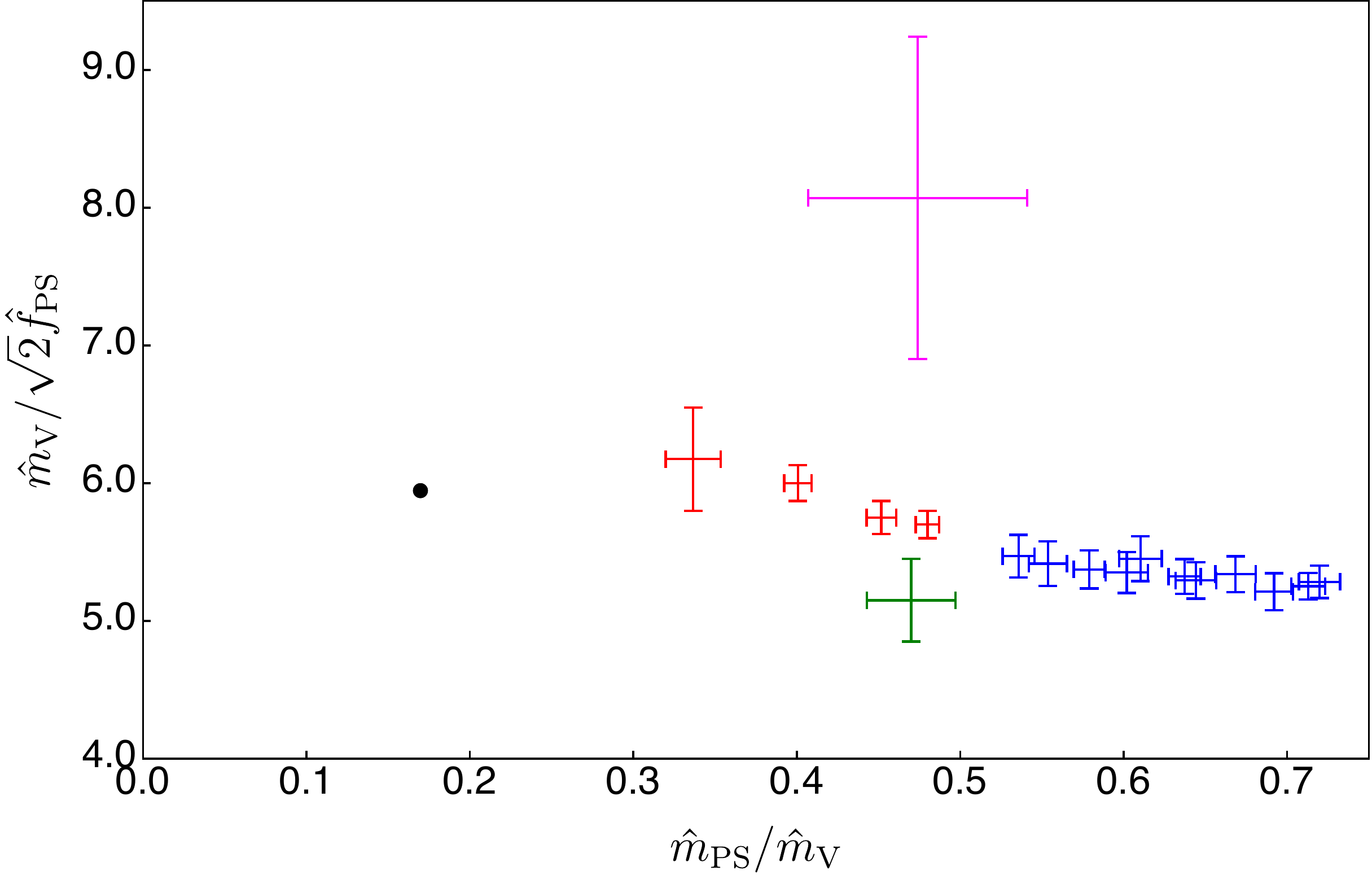}
\caption{%
Vector meson masses in units of the pseudoscalar decay constant obtained from several lattice gauge theories 
coupled to two fundamental Dirac fermions: 
magenta, red, blue, and green colours denote $SU(2)$, $SU(3)$, $Sp(4)$, and $SU(4)$ gauge groups, respectively. 
The black dot denotes the real-world QCD value. 
The plot is taken from Ref.~\cite{Bennett:2019jzz}.
}
\label{Fig:mv_comparison}
\end{center}
\end{figure}

We conclude this Section by comparing several lattice gauge theory calculations taken from the literature,
all with $N_f=2$ (dynamical) fundamental fermions. 
We consider the ratio $m_{\rm V}/\sqrt{2}f_{\rm PS}$, that,
as discussed in Sec.~\ref{sec:quenchedmeson}, appears in the right-hand side of
  the KSRF relation, $g_{\rm VPP}=m_{\rm V}/\sqrt{2}f_{\rm PS}$. 
For $Sp(4)$ one finds that the lightest ensemble yields $m_{\rm V}/\sqrt{2}f_{\rm PS}=5.47(11)$, while
the massless extrapolation is $m_{\rm V}/\sqrt{2}f_{\rm PS}=5.72(18)(13)$. 
The latter is statistically consistent with $g_{\rm VPP}^\chi=6.0(4)(2)$, determined from the global fit of the EFT, 
providing some support for the aforementioned KSRF relation holds. 
For QCD, using experimental values of $m_\pi \simeq 140\,{\rm MeV}$, $m_\rho \simeq 775\,{\rm MeV}$, 
$f_\pi \simeq 93\,{\rm MeV}$, and $\Gamma_\rho \simeq 150\,{\rm MeV}$, one finds $m_\rho/\sqrt{2}f_\pi \simeq 5.9$, 
while one  obtain $g_{\rho\pi\pi} \simeq  6.0$ from the tree-level definition of the decay rate of $\rho$, 
$\Gamma_\rho\equiv \frac{g_{\rho\pi\pi}^2}{48\pi}m_\rho \left(1-\frac{4 m_\pi^2}{m_\rho^2}\right)^{3/2}$.

Figure~\ref{Fig:mv_comparison} displays together the lattice results for $SU(2)$~\cite{Arthur:2016dir}, $SU(3)$~\cite{Jansen:2009hr}, $SU(4)$~\cite{Ayyar:2017qdf}, and 
 $Sp(4)$~\cite{Bennett:2019jzz}, as well as the experimental  QCD value. 
In the case of $SU(4)$,  the result has been obtained by using dynamical ensembles with additional $n_f=2$ 
dynamical (massive) Dirac fermions in the two-index antisymmetric representation. 
Near the threshold of the two-pseudoscalar decay,
the ratio $m_{\rm V}/\sqrt{2}f_{\rm PS}$ in $Sp(4)$ is close to those of $SU(3)$ and $SU(4)$.
Large-$N_c$ arguments suggest that this ratio should be larger for $SU(2)$, as is indeed observed numerically.

\subsection{Antisymmetric and multiple representation dynamical fermions }
\label{sec:beyond_fund}

\begin{figure}
\begin{center}
\includegraphics[width=.5\textwidth]{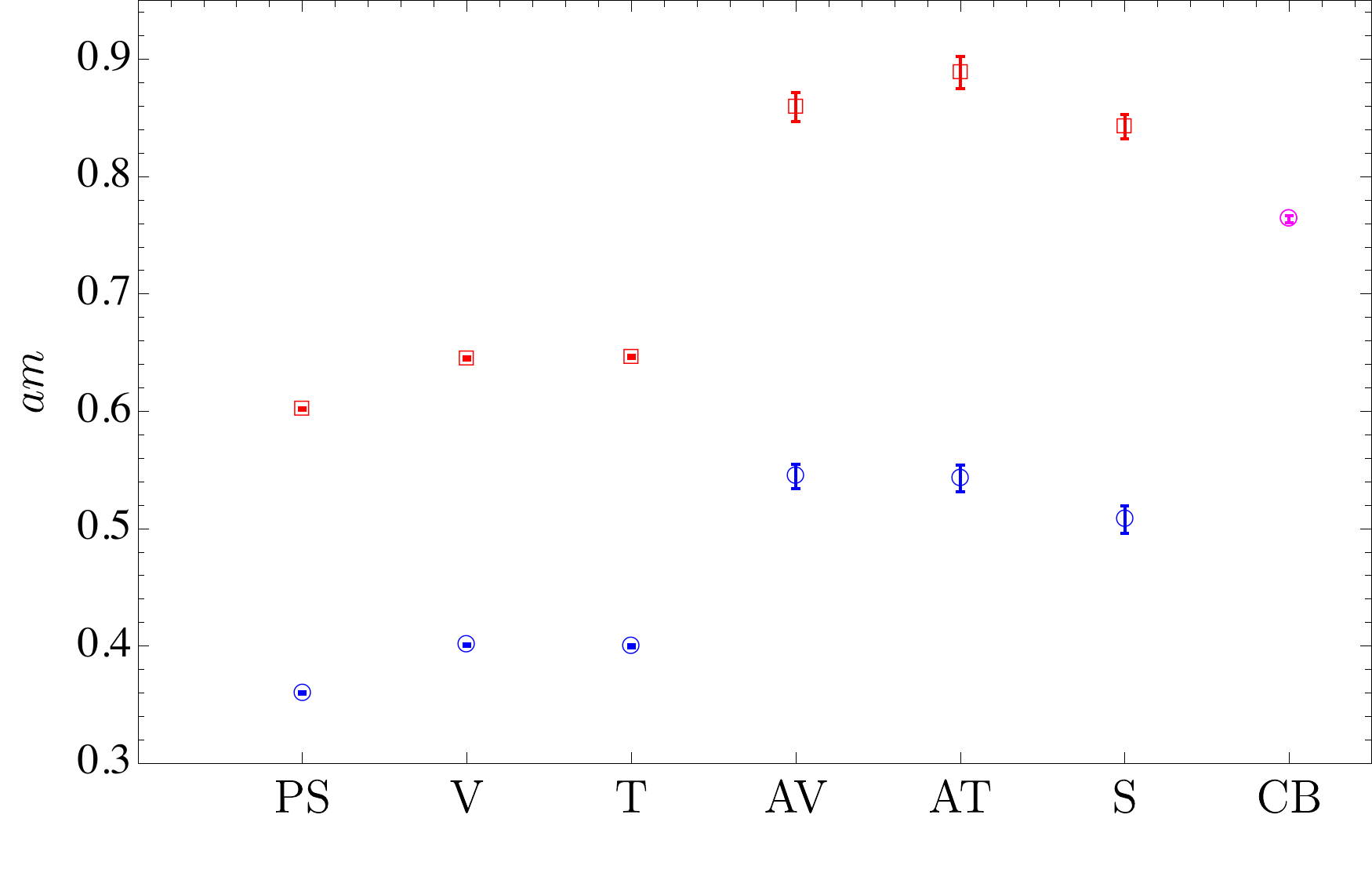}
\caption{%
Masses of composite states in the $Sp(4)$ lattice theory coupled to $N_f=2$  Dirac fermions transforming in the fundamental and $n_f=3$ in the antisymmetric representation. Blue and red colors denote mesons with constituents in the fundamental and antisymmetric representations, respectively. In magenta we display the chimera baryon composed of two constituent fermions in the fundamental and one in the antisymmetric representation. The lattice parameters used 
are $\beta = 6.5$, $am_0^{(as)} = -1.01$, $am_0^{(f)} = -0.71$, and $N_t \times N_s^3 = 54 \times 28^3$.
The plot is taken from Ref.~\cite{Bennett:2022yfa}.
}
\label{Fig:mr_spectrum}
\end{center}
\end{figure}

As discussed in Sec.~\ref{sec:pheno},
the $Sp(4)$ gauge theory with matter consisting of $n_f=3$ Dirac fermions transforming in the antisymmetric representation
(but $N_f=0$ in the fundamental)
is interesting in itself as a completion for alternative CHM and SIMP proposals~\cite{Cacciapaglia:2019ixa},
and it is hence worth studying it in detail. Most importantly, understanding its dynamics is 
a necessary first step towards the study of the theory with multiple species of fermions, transforming in 
different representations of the gauge group, which is relevant to TPC models. A large-scale lattice exploration of the parameter
space of this theory is under way~\cite{AS}. 
We comment briefly on some preliminary results of this exploration that have been presented at the Lattice 2022
Conference~\cite{Lee:2022elf}.
The main focus of the ongoing study is the spectroscopy of the spin-0 and 1 mesons listed in Table~\ref{tab:mesons}, together with the decay constants for pseudoscalar, vector and axial vector mesons. 
Preliminary results for the ratio $m_{\rm PS}/f_{\rm PS}$  indicate that this theory is likely in the broken phase, as evidenced by a sharp drop of the ratio towards the massless limit---see Fig.~2 in Ref.~\cite{Lee:2022elf}.
Yet, the theory also exhibits  a  strong mass dependence in the gradient flow scale, and it is difficult  to
  lower the physical mass of the mesons (expressed in units of $w_0$) in the numerical calculations. These observations
might be explained by the proximity of this theory to the lower edge of conformal window, as suggested by the perturbative analysis in Sec.~\ref{sec:pert}. The long distance dynamical features in this theory in the limit of massless fermions
might show substantial differences from the theory with $N_f=2$ fundamental fermions, or other QCD-like theories, but a dedicated study is needed to  ascertain this.

A main target for the study of lattice gauge theories with $Sp(4)$ gauge group is the theory with $N_f=2$ Dirac fermions transforming in the fundamental representation combined with $n_f=3$ transforming in the 2-index antisymmetric representation.
The literature on lattice calculations with multiple fermionic representations is quite limited~\cite{Ayyar:2017qdf,Ayyar:2018zuk,Ayyar:2018ppa,
 Ayyar:2018glg,Cossu:2019hse,Shamir:2021frg,DelDebbio:2021xlv,Bennett:2022yfa}.
 We have developed the necessary software, adapted from HiRep~\cite{DelDebbio:2008zf}, and performed non-trivial technical 
 tests by studying the bulk phase structure and finite volume effects~\cite{Bennett:2022yfa}. 
The first results characterising the non-perturbative dynamics 
 of phenomenologically interesting regions of parameter space are available.
 
Several species of chimera baryon states with different parity and spin quantum numbers 
have been identified, their spectrum
 for representative examples of parameter choices
in under study~\cite{Bennett:2022yfa,Hsiao:2022kxf}, and future dedicated studies will report on this extensive work.
In Fig.~\ref{Fig:mr_spectrum}, we present the combined mass spectrum of mesons composed of 
fermion constituents in the fundamental and antisymmetric representations, together with the lightest chimera baryon. 
For the one choice of lattice parameters specified in the caption of the figure,  the mass of the chimera baryon with $J^P=\frac{1}{2}^+$ is slightly lighter than the mass of the scalar meson composed of constituent
 fermions in the antisymmetric representation. 
A comprehensive study will be carried out in the lattice parameter space, to determine 
 how the masses of chimera baryons  depend on bare masses of fermion constituents in both representations.

\section{Summary and outlook}
\label{sec:conclusion}

Lattice gauge theories with $Sp(2N)$ gauge group are interesting for a variety of reasons,
both in abstract terms and in view of applications, and this review summaries just the first few steps 
of the systematic programme of explorations of the parameter space of these theories, a programme
that we envision will further develop in the near future.
We listed a number of interesting results, 
and connected them to the ongoing theoretical and phenomenological
developments. We briefly summarise these results and connections in this short section,
and indicate future avenues for further study.

In the case of pure Yang-Mills theories, we collected results for $Sp(2N)$ theories with $N=1,\,\cdots,\,4$,
and the extrapolation to the large-$N$ limit. We presented the measurements of string tension,  masses of glueballs,
and topological susceptibility. All these quantities have primarily a theoretical interest, 
for example because we expect to find agreement in  the large-$N$ extrapolations 
 of the same observables in the $SU(N_c)$ sequence of gauge theories.
There is also an interesting connection with gauge-gravity dualities, in which the non-perturbative 
regime of the large-$N$ theories is captured by perturbative supergravity calculations.
All the quantities we have been able to compute so far show hints of interesting regular patterns when 
extrapolated towards the large-$N$ limit, and furthermore it seems that the convergence is comparatively fast, with $Sp(8)$ 
being close to the continuum limit for several observables.
Applications, for example in the context of dark-matter model building, would benefit from the measurement of additional 
observables, related to interactions between glueballs (such as 3-point functions, decay rates, scattering cross-sections).

The calculation of observables involving quenched fermions provides a good approximation of the complete
dynamical theory if the number of fermion species is small, and their mass is large.
This regime is important for SIMP models, for example, but is also relevant in the CHM context.
We summarised  an extensive number of measurements in the  $Sp(4)$ theory, for
mesons built with fermions transforming either in the fundamental representation, or the 2-index antisymmetric one.
These studies will be extended in three directions: we will consider additional fermion representations
(e.g., fermions transforming in the 2-index symmetric representation of the gauge group),
study the masses of composite states 
containing two fundamental and one antisymmetric fermions (chimera baryons),
and extend the study to $Sp(2N)$ groups with larger $N$.

 The study of theories with dynamical fermions is much more challenging, for a number of reasons. It
 requires specifying the number of species of each type of fermion (in different representations),
 and for each case one has to identify the regime of lattice parameters that is useful in numerical studies.
So far, rather extensive studies of the $Sp(4)$ mesons in ensembles with dynamical fermions in the fundamental
 representation have been performed, so that the continuum limit can be taken.
Masses and decay constants of mesons relevant to CHM phenomenology have been made available.
The masses of the fermions in these studies are large enough that  they preclude decay of the spin-1 states onto 
PNGBs, hence it is not possible yet to measure directly, say, the coupling of a vector and two pseudoscalar mesons. 
Similar studies, but with dynamical matter transforming in the 2-index antisymmetric representation, are under way.
High precision calculations performed with lattice parameters closer to the
massless (chiral) regime 
require a new numerical strategy, which combines smaller fermion masses with larger volumes, and, possibly, 
adopts an improved action, to accelerate the convergence towards the continuum limit.

In the case of $Sp(4)$ with multiple dynamical 
fermion representations (fundamental and 2-index antisymmetric),
 the phase space of the lattice theory 
 is rather complicated, as we have shown in a relevant example, and this observation
 affects the choice of parameters that allows us to approach the continuum limit.
Preliminary results
have been published for one choice of lattice parameters, showing that both meson and chimera baryon
2-point correlation functions can be measured.
This study will be extended, to allow for a systematic study of the continuum and massless extrapolations,
by making use of an extended selection of ensembles.
Work on the observables themselves is also being carried out, to gain access to
an extended set of  composite states and, where possible, their excitations.
These are the first necessary steps towards testing
whether the minimal models combining composite Higgs and top partial compositeness are viable. 
A critical requirement is also to understand how the couplings and dimensionalities of the composite operators
are affected by the presence of many fermions in the dynamics;
the presence of large, non-perturbative anomalous dimensions would have important
model-building implications, but it is not known what theories yield them.

To make contact with  CHM phenomenology,
one would couple  the SM fields to
 the strong coupling sector---the 
 $Sp(2N)$ gauge theories.
For example, this would allow to compute the contributions to the effective potential for the PNGBs,
and to study vacuum (mis-)alignment. 
In this way, one would be able to test directly the properties of 
 the strong coupling sector and its heavy resonances.
Part of this programme can be performed approximating the dynamics of the combined 
system of strong and weak coupling fields by ignoring the back-reaction of the latter on the former,
along the lines of what is routinely done for QCD in the B-mesons system, for instance,
and hence by computing matrix elements of higher-order operators in the strongly
coupled theory.

Last but not least, finite temperature studies are currently being performed, aimed at characterising the 
confinement/deconfinement phase transition of $Sp(2N)$ theories, and hence extending the pioneering work in Ref.~\cite{Holland:2003kg}.
The results of this investigation might play an important role in the context of dark matter,
for example as a source of (detectable) stochastic gravitational wave background.
In general, the complete characterisation of such phase transitions is a topic that has 
great potential to reveal new, theoretical and phenomenological, possible developments.

Lattice studies of $Sp(2N)$ 
gauge theories represent a  lively field of research, which 
 is still in its infancy. We gathered together a  large compilation of results, yet this is but a 
 taster of the wealth of information contained in the original literature~\cite{Holland:2003kg,Bennett:2017kga,
Lee:2018ztv,Bennett:2019jzz,
Bennett:2019cxd,Bennett:2020hqd,
Bennett:2020qtj,Lucini:2021xke,Bennett:2021mbw,
Bennett:2022yfa,Bennett:2022gdz,Bennett:2022ftz,AS,
Lee:2022elf,
Hsiao:2022kxf,
Maas:2021gbf,
Zierler:2021cfa,
Kulkarni:2022bvh}.
This is the first stage  of what  will be a fertile
ground for testing new ideas, and learning new lessons, which are
going to inform further  theoretical developments as well 
as  applications. 


\authorcontributions{
All authors have read and agreed to the published version of the manuscript.

}

\funding{

The work of EB has been supported by the UKRI Science and Technology Facilities Council (STFC)
 Research Software Engineering Fellowship EP/V052489/1, and by the ExaTEPP project EP/X017168/1.

The work of JH at the University of Maryland is partially supported by the Center for Frontier Nuclear Science at Stony Brook University. The work of JH at Michigan State University is supported by NSF grant PHY 2209424 as well as the Research Corporation for Science Advancement through the Cottrell Scholar Award.

The work of DKH was supported by Basic Science Research Program through the National Research Foundation of Korea (NRF) funded by the Ministry of Education (NRF-2017R1D1A1B06033701).

The work of JWL was supported in part by the National Research Foundation of Korea (NRF) grant funded 
by the Korea government(MSIT) (NRF-2018R1C1B3001379) and by IBS under the project code, IBS-R018-D1. 

The work of DKH and JWL was further supported by the National Research Foundation of Korea (NRF) grant funded by the Korea government (MSIT) (2021R1A4A5031460).

The work of HH and CJDL is supported by the Taiwanese MoST grant 109-2112-M-009-006-MY3. 

The work of BL and MP has been supported in part by the STFC 
Consolidated Grants No. ST/P00055X/1 and No. ST/T000813/1.
 BL and MP received funding from
the European Research Council (ERC) under the European
Union’s Horizon 2020 research and innovation program
under Grant Agreement No.~813942. 
The work of BL is further supported in part 
by the Royal Society Wolfson Research Merit Award 
WM170010 and by the Leverhulme Trust Research Fellowship No. RF-2020-4619.

The work of DV is supported in part 
by the Simons Foundation under the program “Targeted 
Grants to Institutes” awarded to 
the Hamilton Mathematics Institute.

Numerical simulations have been performed on the 
Swansea University SUNBIRD cluster (part of the Supercomputing Wales project) and AccelerateAI A100 GPU system,
on the local HPC
clusters in Pusan National University (PNU) and in National Yang Ming Chiao Tung University (NYCU),
and on the DiRAC Data Intensive service at Leicester.
The Swansea University SUNBIRD system and AccelerateAI are part funded by the European Regional Development Fund (ERDF) via Welsh Government.
The DiRAC Data Intensive service at Leicester is operated by 
the University of Leicester IT Services, which forms part of 
the STFC DiRAC HPC Facility (www.dirac.ac.uk). The DiRAC 
Data Intensive service equipment at Leicester was funded 
by BEIS capital funding via STFC capital grants ST/K000373/1 
and ST/R002363/1 and STFC DiRAC Operations grant ST/R001014/1. 
DiRAC is part of the National e-Infrastructure.

\vspace{0.5cm}

\noindent
{\bf Open Access Statement: } 

For the purpose of open access, the authors have applied a Creative Commons 
Attribution (CC BY) licence  to any Author Accepted Manuscript version arising.

}

\institutionalreview{Not applicable}

\informedconsent{Not applicable}

\dataavailability{No new data were generated for this manuscript.
}



\conflictsofinterest{The authors declare no conflict of interest.}
\clearpage
\newpage

\abbreviations{The following abbreviations are used in this manuscript:\\

\noindent 
\begin{tabular}{@{}ll}
(as) & 2-index antisymmetric (representation)\\
(A)T & (Axial-)Tensor (operator, particle)\\
(A)V & (Axial-)Vector (operator, particle)\\
BZ & Banks-Zaks\\
CB & Chimera Baryon\\
CDM & Cold Dark Matter\\
 CERN & European Organisation for Nuclear Research\\
CHM & Composite Higgs Model\\
ChRMT & Chiral Random Matrix Theory\\
CoDM & Composite Dark Matter\\
(E) & Euclidean (space-time)\\
EFT & Effective Field Theory\\
ETC & Extended Technicolor\\
EW(SB) & ElectroWeak (Symmetry Breaking)\\
(f) & fundamental (representation)\\
FCNC & Flavor Changing Neutral Current\\
FV & Finite Volume\\
GIM & Glashow-Iliopoulos-Maiani (mechanism)\\
GMOR & Gell-Mann-Oakes-Renner\\
GW & Gravitational Wave\\
HB & Heat Bath\\
(R)HMC & (Rational) Hybrid Monte Carlo\\
IR & Infra-Red\\
KSRF & Kawarabayashi-Suzuki-Riazuddin-Fayyazuddin (relation)\\
HLS & Hidden Local Symmetry\\
HPC & High Performance Computing\\
LHC & Large Hadron Collider\\
LISA & Laser Interferometer Space Antenna\\
LLR & Logarithmic Linear Relaxation\\
(M) & Minkowski (space-time)\\
MC & Monte Carlo\\
MD & Molecular Dynamics\\
NDA & Naive Dimensional Analysis\\
NLO & Next-to-Leading Order\\ 
OR & Over-Relaxation\\
PNGB & Pseudo-Nambu-Goldstone Boson\\
PS & Pseudoscalar (operator, particle)\\
QCD & Quantum Chromodynamics\\
RG(E) & Renormalisation Group (Equation)\\
S & Scalar (operator, particle)\\
SIMP & Strongly Interacting Massive Particle\\
SM & Standard Model (of particle physics)\\
(W)TC & (Walking) Technicolor\\
TPC & Top Partial Compositeness\\
URL & Uniform Resource Locator\\
UV & Ultra-Violet\\
VEV & Vacuum Expectation Value\\
WIMP & Weakly Interacting Massive Particle\\
WZW & Wess-Zumino-Witten (interaction term)\\
(W)$\chi$PT & (Wilson) Chiral Perturbation Theory\\
\end{tabular}}

\appendixtitles{yes} 
\appendixstart
\appendix

\section{Groups, Algebras and technical details}
\label{sec:algebra}

 We collect in this Appendix technical details, in particular 
 about conventional choices 
and group theory notions, that support the main narrative of the paper.

We start from the generators of the global symmetry group $SU(4)$.
We adopt the convenient parametrisation of the $15$ generators of $SU(4)$ 
in Ref.~\cite{Lee:2017uvl}. 
The generators obey the relations $\Tr \tilde{T}^A\tilde{T}^B=\frac{1}{2}\delta^{AB}$, and  are written as follows---see Eqs.~(\ref{Eq:broken}) and~(\ref{Eq:unbroken}).
\beqs
\label{eq:su4_generators}
\tilde{T}^1&=&\frac{1}{2\sqrt{2}}
\left(
\begin{array}{cccc}
 0 & 1 & 0 & 0 \\
 1 & 0 & 0 & 0 \\
 0 & 0 & 0 & 1 \\
 0 & 0 & 1 & 0
\end{array}
\right)
\,,~~~~~
\tilde{T}^2\,=\,\frac{1}{2\sqrt{2}}
\left(
\begin{array}{cccc}
 0 & -i & 0 & 0 \\
 i & 0 & 0 & 0 \\
 0 & 0 & 0 & i \\
 0 & 0 & -i & 0
\end{array}
\right)
\,,\nonumber\\
\tilde{T}^3&=&\frac{1}{2\sqrt{2}}
\left(
\begin{array}{cccc}
 1 & 0 & 0 & 0 \\
 0 & -1 & 0 & 0 \\
 0 & 0 & 1 & 0 \\
 0 & 0 & 0 & -1
\end{array}
\right)
\,,~~~~~
\tilde{T}^4\,=\,\frac{1}{2\sqrt{2}}
\left(
\begin{array}{cccc}
 0 & 0 & 0 & -i \\
 0 & 0 & i & 0 \\
 0 & -i & 0 & 0 \\
 i & 0 & 0 & 0
\end{array}
\right)
\,,\nonumber\\
\tilde{T}^5&=&\frac{1}{2\sqrt{2}}
\left(
\begin{array}{cccc}
 0 & 0 & 0 & 1 \\
 0 & 0 & -1 & 0 \\
 0 & -1 & 0 & 0 \\
 1 & 0 & 0 & 0
\end{array}
\right)
\,,~~~~~
\tilde{T}^6\,=\,\frac{1}{2\sqrt{2}}
\left(
\begin{array}{cccc}
 0 & 0 & -i & 0 \\
 0 & 0 & 0 & -i \\
 i & 0 & 0 & 0 \\
 0 & i & 0 & 0
\end{array}
\right)
\,,\nonumber\\
\tilde{T}^7&=&\frac{1}{2\sqrt{2}}
\left(
\begin{array}{cccc}
 0 & 0 & 0 & -i \\
 0 & 0 & -i & 0 \\
 0 & i & 0 & 0 \\
 i & 0 & 0 & 0
\end{array}
\right)
\,,~~~~~
\tilde{T}^8\,=\,\frac{1}{2\sqrt{2}}
\left(
\begin{array}{cccc}
 0 & -i & 0 & 0 \\
 i & 0 & 0 & 0 \\
 0 & 0 & 0 & -i \\
 0 & 0 & i & 0
\end{array}
\right)
\,,\nonumber\\
\tilde{T}^9&=&\frac{1}{2\sqrt{2}}
\left(
\begin{array}{cccc}
 0 & 0 & -i & 0 \\
 0 & 0 & 0 & i \\
 i & 0 & 0 & 0 \\
 0 & -i & 0 & 0
\end{array}
\right)
\,,~~~~~
\tilde{T}^{10}\,=\,\frac{1}{2}
\left(
\begin{array}{cccc}
 0 & 0 & 1 & 0 \\
 0 & 0 & 0 & 0 \\
 1 & 0 & 0 & 0 \\
 0 & 0 & 0 & 0
\end{array}
\right)
\,,\nonumber\\
\tilde{T}^{11}&=&\frac{1}{2\sqrt{2}}
\left(
\begin{array}{cccc}
 0 & 0 & 0 & 1 \\
 0 & 0 & 1 & 0 \\
 0 & 1 & 0 & 0 \\
 1 & 0 & 0 & 0
\end{array}
\right)
\,,~~~~~
\tilde{T}^{12}\,=\,\frac{1}{2}
\left(
\begin{array}{cccc}
 0 & 0 & 0 & 0 \\
 0 & 0 & 0 & 1 \\
 0 & 0 & 0 & 0 \\
 0 & 1 & 0 & 0
\end{array}
\right)
\,,\nonumber\\
\tilde{T}^{13}&=&\frac{1}{2\sqrt{2}}
\left(
\begin{array}{cccc}
 0 & 1 & 0 & 0 \\
 1 & 0 & 0 & 0 \\
 0 & 0 & 0 & -1 \\
 0 & 0 & -1 & 0
\end{array}
\right)
\,,~~~~~
\tilde{T}^{14}\,=\,\frac{1}{2\sqrt{2}}
\left(
\begin{array}{cccc}
 1 & 0 & 0 & 0 \\
 0 & -1 & 0 & 0 \\
 0 & 0 & -1 & 0 \\
 0 & 0 & 0 & 1
\end{array}
\right)
\,,\nonumber\\
\tilde{T}^{15}&=&\frac{1}{2\sqrt{2}}
\left(
\begin{array}{cccc}
 1 & 0 & 0 & 0 \\
 0 & 1 & 0 & 0 \\
 0 & 0 & -1 & 0 \\
 0 & 0 & 0 & -1
\end{array}
\right)
\,.\label{Eq:basisSU(4)}
\eeqs

Following Refs.~\cite{Lee:2017uvl,Bennett:2019cxd}, we define the 
unbroken subgroup $SO(4)_0\sim SU(2)_{L,0}\times SU(2)_{R,0}$ 
as the subset of the unbroken global  $Sp(4)\subset SU(4)$ that 
is generated by the following elements of the associated algebra:
\beqs
\label{Eq:SU2L}
\tilde{T}^{1}_{L,0}&=&\frac{1}{2}\left(\begin{array}{cccc}
0 & 0 & 1 & 0\cr
0 & 0 & 0 & 0\cr
1 & 0 & 0 & 0\cr
0 & 0 & 0 & 0\cr
\end{array}\right)\,,\,\,
\tilde{T}^{2}_{L,0}\,=\,\frac{1}{2}\left(\begin{array}{cccc}
0 & 0 & -i & 0\cr
0 & 0 & 0 & 0\cr
i & 0 & 0 & 0\cr
0 & 0 & 0 & 0\cr
\end{array}\right)\,,\nonumber\\
\tilde{T}^{3}_{L,0}&=&\frac{1}{2}\left(\begin{array}{cccc}
1 & 0 & 0 & 0\cr
0 & 0  & 0 & 0\cr
0 & 0 & -1 & 0\cr
0 & 0 & 0 & 0\cr
\end{array}\right)\,,\\
\tilde{T}^{1}_{R,0}&=&\frac{1}{2}\left(\begin{array}{cccc}
0 & 0 & 0 & 0\cr
0 & 0 & 0 & 1\cr
0 & 0 & 0 & 0\cr
0 & 1 & 0 & 0\cr
\end{array}\right)\,,\,\,
\tilde{T}^{2}_{R,0}\,=\,\frac{1}{2}\left(\begin{array}{cccc}
0 & 0 & 0 & 0\cr
0 & 0 & 0 & -i\cr
0 & 0 & 0 & 0\cr
0 & i & 0 & 0\cr
\end{array}\right)\,,\nonumber\\
\tilde{T}^{3}_{R,0}&=&\frac{1}{2}\left(\begin{array}{cccc}
0 & 0 & 0 & 0\cr
0 &1  & 0 & 0\cr
0 & 0 & 0 & 0\cr
0 & 0 & 0 & -1\cr
\end{array}\right)\,.
\label{Eq:SU2R}
\eeqs
The $T_{L,0}$ and $T_{R,0}$ generators satisfy the algebra $\left[ T_L^i\,,\,T_L^j\right]=i\epsilon^{ijk}\,T_L^k$,
$\left[ T_R^i\,,\,T_R^j\right]=i\epsilon^{ijk}\,T_R^k$, and  $\left[T_L^i,T_R^j\right]=0$.
In the vacuum aligned with $\tilde{\Omega}$, this is the natural choice 
of embedding of the $SO(4)_{EW}$ symmetry of the Higgs potential that leaves it unbroken.
These are linear combinations of the generators $\tilde{T}^{10}$, $\tilde{T}^{12}$, $\tilde{T}^6$, $\tilde{T}^9$, $\tilde{T}^{14}$, and $\tilde{T}^{15}$ in
Eqs.~(\ref{Eq:basisSU(4)}).

The following alternative choice of generators 
defines  $SO(4)_{TC}\sim SU(2)_{L,TC}\times SU(2)_{R,TC}$~\cite{Lee:2017uvl}:
\beqs
\tilde{T}^{1}_{L,TC}&=&\frac{1}{2}\left(\begin{array}{cccc}
0 & 1 & 0 & 0\cr
1 & 0 & 0 & 0\cr
0 & 0 & 0 & 0\cr
0 & 0 & 0 & 0\cr
\end{array}\right)\,,\,\,
\tilde{T}^{2}_{L,TC}\,=\,\frac{1}{2}\left(\begin{array}{cccc}
0 & -i & 0 & 0\cr
i & 0 & 0 & 0\cr
0 & 0 & 0 & 0\cr
0 & 0 & 0 & 0\cr
\end{array}\right)\,,\nonumber\\
\tilde{T}^{3}_{L,TC}&=&\frac{1}{2}\left(\begin{array}{cccc}
1 & 0 & 0 & 0\cr
0 & -1  & 0 & 0\cr
0 & 0 & 0 & 0\cr
0 & 0 & 0 & 0\cr
\end{array}\right)\,,
\label{Eq:SU2LTC}\\
\tilde{T}^{1}_{R,TC}&=&-\frac{1}{2}\left(\begin{array}{cccc}
0 & 0 & 0 & 0\cr
0 & 0 & 0 & 0\cr
0 & 0 & 0 & 1\cr
0 & 0 & 1 & 0\cr
\end{array}\right)\,,\,\,
\tilde{T}^{2}_{R,TC}\,=\,-\frac{1}{2}\left(\begin{array}{cccc}
0 & 0 & 0 & 0\cr
0 & 0 & 0 & 0\cr
0 & 0 & 0 & -i\cr
0 & 0 & i & 0\cr
\end{array}\right)\,,\nonumber\\
\tilde{T}^{3}_{R,TC}&=&-\frac{1}{2}\left(\begin{array}{cccc}
0 & 0 & 0 & 0\cr
0 & 0  & 0 & 0\cr
0 & 0 & 1 & 0\cr
0 & 0 & 0 & -1\cr
\end{array}\right)\,.
\label{Eq:SU2RTC}
\eeqs
These are linear combinations of the generators $\tilde{T}^{1}$, $\tilde{T}^{2}$, $\tilde{T}^3$, $\tilde{T}^8$, $\tilde{T}^{13}$, and $\tilde{T}^{14}$ in
Eqs.~(\ref{Eq:basisSU(4)}).
The vacuum $\Sigma_6\propto \tilde{\Omega}$  breaks $SU(2)_{L,TC}\times SU(2)_{R,TC}$ 
to its diagonal subgroup $SU(2)_{V,TC}$.

\section{Data and Analysis code}
\label{sec:open}

Recently, our collaboration has resolved to openly release full datasets
 for the work that goes into our future publications, as well as, where possible,
 the analysis software used, both to obtain these data and to prepare them for publication.
By doing so, we enable other researchers to make maximal use of our results, 
and to fully understand the process by which they are obtained.
Starting from Refs.~\cite{Bennett:2022yfa,Bennett:2022gdz,Bennett:2022ftz}, our analysis
can be fully reproduced,\footnote{In this work we use ``reproduce'' to mean ``perform the same 
analysis on the same data and obtain the same result'', and ``replicate'' to mean ``repeat the 
same or a similar analysis on freshly-obtained data and obtain compatible results'', as suggested
 by the Turing Way~\cite{turing-way-reproduce}.} 
by means of the data and analysis code packages referred to within the publications themselves. 
The intended benefits of this policy include (but are not limited to) the following.

\begin{itemize}
\item A potential reader might be interested in learning how to apply one of the techniques that 
we have used in our work 
 to their own research. Some technical detail 
 might have been omitted from the published paper
 for presentation reasons (length or readability constraints). Said reader will benefit from 
 direct inspection of  the complete procedure we followed, which can be found in the associated code release.
\item A reader, who seeks to replicate independently one of our findings, 
might discover some  tension between the results of our and their own implementation of the analysis.
Direct inspection of the software we used would enable this reader to identify at what point the divergence 
between the two processes occurs, avoiding protracted arguments on reproducibility---see, e.g.,
the case described in Ref.~\cite{supercooled-water}.
\item Lattice studies frequently generate more data than what can be feasible to fully exploit for a single group of researchers. 
The interested readers may perform their own, additional analysis on our data, with alternative methodologies,
without the need to regenerate the data from scratch 
(which might require a significant investment of computer time).
For example, more advanced fitting algorithms may give more detailed or precise results, 
or gain access to additional observables.
\item Phenomenologists and other researchers who look to build on the numerical results of 
lattice computations may import the data from our work directly 
into their computational environment, without the need to resort to copying and pasting from 
published tables (or reading numbers off published plots). By doing so, 
one reduces the risk of introducing  additional uncertainties, and avoids one source of potential 
human error.
\end{itemize}

In the following,  we  discuss our approach to releasing our data, our analysis code, and other components 
of our workflow that affect the reproducibility of our work, before briefly returning to discussing the benefits 
we see in our process for the robustness of our final results. 
This appendix will focus on the approach that has been taken to date by our collaboration,
with specific reference to Refs.~\cite{Bennett:2022yfa,Bennett:2022gdz,Bennett:2022ftz};
 a more general, pedagogical guide to adopting this approach is in preparation~\cite{repro-guide}.

\subsection{Data release}

The primary data we publish are plaintext output files from production of configurations 
and from  subsequent computation of observables (measurements).
We do not  release gauge configurations used for Refs.~\cite{Bennett:2022yfa,Bennett:2022gdz,Bennett:2022ftz}, 
due to the unavailability of a suitable hosting platform with adequate capacity, 
but restrict our release to the measurements.
Even in the case where such capacity were available, 
releasing the measurement output files significantly reduces the barrier to entry 
(in terms of computer time and capability required) for 
those readers who are looking to reproduce the analysis.

To be more specific, we release four primary classes of data:
\begin{description}
\item[Raw data,] such as correlation functions and gradient flow histories,
 are released in their native formats as generated by the HiRep code~\cite{hirep-repo,DelDebbio:2008zf}, in accordance with the principle of ``keeping raw data raw'' \cite{ten-data-rules}. By doing so, we reduce the chances of human error in transcription of data formats, while 
 increasing the opportunity to detect  such type of errors in  a subsequent validation process.

\item[Reformatted raw data,] obtained by taking the output files of raw data, condense the salient 
information in tables stored in HDF5 format~\cite{hdf5}. 
Commonly available library functions can read the data in this format,
so that one does not need to write a parser to interpret the bespoke formats generated by HiRep. 
 Currently this information is  
 generated from the raw log files as part of the analysis process.
 
  \item[Metadata] are collections of parameter values which identify the analysis performed. 
They include physical parameters, such as the lattice coupling $\beta$, algorithmic ones, 
such as the number of trajectories between successive configurations, and analysis ones, such as the start 
and end of plateaux in effective mass plots. The metadata we publish are primarily those which enable the analysis.

\item[Final results,] also presented in tabular form in the corresponding publications,
 are released in CSV format; they are typically compact enough that using a denser format such as HDF5 would
 not yield a significant benefit (in file size, for example), and the use of CSV files makes the data 
 accessible without specialist software tools.

\end{description}

If a data format is not formally defined, 
we include in the release also detailed descriptions to enable the user 
to understand and parse the data. This aids  users who
are unfamiliar with data formats used by bespoke software packages, such as HiRep.

We publish the data to Zenodo~\cite{zenodo}, a general-purpose data repository maintained by the
European Organisation for Nuclear Research (CERN). Each data set 
(or each version of a dataset, in cases where revisions are necessary) 
is allocated a Digital Object Identifier (DOI) which may be used to cite the data directly. 
Unlike a Uniform Resource Locator (URL), typically used to refer to a web page, 
a DOI is designed to avoid ``link rot'', where changes in website structure cause links to become invalid. 
Zenodo, as other dedicated data repositories, is planned to outlast a typical institutional affiliation, 
and its data sets are expected to remain available past the time
when the author has retired, changed institutions, or simply stopped paying a hosting bill. 
The Zenodo DOI is cited from the paper in order to alert the reader to its availability.

\subsection{Analysis code release}

We automate the analysis leading to a publication:
our tooling takes the data and metadata release, and its output consists of the
 full set of plots and tables in the paper.
This analysis kit is not written \emph{a priori} and then run on the data obtained from 
High Performance Computing (HPC) simulations. Rather, any 
manual steps are subsequently translated into data that can be used 
 \emph{a posteriori} to compute the result. 
To provide a concrete example, our choice of plateaux  in mesonic correlation
 functions is not fully algorithmic: the positions of the plateaux 
 are  identified by a member of the collaboration in a semi-algorithmic way, 
 and then the results (start and end time of the plateaux) stored in a text file that is 
 subsequently read in by the analysis code to be released. The end user of the release does not have
  to identify the plateaux manually (which would compromise reproducibility),
  yet they may inspect and test our choices.

Our Collaboration has developed a body of bespoke software, coded in several different computer  languages:
 Python~\cite{python} (in particular the packages Numpy~\cite{numpy}, Scipy~\cite{scipy}, and Matplotlib~\cite{matplotlib}), Mathematica~\cite{mathematica}, and to lesser extent  Bash~\cite{bash}. Individual analysis tools are combined
  together using GNU Make~\cite{gnu-make}, which offers a few significant advantages over using a hard-coded shell script.
\begin{enumerate}
\item Dependencies between steps are automatically managed. The ordering of steps is automatically decided, rather than requiring user's input.
\item Steps can automatically be run in parallel, with Make ensuring that no step runs before its prerequisites are complete. This allows the analysis process to scale with the available compute capacity. 
\item The workflow can be interrupted partway and resumed subsequently, without the need to re-run previously 
completed steps. 
\item  Make is able to re-run only the steps of the analysis that depend on specific  files,
if data are updated, expediting  the debugging cycle.
\end{enumerate}
The moderate  cost to pay for these benefits is that writing and debugging a
 Makefile for the type of workflow we automate is relatively complex.

The workflows for Refs.~\cite{Bennett:2022yfa,Bennett:2022gdz,Bennett:2022ftz} 
were originally run interactively, hence required postproduction reworking and  automation
before release. 
Reformulating our toolchain to be written in an automation-first way is an 
ongoing internal project which will significantly reduce the effort required for future code releases.

We verify that repeated runs of the analysis give identical  output.
Small  fluctuations (within uncertainties) due, for example, to changes in the bootstrap samples
have been removed by fixing the random seeds based on metadata about the files being processed.

Where possible, we specify the full software environment used to 
perform the analysis---for example, the version of Python and all Python packages used. 
Doing so is necessary to enable reproducibility, as
 some libraries give quantitively different results when switching between versions.
 We specify this such via an~\verb|environment.yml| file compatible with the Conda package manager~\cite{conda}.

Analysis tools are held in GitHub~\cite{github} while they are being actively developed, 
and then pushed to Zenodo~\cite{zenodo} when they are ready for publication,
where they acquire a persistent identifier (DOI).
 This process also identifies the specific revision of the code used to generate the publication; as we move to building tools
  that are used for multiple publications and modified or updated in between, this will remove 
  any ambiguity as to software version.
  As with the data release, the analysis code release on Zenodo is cited in the paper to alert the reader to its availability.

\subsection{Closing remarks}

Not all data sets can be released, due to the excessive
 requirements of storage and computer time, but also not all steps of a computation can be automated.
We document  as much of the process as possible 
in the steps of our computations that 
cannot be prepared and published as automated reproducible pipelines.
As an example, while we make use of the open-source tool HiRep~\cite{hirep-upstream},
we have also made a number of customised modifications on it, including adapting the Monte Carlo 
to the $Sp(2N)$ groups
 and implementing the measurement of the chimera baryon correlators. 
 These modifications are  publicly released elsewhere~\cite{hirep-repo}, and the specific branch used is identified 
 in our publications.

In the process of preparing data and analysis software for release, we identified a number of  minor 
inconsistencies in our data sets that otherwise might have been overlooked,
and ultimately did not affect the conclusions of our work.
Said inconsistencies
 originate from the fact that working with large numbers of files and 
data is inherently prone to unavoidable human error.
They might have adversely affected the ability of someone else to replicate our work.
The very  adoption of our open release  policy
ultimately had the serendipitous consequence of adding one more layer of independent consistency checks,
making our scientific output more robust.

We refer the reader to Refs.~\cite{Bennett:2022klt,Athenodorou:2022ixd} and references therein for 
recent surveys on open science.

\end{paracol}




\begin{thebibliography}{999}


\bibitem{Holland:2003kg}
K.~Holland, M.~Pepe and U.~J.~Wiese,
``The Deconfinement phase transition of Sp(2) and Sp(3) Yang-Mills theories in (2+1)-dimensions and (3+1)-dimensions,''
Nucl. Phys. B \textbf{694}, 35-58 (2004)
doi:10.1016/j.nuclphysb.2004.06.026
[arXiv:hep-lat/0312022 [hep-lat]].





\bibitem{Bennett:2017kga}
  E.~Bennett, D.~K.~Hong, J.~W.~Lee, C.-J.~D.~Lin, B.~Lucini, M.~Piai and D.~Vadacchino,
 ``Sp(4) gauge theory on the lattice: towards SU(4)/Sp(4) composite Higgs (and beyond),''
  JHEP {\bf 1803}, 185 (2018)
  doi:10.1007/JHEP03(2018)185
  [arXiv:1712.04220 [hep-lat]].



\bibitem{Lee:2018ztv}
  J.~W.~Lee, E.~Bennett, D.~K.~Hong, C.~J.~D.~Lin, B.~Lucini, M.~Piai and D.~Vadacchino,
  ``Progress in the lattice simulations of Sp(2$N$) gauge theories,''
  PoS LATTICE {\bf 2018}, 192 (2018)
 doi:10.22323/1.334.0192
  [arXiv:1811.00276 [hep-lat]].
 
  

\bibitem{Bennett:2019jzz}
E.~Bennett, D.~K.~Hong, J.~W.~Lee, C.~J.~D.~Lin, B.~Lucini, M.~Piai and D.~Vadacchino,
``Sp(4) gauge theories on the lattice: $N_f=2$ dynamical fundamental fermions,''
JHEP \textbf{12} (2019), 053
doi:10.1007/JHEP12(2019)053
[arXiv:1909.12662 [hep-lat]].

  

\bibitem{Bennett:2019cxd}
E.~Bennett, D.~K.~Hong, J.~W.~Lee, C.~J.~D.~Lin, B.~Lucini, M.~Mesiti, M.~Piai, J.~Rantaharju and D.~Vadacchino,
``$Sp(4)$ gauge theories on the lattice: quenched fundamental and antisymmetric fermions,''
Phys. Rev. D \textbf{101} (2020) no.7, 074516
doi:10.1103/PhysRevD.101.074516
[arXiv:1912.06505 [hep-lat]].

  
  
\bibitem{Bennett:2020hqd}
E.~Bennett, J.~Holligan, D.~K.~Hong, J.~W.~Lee, C.~J.~D.~Lin, B.~Lucini, M.~Piai and D.~Vadacchino,
``Color dependence of tensor and scalar glueball masses in Yang-Mills theories,''
Phys. Rev. D \textbf{102}, no.1, 011501 (2020)
doi:10.1103/PhysRevD.102.011501
[arXiv:2004.11063 [hep-lat]].

\bibitem{Bennett:2020qtj}
E.~Bennett, J.~Holligan, D.~K.~Hong, J.~W.~Lee, C.~J.~D.~Lin, B.~Lucini, M.~Piai and D.~Vadacchino,
``Glueballs and strings in $Sp(2N)$ Yang-Mills theories,''
Phys. Rev. D \textbf{103} (2021) no.5, 054509
doi:10.1103/PhysRevD.103.054509
[arXiv:2010.15781 [hep-lat]].


\bibitem{Lucini:2021xke}
B.~Lucini, E.~ Bennett, J.~Holligan, D.~K.~Hong, H.~Hsiao, J.~W.~Lee, C.~J.~D.~Lin, M.~Mesiti, M.~Piai and D.~Vadacchino,
``Sp(4) gauge theories and beyond the standard model physics,''
EPJ Web Conf. \textbf{258}, 08003 (2022)
doi:10.1051/epjconf/202225808003
[arXiv:2111.12125 [hep-lat]].

\bibitem{Bennett:2021mbw}
E.~Bennett, J.~Holligan, D.~K.~Hong, H.~Hsiao, J.~W.~Lee, C.~J.~D.~Lin, B.~Lucini, M.~Mesiti, M.~Piai and D.~Vadacchino,
``Progress in $Sp(2N)$ lattice gauge theories,''
PoS \textbf{LATTICE2021}, 308 (2022)
doi:10.22323/1.396.0308
[arXiv:2111.14544 [hep-lat]].

\bibitem{Bennett:2022yfa}
E.~Bennett, D.~K.~Hong, H.~Hsiao, J.~W.~Lee, C.~J.~D.~Lin, B.~Lucini, M.~Mesiti, M.~Piai and D.~Vadacchino,
``Lattice studies of the Sp(4) gauge theory with two fundamental and three antisymmetric Dirac fermions,''
Phys. Rev. D \textbf{106}, no.1, 014501 (2022)
doi:10.1103/PhysRevD.106.014501
[arXiv:2202.05516 [hep-lat]].

\bibitem{Bennett:2022gdz}
E.~Bennett, D.~K.~Hong, J.~W.~Lee, C.~J.~D.~Lin, B.~Lucini, M.~Piai and D.~Vadacchino,
``Color dependence of the topological susceptibility in Yang-Mills theories,''
Phys. Lett. B \textbf{835}, 137504 (2022)
doi:10.1016/j.physletb.2022.137504
[arXiv:2205.09254 [hep-lat]].


\bibitem{Bennett:2022ftz}
E.~Bennett, D.~K.~Hong, J.~W.~Lee, C.~J.~D.~Lin, B.~Lucini, M.~Piai and D.~Vadacchino,
``Sp(2N) Yang-Mills theories on the lattice: Scale setting and topology,''
Phys. Rev. D \textbf{106}, no.9, 094503 (2022)
doi:10.1103/PhysRevD.106.094503
[arXiv:2205.09364 [hep-lat]].


\bibitem{AS}
E.~Bennett,  D.~K.~Hong, H.~Hsiao, J.~W.~Lee, C.~J.~D.~Lin,
B.~Lucini,  M.~Piai, and D.~Vadacchino,
``Sp(4) theories on the lattice: dynamical antisymmetric fermions,''
in preparation.

\bibitem{Lee:2022elf}
J.~W.~Lee, E.~Bennett, D.~K.~Hong, H.~Hsiao, C.~J.~D.~Lin, B.~Lucini, M.~Piai and D.~Vadacchino,
``Spectroscopy of $Sp(4)$ lattice gauge theory with $n_f=3$ antisymmetric fermions,''
PoS \textbf{LATTICE2022}, 214 (2023)
doi:10.22323/1.430.0214
[arXiv:2210.08154 [hep-lat]].

\bibitem{Hsiao:2022kxf}
H.~Hsiao, E.~Bennett, D.~K.~Hong, J.~W.~Lee, C.~J.~D.~Lin, B.~Lucini, M.~Piai and D.~Vadacchino,
``Spectroscopy of chimera baryons in a $Sp(4)$ lattice gauge theory,''
PoS \textbf{LATTICE2022}, 211 (2023)
doi:10.22323/1.430.0211
[arXiv:2211.03955 [hep-lat]].


\bibitem{Maas:2021gbf}
A.~Maas and F.~Zierler,
``Strong isospin breaking in Sp(4) gauge theory,''
[arXiv:2109.14377 [hep-lat]].

\bibitem{Zierler:2021cfa}
F.~Zierler and A.~Maas,
``$Sp(4)$ SIMP Dark Matter on the Lattice,''
PoS \textbf{LHCP2021}, 162 (2021)
doi:10.22323/1.397.0162

\bibitem{Kulkarni:2022bvh}
S.~Kulkarni, A.~Maas, S.~Mee, M.~Nikolic, J.~Pradler and F.~Zierler,
``Low-energy effective description of dark $Sp(4)$ theories,''
[arXiv:2202.05191 [hep-ph]].



\bibitem{Peskin:1980gc}
M.~E.~Peskin,
``The Alignment of the Vacuum in Theories of Technicolor,''
Nucl. Phys. B \textbf{175}, 197-233 (1980)
doi:10.1016/0550-3213(80)90051-6


\bibitem{Aad:2012tfa} 
G.~Aad {\it et al.} [ATLAS Collaboration],
``Observation of a new particle in the search for the Standard Model Higgs boson with the ATLAS detector at the LHC,''
Phys.\ Lett.\ B {\bf 716}, 1 (2012)
doi:10.1016/j.physletb.2012.08.020
[arXiv:1207.7214 [hep-ex]].


\bibitem{Chatrchyan:2012xdj} 
S.~Chatrchyan {\it et al.} [CMS Collaboration],
``Observation of a new boson at a mass of 125 GeV with the CMS experiment at the LHC,''
Phys.\ Lett.\ B {\bf 716}, 30 (2012)
doi:10.1016/j.physletb.2012.08.021
[arXiv:1207.7235 [hep-ex]].



\bibitem{Kaplan:1983fs}
D.~B.~Kaplan and H.~Georgi,
``SU(2) x U(1) Breaking by Vacuum Misalignment,''
Phys. Lett. B \textbf{136}, 183-186 (1984)
doi:10.1016/0370-2693(84)91177-8


 \bibitem{Georgi:1984af} 
  H.~Georgi and D.~B.~Kaplan,
   ``Composite Higgs and Custodial SU(2),''
  Phys.\ Lett.\  {\bf 145B}, 216 (1984).
  doi:10.1016/0370-2693(84)90341-1
  


 \bibitem{Dugan:1984hq} 
  M.~J.~Dugan, H.~Georgi and D.~B.~Kaplan,
   ``Anatomy of a Composite Higgs Model,''
  Nucl.\ Phys.\ B {\bf 254}, 299 (1985).
  doi:10.1016/0550-3213(85)90221-4


  



 \bibitem{Panico:2015jxa} 
  G.~Panico and A.~Wulzer,
   ``The Composite Nambu-Goldstone Higgs,''
  Lect.\ Notes Phys.\  {\bf 913}, pp.1 (2016)
  doi:10.1007/978-3-319-22617-0
  [arXiv:1506.01961 [hep-ph]].
  
\bibitem{Witzel:2019jbe} 
  O.~Witzel,
  ``Review on Composite Higgs Models,''
  PoS LATTICE {\bf 2018}, 006 (2019)
  doi:10.22323/1.334.0006
  [arXiv:1901.08216 [hep-lat]].

  
\bibitem{Cacciapaglia:2020kgq}
G.~Cacciapaglia, C.~Pica and F.~Sannino,
``Fundamental Composite Dynamics: A Review,''
Phys. Rept. \textbf{877}, 1-70 (2020)
doi:10.1016/j.physrep.2020.07.002
[arXiv:2002.04914 [hep-ph]].

  
\bibitem{Ferretti:2013kya}
G.~Ferretti and D.~Karateev,
``Fermionic UV completions of Composite Higgs models,''
JHEP \textbf{03}, 077 (2014)
doi:10.1007/JHEP03(2014)077
[arXiv:1312.5330 [hep-ph]].
  
   
\bibitem{Ferretti:2016upr}
G.~Ferretti,
``Gauge theories of Partial Compositeness: Scenarios for Run-II of the LHC,''
JHEP \textbf{06}, 107 (2016)
doi:10.1007/JHEP06(2016)107
[arXiv:1604.06467 [hep-ph]].

     \bibitem{Cacciapaglia:2019bqz} 
  G.~Cacciapaglia, G.~Ferretti, T.~Flacke and H.~Ser\^{o}dio,
  ``Light scalars in composite Higgs models,''
  Front.\ Phys.\  {\bf 7}, 22 (2019)
  doi:10.3389/fphy.2019.00022
  [arXiv:1902.06890 [hep-ph]].



  
 \bibitem{Katz:2005au} 
  E.~Katz, A.~E.~Nelson and D.~G.~E.~Walker,
   ``The Intermediate Higgs,''
  JHEP {\bf 0508}, 074 (2005)
  doi:10.1088/1126-6708/2005/08/074
  [hep-ph/0504252].




 \bibitem{Barbieri:2007bh} 
  R.~Barbieri, B.~Bellazzini, V.~S.~Rychkov and A.~Varagnolo,
   ``The Higgs boson from an extended symmetry,''
  Phys.\ Rev.\ D {\bf 76}, 115008 (2007)
  doi:10.1103/PhysRevD.76.115008
  [arXiv:0706.0432 [hep-ph]].



  
   \bibitem{Lodone:2008yy} 
  P.~Lodone,
   ``Vector-like quarks in a composite Higgs model,''
  JHEP {\bf 0812}, 029 (2008)
  doi:10.1088/1126-6708/2008/12/029
  [arXiv:0806.1472 [hep-ph]].



 \bibitem{Gripaios:2009pe} 
  B.~Gripaios, A.~Pomarol, F.~Riva and J.~Serra,
   ``Beyond the Minimal Composite Higgs Model,''
  JHEP {\bf 0904}, 070 (2009)
  doi:10.1088/1126-6708/2009/04/070
  [arXiv:0902.1483 [hep-ph]].
  
  \bibitem{Mrazek:2011iu}
J.~Mrazek, A.~Pomarol, R.~Rattazzi, M.~Redi, J.~Serra and A.~Wulzer,
``The Other Natural Two Higgs Doublet Model,''
Nucl. Phys. B \textbf{853}, 1-48 (2011)
doi:10.1016/j.nuclphysb.2011.07.008
[arXiv:1105.5403 [hep-ph]].
  
\bibitem{Marzocca:2012zn} 
  D.~Marzocca, M.~Serone and J.~Shu,
   ``General Composite Higgs Models,''
  JHEP {\bf 1208}, 013 (2012)
  doi:10.1007/JHEP08(2012)013
  [arXiv:1205.0770 [hep-ph]].


\bibitem{Grojean:2013qca} 
  C.~Grojean, O.~Matsedonskyi and G.~Panico,
  ``Light top partners and precision physics,''
  JHEP {\bf 1310}, 160 (2013)
  doi:10.1007/JHEP10(2013)160
  [arXiv:1306.4655 [hep-ph]].





   \bibitem{Cacciapaglia:2014uja} 
  G.~Cacciapaglia and F.~Sannino,
   ``Fundamental Composite (Goldstone) Higgs Dynamics,''
  JHEP {\bf 1404}, 111 (2014)
  doi:10.1007/JHEP04(2014)111
  [arXiv:1402.0233 [hep-ph]].


\bibitem{Ferretti:2014qta}
G.~Ferretti,
``UV Completions of Partial Compositeness: The Case for a SU(4) Gauge Group,''
JHEP \textbf{06}, 142 (2014)
doi:10.1007/JHEP06(2014)142
[arXiv:1404.7137 [hep-ph]].

  \bibitem{Arbey:2015exa} 
  A.~Arbey, G.~Cacciapaglia, H.~Cai, A.~Deandrea, S.~Le Corre and F.~Sannino,
   ``Fundamental Composite Electroweak Dynamics: Status at the LHC,''
  Phys.\ Rev.\ D {\bf 95}, no. 1, 015028 (2017)
  doi:10.1103/PhysRevD.95.015028
  [arXiv:1502.04718 [hep-ph]].

  
   
  
   \bibitem{Cacciapaglia:2015eqa} 
  G.~Cacciapaglia, H.~Cai, A.~Deandrea, T.~Flacke, S.~J.~Lee and A.~Parolini,
   ``Composite scalars at the LHC: the Higgs, the Sextet and the Octet,''
  JHEP {\bf 1511}, 201 (2015)
  doi:10.1007/JHEP11(2015)201
  [arXiv:1507.02283 [hep-ph]].

  
    
  
   \bibitem{Feruglio:2016zvt} 
  F.~Feruglio, B.~Gavela, K.~Kanshin, P.~A.~N.~Machado, S.~Rigolin and S.~Saa,
   ``The minimal linear sigma model for the Goldstone Higgs,''
  JHEP {\bf 1606}, 038 (2016)
  doi:10.1007/JHEP06(2016)038
  [arXiv:1603.05668 [hep-ph]].

  
    
 \bibitem{DeGrand:2016pgq} 
  T.~DeGrand, M.~Golterman, E.~T.~Neil and Y.~Shamir,
   ``One-loop Chiral Perturbation Theory with two fermion representations,''
  Phys.\ Rev.\ D {\bf 94}, no. 2, 025020 (2016)
  doi:10.1103/PhysRevD.94.025020
  [arXiv:1605.07738 [hep-ph]].

  

 
   \bibitem{Fichet:2016xvs} 
  S.~Fichet, G.~von Gersdorff, E.~Pont\`{o}n and R.~Rosenfeld,
   ``The Excitation of the Global Symmetry-Breaking Vacuum in Composite Higgs Models,''
  JHEP {\bf 1609}, 158 (2016)
  doi:10.1007/JHEP09(2016)158
  [rXiv:1607.03125 [hep-ph]].

  
    
   \bibitem{Galloway:2016fuo} 
  J.~Galloway, A.~L.~Kagan and A.~Martin,
   ``A UV complete partially composite-pNGB Higgs,''
  Phys.\ Rev.\ D {\bf 95}, no. 3, 035038 (2017)
  doi:10.1103/PhysRevD.95.035038
  [arXiv:1609.05883 [hep-ph]].
 
  
 \bibitem{Agugliaro:2016clv} 
  A.~Agugliaro, O.~Antipin, D.~Becciolini, S.~De Curtis and M.~Redi,
   ``UV complete composite Higgs models,''
  Phys.\ Rev.\ D {\bf 95}, no. 3, 035019 (2017)
  doi:10.1103/PhysRevD.95.035019
  [arXiv:1609.07122 [hep-ph]].

  
 
 \bibitem{Belyaev:2016ftv}
 A.~Belyaev, G.~Cacciapaglia, H.~Cai, G.~Ferretti, T.~Flacke, A.~Parolini and H.~Serodio,
 ``Di-boson signatures as Standard Candles for Partial Compositeness,''
 JHEP \textbf{01}, 094 (2017)
 doi:10.1007/JHEP01(2017)094
 [arXiv:1610.06591 [hep-ph]].



  
  

    
   \bibitem{Csaki:2017cep} 
  C.~Csaki, T.~Ma and J.~Shu,
   ``Maximally Symmetric Composite Higgs Models,''
  Phys.\ Rev.\ Lett.\  {\bf 119}, no. 13, 131803 (2017)
  doi:10.1103/PhysRevLett.119.131803
  [arXiv:1702.00405 [hep-ph]].

 
     \bibitem{Chala:2017sjk} 
  M.~Chala, G.~Durieux, C.~Grojean, L.~de Lima and O.~Matsedonskyi,
   ``Minimally extended SILH,''
  JHEP {\bf 1706}, 088 (2017)
  doi:10.1007/JHEP06(2017)088
  [arXiv:1703.10624 [hep-ph]].

  
  
\bibitem{Golterman:2017vdj} 
  M.~Golterman and Y.~Shamir,
   ``Effective potential in ultraviolet completions for composite Higgs models,''
  Phys.\ Rev.\ D {\bf 97}, no. 9, 095005 (2018)
  doi:10.1103/PhysRevD.97.095005
  [arXiv:1707.06033 [hep-ph]].



\bibitem{Csaki:2017jby} 
  C.~Csaki, T.~Ma and J.~Shu,
   ``Trigonometric Parity for Composite Higgs Models,''
  Phys.\ Rev.\ Lett.\  {\bf 121}, no. 23, 231801 (2018)
  doi:10.1103/PhysRevLett.121.231801
  [arXiv:1709.08636 [hep-ph]].

  
   \bibitem{Alanne:2017rrs} 
  T.~Alanne, D.~Buarque Franzosi and M.~T.~Frandsen,
   ``A partially composite Goldstone Higgs,''
  Phys.\ Rev.\ D {\bf 96}, no. 9, 095012 (2017)
  doi:10.1103/PhysRevD.96.095012
  [arXiv:1709.10473 [hep-ph]].


 
\bibitem{Alanne:2017ymh} 
  T.~Alanne, D.~Buarque Franzosi, M.~T.~Frandsen, M.~L.~A.~Kristensen, A.~Meroni and M.~Rosenlyst,
   ``Partially composite Higgs models: Phenomenology and RG analysis,''
  JHEP {\bf 1801}, 051 (2018)
  doi:10.1007/JHEP01(2018)051
  [arXiv:1711.10410 [hep-ph]].
 


 
 \bibitem{Sannino:2017utc} 
  F.~Sannino, P.~Stangl, D.~M.~Straub and A.~E.~Thomsen,
   ``Flavor Physics and Flavor Anomalies in Minimal Fundamental Partial Compositeness,''
  Phys.\ Rev.\ D {\bf 97}, no. 11, 115046 (2018)
  doi:10.1103/PhysRevD.97.115046
  [arXiv:1712.07646 [hep-ph]].
 

\bibitem{Alanne:2018wtp} 
  T.~Alanne, N.~Bizot, G.~Cacciapaglia and F.~Sannino,
   ``Classification of NLO operators for composite Higgs models,''
  Phys.\ Rev.\ D {\bf 97}, no. 7, 075028 (2018)
  doi:10.1103/PhysRevD.97.075028
  [arXiv:1801.05444 [hep-ph]].







 \bibitem{Bizot:2018tds} 
  N.~Bizot, G.~Cacciapaglia and T.~Flacke,
   ``Common exotic decays of top partners,''
  JHEP {\bf 1806}, 065 (2018)
  doi:10.1007/JHEP06(2018)065
  [arXiv:1803.00021 [hep-ph]].



\bibitem{Cai:2018tet} 
  C.~Cai, G.~Cacciapaglia and H.~H.~Zhang,
  ``Vacuum alignment in a composite 2HDM,''
  JHEP {\bf 1901}, 130 (2019)
  doi:10.1007/JHEP01(2019)130
  [arXiv:1805.07619 [hep-ph]].



\bibitem{Agugliaro:2018vsu} 
  A.~Agugliaro, G.~Cacciapaglia, A.~Deandrea and S.~De Curtis,
  ``Vacuum misalignment and pattern of scalar masses in the SU(5)/SO(5) composite Higgs model,''
  JHEP {\bf 1902}, 089 (2019)
  doi:10.1007/JHEP02(2019)089
  [arXiv:1808.10175 [hep-ph]].
  
\bibitem{Cacciapaglia:2018avr}
G.~Cacciapaglia, T.~Ma, S.~Vatani and Y.~Wu,
``Towards a fundamental safe theory of composite Higgs and Dark Matter,''
Eur. Phys. J. C \textbf{80}, no.11, 1088 (2020)
doi:10.1140/epjc/s10052-020-08648-7
[arXiv:1812.04005 [hep-ph]].


\bibitem{Gertov:2019yqo} 
  H.~Gertov, A.~E.~Nelson, A.~Perko and D.~G.~E.~Walker,
  ``Lattice-Friendly Gauge Completion of a Composite Higgs with Top Partners,''
  JHEP {\bf 1902}, 181 (2019)
  doi:10.1007/JHEP02(2019)181
  [arXiv:1901.10456 [hep-ph]].

  


  
  \bibitem{Ayyar:2019exp} 
  V.~Ayyar, M.~F.~Golterman, D.~C.~Hackett, W.~Jay, E.~T.~Neil, Y.~Shamir and B.~Svetitsky,
  ``Radiative Contribution to the Composite-Higgs Potential in a Two-Representation Lattice Model,''
  Phys.\ Rev.\ D {\bf 99}, no. 9, 094504 (2019)
  doi:10.1103/PhysRevD.99.094504
  [arXiv:1903.02535 [hep-lat]].

  

  
  
\bibitem{Cacciapaglia:2019ixa} 
  G.~Cacciapaglia, H.~Cai, A.~Deandrea and A.~Kushwaha,
  ``Composite Higgs and Dark Matter Model in SU(6)/SO(6),''
  JHEP {\bf 1910}, 035 (2019)
  doi:10.1007/JHEP10(2019)035
  [arXiv:1904.09301 [hep-ph]].



\bibitem{BuarqueFranzosi:2019eee} 
  D.~Buarque Franzosi and G.~Ferretti,
  ``Anomalous dimensions of potential top-partners,''
  SciPost Phys.\  {\bf 7}, no. 3, 027 (2019)
  doi:10.21468/SciPostPhys.7.3.027
  [arXiv:1905.08273 [hep-ph]].

  
\bibitem{Cacciapaglia:2019dsq}
G.~Cacciapaglia, S.~Vatani and C.~Zhang,
``Composite Higgs Meets Planck Scale: Partial Compositeness from Partial Unification,''
Phys. Lett. B \textbf{815}, 136177 (2021)
doi:10.1016/j.physletb.2021.136177
[arXiv:1911.05454 [hep-ph]].

\bibitem{Cacciapaglia:2020vyf}
G.~Cacciapaglia, A.~Deandrea, T.~Flacke and A.~M.~Iyer,
``Gluon-Photon Signatures for color octet at the LHC (and beyond),''
JHEP \textbf{05}, 027 (2020)
doi:10.1007/JHEP05(2020)027
[arXiv:2002.01474 [hep-ph]].


\bibitem{Dong:2020eqy}
Z.~Y.~Dong, C.~S.~Guan, T.~Ma, J.~Shu and X.~Xue,
``UV completed composite Higgs model with heavy composite partners,''
Phys. Rev. D \textbf{104}, no.3, 035013 (2021)
doi:10.1103/PhysRevD.104.035013
[arXiv:2011.09460 [hep-ph]].


\bibitem{Cacciapaglia:2021uqh}
G.~Cacciapaglia, T.~Flacke, M.~Kunkel and W.~Porod,
``Phenomenology of unusual top partners in composite Higgs models,''
[arXiv:2112.00019 [hep-ph]].

\bibitem{Banerjee:2022izw}
A.~Banerjee, D.~B.~Franzosi and G.~Ferretti,
``Modelling vector-like quarks in partial compositeness framework,''
[arXiv:2202.00037 [hep-ph]].


\bibitem{Contino:2003ve}
R.~Contino, Y.~Nomura and A.~Pomarol,
``Higgs as a holographic pseudoGoldstone boson,''
Nucl. Phys. B \textbf{671} (2003), 148-174
doi:10.1016/j.nuclphysb.2003.08.027
[arXiv:hep-ph/0306259 [hep-ph]].

 \bibitem{Agashe:2004rs} 
  K.~Agashe, R.~Contino and A.~Pomarol,
   ``The Minimal composite Higgs model,''
  Nucl.\ Phys.\ B {\bf 719}, 165 (2005)
  doi:10.1016/j.nuclphysb.2005.04.035
  [hep-ph/0412089].
 
  
\bibitem{Agashe:2005dk}
K.~Agashe and R.~Contino,
``The Minimal composite Higgs model and electroweak precision tests,''
Nucl. Phys. B \textbf{742} (2006), 59-85
doi:10.1016/j.nuclphysb.2006.02.011
[arXiv:hep-ph/0510164 [hep-ph]].
  
  \bibitem{Agashe:2006at}
K.~Agashe, R.~Contino, L.~Da Rold and A.~Pomarol,
``A Custodial symmetry for $Zb \bar b$,''
Phys. Lett. B \textbf{641} (2006), 62-66
doi:10.1016/j.physletb.2006.08.005
[arXiv:hep-ph/0605341 [hep-ph]].

 \bibitem{Contino:2006qr} 
  R.~Contino, L.~Da Rold and A.~Pomarol,
   ``Light custodians in natural composite Higgs models,''
  Phys.\ Rev.\ D {\bf 75}, 055014 (2007)
  doi:10.1103/PhysRevD.75.055014
  [hep-ph/0612048].
  
  
    
    \bibitem{Falkowski:2008fz}
A.~Falkowski and M.~Perez-Victoria,
``Electroweak Breaking on a Soft Wall,''
JHEP \textbf{12}, 107 (2008)
doi:10.1088/1126-6708/2008/12/107
[arXiv:0806.1737 [hep-ph]].

\bibitem{Contino:2010rs}
R.~Contino,
``The Higgs as a Composite Nambu-Goldstone Boson,''
doi:10.1142/9789814327183\_0005
[arXiv:1005.4269 [hep-ph]].


\bibitem{Contino:2011np}
R.~Contino, D.~Marzocca, D.~Pappadopulo and R.~Rattazzi,
``On the effect of resonances in composite Higgs phenomenology,''
JHEP \textbf{10} (2011), 081
doi:10.1007/JHEP10(2011)081
[arXiv:1109.1570 [hep-ph]].
  
  
\bibitem{Elander:2023aow}
D.~Elander, A.~Fatemiabhari and M.~Piai,
``Towards composite Higgs: minimal coset from a regular bottom-up holographic model,''
[arXiv:2303.00541 [hep-th]].

 
 

\bibitem{Kaplan:1991dc}
D.~B.~Kaplan,
``Flavor at SSC energies: A New mechanism for dynamically generated fermion masses,''
Nucl. Phys. B \textbf{365} (1991), 259-278
doi:10.1016/S0550-3213(05)80021-5 

  
  
  \bibitem{Grossman:1999ra}
Y.~Grossman and M.~Neubert,
``Neutrino masses and mixings in nonfactorizable geometry,''
Phys. Lett. B \textbf{474}, 361-371 (2000)
doi:10.1016/S0370-2693(00)00054-X
[arXiv:hep-ph/9912408 [hep-ph]].

\bibitem{Gherghetta:2000qt}
T.~Gherghetta and A.~Pomarol,
``Bulk fields and supersymmetry in a slice of AdS,''
Nucl. Phys. B \textbf{586}, 141-162 (2000)
doi:10.1016/S0550-3213(00)00392-8
[arXiv:hep-ph/0003129 [hep-ph]].

\bibitem{Chacko:2012sy}
Z.~Chacko and R.~K.~Mishra,
``Effective Theory of a Light Dilaton,''
Phys. Rev. D \textbf{87}, no.11, 115006 (2013)
doi:10.1103/PhysRevD.87.115006
[arXiv:1209.3022 [hep-ph]].






  
   \bibitem{Hietanen:2014xca} 
  A.~Hietanen, R.~Lewis, C.~Pica and F.~Sannino,
   ``Fundamental Composite Higgs Dynamics on the Lattice: SU(2) with Two Flavors,''
  JHEP {\bf 1407}, 116 (2014)
  doi:10.1007/JHEP07(2014)116
  [arXiv:1404.2794 [hep-lat]].

  
  \bibitem{Detmold:2014kba} 
  W.~Detmold, M.~McCullough and A.~Pochinsky,
   ``Dark nuclei. II. Nuclear spectroscopy in two-color QCD,''
  Phys.\ Rev.\ D {\bf 90}, no. 11, 114506 (2014)
  doi:10.1103/PhysRevD.90.114506
  [arXiv:1406.4116 [hep-lat]].

  
          \bibitem{Arthur:2016dir} 
  R.~Arthur, V.~Drach, M.~Hansen, A.~Hietanen, C.~Pica and F.~Sannino,
   ``SU(2) gauge theory with two fundamental flavors: A minimal template for model building,''
  Phys.\ Rev.\ D {\bf 94}, no. 9, 094507 (2016)
  doi:10.1103/PhysRevD.94.094507
  [arXiv:1602.06559 [hep-lat]].

  
   \bibitem{Arthur:2016ozw} 
  R.~Arthur, V.~Drach, A.~Hietanen, C.~Pica and F.~Sannino,
   ``$SU(2)$ Gauge Theory with Two Fundamental Flavours: Scalar and Pseudoscalar Spectrum,''
  arXiv:1607.06654 [hep-lat].

  
  
     \bibitem{Pica:2016zst} 
  C.~Pica, V.~Drach, M.~Hansen and F.~Sannino,
   ``Composite Higgs Dynamics on the Lattice,''
  EPJ Web Conf.\  {\bf 137}, 10005 (2017)
  doi:10.1051/epjconf/201713710005
  [arXiv:1612.09336 [hep-lat]].


 \bibitem{Lee:2017uvl} 
  J.~W.~Lee, B.~Lucini and M.~Piai,
   ``Symmetry restoration at high-temperature in two-color and two-flavor lattice gauge theories,''
  JHEP {\bf 1704}, 036 (2017)
  doi:10.1007/JHEP04(2017)036
  [arXiv:1701.03228 [hep-lat]].


\bibitem{Drach:2017btk} 
  V.~Drach, T.~Janowski and C.~Pica,
  ``Update on SU(2) gauge theory with NF = 2 fundamental flavours,''
  EPJ Web Conf.\  {\bf 175}, 08020 (2018)
  doi:10.1051/epjconf/201817508020
  [arXiv:1710.07218 [hep-lat]].

  
  \bibitem{Drach:2020wux}
V.~Drach, T.~Janowski, C.~Pica and S.~Prelovsek,
``Scattering of Goldstone Bosons and resonance production in a Composite Higgs model on the lattice,''
JHEP \textbf{04}, 117 (2021)
doi:10.1007/JHEP04(2021)117
[arXiv:2012.09761 [hep-lat]].


\bibitem{Drach:2021uhl}
V.~Drach, P.~Fritzsch, A.~Rago and F.~Romero-L\'opez,
``Singlet channel scattering in a Composite Higgs model on the lattice,''
[arXiv:2107.09974 [hep-lat]].




\bibitem{Ayyar:2017qdf} 
  V.~Ayyar, T.~DeGrand, M.~Golterman, D.~C.~Hackett, W.~I.~Jay, E.~T.~Neil, Y.~Shamir and B.~Svetitsky,
   ``Spectroscopy of SU(4) composite Higgs theory with two distinct fermion representations,''
  Phys.\ Rev.\ D {\bf 97}, no. 7, 074505 (2018)
  doi:10.1103/PhysRevD.97.074505
  [arXiv:1710.00806 [hep-lat]].

  
  \bibitem{Ayyar:2018zuk} 
  V.~Ayyar, T.~Degrand, D.~C.~Hackett, W.~I.~Jay, E.~T.~Neil, Y.~Shamir and B.~Svetitsky,
   ``Baryon spectrum of SU(4) composite Higgs theory with two distinct fermion representations,''
  Phys.\ Rev.\ D {\bf 97}, no. 11, 114505 (2018)
  doi:10.1103/PhysRevD.97.114505
  [arXiv:1801.05809 [hep-ph]].
 


\bibitem{Ayyar:2018ppa} 
  V.~Ayyar, T.~DeGrand, D.~C.~Hackett, W.~I.~Jay, E.~T.~Neil, Y.~Shamir and B.~Svetitsky,
   ``Finite-temperature phase structure of SU(4) gauge theory with multiple fermion representations,''
  Phys.\ Rev.\ D {\bf 97}, no. 11, 114502 (2018)
  doi:10.1103/PhysRevD.97.114502
  [arXiv:1802.09644 [hep-lat]].




\bibitem{Ayyar:2018glg} 
  V.~Ayyar, T.~DeGrand, D.~C.~Hackett, W.~I.~Jay, E.~T.~Neil, Y.~Shamir and B.~Svetitsky,
  ``Partial compositeness and baryon matrix elements on the lattice,''
  Phys.\ Rev.\ D {\bf 99}, no. 9, 094502 (2019)
  doi:10.1103/PhysRevD.99.094502
  [arXiv:1812.02727 [hep-ph]].
  


  
  \bibitem{Cossu:2019hse} 
  G.~Cossu, L.~Del Debbio, M.~Panero and D.~Preti,
  ``Strong dynamics with matter in multiple representations: SU(4) gauge theory with fundamental and sextet fermions,''
  Eur.\ Phys.\ J.\ C {\bf 79}, no. 8, 638 (2019)
  doi:10.1140/epjc/s10052-019-7137-1
  [arXiv:1904.08885 [hep-lat]].
  
  
  \bibitem{Shamir:2021frg}
Y.~Shamir, M.~Golterman, W.~I.~Jay, E.~T.~Neil and B.~Svetitsky,
``$S$ parameter from a prototype composite-Higgs model,''
[arXiv:2110.05198 [hep-lat]].
  
  \bibitem{DelDebbio:2021xlv}
L.~Del Debbio, A.~Lupo, M.~Panero and N.~Tantalo,
``Spectral reconstruction in SU(4) gauge theory with fermions in multiple representations,''
[arXiv:2112.01158 [hep-lat]].








 \bibitem{Vecchi:2015fma}
 L.~Vecchi,
 ``A dangerous irrelevant UV-completion of the composite Higgs,''
 JHEP \textbf{02}, 094 (2017)
 doi:10.1007/JHEP02(2017)094
 [arXiv:1506.00623 [hep-ph]].



\bibitem{Coleman:1985rnk} 
S.~Coleman,
``Aspects of Symmetry : Selected Erice Lectures,''
doi:10.1017/CBO9780511565045


\bibitem{Migdal:1982jp} 
A.~A.~Migdal and M.~A.~Shifman,
``Dilaton Effective Lagrangian in Gluodynamics,''
Phys.\ Lett.\  {\bf 114B}, 445 (1982).
doi:10.1016/0370-2693(82)90089-2




\bibitem{Leung:1985sn} 
C.~N.~Leung, S.~T.~Love and W.~A.~Bardeen,
``Spontaneous Symmetry Breaking in Scale Invariant Quantum Electrodynamics,''
Nucl.\ Phys.\ B {\bf 273}, 649 (1986).
doi:10.1016/0550-3213(86)90382-2




\bibitem{Bardeen:1985sm} 
W.~A.~Bardeen, C.~N.~Leung and S.~T.~Love,
``The Dilaton and Chiral Symmetry Breaking,''
Phys.\ Rev.\ Lett.\  {\bf 56}, 1230 (1986).
doi:10.1103/PhysRevLett.56.1230


\bibitem{Yamawaki:1985zg} 
K.~Yamawaki, M.~Bando and K.~I.~Matumoto,
``Scale Invariant Technicolor Model and a Technidilaton,''
Phys.\ Rev.\ Lett.\  {\bf 56}, 1335 (1986).
doi:10.1103/PhysRevLett.56.1335



\bibitem{Goldberger:2007zk}
W.~D.~Goldberger, B.~Grinstein and W.~Skiba,
``Distinguishing the Higgs boson from the dilaton at the Large Hadron Collider,''
Phys. Rev. Lett. \textbf{100}, 111802 (2008)
doi:10.1103/PhysRevLett.100.111802
[arXiv:0708.1463 [hep-ph]].





\bibitem{Matsuzaki:2013eva} 
S.~Matsuzaki and K.~Yamawaki,
``Dilaton Chiral Perturbation Theory: Determining the Mass and Decay Constant of the Technidilaton on the Lattice,''
Phys.\ Rev.\ Lett.\  {\bf 113}, no. 8, 082002 (2014)
doi:10.1103/PhysRevLett.113.082002
[arXiv:1311.3784 [hep-lat]].


\bibitem{Golterman:2016lsd} 
M.~Golterman and Y.~Shamir,
``Low-energy effective action for pions and a dilatonic meson,''
Phys.\ Rev.\ D {\bf 94}, no. 5, 054502 (2016)
doi:10.1103/PhysRevD.94.054502
[arXiv:1603.04575 [hep-ph]].


\bibitem{Kasai:2016ifi} 
A.~Kasai, K.~i.~Okumura and H.~Suzuki,
``A dilaton-pion mass relation,''
arXiv:1609.02264 [hep-lat].



\bibitem{Hansen:2016fri} 
M.~Hansen, K.~Langaeble and F.~Sannino,
``Extending Chiral Perturbation Theory with an Isosinglet Scalar,''
Phys.\ Rev.\ D {\bf 95}, no. 3, 036005 (2017)
doi:10.1103/PhysRevD.95.036005
[arXiv:1610.02904 [hep-ph]].


\bibitem{Golterman:2016cdd} 
M.~Golterman and Y.~Shamir,
``Effective pion mass term and the trace anomaly,''
Phys.\ Rev.\ D {\bf 95}, no. 1, 016003 (2017)
doi:10.1103/PhysRevD.95.016003
[arXiv:1611.04275 [hep-ph]].


\bibitem{Appelquist:2017wcg} 
T.~Appelquist, J.~Ingoldby and M.~Piai,
``Dilaton EFT Framework For Lattice Data,''
JHEP {\bf 1707}, 035 (2017)
doi:10.1007/JHEP07(2017)035
[arXiv:1702.04410 [hep-ph]].


\bibitem{Appelquist:2017vyy} 
T.~Appelquist, J.~Ingoldby and M.~Piai,
``Analysis of a Dilaton EFT for Lattice Data,''
JHEP {\bf 1803}, 039 (2018)
doi:10.1007/JHEP03(2018)039
[arXiv:1711.00067 [hep-ph]].

\bibitem{Golterman:2018mfm} 
M.~Golterman and Y.~Shamir,
``Large-mass regime of the dilaton-pion low-energy effective theory,''
Phys.\ Rev.\ D {\bf 98}, no. 5, 056025 (2018)
doi:10.1103/PhysRevD.98.056025
[arXiv:1805.00198 [hep-ph]].

\bibitem{Cata:2019edh}
O.~Cata and C.~Muller,
``Chiral effective theories with a light scalar at one loop,''
Nucl. Phys. B \textbf{952}, 114938 (2020)
doi:10.1016/j.nuclphysb.2020.114938
[arXiv:1906.01879 [hep-ph]].

\bibitem{Appelquist:2019lgk}
T.~Appelquist, J.~Ingoldby and M.~Piai,
``Dilaton potential and lattice data,''
Phys. Rev. D \textbf{101}, no.7, 075025 (2020)
doi:10.1103/PhysRevD.101.075025
[arXiv:1908.00895 [hep-ph]].



\bibitem{Golterman:2020tdq}
M.~Golterman, E.~T.~Neil and Y.~Shamir,
``Application of dilaton chiral perturbation theory to $N_f=8$, ${\rm SU}(3)$ spectral data,''
Phys. Rev. D \textbf{102}, no.3, 034515 (2020)
doi:10.1103/PhysRevD.102.034515
[arXiv:2003.00114 [hep-ph]].


\bibitem{Golterman:2020utm}
M.~Golterman and Y.~Shamir,
``Explorations beyond dilaton chiral perturbation theory in the eight-flavor SU(3) gauge theory,''
Phys. Rev. D \textbf{102}, 114507 (2020)
doi:10.1103/PhysRevD.102.114507
[arXiv:2009.13846 [hep-lat]].

\bibitem{Appelquist:2022mjb}
T.~Appelquist, J.~Ingoldby and M.~Piai,
``Dilaton Effective Field Theory,''
Universe \textbf{9}, no.1, 10 (2023)
doi:10.3390/universe9010010
[arXiv:2209.14867 [hep-ph]].



\bibitem{Appelquist:2020bqj}
T.~Appelquist, J.~Ingoldby and M.~Piai,
``Nearly Conformal Composite Higgs Model,''
Phys. Rev. Lett. \textbf{126}, no.19, 191804 (2021)
doi:10.1103/PhysRevLett.126.191804
[arXiv:2012.09698 [hep-ph]].

\bibitem{Appelquist:2022qgl}
T.~Appelquist, J.~Ingoldby and M.~Piai,
``Composite two-Higgs doublet model from dilaton effective field theory,''
[arXiv:2205.03320 [hep-ph]].




 
 
 \bibitem{Ma:2015gra}
 T.~Ma and G.~Cacciapaglia,
 ``Fundamental Composite 2HDM: SU(N) with 4 flavours,''
 JHEP \textbf{03}, 211 (2016)
 doi:10.1007/JHEP03(2016)211
 [arXiv:1508.07014 [hep-ph]].

 
 
 \bibitem{BuarqueFranzosi:2018eaj}
 D.~Buarque Franzosi, G.~Cacciapaglia and A.~Deandrea,
 ``Sigma-assisted low scale composite Goldstone–Higgs,''
 Eur. Phys. J. C \textbf{80}, no.1, 28 (2020)
 doi:10.1140/epjc/s10052-019-7572-z
 [arXiv:1809.09146 [hep-ph]].



\bibitem{Aoki:2014oha} 
Y.~Aoki {\it et al.} [LatKMI Collaboration],
``Light composite scalar in eight-flavor QCD on the lattice,''
Phys.\ Rev.\ D {\bf 89}, 111502 (2014)
doi:10.1103/PhysRevD.89.111502
[arXiv:1403.5000 [hep-lat]].



\bibitem{Aoki:2016wnc} 
Y.~Aoki {\it et al.} [LatKMI Collaboration],
``Light flavor-singlet scalars and walking signals in $N_f=8$ QCD on the lattice,''
Phys.\ Rev.\ D {\bf 96}, no. 1, 014508 (2017)
doi:10.1103/PhysRevD.96.014508
[arXiv:1610.07011 [hep-lat]].

\bibitem{Appelquist:2016viq} 
T.~Appelquist {\it et al.},
``Strongly interacting dynamics and the search for new physics at the LHC,''
Phys.\ Rev.\ D {\bf 93}, no. 11, 114514 (2016)
doi:10.1103/PhysRevD.93.114514
[arXiv:1601.04027 [hep-lat]].




\bibitem{Gasbarro:2017fmi} 
A.~D.~Gasbarro and G.~T.~Fleming,
``Examining the Low Energy Dynamics of Walking Gauge Theory,''
PoS LATTICE {\bf 2016}, 242 (2017)
doi:10.22323/1.256.0242
[arXiv:1702.00480 [hep-lat]].


\bibitem{Appelquist:2018yqe} 
T.~Appelquist {\it et al.} [Lattice Strong Dynamics Collaboration],
``Nonperturbative investigations of SU(3) gauge theory with eight dynamical flavors,''
Phys.\ Rev.\ D {\bf 99}, no. 1, 014509 (2019)
doi:10.1103/PhysRevD.99.014509
[arXiv:1807.08411 [hep-lat]].


  
   \bibitem{Barnard:2013zea} 
  J.~Barnard, T.~Gherghetta and T.~S.~Ray,
   ``UV descriptions of composite Higgs models without elementary scalars,''
  JHEP {\bf 1402}, 002 (2014)
  doi:10.1007/JHEP02(2014)002
  [arXiv:1311.6562 [hep-ph]].






\bibitem{Bizot:2016zyu}
N.~Bizot, M.~Frigerio, M.~Knecht and J.~L.~Kneur,
``Nonperturbative analysis of the spectrum of meson resonances in an ultraviolet-complete composite-Higgs model,''
Phys. Rev. D \textbf{95}, no.7, 075006 (2017)
doi:10.1103/PhysRevD.95.075006
[arXiv:1610.09293 [hep-ph]].


  


\bibitem{Maldacena:1997re} 
J.~M.~Maldacena,
``The Large N limit of superconformal field theories and supergravity,''
Int.\ J.\ Theor.\ Phys.\  {\bf 38}, 1113 (1999)
[Adv.\ Theor.\ Math.\ Phys.\  {\bf 2}, 231 (1998)]
doi:10.1023/A:1026654312961, 10.4310/ATMP.1998.v2.n2.a1
[hep-th/9711200].

\bibitem{Gubser:1998bc} 
S.~S.~Gubser, I.~R.~Klebanov and A.~M.~Polyakov,
``Gauge theory correlators from noncritical string theory,''
Phys.\ Lett.\ B {\bf 428}, 105 (1998)
doi:10.1016/S0370-2693(98)00377-3
[hep-th/9802109].

\bibitem{Witten:1998qj} 
E.~Witten,
``Anti-de Sitter space and holography,''
Adv.\ Theor.\ Math.\ Phys.\  {\bf 2}, 253 (1998)
doi:10.4310/ATMP.1998.v2.n2.a2
[hep-th/9802150].

\bibitem{Aharony:1999ti} 
O.~Aharony, S.~S.~Gubser, J.~M.~Maldacena, H.~Ooguri and Y.~Oz,
``Large N field theories, string theory and gravity,''
Phys.\ Rept.\  {\bf 323}, 183 (2000)
doi:10.1016/S0370-1573(99)00083-6
[hep-th/9905111].


 



\bibitem{Erdmenger:2020lvq}
J.~Erdmenger, N.~Evans, W.~Porod and K.~S.~Rigatos,
``Gauge/gravity dynamics for composite Higgs models and the top mass,''
Phys. Rev. Lett. \textbf{126}, no.7, 071602 (2021)
doi:10.1103/PhysRevLett.126.071602
[arXiv:2009.10737 [hep-ph]].

\bibitem{Erdmenger:2020flu}
J.~Erdmenger, N.~Evans, W.~Porod and K.~S.~Rigatos,
``Gauge/gravity dual dynamics for the strongly coupled sector of composite Higgs models,''
JHEP \textbf{02}, 058 (2021)
doi:10.1007/JHEP02(2021)058
[arXiv:2010.10279 [hep-ph]].

\bibitem{Elander:2020nyd}
D.~Elander, M.~Frigerio, M.~Knecht and J.~L.~Kneur,
``Holographic models of composite Higgs in the Veneziano limit. Part I. Bosonic sector,''
JHEP \textbf{03}, 182 (2021)
doi:10.1007/JHEP03(2021)182
[arXiv:2011.03003 [hep-ph]].

\bibitem{Elander:2021bmt}
D.~Elander, M.~Frigerio, M.~Knecht and J.~L.~Kneur,
``Holographic models of composite Higgs in the Veneziano limit: 2. Fermionic sector,''
[arXiv:2112.14740 [hep-ph]].



\bibitem{Elander:2021kxk}
D.~Elander and M.~Piai,
``Towards top-down holographic composite Higgs: minimal coset from maximal supergravity,''
JHEP \textbf{03}, 049 (2022)
doi:10.1007/JHEP03(2022)049
[arXiv:2110.02945 [hep-th]].




\bibitem{Strassler:2006im}
M.~J.~Strassler and K.~M.~Zurek,
``Echoes of a hidden valley at hadron colliders,''
Phys. Lett. B \textbf{651}, 374-379 (2007)
doi:10.1016/j.physletb.2007.06.055
[arXiv:hep-ph/0604261 [hep-ph]].

\bibitem{Cheung:2007ut}
K.~Cheung and T.~C.~Yuan,
``Hidden fermion as milli-charged dark matter in Stueckelberg Z- prime model,''
JHEP \textbf{03}, 120 (2007)
doi:10.1088/1126-6708/2007/03/120
[arXiv:hep-ph/0701107 [hep-ph]].

\bibitem{Hambye:2008bq}
T.~Hambye,
``Hidden vector dark matter,''
JHEP \textbf{01}, 028 (2009)
doi:10.1088/1126-6708/2009/01/028
[arXiv:0811.0172 [hep-ph]].

\bibitem{Feng:2009mn}
J.~L.~Feng, M.~Kaplinghat, H.~Tu and H.~B.~Yu,
``Hidden Charged Dark Matter,''
JCAP \textbf{07}, 004 (2009)
doi:10.1088/1475-7516/2009/07/004
[arXiv:0905.3039 [hep-ph]].

\bibitem{Cohen:2010kn}
T.~Cohen, D.~J.~Phalen, A.~Pierce and K.~M.~Zurek,
``Asymmetric Dark Matter from a GeV Hidden Sector,''
Phys. Rev. D \textbf{82}, 056001 (2010)
doi:10.1103/PhysRevD.82.056001
[arXiv:1005.1655 [hep-ph]].

\bibitem{Foot:2014uba}
R.~Foot and S.~Vagnozzi,
``Dissipative hidden sector dark matter,''
Phys. Rev. D \textbf{91}, 023512 (2015)
doi:10.1103/PhysRevD.91.023512
[arXiv:1409.7174 [hep-ph]].

\bibitem{Bertone:2016nfn}
G.~Bertone and D.~Hooper,
``History of dark matter,''
Rev. Mod. Phys. \textbf{90}, no.4, 045002 (2018)
doi:10.1103/RevModPhys.90.045002
[arXiv:1605.04909 [astro-ph.CO]].



\bibitem{DelNobile:2011je}
E.~Del Nobile, C.~Kouvaris and F.~Sannino,
``Interfering Composite Asymmetric Dark Matter for DAMA and CoGeNT,''
Phys. Rev. D \textbf{84}, 027301 (2011)
doi:10.1103/PhysRevD.84.027301
[arXiv:1105.5431 [hep-ph]].

\bibitem{Hietanen:2013fya}
A.~Hietanen, R.~Lewis, C.~Pica and F.~Sannino,
``Composite Goldstone Dark Matter: Experimental Predictions from the Lattice,''
JHEP \textbf{12}, 130 (2014)
doi:10.1007/JHEP12(2014)130
[arXiv:1308.4130 [hep-ph]].


\bibitem{Cline:2016nab}
J.~M.~Cline, W.~Huang and G.~D.~Moore,
``Challenges for models with composite states,''
Phys. Rev. D \textbf{94}, no.5, 055029 (2016)
doi:10.1103/PhysRevD.94.055029
[arXiv:1607.07865 [hep-ph]].



\bibitem{Dondi:2019olm}
N.~A.~Dondi, F.~Sannino and J.~Smirnov,
``Thermal history of composite dark matter,''
Phys. Rev. D \textbf{101}, no.10, 103010 (2020)
doi:10.1103/PhysRevD.101.103010
[arXiv:1905.08810 [hep-ph]].


\bibitem{Ge:2019voa}
S.~Ge, K.~Lawson and A.~Zhitnitsky,
``Axion quark nugget dark matter model: Size distribution and survival pattern,''
Phys. Rev. D \textbf{99}, no.11, 116017 (2019)
doi:10.1103/PhysRevD.99.116017
[arXiv:1903.05090 [hep-ph]].

\bibitem{Beylin:2019gtw}
V.~Beylin, M.~Y.~Khlopov, V.~Kuksa and N.~Volchanskiy,
``Hadronic and Hadron-Like Physics of Dark Matter,''
Symmetry \textbf{11}, no.4, 587 (2019)
doi:10.3390/sym11040587
[arXiv:1904.12013 [hep-ph]].




\bibitem{Yamanaka:2019aeq}
N.~Yamanaka, H.~Iida, A.~Nakamura and M.~Wakayama,
``Dark matter scattering cross section and dynamics in dark Yang-Mills theory,''
Phys. Lett. B \textbf{813}, 136056 (2021)
doi:10.1016/j.physletb.2020.136056
[arXiv:1910.01440 [hep-ph]].


\bibitem{Yamanaka:2019yek}
N.~Yamanaka, H.~Iida, A.~Nakamura and M.~Wakayama,
``Glueball scattering cross section in lattice SU(2) Yang-Mills theory,''
Phys. Rev. D \textbf{102}, no.5, 054507 (2020)
doi:10.1103/PhysRevD.102.054507
[arXiv:1910.07756 [hep-lat]].

\bibitem{Cai:2020njb}
H.~Cai and G.~Cacciapaglia,
``Singlet dark matter in the SU(6)/SO(6) composite Higgs model,''
Phys. Rev. D \textbf{103}, no.5, 055002 (2021)
doi:10.1103/PhysRevD.103.055002
[arXiv:2007.04338 [hep-ph]].


\bibitem{Hochberg:2014dra}
Y.~Hochberg, E.~Kuflik, T.~Volansky and J.~G.~Wacker,
``Mechanism for Thermal Relic Dark Matter of Strongly Interacting Massive Particles,''
Phys. Rev. Lett. \textbf{113}, 171301 (2014)
doi:10.1103/PhysRevLett.113.171301
[arXiv:1402.5143 [hep-ph]].

\bibitem{Hochberg:2014kqa}
Y.~Hochberg, E.~Kuflik, H.~Murayama, T.~Volansky and J.~G.~Wacker,
``Model for Thermal Relic Dark Matter of Strongly Interacting Massive Particles,''
Phys. Rev. Lett. \textbf{115}, no.2, 021301 (2015)
doi:10.1103/PhysRevLett.115.021301
[arXiv:1411.3727 [hep-ph]].

\bibitem{Hochberg:2015vrg}
Y.~Hochberg, E.~Kuflik and H.~Murayama,
``SIMP Spectroscopy,''
JHEP \textbf{05}, 090 (2016)
doi:10.1007/JHEP05(2016)090
[arXiv:1512.07917 [hep-ph]].

\bibitem{Berlin:2018tvf}
A.~Berlin, N.~Blinov, S.~Gori, P.~Schuster and N.~Toro,
``Cosmology and Accelerator Tests of Strongly Interacting Dark Matter,''
Phys. Rev. D \textbf{97}, no.5, 055033 (2018)
doi:10.1103/PhysRevD.97.055033
[arXiv:1801.05805 [hep-ph]].


\bibitem{Bernal:2017mqb}
N.~Bernal, X.~Chu and J.~Pradler,
``Simply split strongly interacting massive particles,''
Phys. Rev. D \textbf{95}, no.11, 115023 (2017)
doi:10.1103/PhysRevD.95.115023
[arXiv:1702.04906 [hep-ph]].




\bibitem{Bernal:2019uqr}
N.~Bernal, X.~Chu, S.~Kulkarni and J.~Pradler,
``Self-interacting dark matter without prejudice,''
Phys. Rev. D \textbf{101}, no.5, 055044 (2020)
doi:10.1103/PhysRevD.101.055044
[arXiv:1912.06681 [hep-ph]].



\bibitem{Tsai:2020vpi}
Y.~D.~Tsai, R.~McGehee and H.~Murayama,
``Resonant Self-Interacting Dark Matter from Dark QCD,''
Phys. Rev. Lett. \textbf{128}, no.17, 172001 (2022)
doi:10.1103/PhysRevLett.128.172001
[arXiv:2008.08608 [hep-ph]].

\bibitem{Kondo:2022lgg}
D.~Kondo, R.~McGehee, T.~Melia and H.~Murayama,
``Linear Sigma Dark Matter,''
[arXiv:2205.08088 [hep-ph]].

\bibitem{Bernal:2015xba}
N.~Bernal and X.~Chu,
``$\mathbb {Z}_2$ SIMP Dark Matter,''
JCAP \textbf{01} (2016), 006
doi:10.1088/1475-7516/2016/01/006
[arXiv:1510.08527 [hep-ph]].



\bibitem{Witten:1984rs}
E.~Witten,
``Cosmic Separation of Phases,''
Phys. Rev. D \textbf{30}, 272-285 (1984)
doi:10.1103/PhysRevD.30.272

\bibitem{Kamionkowski:1993fg}
M.~Kamionkowski, A.~Kosowsky and M.~S.~Turner,
``Gravitational radiation from first order phase transitions,''
Phys. Rev. D \textbf{49}, 2837-2851 (1994)
doi:10.1103/PhysRevD.49.2837
[arXiv:astro-ph/9310044 [astro-ph]].

\bibitem{Allen:1996vm}
B.~Allen,
``The Stochastic gravity wave background: Sources and detection,''
[arXiv:gr-qc/9604033 [gr-qc]].

\bibitem{Schwaller:2015tja}
P.~Schwaller,
``Gravitational Waves from a Dark Phase Transition,''
Phys. Rev. Lett. \textbf{115}, no.18, 181101 (2015)
doi:10.1103/PhysRevLett.115.181101
[arXiv:1504.07263 [hep-ph]].

\bibitem{Croon:2018erz}
D.~Croon, V.~Sanz and G.~White,
``Model Discrimination in Gravitational Wave spectra from Dark Phase Transitions,''
JHEP \textbf{08}, 203 (2018)
doi:10.1007/JHEP08(2018)203
[arXiv:1806.02332 [hep-ph]].

\bibitem{Christensen:2018iqi}
N.~Christensen,
``Stochastic Gravitational Wave Backgrounds,''
Rept. Prog. Phys. \textbf{82}, no.1, 016903 (2019)
doi:10.1088/1361-6633/aae6b5
[arXiv:1811.08797 [gr-qc]].




\bibitem{Seto:2001qf}
N.~Seto, S.~Kawamura and T.~Nakamura,
``Possibility of direct measurement of the acceleration of the universe using 0.1-Hz band laser interferometer gravitational wave antenna in space,''
Phys. Rev. Lett. \textbf{87}, 221103 (2001)
doi:10.1103/PhysRevLett.87.221103
[arXiv:astro-ph/0108011 [astro-ph]].

\bibitem{Kawamura:2006up}
S.~Kawamura, T.~Nakamura, M.~Ando, N.~Seto, K.~Tsubono, K.~Numata, R.~Takahashi, S.~Nagano, T.~Ishikawa and M.~Musha, \textit{et al.}
``The Japanese space gravitational wave antenna DECIGO,''
Class. Quant. Grav. \textbf{23}, S125-S132 (2006)
doi:10.1088/0264-9381/23/8/S17

\bibitem{Crowder:2005nr}
J.~Crowder and N.~J.~Cornish,
``Beyond LISA: Exploring future gravitational wave missions,''
Phys. Rev. D \textbf{72}, 083005 (2005)
doi:10.1103/PhysRevD.72.083005
[arXiv:gr-qc/0506015 [gr-qc]].


\bibitem{Corbin:2005ny}
V.~Corbin and N.~J.~Cornish,
``Detecting the cosmic gravitational wave background with the big bang observer,''
Class. Quant. Grav. \textbf{23}, 2435-2446 (2006)
doi:10.1088/0264-9381/23/7/014
[arXiv:gr-qc/0512039 [gr-qc]].

\bibitem{Harry:2006fi}
G.~M.~Harry, P.~Fritschel, D.~A.~Shaddock, W.~Folkner and E.~S.~Phinney,
``Laser interferometry for the big bang observer,''
Class. Quant. Grav. \textbf{23}, 4887-4894 (2006)
[erratum: Class. Quant. Grav. \textbf{23}, 7361 (2006)]
doi:10.1088/0264-9381/23/15/008

\bibitem{Hild:2010id}
S.~Hild, M.~Abernathy, F.~Acernese, P.~Amaro-Seoane, N.~Andersson, K.~Arun, F.~Barone, B.~Barr, M.~Barsuglia and M.~Beker, \textit{et al.}
``Sensitivity Studies for Third-Generation Gravitational Wave Observatories,''
Class. Quant. Grav. \textbf{28}, 094013 (2011)
doi:10.1088/0264-9381/28/9/094013
[arXiv:1012.0908 [gr-qc]].

\bibitem{Yagi:2011wg}
K.~Yagi and N.~Seto,
``Detector configuration of DECIGO/BBO and identification of cosmological neutron-star binaries,''
Phys. Rev. D \textbf{83}, 044011 (2011)
[erratum: Phys. Rev. D \textbf{95}, no.10, 109901 (2017)]
doi:10.1103/PhysRevD.83.044011
[arXiv:1101.3940 [astro-ph.CO]].

\bibitem{Sathyaprakash:2012jk}
B.~Sathyaprakash, M.~Abernathy, F.~Acernese, P.~Ajith, B.~Allen, P.~Amaro-Seoane, N.~Andersson, S.~Aoudia, K.~Arun and P.~Astone, \textit{et al.}
``Scientific Objectives of Einstein Telescope,''
Class. Quant. Grav. \textbf{29}, 124013 (2012)
[erratum: Class. Quant. Grav. \textbf{30}, 079501 (2013)]
doi:10.1088/0264-9381/29/12/124013
[arXiv:1206.0331 [gr-qc]].


\bibitem{Thrane:2013oya}
E.~Thrane and J.~D.~Romano,
``Sensitivity curves for searches for gravitational-wave backgrounds,''
Phys. Rev. D \textbf{88}, no.12, 124032 (2013)
doi:10.1103/PhysRevD.88.124032
[arXiv:1310.5300 [astro-ph.IM]].

\bibitem{Caprini:2015zlo}
C.~Caprini, M.~Hindmarsh, S.~Huber, T.~Konstandin, J.~Kozaczuk, G.~Nardini, J.~M.~No, A.~Petiteau, P.~Schwaller and G.~Servant, \textit{et al.}
``Science with the space-based interferometer eLISA. II: Gravitational waves from cosmological phase transitions,''
JCAP \textbf{04}, 001 (2016)
doi:10.1088/1475-7516/2016/04/001
[arXiv:1512.06239 [astro-ph.CO]].


\bibitem{LISA:2017pwj}
P.~Amaro-Seoane \textit{et al.} [LISA],
``Laser Interferometer Space Antenna,''
[arXiv:1702.00786 [astro-ph.IM]].

\bibitem{LIGOScientific:2016wof}
B.~P.~Abbott \textit{et al.} [LIGO Scientific],
``Exploring the Sensitivity of Next Generation Gravitational Wave Detectors,''
Class. Quant. Grav. \textbf{34}, no.4, 044001 (2017)
doi:10.1088/1361-6382/aa51f4
[arXiv:1607.08697 [astro-ph.IM]].

\bibitem{Isoyama:2018rjb}
S.~Isoyama, H.~Nakano and T.~Nakamura,
``Multiband Gravitational-Wave Astronomy: Observing binary inspirals with a decihertz detector, B-DECIGO,''
PTEP \textbf{2018}, no.7, 073E01 (2018)
doi:10.1093/ptep/pty078
[arXiv:1802.06977 [gr-qc]].

\bibitem{Baker:2019nia}
J.~Baker, J.~Bellovary, P.~L.~Bender, E.~Berti, R.~Caldwell, J.~Camp, J.~W.~Conklin, N.~Cornish, C.~Cutler and R.~DeRosa, \textit{et al.}
``The Laser Interferometer Space Antenna: Unveiling the Millihertz Gravitational Wave Sky,''
[arXiv:1907.06482 [astro-ph.IM]].

\bibitem{Brdar:2018num}
V.~Brdar, A.~J.~Helmboldt and J.~Kubo,
``Gravitational Waves from First-Order Phase Transitions: LIGO as a Window to Unexplored Seesaw Scales,''
JCAP \textbf{02}, 021 (2019)
doi:10.1088/1475-7516/2019/02/021
[arXiv:1810.12306 [hep-ph]].

\bibitem{Reitze:2019iox}
D.~Reitze, R.~X.~Adhikari, S.~Ballmer, B.~Barish, L.~Barsotti, G.~Billingsley, D.~A.~Brown, Y.~Chen, D.~Coyne and R.~Eisenstein, \textit{et al.}
``Cosmic Explorer: The U.S. Contribution to Gravitational-Wave Astronomy beyond LIGO,''
Bull. Am. Astron. Soc. \textbf{51}, no.7, 035 (2019)
[arXiv:1907.04833 [astro-ph.IM]].

\bibitem{Caprini:2019egz}
C.~Caprini, M.~Chala, G.~C.~Dorsch, M.~Hindmarsh, S.~J.~Huber, T.~Konstandin, J.~Kozaczuk, G.~Nardini, J.~M.~No and K.~Rummukainen, \textit{et al.}
``Detecting gravitational waves from cosmological phase transitions with LISA: an update,''
JCAP \textbf{03}, 024 (2020)
doi:10.1088/1475-7516/2020/03/024
[arXiv:1910.13125 [astro-ph.CO]].

\bibitem{Maggiore:2019uih}
M.~Maggiore, C.~Van Den Broeck, N.~Bartolo, E.~Belgacem, D.~Bertacca, M.~A.~Bizouard, M.~Branchesi, S.~Clesse, S.~Foffa and J.~Garc\'\i{}a-Bellido, \textit{et al.}
``Science Case for the Einstein Telescope,''
JCAP \textbf{03}, 050 (2020)
doi:10.1088/1475-7516/2020/03/050
[arXiv:1912.02622 [astro-ph.CO]].


\bibitem{Huang:2020crf}
W.~C.~Huang, M.~Reichert, F.~Sannino and Z.~W.~Wang,
``Testing the dark SU(N) Yang-Mills theory confined landscape: From the lattice to gravitational waves,''
Phys. Rev. D \textbf{104}, no.3, 035005 (2021)
doi:10.1103/PhysRevD.104.035005
[arXiv:2012.11614 [hep-ph]].

\bibitem{Halverson:2020xpg}
J.~Halverson, C.~Long, A.~Maiti, B.~Nelson and G.~Salinas,
``Gravitational waves from dark Yang-Mills sectors,''
JHEP \textbf{05}, 154 (2021)
doi:10.1007/JHEP05(2021)154
[arXiv:2012.04071 [hep-ph]].


\bibitem{Kang:2021epo}
Z.~Kang, J.~Zhu and S.~Matsuzaki,
``Dark confinement-deconfinement phase transition: a roadmap from Polyakov loop models to gravitational waves,''
JHEP \textbf{09}, 060 (2021)
doi:10.1007/JHEP09(2021)060
[arXiv:2101.03795 [hep-ph]].





\bibitem{Lucini:2002ku}
B.~Lucini, M.~Teper and U.~Wenger,
``The Deconfinement transition in SU(N) gauge theories,''
Phys. Lett. B \textbf{545}, 197-206 (2002)
doi:10.1016/S0370-2693(02)02556-X
[arXiv:hep-lat/0206029 [hep-lat]].

\bibitem{Lucini:2003zr}
B.~Lucini, M.~Teper and U.~Wenger,
``The High temperature phase transition in SU(N) gauge theories,''
JHEP \textbf{01}, 061 (2004)
doi:10.1088/1126-6708/2004/01/061
[arXiv:hep-lat/0307017 [hep-lat]].

\bibitem{Lucini:2005vg}
B.~Lucini, M.~Teper and U.~Wenger,
``Properties of the deconfining phase transition in SU(N) gauge theories,''
JHEP \textbf{02}, 033 (2005)
doi:10.1088/1126-6708/2005/02/033
[arXiv:hep-lat/0502003 [hep-lat]].

\bibitem{Panero:2009tv}
M.~Panero,
``Thermodynamics of the QCD plasma and the large-N limit,''
Phys. Rev. Lett. \textbf{103}, 232001 (2009)
doi:10.1103/PhysRevLett.103.232001
[arXiv:0907.3719 [hep-lat]].

\bibitem{Datta:2010sq}
S.~Datta and S.~Gupta,
``Continuum Thermodynamics of the Gluo$N_c$ Plasma,''
Phys. Rev. D \textbf{82}, 114505 (2010)
doi:10.1103/PhysRevD.82.114505
[arXiv:1006.0938 [hep-lat]].

\bibitem{Lucini:2012wq}
B.~Lucini, A.~Rago and E.~Rinaldi,
``SU($N_c$) gauge theories at deconfinement,''
Phys. Lett. B \textbf{712}, 279-283 (2012)
doi:10.1016/j.physletb.2012.04.070
[arXiv:1202.6684 [hep-lat]].





\bibitem{Pepe:2005sz}
M.~Pepe,
``Confinement and the center of the gauge group,''
PoS \textbf{LAT2005}, 017 (2006)
doi:10.1016/j.nuclphysbps.2006.01.045
[arXiv:hep-lat/0510013 [hep-lat]].

\bibitem{Pepe:2006er}
M.~Pepe and U.~J.~Wiese,
``Exceptional Deconfinement in G(2) Gauge Theory,''
Nucl. Phys. B \textbf{768}, 21-37 (2007)
doi:10.1016/j.nuclphysb.2006.12.024
[arXiv:hep-lat/0610076 [hep-lat]].

\bibitem{Cossu:2007dk}
G.~Cossu, M.~D'Elia, A.~Di Giacomo, B.~Lucini and C.~Pica,
``G(2) gauge theory at finite temperature,''
JHEP \textbf{10}, 100 (2007)
doi:10.1088/1126-6708/2007/10/100
[arXiv:0709.0669 [hep-lat]].

\bibitem{Bruno:2014rxa}
M.~Bruno, M.~Caselle, M.~Panero and R.~Pellegrini,
``Exceptional thermodynamics: the equation of state of G$_{2}$ gauge theory,''
JHEP \textbf{03}, 057 (2015)
doi:10.1007/JHEP03(2015)057
[arXiv:1409.8305 [hep-lat]].



\bibitem{Appelquist:2015yfa}
T.~Appelquist, R.~C.~Brower, M.~I.~Buchoff, G.~T.~Fleming, X.~Y.~Jin, J.~Kiskis, G.~D.~Kribs, E.~T.~Neil, J.~C.~Osborn and C.~Rebbi, \textit{et al.}
``Stealth Dark Matter: Dark scalar baryons through the Higgs portal,''
Phys. Rev. D \textbf{92}, no.7, 075030 (2015)
doi:10.1103/PhysRevD.92.075030
[arXiv:1503.04203 [hep-ph]].

\bibitem{Appelquist:2015zfa}
T.~Appelquist, E.~Berkowitz, R.~C.~Brower, M.~I.~Buchoff, G.~T.~Fleming, X.~Y.~Jin, J.~Kiskis, G.~D.~Kribs, E.~T.~Neil and J.~C.~Osborn, \textit{et al.}
``Detecting Stealth Dark Matter Directly through Electromagnetic Polarizability,''
Phys. Rev. Lett. \textbf{115}, no.17, 171803 (2015)
doi:10.1103/PhysRevLett.115.171803
[arXiv:1503.04205 [hep-ph]].

\bibitem{LatticeStrongDynamics:2020jwi}
R.~C.~Brower \textit{et al.} [Lattice Strong Dynamics],
``Stealth dark matter confinement transition and gravitational waves,''
Phys. Rev. D \textbf{103}, no.1, 014505 (2021)
doi:10.1103/PhysRevD.103.014505
[arXiv:2006.16429 [hep-lat]].


\bibitem{Borsanyi:2022xml}
S.~Borsanyi, K.~R., Z.~Fodor, D.~A.~Godzieba, P.~Parotto and D.~Sexty,
``Precision study of the continuum SU(3) Yang-Mills theory: How to use parallel tempering to improve on supercritical slowing down for first order phase transitions,''
Phys. Rev. D \textbf{105}, no.7, 074513 (2022)
doi:10.1103/PhysRevD.105.074513
[arXiv:2202.05234 [hep-lat]].



\bibitem{Langfeld:2012ah}
K.~Langfeld, B.~Lucini and A.~Rago,
``The density of states in gauge theories,''
Phys. Rev. Lett. \textbf{109}, 111601 (2012)
doi:10.1103/PhysRevLett.109.111601
[arXiv:1204.3243 [hep-lat]].

\bibitem{Langfeld:2013xbf}
K.~Langfeld and J.~M.~Pawlowski,
``Two-color QCD with heavy quarks at finite densities,''
Phys. Rev. D \textbf{88}, no.7, 071502 (2013)
doi:10.1103/PhysRevD.88.071502
[arXiv:1307.0455 [hep-lat]].

\bibitem{Langfeld:2015fua}
K.~Langfeld, B.~Lucini, R.~Pellegrini and A.~Rago,
``An efficient algorithm for numerical computations of continuous densities of states,''
Eur. Phys. J. C \textbf{76}, no.6, 306 (2016)
doi:10.1140/epjc/s10052-016-4142-5
[arXiv:1509.08391 [hep-lat]].


\bibitem{Cossu:2021bgn}
G.~Cossu, D.~Lancastera, B.~Lucini, R.~Pellegrini and A.~Rago,
``Ergodic sampling of the topological charge using the density of states,''
Eur. Phys. J. C \textbf{81}, no.4, 375 (2021)
doi:10.1140/epjc/s10052-021-09161-1
[arXiv:2102.03630 [hep-lat]].



\bibitem{Springer:2021liy}
F.~Springer and D.~Schaich,
``Density of states for gravitational waves,''
PoS   {LATTICE2021}, 043 (2022);
doi:10.22323/1.396.0043
[arXiv:2112.11868 [hep-lat]].


\bibitem{Mason:2022trc}
D.~Mason, {\bf B.~Lucini, M.~Piai,}
E.~Rinaldi and D.~Vadacchino,
``The density of states method in Yang-Mills theories and first order phase transitions,''
EPJ Web Conf.   {274}, 08007 (2022);
doi:10.1051/epjconf/202227408007
[arXiv:2211.10373 [hep-lat]].

\bibitem{Mason:2022aka}
D.~Mason, B.~Lucini, M.~Piai, 
E.~Rinaldi and D.~Vadacchino,
``The density of state method for first-order phase transitions in Yang-Mills theories,''
PoS   {LATTICE2022}, 216 (2023).
doi:10.22323/1.430.0216
[arXiv:2212.01074 [hep-lat]].



\bibitem{Springer:2023wok}
F.~Springer \textit{et al.} [Lattice Strong Dynamics (LSD)],
``Advances in using density of states for large-N Yang\textendash{}Mills,''
PoS \textbf{LATTICE2022}, 223 (2023)
doi:10.22323/1.430.0223


\bibitem{Brower:2000rp}
R.~C.~Brower, S.~D.~Mathur and C.~I.~Tan,
``Glueball spectrum for QCD from AdS supergravity duality,''
Nucl. Phys. B \textbf{587}, 249-276 (2000)
doi:10.1016/S0550-3213(00)00435-1
[arXiv:hep-th/0003115 [hep-th]].

\bibitem{Apreda:2003sy}
R.~Apreda, D.~E.~Crooks, N.~J.~Evans and M.~Petrini,
``Confinement, glueballs and strings from deformed AdS,''
JHEP \textbf{05}, 065 (2004)
doi:10.1088/1126-6708/2004/05/065
[arXiv:hep-th/0308006 [hep-th]].

\bibitem{Mueck:2004qg}
W.~Mueck and M.~Prisco,
``Glueball scattering amplitudes from holography,''
JHEP \textbf{04}, 037 (2004)
doi:10.1088/1126-6708/2004/04/037
[arXiv:hep-th/0402068 [hep-th]].

\bibitem{Wen:2004qh}
C.~K.~Wen and H.~X.~Yang,
``QCD(4) glueball masses from AdS(6) black hole description,''
Mod. Phys. Lett. A \textbf{20}, 997-1004 (2005)
doi:10.1142/S0217732305016245
[arXiv:hep-th/0404152 [hep-th]].

\bibitem{Kuperstein:2004yf}
S.~Kuperstein and J.~Sonnenschein,
``Non-critical, near extremal AdS(6) background as a holographic laboratory of four dimensional YM theory,''
JHEP \textbf{11}, 026 (2004)
doi:10.1088/1126-6708/2004/11/026
[arXiv:hep-th/0411009 [hep-th]].


\bibitem{Elander:2013jqa}
D.~Elander, A.~F.~Faedo, C.~Hoyos, D.~Mateos and M.~Piai,
``Multiscale confining dynamics from holographic RG flows,''
JHEP \textbf{05}, 003 (2014)
doi:10.1007/JHEP05(2014)003
[arXiv:1312.7160 [hep-th]].

\bibitem{Athenodorou:2016ndx}
A.~Athenodorou, E.~Bennett, G.~Bergner, D.~Elander, C.~J.~D.~Lin, B.~Lucini and M.~Piai,
``Large mass hierarchies from strongly-coupled dynamics,''
JHEP \textbf{06}, 114 (2016)
doi:10.1007/JHEP06(2016)114
[arXiv:1605.04258 [hep-th]].

\bibitem{Elander:2018aub}
D.~Elander, M.~Piai and J.~Roughley,
``Holographic glueballs from the circle reduction of Romans supergravity,''
JHEP \textbf{02}, 101 (2019)
doi:10.1007/JHEP02(2019)101
[arXiv:1811.01010 [hep-th]].

\bibitem{Elander:2020csd}
D.~Elander, M.~Piai and J.~Roughley,
``Probing the holographic dilaton,''
JHEP \textbf{06}, 177 (2020)
[erratum: JHEP \textbf{12}, 109 (2020)]
doi:10.1007/JHEP06(2020)177
[arXiv:2004.05656 [hep-th]].






\bibitem{Bochicchio:2016toi}
M.~Bochicchio,
``An asymptotic solution of Large-N QCD, for the glueball and meson spectrum and the collinear S-matrix,''
AIP Conf. Proc. \textbf{1735}, no.1, 030004 (2016)
doi:10.1063/1.4949387

\bibitem{Bochicchio:2013sra}
M.~Bochicchio,
``Glueball and meson spectrum in large-N massless QCD,''
[arXiv:1308.2925 [hep-th]].

\bibitem{Hong:2017suj}
D.~K.~Hong, J.~W.~Lee, B.~Lucini, M.~Piai and D.~Vadacchino,
``Casimir scaling and Yang\textendash{}Mills glueballs,''
Phys. Lett. B \textbf{775}, 89-93 (2017)
doi:10.1016/j.physletb.2017.10.050
[arXiv:1705.00286 [hep-th]].






\bibitem{Lucini:2001ej}
B.~Lucini and M.~Teper,
``SU(N) gauge theories in four-dimensions: Exploring the approach to N = infinity,''
JHEP \textbf{06} (2001), 050
doi:10.1088/1126-6708/2001/06/050
[arXiv:hep-lat/0103027 [hep-lat]].


\bibitem{Lucini:2004my}
B.~Lucini, M.~Teper and U.~Wenger,
``Glueballs and k-strings in SU(N) gauge theories: Calculations with improved operators,''
JHEP \textbf{06}, 012 (2004)
doi:10.1088/1126-6708/2004/06/012
[arXiv:hep-lat/0404008 [hep-lat]].

\bibitem{Lucini:2010nv}
B.~Lucini, A.~Rago and E.~Rinaldi,
``Glueball masses in the large N limit,''
JHEP \textbf{08} (2010), 119
doi:10.1007/JHEP08(2010)119
[arXiv:1007.3879 [hep-lat]].



\bibitem{Lucini:2012gg}
B.~Lucini and M.~Panero,
``SU(N) gauge theories at large N,''
Phys. Rept. \textbf{526}, 93-163 (2013)
doi:10.1016/j.physrep.2013.01.001
[arXiv:1210.4997 [hep-th]].

\bibitem{Athenodorou:2015nba}
A.~Athenodorou, R.~Lau and M.~Teper,
``On the weak N -dependence of SO(N) and SU(N) gauge theories in 2+1 dimensions,''
Phys. Lett. B \textbf{749}, 448-453 (2015)
doi:10.1016/j.physletb.2015.08.023
[arXiv:1504.08126 [hep-lat]].


\bibitem{Lau:2017aom}
R.~Lau and M.~Teper,
``SO(N) gauge theories in 2 + 1 dimensions: glueball spectra and confinement,''
JHEP \textbf{10}, 022 (2017)
doi:10.1007/JHEP10(2017)022
[arXiv:1701.06941 [hep-lat]].




\bibitem{Hernandez:2020tbc}
P.~Hern\'andez and F.~Romero-L\'opez,
``The large $N_{c}$ limit of QCD on the lattice,''
Eur. Phys. J. A \textbf{57}, no.2, 52 (2021)
doi:10.1140/epja/s10050-021-00374-2
[arXiv:2012.03331 [hep-lat]].





\bibitem{Athenodorou:2021qvs}
A.~Athenodorou and M.~Teper,
``SU(N) gauge theories in 3+1 dimensions: glueball spectrum, string tensions and topology,''
[arXiv:2106.00364 [hep-lat]].

\bibitem{Yamanaka:2021xqh}
N.~Yamanaka, A.~Nakamura and M.~Wakayama,
``Interglueball potential in lattice SU(N) gauge theories,''
[arXiv:2110.04521 [hep-lat]].

\bibitem{Bonanno:2022yjr}
C.~Bonanno, M.~D'Elia, B.~Lucini and D.~Vadacchino,
``Towards glueball masses of large-$N$$\mathrm{SU}(N)$ pure-gauge theories without topological freezing,''
[arXiv:2205.06190 [hep-lat]].






\bibitem{Aharony:2009gg}
O.~Aharony and E.~Karzbrun,
``On the effective action of confining strings,''
JHEP \textbf{06}, 012 (2009)
doi:10.1088/1126-6708/2009/06/012
[arXiv:0903.1927 [hep-th]].






\bibitem{Witten:1979vv}
E.~Witten,
``Current Algebra Theorems for the U(1) Goldstone Boson,''
Nucl. Phys. B \textbf{156}, 269-283 (1979)
doi:10.1016/0550-3213(79)90031-2

\bibitem{Veneziano:1979ec}
G.~Veneziano,
``U(1) Without Instantons,''
Nucl. Phys. B \textbf{159}, 213-224 (1979)
doi:10.1016/0550-3213(79)90332-8


\bibitem{Witten:1998uka}
E.~Witten,
``Theta dependence in the large N limit of four-dimensional gauge theories,''
Phys. Rev. Lett. \textbf{81} (1998), 2862-2865
doi:10.1103/PhysRevLett.81.2862
[arXiv:hep-th/9807109 [hep-th]].

\bibitem{Vicari:2008jw}
E.~Vicari and H.~Panagopoulos,
``Theta dependence of SU(N) gauge theories in the presence of a topological term,''
Phys. Rept. \textbf{470} (2009), 93-150
doi:10.1016/j.physrep.2008.10.001
[arXiv:0803.1593 [hep-th]].








\bibitem{Luscher:1981zq}
M.~Luscher,
``Topology of Lattice Gauge Fields,''
Commun. Math. Phys. \textbf{85}, 39 (1982)
doi:10.1007/BF02029132

\bibitem{Campostrini:1989dh}
M.~Campostrini, A.~Di Giacomo, H.~Panagopoulos and E.~Vicari,
``Topological Charge, Renormalization and Cooling on the Lattice,''
Nucl. Phys. B \textbf{329}, 683-697 (1990)
doi:10.1016/0550-3213(90)90077-Q




\bibitem{DelDebbio:2002xa}
L.~Del Debbio, H.~Panagopoulos and E.~Vicari,
``theta dependence of SU(N) gauge theories,''
JHEP \textbf{08}, 044 (2002)
doi:10.1088/1126-6708/2002/08/044
[arXiv:hep-th/0204125 [hep-th]].


\bibitem{Lucini:2004yh}
B.~Lucini, M.~Teper and U.~Wenger,
``Topology of SU(N) gauge theories at T =\textasciitilde{} 0 and T =\textasciitilde{} T(c),''
Nucl. Phys. B \textbf{715} (2005), 461-482
doi:10.1016/j.nuclphysb.2005.02.037
[arXiv:hep-lat/0401028 [hep-lat]].



\bibitem{DelDebbio:2004ns}
L.~Del Debbio, L.~Giusti and C.~Pica,
``Topological susceptibility in the SU(3) gauge theory,''
Phys. Rev. Lett. \textbf{94}, 032003 (2005)
doi:10.1103/PhysRevLett.94.032003
[arXiv:hep-th/0407052 [hep-th]].








\bibitem{Luscher:2010ik}
M.~Luscher and F.~Palombi,
``Universality of the topological susceptibility in the SU(3) gauge theory,''
JHEP \textbf{09}, 110 (2010)
doi:10.1007/JHEP09(2010)110
[arXiv:1008.0732 [hep-lat]].




\bibitem{Panagopoulos:2011rb}
H.~Panagopoulos and E.~Vicari,
``The 4D SU(3) gauge theory with an imaginary $\theta$ term,''
JHEP \textbf{11}, 119 (2011)
doi:10.1007/JHEP11(2011)119
[arXiv:1109.6815 [hep-lat]].

\bibitem{Bonati:2015sqt}
C.~Bonati, M.~D'Elia and A.~Scapellato,
``$\theta$ dependence in $SU(3)$ Yang-Mills theory from analytic continuation,''
Phys. Rev. D \textbf{93}, no.2, 025028 (2016)
doi:10.1103/PhysRevD.93.025028
[arXiv:1512.01544 [hep-lat]].



\bibitem{Bonati:2016tvi}
C.~Bonati, M.~D'Elia, P.~Rossi and E.~Vicari,
``$\theta$ dependence of 4D $SU(N)$ gauge theories in the large-$N$ limit,''
Phys. Rev. D \textbf{94}, no.8, 085017 (2016)
doi:10.1103/PhysRevD.94.085017
[arXiv:1607.06360 [hep-lat]].



\bibitem{Ce:2016awn}
M.~C\`e, M.~Garc\'\i{}a Vera, L.~Giusti and S.~Schaefer,
``The topological susceptibility in the large-$N$ limit of SU($N$) Yang\textendash{}Mills theory,''
Phys. Lett. B \textbf{762}, 232-236 (2016)
doi:10.1016/j.physletb.2016.09.029
[arXiv:1607.05939 [hep-lat]].





\bibitem{Alexandrou:2017hqw}
C.~Alexandrou, A.~Athenodorou, K.~Cichy, A.~Dromard, E.~Garcia-Ramos, K.~Jansen, U.~Wenger and F.~Zimmermann,
``Comparison of topological charge definitions in Lattice QCD,''
Eur. Phys. J. C \textbf{80} (2020) no.5, 424
doi:10.1140/epjc/s10052-020-7984-9
[arXiv:1708.00696 [hep-lat]].


\bibitem{Bonanno:2020hht}
C.~Bonanno, C.~Bonati and M.~D'Elia,
``Large-$N$ $SU(N)$ Yang-Mills theories with milder topological freezing,''
JHEP \textbf{03}, 111 (2021)
doi:10.1007/JHEP03(2021)111
[arXiv:2012.14000 [hep-lat]].



\bibitem{Borsanyi:2021gqg} 
S.~Borsanyi and D.~Sexty, 
``Topological susceptibility
 of pure gauge theory using Density of States,'' 
 Phys. Lett. B \textbf{815}, 136148 (2021)
 doi:10.1016/j.physletb.2021.136148 
 [arXiv:2101.03383 [hep-lat]]. 




\bibitem{Teper:2022mmj}
M.~Teper,
``More methods for calculating the topological charge (density) of SU(N) lattice gauge fields in 3+1 dimensions,''
[arXiv:2202.02528 [hep-lat]].


\bibitem{Bonanno:2022vot}
C.~Bonanno, M.~D'Elia, B.~Lucini and D.~Vadacchino,
``Towards glueball masses of large-$N$ $\mathrm{SU}(N)$ Yang-Mills theories without topological freezing via parallel tempering on boundary conditions,''
PoS \textbf{LATTICE2022}, 392 (2023)
doi:10.22323/1.430.0392
[arXiv:2210.07622 [hep-lat]].

\bibitem{Bonanno:2022hmz}
C.~Bonanno,
``Lattice determination of the topological susceptibility slope \ensuremath{\chi}' of 2d CPN-1 models at large N,''
Phys. Rev. D \textbf{107}, no.1, 014514 (2023)
doi:10.1103/PhysRevD.107.014514
[arXiv:2212.02330 [hep-lat]].



 \bibitem{Bando:1984ej} 
  M.~Bando, T.~Kugo, S.~Uehara, K.~Yamawaki and T.~Yanagida,
   ``Is rho Meson a Dynamical Gauge Boson of Hidden Local Symmetry?,''
  Phys.\ Rev.\ Lett.\  {\bf 54}, 1215 (1985).
  doi:10.1103/PhysRevLett.54.1215
  
 \bibitem{Casalbuoni:1985kq} 
  R.~Casalbuoni, S.~De Curtis, D.~Dominici and R.~Gatto,
   ``Effective Weak Interaction Theory with Possible New Vector Resonance from a Strong Higgs Sector,''
  Phys.\ Lett.\  {\bf 155B}, 95 (1985).
  doi:10.1016/0370-2693(85)91038-X
  
 \bibitem{Bando:1987br} 
  M.~Bando, T.~Kugo and K.~Yamawaki,
   ``Nonlinear Realization and Hidden Local Symmetries,''
  Phys.\ Rept.\  {\bf 164}, 217 (1988).
  doi:10.1016/0370-1573(88)90019-1
  
   \bibitem{Casalbuoni:1988xm} 
  R.~Casalbuoni, S.~De Curtis, D.~Dominici, F.~Feruglio and R.~Gatto,
   ``Vector and Axial Vector Bound States From a Strongly Interacting Electroweak Sector,''
  Int.\ J.\ Mod.\ Phys.\ A {\bf 4}, 1065 (1989).
  doi:10.1142/S0217751X89000492
  
 \bibitem{Harada:2003jx} 
  M.~Harada and K.~Yamawaki,
   ``Hidden local symmetry at loop: A New perspective of composite gauge boson and chiral phase transition,''
  Phys.\ Rept.\  {\bf 381}, 1 (2003)
  doi:10.1016/S0370-1573(03)00139-X
  [hep-ph/0302103].


 \bibitem{Georgi:1989xy} 
  H.~Georgi,
   ``Vector Realization of Chiral Symmetry,''
  Nucl.\ Phys.\ B {\bf 331}, 311 (1990).
  doi:10.1016/0550-3213(90)90210-5


\bibitem{Appelquist:1999dq}
T.~Appelquist, P.~S.~Rodrigues da Silva and F.~Sannino,
``Enhanced global symmetries and the chiral phase transition,''
Phys. Rev. D \textbf{60}, 116007 (1999)
doi:10.1103/PhysRevD.60.116007
[arXiv:hep-ph/9906555 [hep-ph]].

 \bibitem{Piai:2004yb} 
  M.~Piai, A.~Pierce and J.~G.~Wacker,
   ``Composite vector mesons from QCD to the little Higgs,''
  hep-ph/0405242.

   \bibitem{Franzosi:2016aoo} 
  D.~Buarque Franzosi, G.~Cacciapaglia, H.~Cai, A.~Deandrea and M.~Frandsen,
   ``Vector and Axial-vector resonances in composite models of the Higgs boson,''
  JHEP {\bf 1611}, 076 (2016)
  doi:10.1007/JHEP11(2016)076
  [arXiv:1605.01363 [hep-ph]].
  
  
  

\bibitem{Cabibbo:1982zn}
N.~Cabibbo and E.~Marinari,
``A New Method for Updating SU(N) Matrices in Computer Simulations of Gauge Theories,''
Phys. Lett. B \textbf{119}, 387-390 (1982)
doi:10.1016/0370-2693(82)90696-7

  
  \bibitem{Lewis:2011zb}
R.~Lewis, C.~Pica and F.~Sannino,
``Light Asymmetric Dark Matter on the Lattice: SU(2) Technicolor with Two Fundamental Flavors,''
Phys. Rev. D \textbf{85}, 014504 (2012)
doi:10.1103/PhysRevD.85.014504
[arXiv:1109.3513 [hep-ph]].
  
  
  \bibitem{Slansky:1981yr}
R.~Slansky,
``Group Theory for Unified Model Building,''
Phys. Rept. \textbf{79}, 1-128 (1981)
doi:10.1016/0370-1573(81)90092-2
  

  
  
  
 \bibitem{Caswell:1974gg}
W.~E.~Caswell,
``Asymptotic Behavior of Nonabelian Gauge Theories to Two Loop Order,''
Phys. Rev. Lett. \textbf{33}, 244 (1974)
doi:10.1103/PhysRevLett.33.244

\bibitem{Banks:1981nn}
T.~Banks and A.~Zaks,
``On the Phase Structure of Vector-Like Gauge Theories with Massless Fermions,''
Nucl. Phys. B \textbf{196}, 189-204 (1982)
doi:10.1016/0550-3213(82)90035-9


\bibitem{Chivukula:2000mb}
R.~S.~Chivukula,
``Lectures on technicolor and compositeness,''
[arXiv:hep-ph/0011264 [hep-ph]].

\bibitem{Lane:2002wv}
K.~Lane,
``Two Lectures on Technicolor,''
[arXiv:hep-ph/0202255 [hep-ph]].

\bibitem{Hill:2002ap}
C.~T.~Hill and E.~H.~Simmons,
``Strong Dynamics and Electroweak Symmetry Breaking,''
Phys. Rept. \textbf{381}, 235-402 (2003)
[erratum: Phys. Rept. \textbf{390}, 553-554 (2004)]
doi:10.1016/S0370-1573(03)00140-6
[arXiv:hep-ph/0203079 [hep-ph]].

\bibitem{Martin:2008cd}
A.~Martin,
``Predicted Signals at the LHC from Technicolor: Erice Lecture,''
Subnucl. Ser. \textbf{46}, 135-159 (2011)
doi:10.1142/9789814340212\_0004
[arXiv:0812.1841 [hep-ph]].

\bibitem{Sannino:2009za}
F.~Sannino,
``Conformal Dynamics for TeV Physics and Cosmology,''
Acta Phys. Polon. B \textbf{40}, 3533-3743 (2009)
[arXiv:0911.0931 [hep-ph]].

\bibitem{Piai:2010ma}
M.~Piai,
``Lectures on walking technicolor, holography and gauge/gravity dualities,''
Adv. High Energy Phys. \textbf{2010}, 464302 (2010)
doi:10.1155/2010/464302
[arXiv:1004.0176 [hep-ph]].


\bibitem{Pica:2010xq}
C.~Pica and F.~Sannino,
``UV and IR Zeros of Gauge Theories at The Four Loop Order and Beyond,''
Phys. Rev. D \textbf{83}, 035013 (2011)
doi:10.1103/PhysRevD.83.035013
[arXiv:1011.5917 [hep-ph]].


\bibitem{Baikov:2016tgj}
P.~A.~Baikov, K.~G.~Chetyrkin and J.~H.~K\"uhn,
``Five-Loop Running of the QCD coupling constant,''
Phys. Rev. Lett. \textbf{118}, no.8, 082002 (2017)
doi:10.1103/PhysRevLett.118.082002
[arXiv:1606.08659 [hep-ph]].


\bibitem{Herzog:2017ohr}
F.~Herzog, B.~Ruijl, T.~Ueda, J.~A.~M.~Vermaseren and A.~Vogt,
``The five-loop beta function of Yang-Mills theory with fermions,''
JHEP \textbf{02}, 090 (2017)
doi:10.1007/JHEP02(2017)090
[arXiv:1701.01404 [hep-ph]].

\bibitem{Ryttov:2016ner}
T.~A.~Ryttov and R.~Shrock,
``Infrared Zero of $\beta$ and Value of $\gamma_m$ for an SU(3) Gauge Theory at the Five-Loop Level,''
Phys. Rev. D \textbf{94}, no.10, 105015 (2016)
doi:10.1103/PhysRevD.94.105015
[arXiv:1607.06866 [hep-th]].




\bibitem{Appelquist:1988yc}
T.~Appelquist, K.~D.~Lane and U.~Mahanta,
``On the Ladder Approximation for Spontaneous Chiral Symmetry Breaking,''
Phys. Rev. Lett. \textbf{61}, 1553 (1988)
doi:10.1103/PhysRevLett.61.1553


\bibitem{Cohen:1988sq}
A.~G.~Cohen and H.~Georgi,
``Walking Beyond the Rainbow,''
Nucl. Phys. B \textbf{314}, 7-24 (1989)
doi:10.1016/0550-3213(89)90109-0

\bibitem{Ryttov:2007cx}
T.~A.~Ryttov and F.~Sannino,
``Supersymmetry inspired QCD beta function,''
Phys. Rev. D \textbf{78}, 065001 (2008)
doi:10.1103/PhysRevD.78.065001
[arXiv:0711.3745 [hep-th]].


\bibitem{Pica:2010mt}
C.~Pica and F.~Sannino,
``Beta Function and Anomalous Dimensions,''
Phys. Rev. D \textbf{83}, 116001 (2011)
doi:10.1103/PhysRevD.83.116001
[arXiv:1011.3832 [hep-ph]].


\bibitem{Intriligator:1995au}
K.~A.~Intriligator and N.~Seiberg,
``Lectures on supersymmetric gauge theories and electric-magnetic duality,''
Nucl. Phys. B Proc. Suppl. \textbf{45BC}, 1-28 (1996)
doi:10.1016/0920-5632(95)00626-5
[arXiv:hep-th/9509066 [hep-th]].


\bibitem{Ryttov:2016hdp}
T.~A.~Ryttov,
``Consistent Perturbative Fixed Point Calculations in QCD and Supersymmetric QCD,''
Phys. Rev. Lett. \textbf{117}, no.7, 071601 (2016)
doi:10.1103/PhysRevLett.117.071601
[arXiv:1604.00687 [hep-th]].


\bibitem{Ryttov:2016asb}
T.~A.~Ryttov and R.~Shrock,
``Scheme-independent calculation of $\gamma_{\bar\psi\psi,IR}$ for an SU(3) gauge theory,''
Phys. Rev. D \textbf{94}, no.10, 105014 (2016)
doi:10.1103/PhysRevD.94.105014
[arXiv:1608.00068 [hep-th]].


\bibitem{Ryttov:2016hal}
T.~A.~Ryttov and R.~Shrock,
``Scheme-Independent Series Expansions at an Infrared Zero of the Beta Function in Asymptotically Free Gauge Theories,''
Phys. Rev. D \textbf{94}, no.12, 125005 (2016)
doi:10.1103/PhysRevD.94.125005
[arXiv:1610.00387 [hep-th]].



\bibitem{Ryttov:2017toz}
T.~A.~Ryttov and R.~Shrock,
``Higher-Order Scheme-Independent Calculations of Physical Quantities in the Conformal Phase of a Gauge Theory,''
Phys. Rev. D \textbf{95}, no.8, 085012 (2017)
doi:10.1103/PhysRevD.95.085012
[arXiv:1701.06083 [hep-th]].



\bibitem{Ryttov:2017kmx}
T.~A.~Ryttov and R.~Shrock,
``Higher-order scheme-independent series expansions of $\gamma_{\bar\psi\psi,IR}$ and $\beta'_{IR}$ in conformal field theories,''
Phys. Rev. D \textbf{95}, no.10, 105004 (2017)
doi:10.1103/PhysRevD.95.105004
[arXiv:1703.08558 [hep-th]].

\bibitem{Ryttov:2017dhd}
T.~A.~Ryttov and R.~Shrock,
``Infrared fixed point physics in SO($N_c$) and Sp($N_c$) gauge theories,''
Phys. Rev. D \textbf{96}, no.10, 105015 (2017)
doi:10.1103/PhysRevD.96.105015
[arXiv:1709.05358 [hep-th]].


\bibitem{Gracey:2018oym}
J.~A.~Gracey, T.~A.~Ryttov and R.~Shrock,
``Scheme-Independent Calculations of Anomalous Dimensions of Baryon Operators in Conformal Field Theories,''
Phys. Rev. D \textbf{97}, no.11, 116018 (2018)
doi:10.1103/PhysRevD.97.116018
[arXiv:1805.02729 [hep-th]].


\bibitem{Ryttov:2018uue}
T.~A.~Ryttov and R.~Shrock,
``Scheme-Independent Calculations of Properties at a Conformal Infrared Fixed Point in Gauge Theories with Multiple Fermion Representations,''
Phys. Rev. D \textbf{98}, no.9, 096003 (2018)
doi:10.1103/PhysRevD.98.096003
[arXiv:1809.02242 [hep-th]].




\bibitem{Ryttov:2020scx}
T.~A.~Ryttov and R.~Shrock,
``Scheme-Independent Series for Anomalous Dimensions of Higher-Spin Operators at an Infrared Fixed Point in a Gauge Theory,''
Phys. Rev. D \textbf{101}, 076018 (2020)
doi:10.1103/PhysRevD.101.076018
[arXiv:2002.08979 [hep-ph]].

\bibitem{Kaplan:2009kr}
D.~B.~Kaplan, J.~W.~Lee, D.~T.~Son and M.~A.~Stephanov,
``Conformality Lost,''
Phys. Rev. D \textbf{80}, 125005 (2009)
doi:10.1103/PhysRevD.80.125005
[arXiv:0905.4752 [hep-th]].



\bibitem{Kim:2020yvr}
B.~S.~Kim, D.~K.~Hong and J.~W.~Lee,
``Into the conformal window: Multirepresentation gauge theories,''
Phys. Rev. D \textbf{101} (2020) no.5, 056008
doi:10.1103/PhysRevD.101.056008
[arXiv:2001.02690 [hep-ph]].


\bibitem{Lee:2020ihn}
J.~W.~Lee,
``Conformal window from conformal expansion,''
Phys. Rev. D \textbf{103} (2021) no.7, 076006
doi:10.1103/PhysRevD.103.076006
[arXiv:2008.12223 [hep-ph]].


\bibitem{Appelquist:1998rb}
T.~Appelquist, A.~Ratnaweera, J.~Terning and L.~C.~R.~Wijewardhana,
``The Phase structure of an SU(N) gauge theory with N(f) flavors,''
Phys. Rev. D \textbf{58}, 105017 (1998)
doi:10.1103/PhysRevD.58.105017
[arXiv:hep-ph/9806472 [hep-ph]].


  
  
    
  \bibitem{Gell-Mann:1968hlm}
M.~Gell-Mann, R.~J.~Oakes and B.~Renner,
``Behavior of current divergences under SU(3) x SU(3),''
Phys. Rev. \textbf{175}, 2195-2199 (1968)
doi:10.1103/PhysRev.175.2195
  
  
  

 \bibitem{Coleman:1973jx} 
  S.~R.~Coleman and E.~J.~Weinberg,
  ``Radiative Corrections as the Origin of Spontaneous Symmetry Breaking,''
  Phys.\ Rev.\ D {\bf 7}, 1888 (1973).
  doi:10.1103/PhysRevD.7.1888
  

\bibitem{Einhorn:1992um}
M.~B.~Einhorn and D.~R.~T.~Jones,
``The Effective potential and quadratic divergences,''
Phys. Rev. D \textbf{46}, 5206-5208 (1992)
doi:10.1103/PhysRevD.46.5206

  

\bibitem{Weinberg:1975gm}
S.~Weinberg,
``Implications of Dynamical Symmetry Breaking,''
Phys. Rev. D \textbf{13}, 974-996 (1976)
Phys. Rev. D \textbf{19} 1277-1280 (1979, addendum)
doi:10.1103/PhysRevD.19.1277

\bibitem{Susskind:1978ms}
L.~Susskind,
``Dynamics of Spontaneous Symmetry Breaking in the Weinberg-Salam Theory,''
Phys. Rev. D \textbf{20}, 2619-2625 (1979)
doi:10.1103/PhysRevD.20.2619




\bibitem{Holdom:1984sk}
B.~Holdom,
``Techniodor,''
Phys. Lett. B \textbf{150}, 301-305 (1985)
doi:10.1016/0370-2693(85)91015-9




\bibitem{Appelquist:1986an}
T.~W.~Appelquist, D.~Karabali and L.~C.~R.~Wijewardhana,
``Chiral Hierarchies and the Flavor Changing Neutral Current Problem in Technicolor,''
Phys. Rev. Lett. \textbf{57}, 957 (1986)
doi:10.1103/PhysRevLett.57.957


\bibitem{Dimopoulos:1979es}
S.~Dimopoulos and L.~Susskind,
``Mass Without Scalars,''
Nucl. Phys. B \textbf{155}, 237-252 (1979)
doi:10.1016/0550-3213(79)90364-X


\bibitem{Eichten:1979ah}
E.~Eichten and K.~D.~Lane,
``Dynamical Breaking of Weak Interaction Symmetries,''
Phys. Lett. B \textbf{90}, 125-130 (1980)
doi:10.1016/0370-2693(80)90065-9




\bibitem{Peskin:1991sw}
M.~E.~Peskin and T.~Takeuchi,
``Estimation of oblique electroweak corrections,''
Phys. Rev. D \textbf{46}, 381-409 (1992)
doi:10.1103/PhysRevD.46.381


  
\bibitem{Barbieri:2004qk}
R.~Barbieri, A.~Pomarol, R.~Rattazzi and A.~Strumia,
``Electroweak symmetry breaking after LEP-1 and LEP-2,''
Nucl. Phys. B \textbf{703}, 127-146 (2004)
doi:10.1016/j.nuclphysb.2004.10.014
[arXiv:hep-ph/0405040 [hep-ph]].


\bibitem{Appelquist:1980vg}
T.~Appelquist and C.~W.~Bernard,
``Strongly Interacting Higgs Bosons,''
Phys. Rev. D \textbf{22}, 200 (1980)
doi:10.1103/PhysRevD.22.200

\bibitem{Longhitano:1980iz}
A.~C.~Longhitano,
``Heavy Higgs Bosons in the Weinberg-Salam Model,''
Phys. Rev. D \textbf{22}, 1166 (1980)
doi:10.1103/PhysRevD.22.1166

\bibitem{Longhitano:1980tm}
A.~C.~Longhitano,
``Low-Energy Impact of a Heavy Higgs Boson Sector,''
Nucl. Phys. B \textbf{188}, 118-154 (1981)
doi:10.1016/0550-3213(81)90109-7

\bibitem{Appelquist:1993ka}
T.~Appelquist and G.~H.~Wu,
``The Electroweak chiral Lagrangian and new precision measurements,''
Phys. Rev. D \textbf{48}, 3235-3241 (1993)
doi:10.1103/PhysRevD.48.3235
[arXiv:hep-ph/9304240 [hep-ph]].

\bibitem{Appelquist:1994qz}
T.~Appelquist and G.~H.~Wu,
``The Electroweak chiral Lagrangian and CP violating effects in technicolor theories,''
Phys. Rev. D \textbf{51}, 240-250 (1995)
doi:10.1103/PhysRevD.51.240
[arXiv:hep-ph/9406416 [hep-ph]].




\bibitem{Glashow:1970gm}
S.~L.~Glashow, J.~Iliopoulos and L.~Maiani,
``Weak Interactions with Lepton-Hadron Symmetry,''
Phys. Rev. D \textbf{2}, 1285-1292 (1970)
doi:10.1103/PhysRevD.2.1285


\bibitem{Appelquist:1993sg}
T.~Appelquist and J.~Terning,
``An Extended technicolor model,''
Phys. Rev. D \textbf{50}, 2116-2126 (1994)
doi:10.1103/PhysRevD.50.2116
[arXiv:hep-ph/9311320 [hep-ph]].

\bibitem{Appelquist:2002me}
T.~Appelquist and R.~Shrock,
``Neutrino masses in theories with dynamical electroweak symmetry breaking,''
Phys. Lett. B \textbf{548}, 204-214 (2002)
doi:10.1016/S0370-2693(02)02854-X
[arXiv:hep-ph/0204141 [hep-ph]].

\bibitem{Appelquist:2003uu}
T.~Appelquist and R.~Shrock,
``Dynamical symmetry breaking of extended gauge symmetries,''
Phys. Rev. Lett. \textbf{90}, 201801 (2003)
doi:10.1103/PhysRevLett.90.201801
[arXiv:hep-ph/0301108 [hep-ph]].

\bibitem{Appelquist:2003hn}
T.~Appelquist, M.~Piai and R.~Shrock,
``Fermion masses and mixing in extended technicolor models,''
Phys. Rev. D \textbf{69}, 015002 (2004)
doi:10.1103/PhysRevD.69.015002
[arXiv:hep-ph/0308061 [hep-ph]].

\bibitem{Appelquist:2004mn}
T.~Appelquist, M.~Piai and R.~Shrock,
``Lepton dipole moments in extended technicolor models,''
Phys. Lett. B \textbf{593}, 175-180 (2004)
doi:10.1016/j.physletb.2004.04.062
[arXiv:hep-ph/0401114 [hep-ph]].

\bibitem{Appelquist:2004es}
T.~Appelquist, M.~Piai and R.~Shrock,
``Quark dipole operators in extended technicolor models,''
Phys. Lett. B \textbf{595}, 442-452 (2004)
doi:10.1016/j.physletb.2004.06.066
[arXiv:hep-ph/0406032 [hep-ph]].

\bibitem{Appelquist:2004ai}
T.~Appelquist, N.~D.~Christensen, M.~Piai and R.~Shrock,
``Flavor-changing processes in extended technicolor,''
Phys. Rev. D \textbf{70}, 093010 (2004)
doi:10.1103/PhysRevD.70.093010
[arXiv:hep-ph/0409035 [hep-ph]].  


\bibitem{Georgi:1992dw}
H.~Georgi,
``Generalized dimensional analysis,''
Phys. Lett. B \textbf{298}, 187-189 (1993)
doi:10.1016/0370-2693(93)91728-6
[arXiv:hep-ph/9207278 [hep-ph]].
  



\bibitem{Leung:1989hw}
C.~N.~Leung, S.~T.~Love and W.~A.~Bardeen,
``Aspects of Dynamical Symmetry Breaking in Gauge Field Theories,''
Nucl. Phys. B \textbf{323}, 493-512 (1989)
doi:10.1016/0550-3213(89)90121-1


\bibitem{Luty:2004ye}
M.~A.~Luty and T.~Okui,
``Conformal technicolor,''
JHEP \textbf{09}, 070 (2006)
doi:10.1088/1126-6708/2006/09/070
[arXiv:hep-ph/0409274 [hep-ph]].

  

\bibitem{Spergel:1999mh}
D.~N.~Spergel and P.~J.~Steinhardt,
``Observational evidence for selfinteracting cold dark matter,''
Phys. Rev. Lett. \textbf{84}, 3760-3763 (2000)
doi:10.1103/PhysRevLett.84.3760
[arXiv:astro-ph/9909386 [astro-ph]].

\bibitem{deBlok:2009sp}
W.~J.~G.~de Blok,
``The Core-Cusp Problem,''
Adv. Astron. \textbf{2010}, 789293 (2010)
doi:10.1155/2010/789293
[arXiv:0910.3538 [astro-ph.CO]].

\bibitem{Boylan-Kolchin:2011qkt}
M.~Boylan-Kolchin, J.~S.~Bullock and M.~Kaplinghat,
``Too big to fail? The puzzling darkness of massive Milky Way subhaloes,''
Mon. Not. Roy. Astron. Soc. \textbf{415}, L40 (2011)
doi:10.1111/j.1745-3933.2011.01074.x
[arXiv:1103.0007 [astro-ph.CO]].



\bibitem{Wess:1971yu}
J.~Wess and B.~Zumino,
``Consequences of anomalous Ward identities,''
Phys. Lett. B \textbf{37}, 95-97 (1971)
doi:10.1016/0370-2693(71)90582-X

\bibitem{Witten:1983tw}
E.~Witten,
``Global Aspects of Current Algebra,''
Nucl. Phys. B \textbf{223}, 422-432 (1983)
doi:10.1016/0550-3213(83)90063-9

\bibitem{Witten:1983tx}
E.~Witten,
``Current Algebra, Baryons, and Quark Confinement,''
Nucl. Phys. B \textbf{223}, 433-444 (1983)
doi:10.1016/0550-3213(83)90064-0



\bibitem{McDonald:2001vt}
J.~McDonald,
``Thermally generated gauge singlet scalars as selfinteracting dark matter,''
Phys. Rev. Lett. \textbf{88}, 091304 (2002)
doi:10.1103/PhysRevLett.88.091304
[arXiv:hep-ph/0106249 [hep-ph]].

\bibitem{Hall:2009bx}
L.~J.~Hall, K.~Jedamzik, J.~March-Russell and S.~M.~West,
``Freeze-In Production of FIMP Dark Matter,''
JHEP \textbf{03}, 080 (2010)
doi:10.1007/JHEP03(2010)080
[arXiv:0911.1120 [hep-ph]].

\bibitem{Yaguna:2011qn}
C.~E.~Yaguna,
``The Singlet Scalar as FIMP Dark Matter,''
JHEP \textbf{08}, 060 (2011)
doi:10.1007/JHEP08(2011)060
[arXiv:1105.1654 [hep-ph]].

\bibitem{Campbell:2015fra}
R.~Campbell, S.~Godfrey, H.~E.~Logan, A.~D.~Peterson and A.~Poulin,
``Implications of the observation of dark matter self-interactions for singlet scalar dark matter,''
Phys. Rev. D \textbf{92}, no.5, 055031 (2015)
[erratum: Phys. Rev. D \textbf{101}, no.3, 039905 (2020)]
doi:10.1103/PhysRevD.92.055031
[arXiv:1505.01793 [hep-ph]].

\bibitem{Kang:2015aqa}
Z.~Kang,
``View FImP miracle (by scale invariance) \`a la self-interaction,''
Phys. Lett. B \textbf{751}, 201-204 (2015)
doi:10.1016/j.physletb.2015.10.031
[arXiv:1505.06554 [hep-ph]].


\bibitem{Espinosa:2010hh}
J.~R.~Espinosa, T.~Konstandin, J.~M.~No and G.~Servant,
``Energy Budget of Cosmological First-order Phase Transitions,''
JCAP \textbf{06}, 028 (2010)
doi:10.1088/1475-7516/2010/06/028
[arXiv:1004.4187 [hep-ph]].




\bibitem{Bigazzi:2020phm}
F.~Bigazzi, A.~Caddeo, A.~L.~Cotrone and A.~Paredes,
``Fate of false vacua in holographic first-order phase transitions,''
JHEP   {12}, 200 (2020);
doi:10.1007/JHEP12(2020)200
[arXiv:2008.02579 [hep-th]].

\bibitem{Ares:2020lbt}
F.~R.~Ares, M.~Hindmarsh, C.~Hoyos and N.~Jokela,
``Gravitational waves from a holographic phase transition,''
JHEP   {21}, 100 (2020);
doi:10.1007/JHEP04(2021)100
[arXiv:2011.12878 [hep-th]].

\bibitem{Bea:2021zsu}
Y.~Bea, J.~Casalderrey-Solana, T.~Giannakopoulos, 
D.~Mateos, 
M.~Sanchez-Garitaonandia and M.~Zilh\~ao,
``Bubble wall velocity from holography,''
Phys. Rev. D   {104}, no.12, L121903 (2021);
doi:10.1103/PhysRevD.104.L121903
[arXiv:2104.05708 [hep-th]].

\bibitem{Bigazzi:2021ucw}
F.~Bigazzi, A.~Caddeo, T.~Canneti and A.~L.~Cotrone,
``Bubble wall velocity at strong coupling,''
JHEP   {08}, 090 (2021)
doi:10.1007/JHEP08(2021)090
[arXiv:2104.12817 [hep-ph]].

\bibitem{Henriksson:2021zei}
O.~Henriksson,
``Black brane evaporation through D-brane bubble nucleation,''
Phys. Rev. D   {105}, no.4, L041901 (2022);
doi:10.1103/PhysRevD.105.L041901
[arXiv:2106.13254 [hep-th]].

\bibitem{Ares:2021ntv}
F.~R.~Ares, O.~Henriksson, M.~Hindmarsh, 
C.~Hoyos and N.~Jokela,
``Effective actions and bubble nucleation from holography,''
Phys. Rev. D   {105}, no.6, 066020 (2022);
doi:10.1103/PhysRevD.105.066020
[arXiv:2109.13784 [hep-th]].

\bibitem{Ares:2021nap}
F.~R.~Ares, O.~Henriksson, M.~Hindmarsh, 
C.~Hoyos and N.~Jokela,
``Gravitational Waves at Strong Coupling from an Effective Action,''
Phys. Rev. Lett.   {128}, no.13, 131101 (2022);
doi:10.1103/PhysRevLett.128.131101
[arXiv:2110.14442 [hep-th]].

\bibitem{Morgante:2022zvc}
E.~Morgante, N.~Ramberg and P.~Schwaller,
``Echo of the Dark: Gravitational Waves from Dark SU(3) Yang-Mills Theory,''
[arXiv:2210.11821 [hep-ph]].




\bibitem{Verbaarschot:1994qf}
J.~J.~M.~Verbaarschot,
``The Spectrum of the QCD Dirac operator and chiral random matrix theory: The Threefold way,''
Phys. Rev. Lett. \textbf{72}, 2531-2533 (1994)
doi:10.1103/PhysRevLett.72.2531
[arXiv:hep-th/9401059 [hep-th]].

\bibitem{DelDebbio:2008zf}
L.~Del Debbio, A.~Patella and C.~Pica,
``Higher representations on the lattice: Numerical simulations. SU(2) with adjoint fermions,''
Phys. Rev. D \textbf{81} (2010), 094503
doi:10.1103/PhysRevD.81.094503
[arXiv:0805.2058 [hep-lat]].


\bibitem{hirep-upstream}
  \url{https://github.com/claudiopica/HiRep}


\bibitem{hirep-repo}
\url{https://github.com/sa2c/HiRep}.





\bibitem{Clark:2003na}
M.~A.~Clark and A.~D.~Kennedy,
``The RHMC algorithm for two flavors of dynamical staggered fermions,''
Nucl. Phys. B Proc. Suppl. \textbf{129}, 850-852 (2004)
doi:10.1016/S0920-5632(03)02732-4
[arXiv:hep-lat/0309084 [hep-lat]].


\bibitem{Takaishi:2005tz}
T.~Takaishi and P.~de Forcrand,
``Testing and tuning new symplectic integrators for hybrid Monte Carlo algorithm in lattice QCD,''
Phys. Rev. E \textbf{73}, 036706 (2006)
doi:10.1103/PhysRevE.73.036706
[arXiv:hep-lat/0505020 [hep-lat]].
 

\bibitem{DeGrand:1990dk}
T.~A.~DeGrand and P.~Rossi,
``Conditioning Techniques for Dynamical Fermions,''
Comput. Phys. Commun. \textbf{60}, 211-214 (1990)
doi:10.1016/0010-4655(90)90006-M




\bibitem{Luscher:2010iy}
M.~L\"uscher,
``Properties and uses of the Wilson flow in lattice QCD,''
JHEP \textbf{08}, 071 (2010)
[erratum: JHEP \textbf{03}, 092 (2014)]
doi:10.1007/JHEP08(2010)071
[arXiv:1006.4518 [hep-lat]].

\bibitem{Luscher:2013vga}
M.~L\"uscher,
``Future applications of the Yang-Mills gradient flow in lattice QCD,''
PoS \textbf{LATTICE2013}, 016 (2014)
doi:10.22323/1.187.0016
[arXiv:1308.5598 [hep-lat]].

\bibitem{Luscher:2011bx}
M.~Luscher and P.~Weisz,
``Perturbative analysis of the gradient flow in non-abelian gauge theories,''
JHEP \textbf{02}, 051 (2011)
doi:10.1007/JHEP02(2011)051
[arXiv:1101.0963 [hep-th]].




\bibitem{Borsanyi:2012zs}
S.~Borsanyi, S.~Durr, Z.~Fodor, C.~Hoelbling, S.~D.~Katz, S.~Krieg, T.~Kurth, L.~Lellouch, T.~Lippert and C.~McNeile, \textit{et al.}
``High-precision scale setting in lattice QCD,''
JHEP \textbf{09} (2012), 010
doi:10.1007/JHEP09(2012)010
[arXiv:1203.4469 [hep-lat]].




\bibitem{Sheikholeslami:1985ij}
B.~Sheikholeslami and R.~Wohlert,
``Improved Continuum Limit Lattice Action for QCD with Wilson Fermions,''
Nucl. Phys. B \textbf{259}, 572 (1985)
doi:10.1016/0550-3213(85)90002-1


\bibitem{Hasenbusch:2002ai}
M.~Hasenbusch and K.~Jansen,
``Speeding up lattice QCD simulations with clover improved Wilson fermions,''
Nucl. Phys. B \textbf{659}, 299-320 (2003)
doi:10.1016/S0550-3213(03)00227-X
[arXiv:hep-lat/0211042 [hep-lat]].



\bibitem{Luscher:2011kk}
M.~Luscher and S.~Schaefer,
``Lattice QCD without topology barriers,''
JHEP \textbf{07}, 036 (2011)
doi:10.1007/JHEP07(2011)036
[arXiv:1105.4749 [hep-lat]].




\bibitem{Endres:2015yca}
M.~G.~Endres, R.~C.~Brower, W.~Detmold, K.~Orginos and A.~V.~Pochinsky,
``Multiscale Monte Carlo equilibration: Pure Yang-Mills theory,''
Phys. Rev. D \textbf{92}, no.11, 114516 (2015)
doi:10.1103/PhysRevD.92.114516
[arXiv:1510.04675 [hep-lat]].

\bibitem{Luscher:2017cjh}
M.~L\"uscher,
``Stochastic locality and master-field simulations of very large lattices,''
EPJ Web Conf. \textbf{175}, 01002 (2018)
doi:10.1051/epjconf/201817501002
[arXiv:1707.09758 [hep-lat]].




\bibitem{Boyle:2008rh}
P.~A.~Boyle, A.~Juttner, C.~Kelly and R.~D.~Kenway,
``Use of stochastic sources for the lattice determination of light quark physics,''
JHEP \textbf{08}, 086 (2008)
doi:10.1088/1126-6708/2008/08/086
[arXiv:0804.1501 [hep-lat]].




\bibitem{Martinelli:1982mw}
G.~Martinelli and Y.~C.~Zhang,
``The Connection Between Local Operators on the Lattice and in the Continuum and Its Relation to Meson Decay Constants,''
Phys. Lett. B \textbf{123} (1983), 433
doi:10.1016/0370-2693(83)90987-5




\bibitem{Luscher:1980fr}
M.~Luscher, K.~Symanzik and P.~Weisz,
``Anomalies of the Free Loop Wave Equation in the WKB Approximation,''
Nucl. Phys. B \textbf{173}, 365 (1980)
doi:10.1016/0550-3213(80)90009-7

\bibitem{Polchinski:1991ax}
J.~Polchinski and A.~Strominger,
``Effective string theory,''
Phys. Rev. Lett. \textbf{67}, 1681-1684 (1991)
doi:10.1103/PhysRevLett.67.1681



\bibitem{Luscher:1980ac}
M.~Luscher,
``Symmetry Breaking Aspects of the Roughening 
Transition in Gauge Theories,''
Nucl. Phys. B \textbf{180}, 317-329 (1981)
doi:10.1016/0550-3213(81)90423-5

\bibitem{Luscher:2004ib}
M.~Luscher and P.~Weisz,
``String excitation energies in SU(N) gauge theories beyond the free-string approximation,''
JHEP \textbf{07}, 014 (2004)
doi:10.1088/1126-6708/2004/07/014
[arXiv:hep-th/0406205 [hep-th]].

\bibitem{Aharony:2009gg}
O.~Aharony and E.~Karzbrun,
``On the effective action of confining strings,''
JHEP \textbf{06}, 012 (2009)
doi:10.1088/1126-6708/2009/06/012
[arXiv:0903.1927 [hep-th]].

\bibitem{Drummond:2004yp}
J.~M.~Drummond,
``Universal subleading spectrum of effective string theory,''
[arXiv:hep-th/0411017 [hep-th]].


\bibitem{HariDass:2006sd}
N.~D.~Hari Dass and P.~Matlock,
``Universality of correction to Luescher term in Polchinski-Strominger effective string theories,''
[arXiv:hep-th/0606265 [hep-th]].


\bibitem{Drummond:2006su}
J.~M.~Drummond,
``Reply to hep-th/0606265,''
[arXiv:hep-th/0608109 [hep-th]].


\bibitem{Dass:2006ud}
N.~D.~H.~Dass and P.~Matlock,
``Our response to the response hep-th/0608109 by Drummond,''
[arXiv:hep-th/0611215 [hep-th]].

\bibitem{Aharony:2013ipa}
O.~Aharony and Z.~Komargodski,
``The Effective Theory of Long Strings,''
JHEP \textbf{05}, 118 (2013)
doi:10.1007/JHEP05(2013)118
[arXiv:1302.6257 [hep-th]].

\bibitem{Dubovsky:2015zey}
S.~Dubovsky and V.~Gorbenko,
``Towards a Theory of the QCD String,''
JHEP \textbf{02}, 022 (2016)
doi:10.1007/JHEP02(2016)022
[arXiv:1511.01908 [hep-th]].


\bibitem{Bijnens:2009qm}
J.~Bijnens and J.~Lu,
``Technicolor and other QCD-like theories at next-to-next-to-leading order,''
JHEP \textbf{11}, 116 (2009)
doi:10.1088/1126-6708/2009/11/116
[arXiv:0910.5424 [hep-ph]].



\bibitem{Athenodorou:2016ebg}
A.~Athenodorou and M.~Teper,
``SU(N) gauge theories in 2+1 dimensions: glueball spectra and k-string tensions,''
JHEP \textbf{02}, 015 (2017)
doi:10.1007/JHEP02(2017)015
[arXiv:1609.03873 [hep-lat]].

\bibitem{Elander:2018gte}
D.~Elander, A.~F.~Faedo, D.~Mateos, D.~Pravos and J.~G.~Subils,
``Mass spectrum of gapped, non-confining theories with multi-scale dynamics,''
JHEP \textbf{05}, 175 (2019)
doi:10.1007/JHEP05(2019)175
[arXiv:1810.04656 [hep-th]].

\bibitem{Leigh:2006vg}
R.~G.~Leigh, D.~Minic and A.~Yelnikov,
``On the Glueball Spectrum of Pure Yang-Mills Theory in 2+1 Dimensions,''
Phys. Rev. D \textbf{76}, 065018 (2007)
doi:10.1103/PhysRevD.76.065018
[arXiv:hep-th/0604060 [hep-th]].







\bibitem{Rupak:2002sm}
G.~Rupak and N.~Shoresh,
``Chiral perturbation theory for the Wilson lattice action,''
Phys. Rev. D \textbf{66}, 054503 (2002)
doi:10.1103/PhysRevD.66.054503
[arXiv:hep-lat/0201019 [hep-lat]].

\bibitem{Sharpe:1998xm}
S.~R.~Sharpe and R.~L.~Singleton, Jr,
``Spontaneous flavor and parity breaking with Wilson fermions,''
Phys. Rev. D \textbf{58}, 074501 (1998)
doi:10.1103/PhysRevD.58.074501
[arXiv:hep-lat/9804028 [hep-lat]].

\bibitem{Symanzik:1983dc}
K.~Symanzik,
``Continuum Limit and Improved Action in Lattice Theories. 1. Principles and phi**4 Theory,''
Nucl. Phys. B \textbf{226}, 187-204 (1983)
doi:10.1016/0550-3213(83)90468-6

\bibitem{Luscher:1996sc}
M.~Luscher, S.~Sint, R.~Sommer and P.~Weisz,
``Chiral symmetry and O(a) improvement in lattice QCD,''
Nucl. Phys. B \textbf{478}, 365-400 (1996)
doi:10.1016/0550-3213(96)00378-1
[arXiv:hep-lat/9605038 [hep-lat]].



\bibitem{Bar:2013ora}
O.~Bar and M.~Golterman,
``Chiral perturbation theory for gradient flow observables,''
Phys. Rev. D \textbf{89}, no.3, 034505 (2014)
[erratum: Phys. Rev. D \textbf{89}, no.9, 099905 (2014)]
doi:10.1103/PhysRevD.89.034505
[arXiv:1312.4999 [hep-lat]].

\bibitem{Jansen:2009hr}
K.~Jansen \textit{et al.} [ETM],
``Meson masses and decay constants from unquenched lattice QCD,''
Phys. Rev. D \textbf{80}, 054510 (2009)
doi:10.1103/PhysRevD.80.054510
[arXiv:0906.4720 [hep-lat]].





\bibitem{turing-way-reproduce}
  The Turing Way Community. (2021). The Turing Way: A handbook for reproducible, ethical and collaborative research (1.0.1). Zenodo. \url{https://doi.org/10.5281/zenodo.5671094}

\bibitem{supercooled-water}
  The war over supercooled water. (2018). Physics Today. \url{https://doi.org/10.1063/pt.6.1.20180822a}

\bibitem{repro-guide}
  E.~Bennett, J.~Lenz, ``Recommendations for reproducibility in analysis of lattice data,'' in preparation.







\bibitem{ten-data-rules}
  E.~M.~Hart, P.~Barmby, D.~LeBauer, F.~Michonneau, S.~Mount, \textit{et al.},
  ``Ten Simple Rules for Digital Data Storage,'' PLOS Computational Biology 12(10): e1005097 (2016) doi:10.1371/journal.pcbi.1005097.
 
\bibitem{hdf5}
  \url{https://www.hdfgroup.org/solutions/hdf5}

\bibitem{zenodo}
  \url{https://www.zenodo.org}

\bibitem{python}
  G.~Van Rossum, \& F.~L.~Drake (2009), Python 3 Reference Manual. Scotts Valley, CA: CreateSpace.

\bibitem{numpy}
  C.~R.~Harris, K.~J.~Millman, S.~J.~van der Walt, \textit{et al.}, ``Array programming with NumPy'', Nature 585, 357–362 (2020). doi:10.1038/s41586-020-2649-2.

\bibitem{scipy}
  P.~Virtanen, R.~Gommers, T.~E.~Oliphant, M.~Haberland, \textit{et al.}, ``SciPy 1.0: Fundamental Algorithms for Scientific Computing in Python,'' Nature Methods, 17(3), 261-272 (2020) doi:10.1038/s41592-019-0686-2.

\bibitem{matplotlib}
  J.~D.~Hunter, ``Matplotlib: A 2D Graphics Environment'', Computing in Science \& Engineering, vol. 9, no. 3, pp. 90-95, 2007 doi:10.1109/MCSE.2007.55.

\bibitem{mathematica}
  Wolfram Research, Inc., Mathematica, Version 13.0, Champaign, IL (2021).

\bibitem{bash}
  \url{https://www.gnu.org/software/bash/}

\bibitem{gnu-make}
  \url{https://www.gnu.org/software/make/}

\bibitem{conda}
  \url{https://conda.io/projects/conda/en/latest/}

\bibitem{github}
  \url{https://github.com}





\bibitem{Bennett:2022klt}
E.~Bennett,
``Status of reproducibility and open science in hep-lat in 2021,''
PoS \textbf{LATTICE2022}, 337 (2023)
doi:10.22323/1.430.0337
[arXiv:2211.15547 [hep-lat]].

\bibitem{Athenodorou:2022ixd}
A.~Athenodorou, E.~Bennett, J.~Lenz and E.~Papadopoullou,
``Open Science in Lattice Gauge Theory community,''
PoS \textbf{LATTICE2022}, 341 (2023)
doi:10.22323/1.430.0341
[arXiv:2212.04853 [hep-lat]].




\end{thebibliography}
\end{document}